\documentclass[12pt]{article}

\pdfoutput=1 
\usepackage{graphicx} 
\usepackage{color}                                                                                                                                                                                                                                                                                          
\usepackage{cancel,tabularx,moreverb,fancybox,amsmath,float,bm,braket,txfonts,amssymb,bm,accents}
\usepackage[top=30truemm,bottom=30truemm,left=25truemm,right=25truemm]{geometry}
\usepackage{latexsym}
\usepackage{here}
\usepackage{cite}
\usepackage{hyperref}
\usepackage{tcolorbox}

\newcommand{\ex}[1]{\mathrm{e}^{#1}}

\newcommand{\pa}[1]{\left(#1 \right)}
\newcommand{\PA}[1]{\biggl(#1 \biggr)}

\newcommand{\br}[1]{\left[#1 \right]}

\newcommand{\bb}[1]{\mathbb{#1}}

\newcommand{\ca}[1]{\mathcal{#1}}

\newcommand{\abs}[1]{\left|#1\right|}

\newcommand{\ar}[1]{\xrightarrow[#1]{}}

\newcommand{\comm}[2]{\lbrack#1,#2 \rbrack}

\newcommand{\Res}[1]{\mathop{\mathrm{Res}}_{#1}}

\newcommand{\ti}[1]{\tilde{#1}}

\newcommand{\fr}{\frac}

\newcommand{\und}[1]{{\it \textbf{#1}}}

\def\del{{\partial}}

 \def\d{{\delta}}

 \def\a{{\alpha}}

 \def\l{{\lambda}}
 \def\G{{\Gamma}}
 \def\D{{\Delta}}
 \def\g{{\gamma}}

 \def\e{{\epsilon}}

\def\tr{{\text{tr}}}
\def\bz{{\bar{z}}}

  \makeatletter
    
    \@addtoreset{equation}{section}
  \makeatother

\begin{document}

\begin{titlepage}
\thispagestyle{empty}

\begin{flushright}
KYUSHU-HET-306,
\\
RIKEN-iTHEMS-Report-24,
\\

\end{flushright}

\bigskip

\begin{center}
\noindent{{\large \textbf{
Modern Approach to 2D Conformal Field Theory
}}}\\
\vspace{2cm}
Yuya Kusuki ${}^{1,2,3}$
\vspace{1cm}

${}^{1}${\small \sl 
Institute for Advanced Study, \\
Kyushu University, Fukuoka 819-0395, Japan
}

${}^{2}${\small \sl 
Department of Physics, \\
Kyushu University, Fukuoka 819-0395, Japan
}

${}^{3}${\small \sl RIKEN Interdisciplinary Theoretical and Mathematical Sciences (iTHEMS), \\Wako, Saitama 351-0198, Japan}

\vskip 2em
\end{center}

\begin{abstract}
The primary aim of these lecture notes is to introduce the modern approach to two-dimensional conformal field theory (2D CFT).
The study of analytical methods in two-dimensional conformal field theory has developed over several decades, starting with BPZ.
The development of analytical methods, particularly in rational conformal field theory (RCFT), has been remarkable, with complete classifications achieved for certain model groups.
One motivation for studying CFT comes from its ability to describe quantum critical systems.
Given that realistic quantum critical systems are fundamentally RCFTs, it is somewhat natural that the analytical methods of RCFT have evolved significantly.

CFTs other than RCFTs are called irrational conformal field theories (ICFTs).
Compared to RCFTs, the study of ICFTs has not progressed as much.
Leaving aside whether there is physical motivation or not, ICFTs inherently possess a difficulty that makes them challenging to approach.
However, with the development of quantum gravity, the advancement of analytical methods for ICFTs has become essential.
The reason lies in the AdS/CFT correspondence.
The AdS/CFT correspondence refers to the relationship between $d+1$ dimensional quantum gravity and $d$ dimensional CFT.
Within this correspondence, the CFT appears as a non-perturbative formulation of quantum gravity.
Except in special cases, this CFT belongs to ICFT.
Against this backdrop, the methods for ICFTs have developed rapidly in recent years.
Many of these ICFT methods are indispensable for modern quantum gravity research.
Unfortunately, these cannot be learned from textbooks on 2D CFTs, such as the Yellow book \cite{Francesco2012}.
These lecture notes aim to fill this gap.
Specifically,
we will cover techniques that have already been applied in many studies, such as {\it Heavy-Light block} and {\it monodromy method},
and significant results that have become proper nouns, such as {\it Hellerman bound} and {\it HKS bound}.

 \end{abstract}

\end{titlepage}

\restoregeometry

\tableofcontents

\section{Introduction}

This lecture focuses on two-dimensional conformal field theory (CFT).
A conformal field theory is a field theory with a high degree of symmetry called conformal symmetry.
When one hears about analytical methods in field theory, one generally thinks of perturbative methods.
However, there are many situations where perturbative analysis is not very useful.
One characteristic of conformal field theory is that non-perturbative analysis can be easily realized due to conformal symmetry.
Conformal field theory is an ideal model for understanding non-perturbative effects.

\subsection{Why Consider CFT?}
CFT is not just an idealized field theory
but can actually be used to analyze real physical systems (quantum many-body systems).
For example, the critical Ising model is an example of a quantum many-body system described by CFT.
In particular, the three-dimensional critical Ising model is primarily analyzed using CFT.
In general, CFT is known to describe critical systems.
This is for the following reasons.
At the critical point, the correlation length diverges, and there is no characteristic scale.
Such a field theory is called scale-invariant field theory.
Moreover, under certain basic assumptions, scale invariance extends to conformal symmetry.
Thus, critical systems are reduced to CFT.

CFT also plays an essential role in understanding QFT through Wilsonian renormalization.
A major problem in dealing with QFT is the infinite degrees of freedom.
We approximate QFT by averaging some degrees of freedom and considering them as a single degree of freedom.
In other words, we construct an ``effective theory" that drops information on short-range degrees of freedom below a specific reference scale,
thus approximating the infinite degrees of freedom.
Naturally, this effective theory depends on the choice of the reference scale.
The series of QFTs obtained by increasing this reference scale (often referred to as ``zooming out") is called the Renormalization Group (RG) flow.
Since we are generally interested in long-range physics (i.e. macroscopic behavior),
QFT, as an effective theory, is useful in both particle physics and condensed matter physics.
Now, let us consider treating QFT mathematically rigorously.
For mathematical treatment, it is desirable to ensure the existence of effective theories at all scales.
In particular, we hope that the limit where the reference scale is taken to be small (the short-distance limit, the limit where the UV cutoff is taken to infinity) exists (such QFTs are called well-defined QFTs).
For this limit to exist, the QFT must be an RG flow from the vicinity of a scale-invariant theory.
\footnote{
This does not deny the existence of effective theories with physical cutoffs.
In fact, some QFTs as effective theories that appear in condensed matter physics have physical cutoffs.
In such theories, it is unknown whether the short-distance limit leads to a QFT,
but this does not diminish their value as effective theories.
}
This scale-invariant theory, in other words, a theory that does not change even when zoomed out, is precisely CFT.
In other words, a well-defined QFT is interpreted as a point on the RG flow from a UV CFT to an IR CFT.
In this sense, the classification of well-defined QFTs is equivalent to the classification of CFTs,
and the study of CFTs plays an important role in understanding QFTs.

\subsection{Characteristics of CFT}
The task of ``solving" a field theory can be said to be the calculation of all correlation functions.
A correlation function is the expectation value of a product of $N$ local operators,
\begin{equation}
\braket{O_1(x_1) O_2(x_2) \cdots O_N(x_N)}.
\end{equation}
Here, a ``local operator" refers to a local operation.
For example, ``measuring" the spin at a single site in a spin system is an example of a local operator.
In field theory, local operators are called fields (local fields).
One of the characteristics of CFT is that these correlation functions are completely determined solely by scalar quantities called \und{OPE coefficients}.
OPE coefficients are the coefficients ${ C_{ijk} }$ that appear in the Operator Product Expansion (OPE),
\begin{equation}\label{eq
}
O_i \times O_j = \sum_k C_{ijk} O_k.
\end{equation}
In general field theories, OPE may not be helpful because of its limited applicability and structural complexity.
However, in CFT, this OPE can always reduce an $N$-point correlation function to a sum of $N-1$-point correlation functions,
and by repeating this process, any correlation function can be determined solely from the OPE coefficients.
That is, in CFT, the parameters that determine a given theory are just the spectrum and the OPE coefficients.

Furthermore, this spectrum and OPE coefficients are highly constrained by the "``consistency" of the correlation functions.
This consistency refers to the associativity of the OPE,
\begin{equation}
(O_i \times O_j) \times O_k = O_i \times (O_j \times O_k).
\end{equation}
In other words, the calculation results do not depend on the order in which the ``products" in the OPE are taken.
The program to determine the spectrum and OPE coefficients using this consistency is called \und{ conformal bootstrap}.
This is entirely different from the usual process of solving a field theory.
In CFT, there is no need to compute path integrals, and the Lagrangian does not even appear.
In CFT, one can avoid all the complex tasks in general field theory.

\subsection{Contents Covered in This Lecture: Bootstrapping ICFT}
There are broadly two types of CFT.
One is called {\it rational conformal field theory} (RCFT).
A precise definition will be given later, but, roughly speaking, RCFTs are CFTs with a finite number of field types.
The critical phenomena of exactly solvable models, such as the critical Ising model, belong to this RCFT category.
Since critical phenomena are one of the most important topics in condensed matter theory,
RCFTs, directly related to critical phenomena, have been extensively studied, and numerous analytical methods are available.
Many of these can be learned from the Yellow book \cite{Francesco2012}.
However, the methods developed for studying RCFTs heavily utilize the unique nature of RCFTs, namely, that there are only a finite number of field types.
If all CFTs were RCFTs, there would be no problem, but in reality, there are CFTs other than RCFTs, called {\it irrational conformal field theories} (ICFTs).
For example, some of the free theories and Liouville CFT belong to ICFT.
When viewed as quantum many-body systems, ICFTs may not seem as important as RCFTs, as they are not realistic or interesting systems.
However, ICFTs play a very important role in modern physics.

The motivation for studying ICFTs in recent years comes from particle physics.
One of the major goals of particle physics is to complete a theory that unifies gravity and quantum mechanics.
Specifically, the goal is to construct a theory to understand issues like the black hole information paradox (the problem where the existence of black holes violates the information preservation law, a cornerstone of quantum mechanics).
For this purpose, non-perturbative analysis of quantum gravity is essential.
This is where the \und{AdS/CFT correspondence (holography principle)} comes into play.
The AdS/CFT correspondence is a conjecture that a $d+1$ dimensional quantum gravity in AdS space is equivalent to a $d$ dimensional CFT.
In short, CFT appears as a non-perturbative formulation of quantum gravity.
Since the CFT that appears here belongs to ICFT, understanding ICFT has become essential in modern particle physics research.

\begin{figure}[t]
\begin{center}
\includegraphics[width=10.0cm,clip]{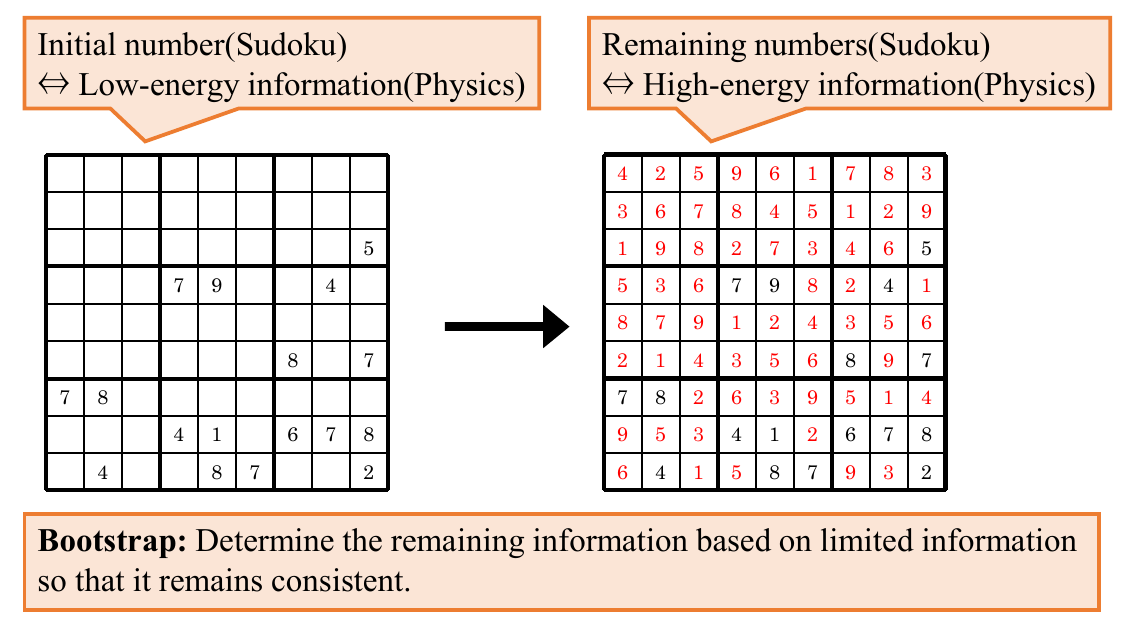}
\end{center}
\caption{Sudoku as an analogy for the conformal bootstrap}
\label{fig:sudoku}
\end{figure}

In the analysis of ICFT, the conformal bootstrap plays an important role.
As briefly mentioned above, conformal bootstrap is a theoretical tool that determines the remaining information to make the theory consistent based on limited initial information.
The process of ``using the consistency of the theory" might seem unfamiliar, but this method is used in puzzles such as Sudoku (see Figure \ref{fig:sudoku}).
When constructing a theory, we start with assumptions such as ``there is a particle with mass $m$," which corresponds to the initial placement of numbers in Sudoku.
Then, similar to how the remaining numbers are determined in Sudoku, the conformal bootstrap determines other properties, such as the masses of other particles, using the consistency of the theory.
While the specific ``rules" of the ``game" of CFT will be explained later,
here we explain what can be learned through the conformal bootstrap.

What we know about quantum gravity is very limited.
In such situations, the conformal bootstrap is particularly useful.
As a first example, let us touch on black hole thermodynamics.
It is known that the entropy of a black hole is proportional to the area of its event horizon.
To provide a statistical mechanical understanding of this black hole thermodynamics,
we need to determine the number of microscopic states at high energy.
Using the conformal bootstrap, we can easily determine this number of states (known as the Cardy formula \cite{Cardy1986a}), and confirm that it indeed reproduces black hole thermodynamics.
In short, the conformal bootstrap shows that black hole thermodynamics is truly thermodynamics.
Section \ref{subsec:Cardy} will explain this in detail.

\begin{figure}[t]
\begin{center}
\includegraphics[width=10.0cm,clip]{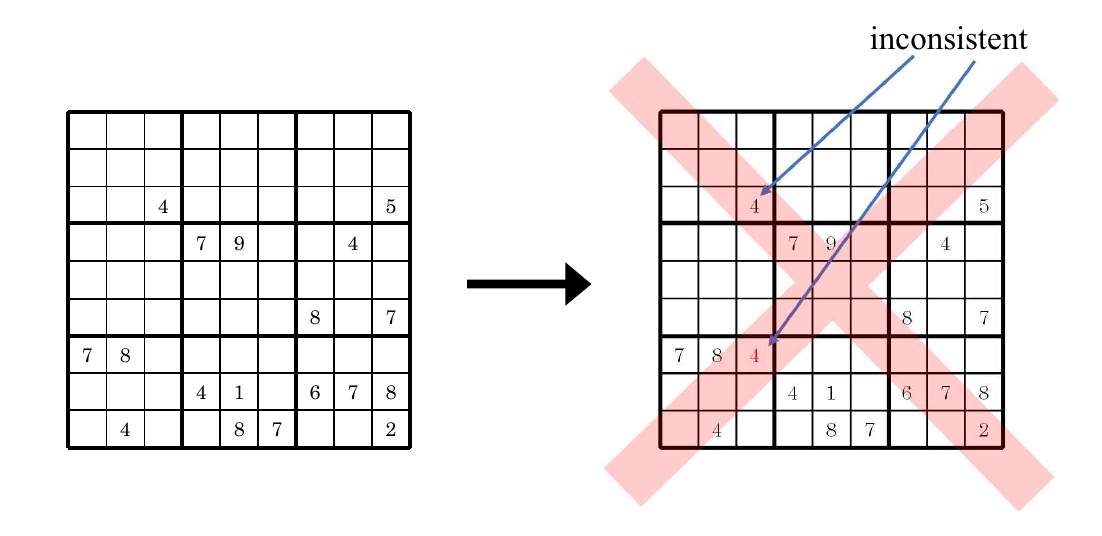}
\end{center}
\caption{Sudoku demonstrating the non-existence of a solution}
\label{fig:sudoku2}
\end{figure}

Additionally, the conformal bootstrap allows us to discuss the existence or non-existence of a particular theory.
The process of physics research involves constructing toy models under several natural assumptions and understanding the physics from their qualitative behavior.
However, if the assumptions are inconsistent, the toy model will not have predictive power.
Therefore, we must always be careful whether the toy model we consider truly exists.
For example, when quantizing a classical model, it may turn out that the quantization is impossible.
In verifying the existence of such theories, the conformal bootstrap is useful.
To explain this, we again use the analogy of Sudoku.
Consider imposing multiple conditions such as ``there exist both particles X and Y".
This corresponds to placing multiple numbers in Sudoku,
but depending on the initial placement, a solution may not exist.
As in the Sudoku example in Figure \ref{fig:sudoku2},
the conformal bootstrap can be used to discuss the non-existence of a theory.
One of the significant questions in quantum gravity is whether
``quantum gravity in AdS space without any matter exists."
Many studies address this question using conformal bootstrap and have achieved significant results (see, for example, \cite{Hellerman2009, Friedan2013}).
Section \ref{subsec:Hellerman} will explain this in detail.

In the following, we briefly summarize the application methods of conformal bootstrap using analogies with Sudoku.

\begin{tabular}{|p{3cm}||p{5cm}|p{6cm}|}
\hline
Problem Setup & Corresponding Theory & Application \\
\hline
\hline
Standard Sudoku & Theory with sufficient initial conditions & The theory can be specified (all parameters except the initial conditions can be determined) \\
\hline
Sudoku with missing numbers & Theory with insufficient assumptions, multiple candidates exist & Common properties among all candidates can be identified (universal behavior under certain assumptions can be extracted) \\
\hline
Incorrect Sudoku & No theory satisfies the given assumptions & Impossible theories can be immediately excluded \\
\hline
\end{tabular}
\\
\\

\section{Conformal Field Theory}\label{sec:CFT}

\begin{figure}[t]
\begin{center}
\includegraphics[width=14.0cm,clip]{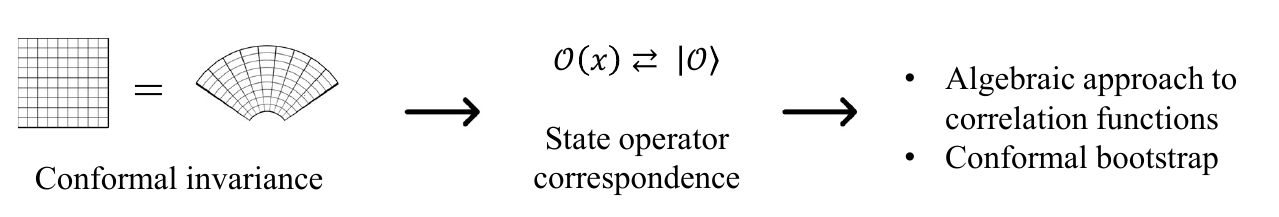}
\end{center}
\caption{Consequences of conformal symmetry}
\label{fig:col}
\end{figure}

The computational methods in CFT differ significantly from those in general field theory.
One reason is that the calculations in CFT are much simpler when using the Operator Product Expansion (OPE).
Although OPE can be defined in general field theories,
it is not mathematically tractable.
However,
with conformal symmetry,
it becomes mathematically tractable in the sense that one can always move to coordinates where the OPE converges,
acquiring useful properties such as the associativity of the OPE.

What bridges conformal symmetry and the convergence of the OPE is the state-operator correspondence (see Figure \ref{fig:col}).
The state-operator correspondence refers to the one-to-one correspondence between {\it energy eigenstates in natural quantization called radial quantization} and {\it local operators}.
In general field theories,
states can easily be created from local operators,
but creating local operators from states requires conformal symmetry.
This state-operator correspondence plays an essential role in the analysis of CFT.
It allows local operators to construct the Hilbert space.
This forms representations of the conformal algebra (more precisely,
the Virasoro algebra).
Thus, the analysis of multi-point functions of local operators reduces to the representation theory of the Virasoro algebra.

The associativity of the OPE is also a consequence of the state-operator correspondence.
This associativity of the OPE corresponds to the ``rules" in Sudoku (see Figure \ref{fig:sudoku}).
This chapter will explain these topics in detail.

\subsection{Conformal Symmetry}\label{subsec:ct}
In two-dimensional Euclidean space,
it is convenient to consider complex coordinates $(z,\bar{z})$ instead of $(x,\tau)$ coordinates,
\begin{equation}
z=x+i\tau, \ \ \ \bar{z}=x-i\tau.
\end{equation}
For convenience,
we will treat $z$ and $\bar{z}$ as independent variables by complexifying $\bb{R}^2 \to \bb{C}^2$.
This allows for the use of convenient methods for complex analysis.
To obtain physical results,
we can restrict the results from complex analysis to $\bb{R}^2 \subset \bb{C}^2$ by setting $\bar{z}=z^*$.

In this complex Euclidean space,
conformal transformations are defined by the following coordinate transformations:
\begin{tcolorbox}[title=Conformal transformation]
\begin{equation}\label{eq:ct2}
z \to w=f(z), \ \ \ \ \ \ \bz  \to \bar{w} =\bar{f}(\bz ).
\end{equation}
\end{tcolorbox}
Here,
$f(z)$ is an arbitrary holomorphic function.
This coordinate transformation has the property of preserving the metric up to a scale factor.
\footnote{
Note that,
unlike general coordinate transformations, where the metric is treated dynamically,
conformal transformations are defined in a theory with a fixed metric.
}
\begin{equation}\label{eq:ct}
dw d\bar{w} = \abs{f'(z)}^2 dz d\bar{z}.
\end{equation}
On the contrary,
it can be shown that the only transformations that preserve the metric up to a scale factor are given by (\ref{eq:ct2}).
Thus,
(\ref{eq:ct}) can also be taken as the definition of conformal transformations.

To get an intuitive understanding of conformal transformations,
consider the map from the Riemann sphere $\hat{\bb{C}} = \bb{C} \cup \{\infty \}$ to itself, which is called {\it global conformal transformations}.
Global conformal transformations can be expressed using four complex numbers $a,b,c,d$ as
\begin{equation}
f(z) = \fr{az+b}{cz+d}, \ \ \ \ \ ad-bc=1,
\end{equation}
which is also known as the Möbius transformation.
In fact,
the global conformal transformations are generated by the following familiar four transformations:
\begin{equation}
\begin{array}{ll}
\text{(translation)} & f(z) = z+a, \\
\text{(dilation)} & f(z) = cz, \\
\text{(rotation)}& f(z) = \ex{i \theta} z, \\
\text{(inversion)} & f(z) = \fr{1}{z}.
\end{array}
\end{equation}
Rotational and translational symmetries are symmetries possessed by many theories.
The symmetry specific to CFT is scale invariance (dilation).
The local extension of this transformation (which is holomorphic locally but not necessarily globally) is the local conformal transformation (\ref{eq:ct2}).

Laurent expansion allows us to parametrize infinitesimal conformal transformations with small constants $\epsilon_n$ as follows.
\begin{equation}\label{eq:epsilon}
z \mapsto z+\epsilon(z) = z - \sum_{n \in \bb{Z}} \epsilon_n z^{n+1}.
\end{equation}
The generators of this infinitesimal transformation are represented using differential operators as
\begin{equation}
l_n \equiv -z^{n+1} \del_z.
\end{equation}
These generators satisfy the following commutation relations,
\begin{equation}\label{eq:witt}
[l_n, l_m] = (n-m) l_{n+m}.
\end{equation}
This algebra is called \und{ Witt algebra}.
The generators of the global conformal transformations ${l_{-1}, l_0, l_1 }$ form its subalgebra.
So far, we have focused only on the holomorphic part (the $z$ coordinate),
but a similar discussion holds for the antiholomorphic part (the $\bar{z}$ coordinate).
That is, if we denote the generators of the coordinate transformation $\bar{w} = \bar{f}(\bar{z})$ by $\bar{l}_n$,
then $\bar{l}_n$ satisfy the following commutation relations.
\begin{equation}\label{eq:ind}
[\bar{l}_n, \bar{l}_m] = (n-m) \bar{l}_{n+m}.
\end{equation}
Moreover, since the holomorphic and antiholomorphic parts are independent,
\begin{equation}
\del_z \bar{z} = 0, \ \ \ \ \del_{\bar{z}} z = 0,
\end{equation}
we obtain the following commutation relation,
\begin{equation}
[l_n, \bar{l}_m] = 0.
\end{equation}
The interpretation of each generator is as follows,
\begin{equation}
\begin{array}{ll}
\text{(translation)} & l_{-1} \text{ and }  \bar{l}_{-1} , \\
\text{(dilation)} &  l_0 + \bar{l}_0,  \\
\text{(rotation)}&  l_0 - \bar{l}_0,    \\
\text{(SCT)} & l_{1} \text{ and }  \bar{l}_{1}.
\end{array}
\end{equation}
SCT (Special Conformal Transformation) is related to inversion.

\subsection{Virasoro Algebra}
So far, we have looked at the generators of groups acting on the geometry.
Let us upgrade this to the generators in quantum theory.
In quantum mechanics,
symmetry is described by linear representations acting on states in the Hilbert space.
Here, a linear representation of a group $G$
is a map $\rho : G \to GL(V)$ from the group $G$ to a general linear group on a linear space $V$,
satisfying the following conditions:
\begin{equation}\label{eq:GL}
\rho(g_1) \rho(g_2) = \rho(g_1 g_2), \ \ \ \ \ g_1, g_2 \in G.
\end{equation}
In quantum theory,
the states correspond to rays in the Hilbert space,
rather than the vectors themselves.
Here, a ray is an equivalence class under the following relation:
\begin{equation}
\ket{\phi} \sim \ket{\psi} :\Leftrightarrow \ket{\phi} = c \ket{\psi}, \ \ \ \ c \in \bb{C} - \{0 \}.
\end{equation}
This is because multiplying a vector by a constant does not change the following expectation value:
\begin{equation}
\braket{A}_\phi = \fr{\braket{\phi|A|\phi}}{\braket{\phi|\phi}}.
\end{equation}
In other words,
we cannot distinguish between $\ket{\phi}$ and its multiple $c\ket{\phi}$ $(c \in \bb{C})$,
so they should be considered the same state.
This identification is realized by introducing equivalence classes under constant multiplication.
All states are truly distinct in this projective Hilbert space,
which is convenient.
Now,
by identifying differences by constant multiples,
the symmetry becomes a projective unitary representation $\ti{\rho}: G \to PU(\ca{H})$,
\begin{equation}
\ti{\rho}(g_1) \ti{\rho}(g_2) = \ti{\rho}(g_1 g_2), \ \ \ \ \ g_1, g_2 \in G.
\end{equation}
For each $g \in G$,
we can choose a representative $\rho(g) \in \ti{\rho}(g)$,
and the projective representation can be expressed as a homomorphism that allows for differences by constants,
\begin{equation}
\rho(g_1) \rho(g_2) = c(g_1, g_2) \rho(g_1 g_2), \ \ \ \ \ \ \ \ g_1, g_2 \in G.
\end{equation}
Here,
$c: G \times G \to U(1)$.
From the associativity of multiplication $\rho(g_1) (\rho(g_2) \rho(g_3)) = (\rho(g_1) \rho(g_2)) \rho(g_3)$,
$c$ satisfies the following cocycle condition:
\begin{equation}\label{eq:cocycle}
c(g_1, g_2 g_3)c(g_2, g_3) = c(g_1, g_2)c(g_1 g_2,g_3), \ \ \ g_1, g_2, g_3 \in G.
\end{equation}
There are many different ways to choose the representatives for the same projective representation.
This arbitrariness can be expressed using $f: G \to U(1)$ as follows:
\footnote{
This means that projective representations of $G$ are classified by the following set:
\begin{equation}
H^2(G,U(1)) := \{ c: G\times G \to U(1) \mid c \text{ satisfies } (\ref{eq:cocycle}) \}/c \sim c' : \Leftrightarrow \exists f, (\ref{eq:cequiv}) \text{ holds }.
\end{equation}
}
\begin{equation}\label{eq:cequiv}
\rho'(g) = f(g) \rho(g), \ \ \  g \in G.
\end{equation}
By choosing the representatives appropriately,
if
\begin{equation}
\rho'(g)\rho'(h) = \rho'(gh), \ \ \ \ \Leftrightarrow c=1,
\end{equation}
is satisfied,
then $\ti{\rho}$ is said to be {\it de-projectivized}.
In physics,
when it is not de-projectivized,
it is said to have an anomaly.

Is it inevitable to give up unitary representations when it is anomalous?
Even if it is anomalous,
the \und{central extension} allows us to realize projective representations as unitary representations.
\footnote{
In 4D CFT,
there is a trace anomaly,
but no central extension exists.
The conformal symmetry in 4D CFT is much smaller than that in 2D CFT,
where it corresponds to the Möbius transformations.
Here, we notice that the projective representations of the Möbius group are given by linear representations.
In other words,
the trace anomaly is an anomaly of local conformal transformations and does not contribute to global conformal transformations in 4D CFT.
}
The central extension of $G$ is defined as follows:
\begin{equation}\label{eq:GU}
H := G \times_c U(1).
\end{equation}
Here,
$\times_c$ is a direct product with the following multiplication:
\begin{equation}
(g,\alpha) \cdot (h,\beta) = (gh, c(g,h)\alpha \beta), \ \ \ \ \ (g,\alpha), (h,\beta) \in G\times U(1).
\end{equation}
In this case,
a unitary representation $\pi: H \to U(\ca{H})$ can be defined by $\pi((g,\alpha))=\alpha \rho(g) \ \ \ ((g, \alpha) \in H)$,
thus realizing the projective representation as a unitary representation.

It is known that conformal symmetry has a {\it conformal anomaly} (also called trace anomaly).
The details of the conformal anomaly will be discussed later,
so for now, accept the fact that conformal symmetry is anomalous.
In Section \ref{subsec:ct},
we dealt with the conformal algebra (Witt algebra),
so instead of the ``quantization" of the group,
we consider the ``quantization" of the algebra.
The central extension of a Lie group induces a central extension of the Lie algebra.
In this case,
the direct product of the Lie group (\ref{eq:GU}) leads to the direct sum of the Lie algebra.
Similarly,
for the linear representation $\pi((g,\alpha))=\alpha \rho(g) \ \ \ ((g, \alpha) \in H)$ of the central extension of the Lie group,
the linear representation $\phi$ of the central extension of the Lie algebra is given by $\phi((u\oplus a)) = \Phi(u)+a \ \ \ \ (u\oplus a \in \mathfrak{g}\oplus \bb{R})$.

The above discussion became somewhat technical,
but can be summarized in plain terms as follows:
\begin{tcolorbox}
Even if we add a suitable constant to the Witt algebra,
the commutation relation (\ref{eq:witt}) holds.
The quantization of Witt algebra (i.e. central extension) is realized by allowing this freedom of constants.
\end{tcolorbox}
For Witt algebra,
the central extension is uniquely determined.
We demonstrate this below.
Let $c,p(n,m) \in \bb{C}$,
the central extension can be written as
\begin{equation}
[L_n,L_m]=(n-m)L_{n+m}+c p(n,m).
\end{equation}
First,
from the antisymmetry of the commutation relation, we obtain
\begin{equation}
p(n,m)=-p(m,n).
\end{equation}
Next,
for CFT to be defined consistently,
there must exist a vacuum invariant under global conformal transformations.
This requires that the central terms do not appear in the commutation relations among $\{ L_{-1}, L_{0}, L_{1}\}$.
\begin{equation}
p(1,-1)=p(1,0)=p(0,0)=p(-1,0)=0.
\end{equation}
Next,
consider the Jacobi identity
\begin{equation}
[L_n,L_m],L_0]+[[L_m,L_0],L_n]+[[L_0,L_n],L_m]=0,
\end{equation}
which leads to
\begin{equation}
(n+m)c p(n,m)=0.
\end{equation}
Therefore,
$p(n,m)=0$ when $n + m \neq 0$.
Also,
\begin{equation}
[L_{-n+1},L_n],L_{-1}]+[[L_{n},L_{-1}],L_{-n+1}]+[[L_{-1},L_{-n+1}],L_n]=0,
\end{equation}
leads to the following recurrence relation,
\begin{equation}
(n-2)p(n,-n)=(n+1)p(n-1,-(n-1)).
\end{equation}
Solving this,
we obtain
\begin{equation}
p(n,-n)=\fr{1}{6}n(n^2-1)p(2,-2).
\end{equation}
Letting $p(2,-2)=\fr{1}{2}$, we obtain
\begin{equation}
p(n,-n)=\fr{1}{12}n(n^2-1).
\end{equation}
Finally,
we obtain the following algebraic relation:
\begin{tcolorbox}[title=Virasoro algebra]
\begin{equation}\label{eq:Virasoro}
[L_n,L_m]=(n-m)L_{n+m}+\fr{c}{12}n(n^2-1)\d_{n+m,0}.
\end{equation}
\end{tcolorbox}
\noindent
This is called the \und{Virasoro algebra}.
$c$ is called the \und{central charge}.
It is a constant not determined by symmetry and is one of the parameters of a theory.
For example,
in the free boson CFT,
$c=1$,
and in the critical Ising model,
$c=\fr{1}{2}$.
This central charge has various physical interpretations.
For instance,
when placing CFT on a curved spacetime,
if the scale symmetry is broken,
$c$ appears directly as the magnitude of the anomaly.
Indeed,
when $c=0$,
the Virasoro algebra becomes the classical Witt algebra.
The factor of $1/12$ is simply a convention so that $c = 1$ for the free boson.

To summarize the key points of this section,
\begin{itemize}
\item
We identified how conformal symmetry acts on states in quantum theories (Virasoro algebra).
\item
Conformal symmetry is anomalous,
but it can be treated similarly to ordinary symmetry through the central extension.
\end{itemize}

\subsection{Representation of Virasoro Algebra}

In studying quantum mechanics,
the operator formalism (treating it as a linear algebra on the Hilbert space) is extremely useful.
For example,
its convenience can be seen in the quantization of angular momentum.
The procedure for examining the Hilbert space of a CFT is similar to the procedure for using the representation theory of $su(2)$ in systems with rotational symmetry,
so let us first recall the case of $su(2)$.

The algebraic relations of $su(2)$ are given by
\begin{equation}
\comm{J_0}{J_{\pm}}=\pm J_{\pm}, \ \ \ \ \comm{J_+}{J_-}=2J_0.
\end{equation}
The procedure of classifying states (particles) by symmetry is mathematically done by the irreducible decomposition of representations.
Here, a representation is said to be {\it irreducible} if it does not have any non-trivial invariant subspaces.
A subspace $W$ is said to be {\it invariant} if for any element $g$ of the group $G$ and any element $w$ of the subspace $W$, the following holds:
\begin{equation}
gw \in W.
\end{equation}
A representation that is not irreducible is called a reducible representation,
and the process of transforming a {\it reducible} representation into a direct sum of irreducible representations is called {\it irreducible decomposition}.
An irreducible decomposition is analogous to factoring an integer into its prime factors, allowing us to handle the object as a collection of simpler constituents.
When classified in this way,
states belonging to the same irreducible representation can transform into each other under the action of symmetry.
\footnote{
One benefit of the irreducible representation comes from Schur's lemma. For example, the fact that particles are characterized by only two real numbers -- mass and spin -- is a consequence of Schur's lemma. Moreover, the orthogonality of characters, which is a corollary of Schur's lemma, also plays an important role in physics research.
}

The irreducible representation of $su(2)$ is obtained by acting $J_-$ on the highest weight state (h.w.s.)
\begin{equation}\label{eq:mstate}
\ket{m}=\pa{J_-}^{j-m}\ket{j},
\end{equation}
where $\ket{j}$ is the highest weight state defined by
\begin{equation}
J_0\ket{j}=j\ket{j}, \ \ \ \ J_+\ket{j}=0.
\end{equation}
The inner product of each state can be obtained recursively by
\begin{equation}\label{eq:su2}
\braket{m-1|m-1}=\pa{j(j+1)-m(m-1)}\braket{m|m}.
\end{equation}
According to this recurrence relation,
states with negative norms generally appear.
To avoid negative norm states,
$j$ must be an integer or a half-integer.
When $j$ is an integer or a half-integer,
$\ket{m} \ \ \ \ (m \leq -j-1)$ has zero norm.
States with zero norms do not contribute to expectation values,
so they can be excluded from the Hilbert space.
Thus,
only the first $2j+1$ states in (\ref{eq:mstate}) remain,
resulting in a finite-dimensional representation.
This is the quantization of angular momentum derived from the representation theory of $su(2)$.

In a similar procedure,
let us examine the Hilbert space of CFT.
We consider constructing irreducible representations of the Virasoro algebra from the highest weight state.
Here,
the highest weight state for the Virasoro algebra is defined as follows:
\begin{equation}\label{eq:hws}
L_0\ket{h} = h\ket{h}, \ \ \ L_{n>0}\ket{h} = 0.
\end{equation}
This corresponds to spanning the Hilbert space with the eigenstates of $L_0$.
The eigenvalue $h$ is called the \und{conformal dimension}.
It should be emphasized here that $L_{n\neq0}$ acts as raising and lowering operators.
This is evident from the following commutation relation:
\begin{equation}
[L_0, L_{-m}]=mL_{-m}.
\end{equation}
In CFT,
the vacuum state corresponds to the highest weight state $\ket{0}$ with $h=0$.
Although it is invariant under global conformal transformations $\{ L_{-1}, L_{0}, L_{1}\}$,
it does not vanish for all Virasoro algebra elements due to the presence of the conformal anomaly.

\begin{figure}[t]
 \begin{center}
  \includegraphics[width=10.0cm,clip]{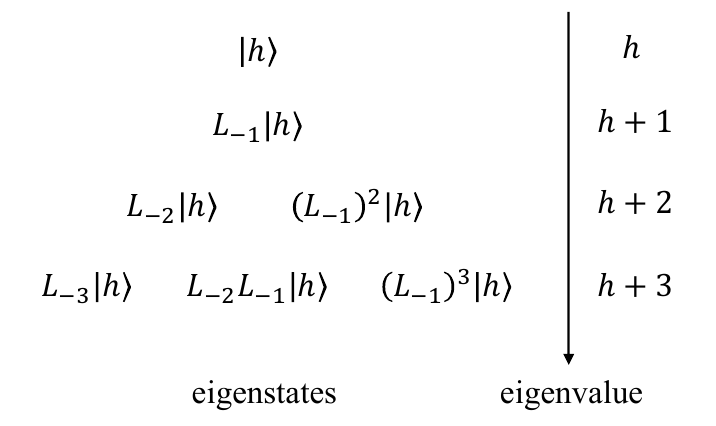}
 \end{center}
 \caption{Verma Module}
 \label{fig:verma}
\end{figure}

In CFT,
the highest weight state is called a \und{primary state}.
A state obtained by acting $L_{n<0}$ on the highest weight state is called a \und{descendant state}.
\begin{equation}
L_{-n_1} L_{-n_2}\cdots L_{-n_k} \ket{h}, \ \ \ \ 0\leq n_1 \leq n_2 \leq \cdots \leq n_k.
\end{equation}
$N\equiv \sum^k_{i=1} n_i$ is called the level,
and a descendant state of level $N$ is an eigenstate of $L_0$ with eigenvalue $h+N$.
The vector space constructed from the highest weight state is called the {\it Verma module} $\ca{V}(c,h)$.
The first few states of the Verma module are shown in Figure \ref{fig:verma}.
As can be imagined from this table,
the number of states at each level of the Verma module is given by the partition number $p(N)$.
Although $p(N)$ cannot be written as a simple formula in terms of $N$,
the following asymptotic formula (Hardy-Ramanujan formula) is known:
\begin{equation}\label{eq:HR}
p(N) \simeq \fr{1}{4N\sqrt{3}}\ex{\pi\sqrt{\fr{2N}{3}}}, \ \ \ \ \  N \to \infty.
\end{equation}
In this and subsequent equations,
the symbol $\simeq$ is used for asymptotically valid expressions.
This asymptotic formula concerning the ``number of descendant states" is useful for intuitive understanding later on,
so remember it.
\footnote{Strictly speaking,
this holds when the Verma module is irreducible.
In the context of AdS/CFT correspondence,
the Verma modules of CFT that appear are often irreducible,
so it can be shown to be correct for the CFT of our interest.
This will be explained in more detail later.}

The above discussion also applies to the anti-holomorphic part.
Considering both the holomorphic and anti-holomorphic parts together,
since $L_0+\bar{L}_0$ is the generator of scale transformation,
its eigenvalue $\Delta\equiv h+\bar{h}$ is called the scaling dimension (also referred to as energy, as explained later).
Similarly,
the eigenvalue $s \equiv \abs{h-\bar{h}}$ of the generator of rotations $L_0-\bar{L}_0$ is called the spin.

\subsection{Hilbert Space}\label{subsec:Hilbert}

So far,
we have seen the representation of the Virasoro algebra described as a direct sum of Verma modules,
but is the Verma module itself irreducible?
This is a non-trivial problem,
and depending on the values of $(c,h)$,
it can be reducible.
Here, we explain how to obtain the irreducible representation of the Virasoro algebra.

A Verma module being reducible means that there is a subspace closed under conformal transformations (i.e. the action of the Virasoro algebra).
Such a subspace is formed from descendant states $\ket{\chi}$ that satisfy the following condition:
\begin{equation}\label{eq:null}
L_{n>0}\ket{\chi} = 0.
\end{equation}
This is the same condition as the highest weight state (\ref{eq:hws}),
so the subspace created from this becomes an invariant subspace.
These descendant states are called {\it null states} or {\it singular vectors}.
The characteristic of this null state is that its norm is zero.
Given the Hermitian conjugate of the Virasoro algebra
\footnote{
Recall that the raising and lowering operators of $su(2)$ satisfy $J_{\pm}^\dagger = J_{\mp}$.
This comes from the fact that $J_0$ is Hermitian and $[J_0,J_\pm^\dagger]=\mp J_\pm^\dagger$.
Similarly,
(\ref{eq:hermite}) is derived by assuming that $L_0$ is Hermitian.
This assumption comes from the fact that $L_0$ corresponds to the Hamiltonian.
The state-operator correspondence is useful when discussing the interpretation and Hermitian conjugates of the Virasoro algebra,
so a detailed explanation will be given in the section on the state-operator correspondence (Section \ref{subsec:stateoperator}).
}
\begin{equation}\label{eq:hermite}
L_n^\dagger = L_{-n},
\end{equation}
the norm of the null state takes the following form:
\begin{equation}
\braket{\chi|\chi} = \bra{h}L_{n_1} \cdots L_{n_k}\ket{\chi}.
\end{equation}
However, this vanishes due to the definition of the null state (\ref{eq:null}).
Similarly,
it can be proven that descendant states of null states also vanish.
Therefore,
when the subspace formed by a null state is denoted as $\ca{J}_h$,
an irreducible representation can be constructed by the quotient space $\ca{M}_h \equiv \ca{V}_h/\ca{J}_h$,
removing this invariant subspace.
\footnote{
Recall that zero-norm states were removed in the process of quantizing angular momentum.
What we are doing here is precisely that.
}
Note that the number of descendant states at each level is no longer given by the partition number $p(N)$.
Following convention,
we refer to $\ca{M}_h$ as a Verma module and denote it by $\ca{V}_h$.
Since we essentially do not deal with the reducible Verma module itself,
there is unlikely to be any confusion between the two.

Finally, the Hilbert space of the CFT is described as follows, including the anti-holomorphic part:
\begin{tcolorbox}[title=Hilbert space of CFT]
    \begin{equation}\label{eq:hilbert}
\ca{H}_{CFT} = \bigoplus_{(h,\bar{h}) \in \ca{S}} M_{h,\bar{h}} \ca{V}_h \otimes \bar{\ca{V}}_{\bar{h}},
\end{equation}
\end{tcolorbox}
\noindent
Here, the set $\ca{S}$ is called the {\it spectrum}, and $M_{h,\bar{h}}$ is called the {\it multiplicity}.
The reason why the Hilbert space is constructed from the product of holomorphic and anti-holomorphic Verma modules is that the holomorphic and anti-holomorphic parts of the conformal symmetry act independently (refer to (\ref{eq:ind}) or the Ward-Takahashi identity (\ref{eq:WT})).
This decomposition into holomorphic and anti-holomorphic parts is one of the important properties of 2D CFT.
However, as seen in (\ref{eq:hilbert}), it should be noted that while the algebra decomposes, the Hilbert space itself does not.
\footnote{
Since the Virasoro algebra completely separates into holomorphic and anti-holomorphic parts,
the representation theory of the Virasoro algebra tells us nothing about the possible pairs $(h,\bar{h})$.
To investigate the possible pairs $(h,\bar{h})$, we need to use the power of the conformal bootstrap.
This will be explained again in the section on conformal bootstrap.
}
In general, the central charges $c$ of the holomorphic part and $\bar{c}$ of the anti-holomorphic part do not have to be the same.
However, for simplicity, we assume $c=\bar{c}$ here.

Finally, let us introduce some terms about the spectrum.
\begin{description}

\item[Diagonal CFT]\mbox{}\\
When $\bar{i}=i$ holds for all irreducible representations $\ca{V}_i\times\bar{\ca{V}}_{\bar{i}}$ appearing in the spectrum,
the CFT is said to be {\it diagonal CFT}.
Its partition function is called the {\it diagonal modular invariant}.

\item[Charge-conjugate CFT]\mbox{}\\
A CFT is called a {\it charge-conjugate CFT} if, for every irreducible representation $\ca{V}_i \times \bar{\ca{V}}_{\bar{i}}$ that appears in the spectrum, the condition $\bar{i} = i^*$ holds. Here, $i^*$ denotes the charge conjugate of $i$. 
Its partition function is called the {\it charge conjugate modular invariant}.
One example can be obtained by gauging $\bb{Z}_2$ of a diagonal free boson CFT.

\item[Chiral CFT]\mbox{}\\
When the Hilbert space itself decomposes into holomorphic and anti-holomorphic parts,
the CFT is said to be chiral CFT.

\item[RCFT and ICFT]\mbox{}\\
When the spectrum is represented by a finite number of irreducible representations,
the CFT is called an RCFT (Rational CFT).
\footnote{
When a CFT has other symmetries,
the Virasoro algebra may be a subalgebra of a larger algebra $\ca{A}$.
In this case,
even if there are infinitely many irreducible representations of $Vir\otimes \overline{Vir}$ in the spectrum,
there can be a finite number of irreducible representations of $\ca{A} \otimes \overline{\ca{A}}$.
This will be explained again later.
}
CFTs other than RCFT are called ICFT (Irrational CFT).

\end{description}
{\color{gray} Exercise: Many textbooks state that `` every field in a CFT is a primary or a descendant field (or a linear combination of primaries and descendants)," but this is not always true; explain in what cases this fact breaks down.}

\subsection{Irreducible Representation and Gram Matrix}\label{subsec:Gram}

How can we determine whether a Verma module is irreducible?
For this,
we use the Gram matrix for the subspace of the Verma module at level $N$.
For simplicity,
we introduce the following notation:
\begin{equation}
\ket{h,\bm{n}} \equiv L_{-\bm{n}}\ket{h} \equiv L_{-n_1} L_{-n_2}\cdots L_{-n_k} \ket{h},
\end{equation}
where $\bm{n}$ represents the vector $(n_1, n_2, \cdots, n_k )$.
To avoid complexity in notation,
a general state is sometimes abbreviated as:
\begin{equation}
\ket{\nu_{i,\bm{n}}} \equiv \ket{h_i, \bm{n}}.
\end{equation}
Consider the Gram matrix composed of all states at level $N$ of the Verma module $\ca{V}_h$,
\begin{equation}
[G^{(N)}_{c,h}]_{\bm{n} \bm{m}} \equiv \braket{h, \bm{n}| h, \bm{m} }.
\end{equation}
When there exists a null state,
the Gram matrix must be degenerate.
To check whether the Gram matrix is degenerate,
we look at whether the determinant is zero.
Let us see an example to understand when the determinant becomes zero.
The simplest example is at level $1$,
where the Gram matrix is a $1\times1$ matrix,
\begin{equation}
G^{(1)}_{c,h} = \bra{h}L_1 L_{-1} \ket{h} = 2h.
\end{equation}
Here,
we used the Virasoro algebra (\ref{eq:Virasoro}).
From this result,
the vacuum state (i.e. a primary state with $h=0$) of any CFT has a null state at level $1$.
Next,
let us consider the Gram matrix at level $2$.
\begin{equation}
G^{(2)}_{c,h} = \left(
\begin{array}{cc}
\bra{h}L_1^2 L_{-1}^2 \ket{h} & \bra{h}L_1^2 L_{-2} \ket{h} \\
\bra{h}L_2 L_{-1}^2 \ket{h} & \bra{h}L_2 L_{-2} \ket{h} \\
\end{array}
\right)
=
\left(
\begin{array}{cc}
4h(2h+1) & 6h \\
6h & 4h+\frac{c}{2} \\
\end{array}
\right).
\end{equation}
The determinant of this matrix is
\begin{equation}
\text{det} G^{(2)}_{c,h} = 4h\left(8h^2+(c-5)h+\frac{c}{2}\right).
\end{equation}
There is a trivial solution $h=0$,
but this has already been removed at level $1$.
The other conformal dimensions for which the determinant becomes zero are
\begin{equation}
h=\frac{5-c\pm\sqrt{(c-1)(c-25)}}{16}.
\end{equation}
To confirm,
let us construct the null state $\chi$ explicitly.
Note that, in general, null states are often assigned the symbol $\chi$.
The states at level $2$ take the following form (up to a normalization factor):
\begin{equation}
\ket{\chi}=\left(L_{-2}+a L_{-1}^2\right)\ket{h}.
\end{equation}
When this is a null state,
we get the following system of equations:
\begin{equation}
\begin{aligned}
\left\{
\begin{array}{l}
L_1\ket{\chi} = (3+2a+4ah)L_{-1}\ket{h}=0, \\
L_2\ket{\chi} = \left(4h+3ahc\right)\ket{h}=0. \\
\end{array}
\right.\\
\end{aligned}
\end{equation}
Solving these,
we get
\begin{equation}
\begin{aligned}
\left\{
\begin{array}{l}
a=-\frac{3}{4h+2}, \\
h=\frac{5-c\pm\sqrt{(c-1)(c-25)}}{16}. \\
\end{array}
\right.\\
\end{aligned}
\end{equation}
Therefore,
the null state at level $2$ is
\begin{equation}
\ket{\chi} = \left(L_{-2}-\frac{3}{4h+2} L_{-1}^2\right)\ket{h}.
\end{equation}

How can we determine the existence of a null state at a general level $N$?
In fact, for a general $N$, the following formula, Kac's formula, is known.
\begin{tcolorbox}[title=Kac formula]
\begin{equation}
\det   G^{(N)}_{c,h} = K_N \prod^N_{\substack{r,s=1\\ 1\leq rs \leq N  }} \pa{h-h_{rs}}^{p(N-rs)}.
\end{equation}
\end{tcolorbox}
\noindent
Here, $p(N)$ is the partition number, and $K_N$ is a constant that does not depend on $(c,h)$,
\begin{equation}
K_N \equiv \prod^N_{\substack{r,s=1\\ 1\leq rs \leq N  }}  \pa{(2r)^s s!}^{p(N-rs)-p(N-r(s+1))}.
\end{equation}
The zeros $\{ h_{rs} \}$ of the Kac determinant are given by
\begin{equation}\label{eq:kaczero}
\begin{aligned}
h_{rs}\equiv \fr{Q^2}{4} - \fr{\pa{rb + \fr{s}{b}  }^2 }{4},
\end{aligned}
\end{equation}
where we used the Liouville notation,
\begin{equation}
c\equiv 1+6Q^2, \ \ \ \ Q\equiv \pa{b+\fr{1}{b}}.
\end{equation}
Examining whether this Kac determinant is zero can determine whether a general Verma module is irreducible.
Also, as immediately apparent from the Kac formula,
it can be seen that the level $rs$ descendant state on the primary state with conformal dimension $h_{rs}$ (called degenerate primary field, degenerate field, etc.) contains a null state.
For example, the null states at levels 1 and 2 above are expressed as follows.
\begin{equation}\label{eq:LevelTwo}
\begin{aligned}
\ket{ \chi_{11} } &= L_{-1}\ket{h_{11}},    \\
\ket{ \chi_{21} } &= \pa{L_{-2} - \fr{3}{3h_{21}+2}L_{-1}^2}    \ket{h_{21}} =  \pa{L_{-2} + \fr{1}{b^2}L_{-1}^2}    \ket{h_{21}},    \\
\ket{ \chi_{12} } &= \pa{L_{-2} - \fr{3}{3h_{12}+2}L_{-1}^2}    \ket{h_{12}} =   \pa{L_{-2} + b^2 L_{-1}^2}\ket{h_{12}}.    \\
\end{aligned}
\end{equation}
In addition, for $N>rs$, the Kac determinant has a zero of order $p(N-rs)$,
which reflects the number of null states, i.e. $p(N-rs)$.
As already mentioned, all descendant states of null states are null states.

A significant result obtained by examining the Kac determinant in detail is as follows.

\begin{description}

\item[Representations with $c>1, h>0$ are irreducible unitary representations]\mbox{}\\
It must be at least $c\geq 0, h \geq0$ for the representation to be unitary.
The following inequality can easily show this:
\begin{equation}\label{eq:lnln}
\braket{h|L_n L_{-n}|h} =  2nh + \fr{1}{12} cn(n^2-1) >0.
\end{equation}
First, setting $n=1$, it can be shown that $h\geq0$.
Next, considering the limit as $n\to \infty$, it can be shown that $c\geq 0$.
However, to know at which points in the region $c\geq 0, h \geq0$ the representation is unitary,
we need the help of the Kac determinant.

In the two-dimensional plane $(c,h)$, equation (\ref{eq:kaczero}) gives the curves where the sign of the Kac determinant can change.
These curves do not cross the region $c>1, h>0$, so it can be shown that the sign of the Kac determinant does not change in this region.
Therefore, if we know the sign of the Kac determinant at a particular point in this region, we can determine the sign of the Kac determinant over the entire region $c>1, h>0$.
Concrete examples of $c>1$ unitary CFTs can be easily constructed (such as a tensor product of free bosons),
so it can be shown that the Kac determinant is always positive for $c>1, h>0$.

\item[Representations with $0<c<1, h>0$ are unitary only at specific discrete values]\mbox{}\\

By examining the curves (\ref{eq:kaczero}) in detail in the two-dimensional plane $(c,h)$,
it can be shown that the unitarity cannot be achieved except at the following points,
\footnote{For details of the proof, see Section 7.2.3 of the Yellow book \cite{Francesco2012}.}
\begin{equation}\label{eq:unitarypoint}
\begin{aligned}
c&=1-\fr{6}{m(m+1)}, \\
h_{rs} &=  \fr{\pa{(m+1)r-ms}^2-1 }{4m(m+1)}, \ \ \ \ \ 1\leq r < m, \ \ 1\leq s < r. 
\end{aligned}
\end{equation}
Here, $h_{rs}$ is rewritten using the variable $m$ from (\ref{eq:kaczero}).
What can be understood from this is that for $0<c<1$, unitary CFTs can only be constructed for specific rational values of $c$.
\footnote{
The unitary CFT with $c=0$ is called the {\it trivial CFT}
because there are no primary states other than the vacuum state in the unitary CFT with $c=0$.
}
Also, the fact that the conformal dimensions are always given by rational numbers is noteworthy.
When associating CFTs with critical systems, the conformal dimensions correspond to critical exponents.
Thus, the above fact indicates that critical exponents are always rational numbers.
Just as the quantization of spin was derived from the representation theory of $su(2)$,
the quantization of conformal dimensions and the rationality of critical exponents are derived from the representation theory of the Virasoro algebra.

Finally, it should be noted that the proof using the curves (\ref{eq:kaczero}) in the two-dimensional plane $(c,h)$
cannot prove that all representations (\ref{eq:unitarypoint}) are unitary.
In fact, (\ref{eq:unitarypoint}) does represent unitary representations,
but proving this requires another technique (i.e. the coset construction).
These lecture notes do not explain the coset construction,
but those interested may refer to Section 18 of the Yellow book \cite{Francesco2012}.

\end{description}

\subsection{State-Operator Correspondence}\label{subsec:stateoperator}

In all previous sections, we have been examining the state space.
However, there should also be ``fields" in field theory.
Where have they gone?
In fact, in CFT, fields and states are equivalent.
In this section, we will explain this ``state-operator correspondence."

We have described the Hilbert space using the eigenstates of $L_0$,
but what physical role does $L_0$ play?
In the operator formalism, choosing a time direction is essential in the quantization process.
Generally, we use Cartesian coordinates, wherein the current two-dimensional complex plane, the real axis is taken as the spatial direction and the imaginary axis as the time direction.
However, there is a more convenient quantization in CFT.
When the two-dimensional complex plane is expressed in polar coordinates $z=r \ex{i\theta}$, quantization takes the angular direction $\theta$ as the spatial direction and the radial direction $r$ as the time direction.
This type of quantization is called \und{radial quantization}.
What is the Hamiltonian, or the time evolution generator, in radial quantization?
Since time evolution corresponds to dilatation,
the generator $L_0+\bar{L}_0$ is the Hamiltonian.
We have seen through the representation theory of the Virasoro algebra the quantization of the eigenvalues of $L_0$,
and it turns out that the eigenvalues of $L_0$ are indeed the energy.
We assumed that $L_0$ is Hermitian because we wanted to regard the eigenvalues of $L_0$ as the energy of the physical system.
Moreover, if we interpret the eigenvalues of $L_0$ as energy,
those values must be bounded from below,
and as discussed in (\ref{eq:lnln}),
unitarity guarantees the lower bound of the eigenvalues of $L_0+\bar{L}_0$.
Additionally, the eigenvalues of $L_0-\bar{L}_0$ are interpreted as spin.
Thus, it is natural to assume $h-\bar{h}\in\bb{Z}$.
\footnote{
From the single-valuedness of correlation functions, we can show that $h-\bar{h} \in \fr{1}{2}\bb{Z}$.
Furthermore, imposing modular $T$ invariance yields $h-\bar{h} \in \bb{Z}$,
which will be explained later.
}

In radial quantization, the infinite past corresponds to the origin.
This suggests that the state is determined by the local information at the origin.
This is called the \und{state-operator correspondence}.

\newsavebox{\boxToOperator}
\sbox{\boxToOperator}{\includegraphics[width=9cm]{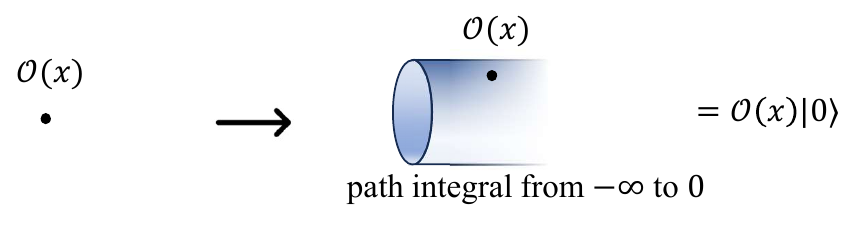}}
\newlength{\ToOperatorw}
\settowidth{\ToOperatorw}{\usebox{\boxToOperator}} 

\newsavebox{\boxToState}
\sbox{\boxToState}{\includegraphics[width=10cm]{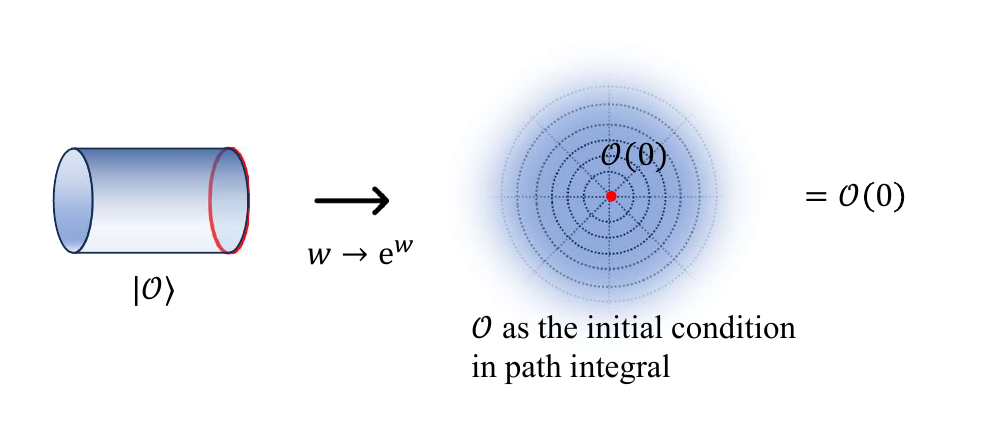}}
\newlength{\ToStatew}
\settowidth{\ToStatew}{\usebox{\boxToState}} 

\begin{description}
\item[Operator $\Rightarrow$ State]\mbox{}\\
A state at a particular time $t$ is defined by the path integral from $-\infty$ to $t$.
By inserting a local operator into this path integral,
various states can be created.
In the operator formalism,
when the vacuum state is written as $\ket{0}$,
a local operator $\ca{O}(x)$ corresponds to creating the following state.
\begin{equation}
\parbox{\ToOperatorw}{\usebox{\boxToOperator}}.
\end{equation}
Defining states from operators is possible not only in CFT but also in general QFT.

\item[State $\Rightarrow$ Operator]\mbox{}\\
In the path integral formalism, the initial state is defined by the boundary condition for the fields in the infinite past.
In radial quantization, this infinite past is localized at the origin.
Consequently, specifying a state is equivalent to a local operation, namely, inserting a local operator.
\begin{equation}
\parbox{\ToStatew}{\usebox{\boxToState}}.
\end{equation}
This is a property unique to CFT.
We will see this more concretely in Section \ref{subsec:radial}.

\end{description}
Combining these two, we conclude
\footnote{
The term "state-operator correspondence" may be somewhat misleading.
As can be understood from the construction, it is precisely restricted to ``eigen" states and ``local" operators.
}
\begin{tcolorbox}[title=State-operator correspondence]
There is a one-to-one correspondence between eigenstates of the dilatation $L_0+\bar{L}_0$ and local operators
\end{tcolorbox}
\noindent
In the first half of this section, we examined the structure of the state space through the representation theory of the Virasoro algebra,
which was equivalent to identifying the Hilbert space of local operators (or fields) existing in the field theory.

\subsection{Relationship Between Radial Quantization and Vertical Quantization}\label{subsec:radial}

\begin{figure}[t]
 \begin{center}
  \includegraphics[width=12.0cm,clip]{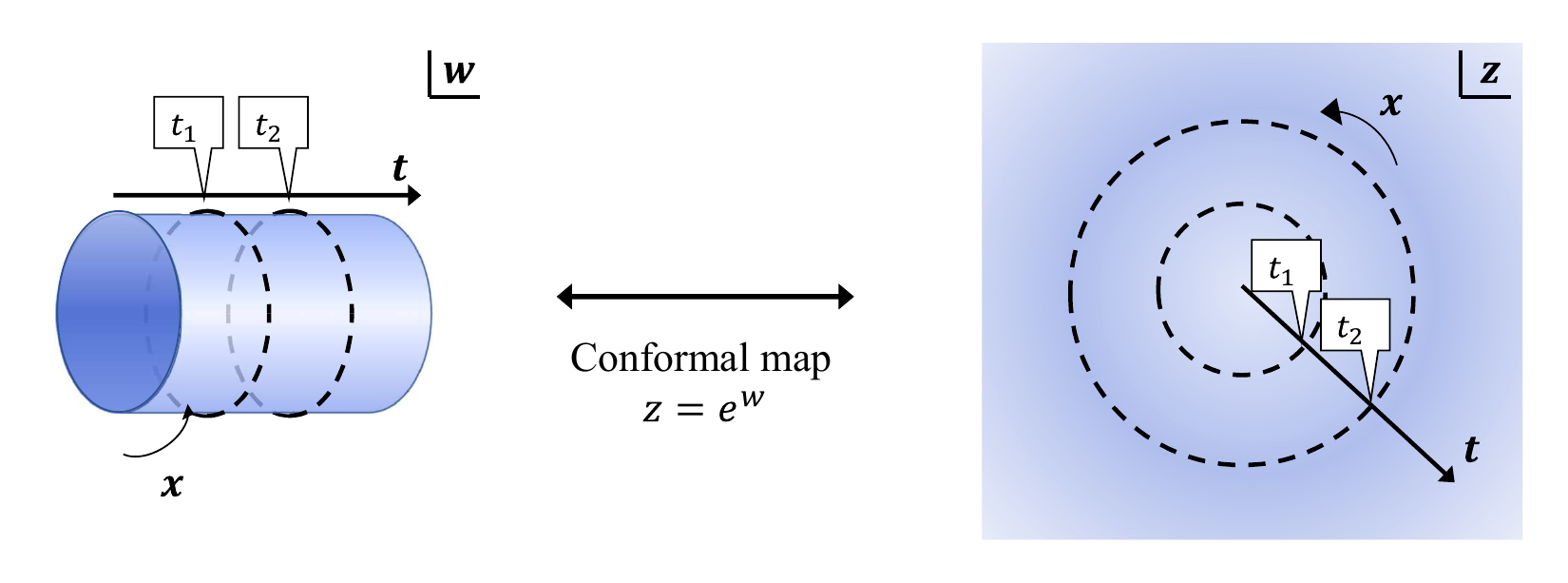}
 \end{center}
 \caption{Relation between the vertical quantization and the radial quantization.}
 \label{fig:radial}
\end{figure}

Consider the Hilbert space of a CFT defined on a cylinder $\bb{R}\times S^1$.
Typically, a different quantization gives an entirely different structure of the Hilbert space and a completely different action on operators.
However, in CFT,
it is possible to directly relate the vertical quantization on the cylinder to the radial quantization on the Riemann sphere by the conformal transformation $z = \ex{w}$ (see Figure \ref{fig:radial}).
Since we are more familiar with the vertical quantization on the cylinder,
it is common to consider the state space of a CFT after mapping it onto the cylinder.

Using this relationship with quantization on the cylinder, we can understand the state-operator correspondence more concretely.
A state can be represented in the path integral with boundary conditions on the fields and a formal cut as follows.
\footnote{
Refer to Chapter 4 of Thomas Hartman's \href{http://www.hartmanhep.net/topics2015/}{lecture notes}.}

\newsavebox{\boxaaa}
\sbox{\boxaaa}{\includegraphics[width=4cm]{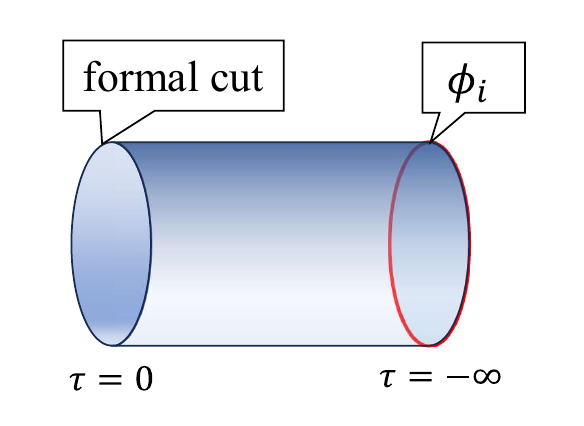}}
\newlength{\aaaw}
\settowidth{\aaaw}{\usebox{\boxaaa}} 

\begin{equation}
\ket{\Psi} = \ex{-\beta H} \ket{\phi_i}
= \int^{\phi(\tau=0)=?}_{\phi(\tau=-\infty)=\phi_i} D\phi \ \ex{S[\phi]} = \parbox{\aaaw}{\usebox{\boxaaa}}.
\end{equation}
The formal cut introduced at $\tau = 0$ is similar to labeling without specifying the indices of the matrix.
More generally, it is possible to have a weighted sum,
\begin{equation}
\ket{\Psi} 
= \int D \phi_i \ \int^{\phi(\tau=0)=?}_{\phi(\tau=-\infty)=\phi_i} D\phi \ \ex{S[\phi]}  \Psi[\phi_i] .
\end{equation}
By moving to radial quantization through the conformal transformation $z=\ex{w}$,
specifying the initial state is replaced by the local operation of changing the weight at the origin of the path integral,
which is nothing but the insertion of a local operator,
\begin{equation}
\ket{\Psi} =  \ \int^{\phi(\tau=0)=?} D\phi   \ \ex{S[\phi]}   \ca{O}_i (0).
\end{equation}
Thus, it is understood that states and operators correspond to each other.
\footnote{
This discussion references Chapter 4 of David Tong's \href{https://www.damtp.cam.ac.uk/user/tong/string.html}{lecture notes}.
}

Finally, one comment to make is that the state-operator correspondence also holds in higher dimensions.
However, what corresponds to a local operator is a state of the CFT on $\bb{R}\times S^{d-1}$.
In the context of the conformal bootstrap discussed later, the CFT on $T^d$ is the most manageable,
but considering quantization on $T^d$, the states correspond to non-local operators,
making the situation much more complicated.
\footnote{
This is briefly mentioned in \cite{Luo2022}.
}

\subsection{Transformation Law of Fields}

We have explained that states and fields are equivalent,
but there are many situations where the field picture is useful.
In this subsection, we will explore the analysis of CFT through fields (or equivalently, correlation functions).
Additionally, we will show how the analytical methods based on fields are related to those based on algebra.

Just as we classified states using the representation theory of the Virasoro algebra,
fields are classified by their transformation laws under conformal transformations.
Therefore, let us first explain the method of calculating the transformation laws of the fields.
By Noether's theorem, the conserved current for the conformal transformation $x^\mu \to x^\mu + \epsilon^\mu (x)$ is given by
\begin{equation}
j^\mu = T_{\mu \nu} \epsilon^\nu,
\end{equation}
where $T_{\mu \nu}$ is the energy-momentum tensor, satisfying $T_{\mu \nu} = T_{\nu \mu}$.
\footnote{
The energy-stress tensor can also be defined by the variation of the action $S$ with respect to the metric,
\begin{equation}
T_{\mu \nu} \equiv -\fr{2}{\sqrt{g}}\fr{\delta S}{\delta g^{\mu \nu}}.
\end{equation}
This definition is equivalent to its definition as the Noether current associated with translations (see the details in Section 2.5 of the Yellow book \cite{Francesco2012}).
}
The energy-momentum tensor is the generator of translations, and the conformal transformation is its generalization.
In fact, assuming $\epsilon^\mu$ is a constant, $x^\mu \to x^\mu + \epsilon^\mu$ becomes a translation,
and the conservation law $\del^\mu j_\mu=0$ reduces to the conservation law of the energy-momentum tensor,
\begin{equation}
\del^\mu T_{\mu \nu}=0.
\end{equation}
For convenience, we move to the complex coordinates again.
Remember that a tensor transforms under diffeomorphism as
\begin{equation}
T'_{\mu \nu}(x') =  \fr{\del x^a}{\del x'^\mu} \fr{\del x^b}{\del x'^\nu}   T_{a b}(x).
\end{equation}
We find that the energy-momentum tensor in vertical coordinates is related to that in complex coordinates as follows:
\begin{equation}
\begin{aligned}
T_{zz} &= \fr{1}{4}\pa{T_{xx} - T_{tt} - iT_{xt} - iT_{tx} },  \\
T_{\bz \bz } &= \fr{1}{4}\pa{T_{xx} - T_{tt} + iT_{xt} + iT_{tx} },  \\
T_{z \bz } &= T_{\bz  z} = \fr{1}{4}\pa{T_{xx} + T_{tt}}.
\end{aligned}
\end{equation}
Thus, we obtain the current conservation law as
\begin{equation}\label{eq:trace}
\begin{aligned}
\del^\mu \pa{T_{\mu \nu} \epsilon^\nu}
&=(\del^\mu T_{\mu \nu}) \epsilon^\mu + T_{\mu \nu} \del^\mu \epsilon^\nu \\
&=2 \pa{T_{zz} \del_{\bz } {\epsilon}(z)   + T_{\bz \bz } \del_{z}  \bar{\epsilon}(\bz )  
+ T_{\bz  z} \del_z \epsilon(z) +  T_{z \bz }\del_{\bz } \bar{\epsilon}(\bz ) }  \\
&= 2 T_{z\bz }\pa{  \del_z \epsilon(z) + \del_{\bz } \bar{\epsilon}(\bz ) } = 0.
\end{aligned}
\end{equation}
Here, from the first to the second line, we used the conservation law of the energy-momentum tensor $\del^\mu T_{\mu \nu}=0$.
In the second line, we use the following metric to raise and lower indices (i.e. $\del^z = 2\del_{\bz }$), and we denote $\epsilon^z$ as $\epsilon(z)$.
\begin{equation}
 g^{\mu \nu} =\left(
    \begin{array}{cc}
      0 & 2  \\
      2 & 0  \\
    \end{array}
  \right).
\end{equation}
From the second to the third line, we used $\del_{\bz } \epsilon(z) = 0$.
Also, in the third line, we used the fact that the energy-momentum tensor is symmetric, $T_{z\bz } = T_{\bz z}$.
Finally, the equation (\ref{eq:trace}) must hold for any $\epsilon^\mu$, so the trace of the energy-momentum tensor must be zero.
\begin{equation}
T_{z\bz } = \fr{1}{4} (T_{xx} +T_{tt}) = 0.
\end{equation}
Consequently, we find that the nonzero components of the energy-momentum tensor are only two components $T_{zz}$ and $T_{\bz \bz}$.
Recalling the conservation law of the energy-momentum tensor again, we obtain
\begin{equation}
\del^z T_{z z} + \del^{\bz } T_{\bz  z} = \del_{\bz } T_{zz} =0.
\end{equation}
In other words, $T_{zz}$ is a holomorphic function. Similarly, $T_{\bz  \bz }$ is an anti-holomorphic function.
For convenience, we redefine these as follows:
\begin{equation}
T(z)\equiv -2\pi T_{zz}, \ \ \ \ \ \bar{T}(\bz ) \equiv -2\pi T_{\bz  \bz }.
\end{equation}

\begin{figure}[t]
 \begin{center}
  \includegraphics[width=6.0cm,clip]{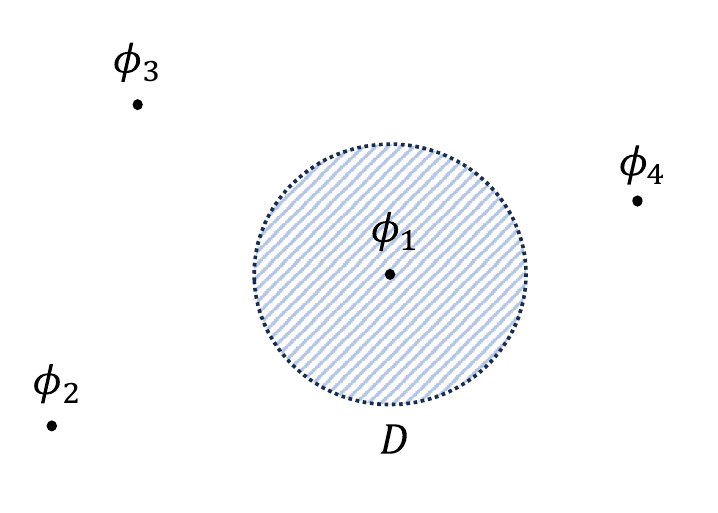}
 \end{center}
 \caption{Conformal transformation restricted to a region $D$, which is a {\it local} conformal transformation.}
 \label{fig:D}
\end{figure}

Next, let us quantize the current conservation law $\del^\mu j_\mu=0$. 
In a field theory, fields are affected by symmetry transformations,
and we describe the change of the fields at the same point by
\begin{equation}
\phi(x) \to \phi'(x) = \phi(x) + \delta\phi(x).
\end{equation}
We consider local symmetry transformations to obtain a quantum version of Noether's theorem.
Specifically, let us apply the symmetry transformation only within an arbitrary region $D$ surrounding $\phi_1$ (see Figure \ref{fig:D}).
\footnote{
We do not think of this local transformation as a symmetry of the theory but introduce it as a tool to derive the quantum version of Noether's theorem.
}
It is possible to include multiple fields within the region $D$, but for simplicity, we consider the case where only $\phi_1$ is in the region $D$.
The change in the action under this transformation is described as follows,
\begin{equation}
 S[\phi'] = S[\phi] + \delta S.
\end{equation}
The change in the action can be written using the Noether current as
\begin{equation}\label{eq:dS}
\delta S = \int_D d^2 x \del^\mu j_\mu.
\end{equation}
By Gauss's theorem, this reduces to a surface integral.
Intuitively, the reason it reduces to a surface integral is that the action inside $D$ (and obviously outside $D$) remains invariant under the global transformation, therefore, only the boundary part $\del D$ contributes to the change in the action.
Since we are in the complex coordinates, Stokes's theorem implies
\begin{equation}
\delta S = \oint j_z dz - \oint j_{\bz } d\bz .
\end{equation}
Consider replacing the variables of the path integral for the correlation function with fields transformed by the local transformation.
Since this is merely a change of integral variables, it does not alter the value of the correlation function,
\begin{equation}
\int D\phi  X \ex{-S[\phi]}  = \int D\phi  (X+\delta X) \ex{-S[\phi]-\delta S}.
\end{equation}
Here, since we assume that only $\phi_1$ is included in the region $D$, 
if we denote $\phi_1 \phi_2 \cdots $ by $X$, then $\delta X = (\delta \phi_1) \phi_2 \cdots$.
For simplicity, we assume the invariance of the integration measure $D\phi=D\phi'$ under the local transformation.
It is important to note that the action is not invariant under the local transformation.
Looking at the first-order perturbation of this equation under an infinitesimal transformation, we obtain
\footnote{
More generally, when the surface operator defined by the surface integral is expressed as $\mathrm{D}^{-1} \equiv \ex{-\del S}$, the Ward-Takahashi identity can be written as
 \begin{equation}
\braket{X'} = \braket{\mathrm{D} X}.
\end{equation}
Here, $\mathrm{D}$ is specifically a topological surface operator.
This identity can be represented diagrammatically as follows,
\newsavebox{\boxTDL}
\sbox{\boxTDL}{\includegraphics[width=6cm]{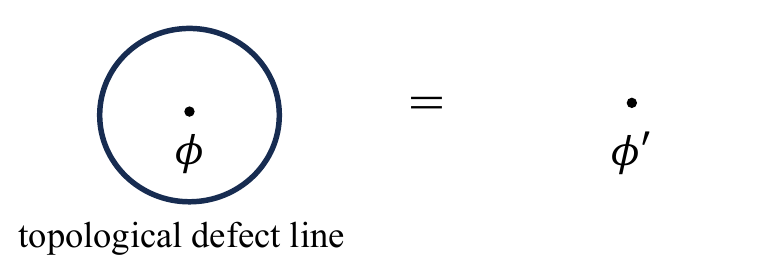}}
\newlength{\TDLw}
\settowidth{\TDLw}{\usebox{\boxTDL}} 

\begin{equation}
\parbox{\TDLw}{\usebox{\boxTDL}}.
\end{equation}
}
\begin{equation}\label{eq:WT0}
\braket{ \delta X } = \braket{\delta S X}.
\end{equation}
By substituting (\ref{eq:dS}) into this equation,
we obtain an identity for the expectation value of the Noether current.
This identity is the quantum version of Noether's theorem that we have wanted,
and it is referred to as the \und{Ward-Takahashi identity}.
By substituting the current for the conformal transformation $z \to z+\varepsilon(z)$ and using Stokes's theorem,
we obtain the following conformal Ward-Takahashi identity:
\begin{tcolorbox}[title=Conformal Ward-Takahashi identity]
\begin{equation}\label{eq:WT}
\delta_{\varepsilon,\overline{\varepsilon}}\left< X\right>
=-\frac{1}{2\pi i}\oint_C dz\varepsilon(z)\left< T(z)X\right>
+\frac{1}{2\pi i}\oint_C d \bz  \bar{\varepsilon}(\bz )\left< \bar{T}(\bz )X\right>.
\end{equation}
\end{tcolorbox}
\noindent
Here, the integration contour $C$ is any path surrounding the insertion points of all transformed local fields.
\footnote{
By Cauchy's theorem, the integral result does not change under contour deformations that do not cross poles.
}
The decomposition into holomorphic and anti-holomorphic parts explained in Section \ref{subsec:Hilbert} can be explicitly read from this Ward-Takahashi identity.

From the Ward-Takahashi identity, the generator of the transformation is
\begin{equation}
Q_\varepsilon \equiv -\frac{1}{2\pi i}\oint_C dz\varepsilon(z) T(z),
\end{equation}
which is a surface operator.
The surface operators as generators are topological.
This can be shown as follows.
Let $S_1$ and $S_2$ be two surfaces,
and let $D$ be the region enclosed by these two surfaces.
If no operators are contained in $D$, $\delta X=0$,
thus, from the Ward-Takahashi identity,
we obtain
\begin{equation}
\int_{S_1} dS_\mu j^\mu -  \int_{S_2} dS_\mu j^\mu=0.
\end{equation}
In other words, the generator is invariant under deformations that do not cross the insertion points of the operators.
\footnote{
Not all topological surface operators correspond to symmetries.
For example, there exist topological surface operators without group structure, such as the Krammers-Wannier duality defect.
However, recent research shows that general topological surface operators play a role similar to symmetries.
These topological operators beyond group structure are called {\it generalized symmetries}.
}
The generator of the conformal transformation can be decomposed as follows on the basis of $-z^{n+1}\del_z$:
\begin{equation}
Q_\varepsilon = \sum_{n \in \bb{Z}} \varepsilon_n L_n,
\end{equation}
and
\begin{equation}\label{eq:taylorT}
L_n \equiv \fr{1}{2\pi i} \oint_C dz \ z^{n+1} T(z)
\end{equation}
is the generator of local conformal transformations.

The Ward-Takahashi identity can be expressed in the following form:
\begin{equation}
\delta \phi(x) = [Q, \phi(x)].
\end{equation}
This commutation relation is introduced as follows.
Let $t$ be the time at which $\phi(x)$ is inserted.
Let $S_1$ and $S_2$ be the constant time slices at times $t_1$ and $t_2$, respectively, with $t_1<t<t_2$.
Since correlation functions follow the time ordering, $Q(S_2)-Q(S_1)$ becomes a commutation relation,
\begin{equation}
\pa{Q(S_2)-Q(S_1)}\phi(x) = [Q,\phi],
\end{equation}
where the integral region $S$ is explicitly written to make the explanation clearer,
but $Q$ is topological, so $S$ is generally omitted from expressions.
Finally, using the Ward-Takahashi identity (\ref{eq:WT0}),
we obtain
\begin{equation}\label{eq:gen}
\delta \phi = [Q,\phi].
\end{equation}
Since this argument does not depend on the choice of the time direction, (\ref{eq:gen}) holds for any quantization.
It should be emphasized that the algebraic structure obtained independent of the quantization is thanks to the topological nature of the operator.

\begin{figure}[t]
 \begin{center}
  \includegraphics[width=10.0cm,clip]{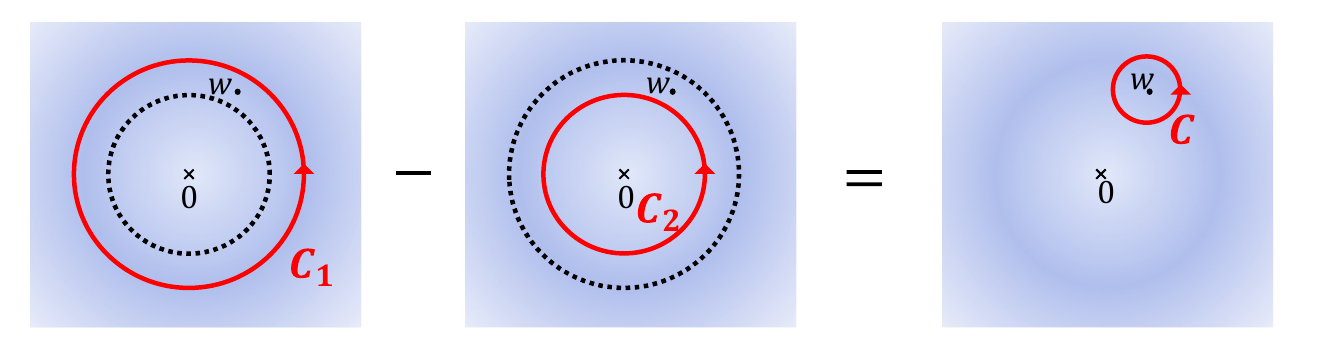}
 \end{center}
 \caption{Integration contours corresponding to the commutation relation and their deformation.}
 \label{fig:comm}
\end{figure}

The commutation relation (\ref{eq:gen}) holds in any QFT.
Let us specifically focus on CFT.
CFT was quantized using radial quantization.
In radial quantization, instead of time ordering, we impose the following rule called {\it radial ordering}:
\begin{equation}
\begin{aligned}
\ca{R} \phi_1(z) \phi_2(w)  &=\left\{
    \begin{array}{ll}
    \phi_1(z)\phi_2(w)   ,& \text{if }  |z|>|w| ,\\
    \varepsilon\phi_2(w)\phi_1(z)   ,& \text{if } |z|<|w|  .\\
    \end{array}
  \right.\\
\end{aligned}
\end{equation}
Here, if both $\phi_1$ and $\phi_2$ are fermions, $\varepsilon=-1$,
and otherwise, $\varepsilon=1$.
Next, let us introduce the commutation relation in radial quantization.
Consider calculating the commutation relation between a topological surface operator
\begin{equation}
A = \oint_C \fr{dz}{2\pi i} a(z)
\end{equation}
and a local operator $b(w)$,
\begin{equation}
[A,b(w)].
\end{equation}
Since the order of operators is given by radial ordering,
we can achieve the desired commutation relation by slightly shifting the contour of $A$ past and future relative to the time $\abs{w}$.
By setting the contour $C_1$ as $\abs{z}>\abs{w}$ and the contour $C_2$ as $\abs{z}<\abs{w}$, the commutation relation can be expressed as follows:
\begin{equation}
[A,b(w)] = \oint_{C_1} \fr{dz}{2\pi i} a(z) b(w) -  \oint_{C_2} \fr{dz}{2\pi i} b(w) a(z). 
\end{equation}
Recalling that the contour can be freely deformed except around the insertion point of the local operator,
we take a sufficiently small closed contour $C$ around the insertion point $w$, which ultimately becomes the following complex integral (see Figure \ref{fig:comm}):
\begin{equation}
[A,b(w)] = \oint_{C} \fr{dz}{2\pi i} a(z) b(w).
\end{equation}
From the residue theorem, it can be seen that the poles of the operator product determine the commutation relation.
In the following, we denote the integral around the vicinity of the point $w$ as $\oint_w$.

\subsection{Primary Field $=$ Primary State}

In the Hilbert space, the fundamental states are primary states.
Using the state-operator correspondence, we can define a primary field using the primary state $\ket{h,\bar{h}}$ as follows:
\begin{equation}
\lim_{z,\bz  \to 0} \phi_{h,\bar{h}}(z,\bz ) \ket{0} \equiv \ket{h,\bar{h}}.
\end{equation}
Since the classification of fields is generally based on their transformation laws,
let us see in what sense primary fields are characteristic.
We can show that primary fields transform under conformal transformations as follows:
\begin{equation}
\phi_{h,\bar{h}}(z,\bz ) \to \phi'_{h,\bar{h}}(w,\bar{w}) = \pa{ \fr{dw}{dz}}^{-h}  \pa{ \fr{d\bar{w}}{d\bz }}^{-\bar{h}} \phi_{h,\bar{h}}(z,\bz ). 
\end{equation}
Let us see that this definition of primary fields is equivalent to the definition of primary states (\ref{eq:hws}).
Under the conformal transformation $z \to w=z+\epsilon(z)$, $\bz  \to \bar{w} = \bz +\bar{\epsilon}(\bz )$, the infinitesimal transformation of a primary field to first order in $\epsilon, \bar{\epsilon}$ is
\begin{equation}\label{eq:primarytrans}
\begin{aligned}
\delta \phi_{h,\bar{h}}(z,\bz ) &\equiv \phi'_{h,\bar{h}}(z,\bz ) - \phi_{h,\bar{h}}(z,\bz ) \\
&= \phi'_{h,\bar{h}}(w,\bar{w}) - \phi_{h,\bar{h}}(z,\bz ) - \pa{ \phi'_{h,\bar{h}}(w,\bar{w}) - \phi'_{h,\bar{h}}(z,\bz )   } \\
&= -h\del_z \epsilon(z) - \bar{h} \del_{\bz } \bar{\epsilon}(\bz ) - \epsilon(z) \del_z \phi_{h,\bar{h}}(z,\bz ) -  \bar{\epsilon}(\bz ) \del_{\bz } \phi_{h,\bar{h}}(z,\bz ).
\end{aligned}
\end{equation}
For this transformation law to be consistent with the Ward-Takahashi identity (\ref{eq:WT}), the following equations must hold:
\begin{equation}\label{eq:primaryope}
\begin{aligned}
T(z)\phi_{h,\bar{h}}(w,\bar{w})  &= \fr{ h \phi_{h,\bar{h}}(w,\bar{w})}{(z-w)^2} + \fr{ \del_w \phi_{h,\bar{h}}(w,\bar{w})}{z-w} + \text{regular terms}, \\
\bar{T}(\bz )\phi_{h,\bar{h}}(w,\bar{w})  &= \fr{ \bar{h} \phi_{h,\bar{h}}(w,\bar{w})}{(\bar{z}-\bar{w})^2} + \fr{ \del_{\bar{w}} \phi_{h,\bar{h}}(w,\bar{w})}{\bz -\bar{w}} + \text{regular terms}.
\end{aligned}
\end{equation}
The equation (\ref{eq:primarytrans}) particularly leads to the absence of higher-order terms than $(z-w)^{-2}$,
which is equivalent to the definition of primary states $L_{n>0}\ket{h,\bar{h}}=0$.
Indeed,
\begin{equation}
\begin{aligned}
L_n\ket{h,\bar{h}} &= L_n \phi_{h,\bar{h}}(0,0)\ket{0} \\
&=[L_n, \phi_{h,\bar{h}}(0,0)] \ket{0} \\
&=\oint_0 \fr{dz}{2\pi i} z^{n+1}T(z) \phi_{h,\bar{h}}(0,0) \ket{0}  \\
&=\oint_0 \fr{dz}{2\pi i} z^{n+1}  \pa{ \fr{ h \phi_{h,\bar{h}}(0,0)}{z^2} + \fr{ \del \phi_{h,\bar{h}}(0,0)}{z} }  \ket{0} \\
&=0, \ \ \ \ \ \ \text{if } n>0.
\end{aligned}
\end{equation}
For simplicity, we denote the holomorphic and anti-holomorphic derivatives by $\del$ and $\bar{\del}$, respectively.
Similarly, we can show that
\begin{equation}
L_0\ket{h,\bar{h}}=h\ket{h,\bar{h}}.
\end{equation}

Let us define the Hermitian conjugate of a primary field.
In Lorentz coordinates, the Hermitian conjugate is defined as an action on operators in the Hilbert space.
On the other hand, in Euclidean coordinates, we must be careful about the fact that the Hermitian conjugate is associated with time-reversal $\tau \to -\tau$ of Euclidean time $\tau \equiv it$.
In radial quantization, this corresponds to $z \to \frac{1}{z^*}$.
Thus, the Hermitian conjugate of a primary field is defined as follows:
\begin{equation}
[\phi_{h,\bar{h}}(z,\bz )]^\dagger \equiv \bz ^{-2h} z^{-2\bar{h}} \phi_{h,\bar{h}}(1/\bz , 1/z).
\end{equation}
Also, using the state-operator correspondence, the Hermitian conjugate of a primary state is defined as
\begin{equation}
\bra{h,\bar{h}} = \lim_{z,\bz  \to 0} [\phi_{h,\bar{h}}(z,\bz ) \ket{0}]^\dagger \equiv  \lim_{\xi, \bar{\xi} \to \infty} \bar{\xi}^{2h} \xi^{2\bar{h}}  \bra{0} \phi_{h,\bar{h}}(\bar{\xi},\xi).
\end{equation}
Since different highest weight states are orthogonal, we have
\begin{equation}
\braket{h_i,\bar{h}_i| h_j, \bar{h}_j } = \delta_{ij}.
\end{equation}
Here, due to the arbitrariness of the normalization of the primary fields, we normalize them such that their norm is 1.
Similarly, the Hermitian conjugate of the Virasoro algebra is given by
\begin{equation}
L_n^\dagger = L_{-n}.
\end{equation}
In particular, since $L_0^\dagger = L_0$, its eigenvalues, the conformal dimensions, are real numbers.
\footnote{
It is possible to consider a CFT whose Hamiltonian is not Hermitian (i.e. non-unitary CFT).
In such a case, it has peculiar properties, such as complex conformal dimensions.
Nevertheless, it is a well-defined CFT in the sense that it is a solution to the conformal bootstrap equations.
Specifically, CFTs with complex CFT data are called {\it complex CFTs}.
}

\subsection{Energy-Momentum Tensor}

Let us rederive the Virasoro algebra from the transformation law of the energy-momentum tensor.
The simplest way to derive the transformation law of the energy-momentum tensor is to calculate $T(z)T(0)$ in specific CFTs (e.g. free CFTs) and find universal features from them.
In fact, such a derivation is given in many textbooks like \cite{Francesco2012}.
Here, instead, we will consider predicting $T(z)T(0)$ from physical considerations.
First, from the fact that the integral of the energy-momentum tensor is energy $E \sim \int T$ and that the energy has mass dimension one, we obtain by dimensional analysis,
\begin{equation}
T'(\lambda z) = \lambda^{-2}T(z).
\end{equation}
Also, from Noether's theorem, the Noether current for dilatation is
\begin{equation}
j_z = zT(z).
\end{equation}
Using these facts, we consider the Ward-Takahashi identity for dilatation and obtain,
\begin{equation}
T(z)T(w) = \fr{2T(w)}{(z-w)^2} + \fr{\del T(w)}{z-w} + \text{regular terms}.
\end{equation}
However, the dilatation alone cannot rule out the presence of higher-order terms.
What kind of higher-order terms are possible?
Consider writing the general term as follows:
\begin{equation}
\fr{\ca{O}_n}{(z-w)^n}.
\end{equation}
By dimensional analysis, $\ca{O}_n$ should have conformal dimension $4-n$.
From unitarity, negative conformal dimensions are not allowed (recall the statements around (\ref{eq:lnln})), so we obtain $n \leq 4$.
Furthermore, from the commutativity $T(z)T(w)=T(w)T(z)$, a term of $(z-w)^{-3}$ cannot exist.
One might think that a term of $(z-w)^{-1}$ should not exist for the same reason.
However, if we expand the $(z-w)^{-2}$ term around $z$, another first-order term appears as follows, ensuring $T(z)T(w)=T(w)T(z)$ is satisfied:
\begin{equation}
\fr{T(w)}{(z-w)^2} = \fr{T(z)}{(z-w)^2} - \fr{\del T(z)}{(z-w)}.
\end{equation}
However, this mechanism does not apply to a term of $(z-w)^{-4}$ because $\ca{O}_4$ is the identity operator of conformal dimension 0 and does not depend on $w$.

Finally, we can conclude that the possible form for $T(z)T(w)$ is as follows:
\begin{equation}
T(z)T(w) = \fr{c}{2(z-w)^4} + \fr{2T(w)}{(z-w)^2} + \fr{\del T(w)}{z-w} + \text{regular terms}.
\end{equation}
In fact, in concrete calculations for a free boson ($c=1$) or a free fermion ($c=1/2$), this form can be confirmed.
Furthermore, it can be verified that this result for $T(z)T(w)$ is equivalent to the Virasoro algebra,
\begin{equation}
\begin{aligned}
[L_n, L_m] = \oint \fr{dw}{2\pi i} \oint_w \fr{dz}{2\pi i} z^{n+1}w^{m+1} T(z)T(w)
= (n-m)L_{n+m} + \fr{c}{12}n(n^2-1)\delta_{n+m,0}.
\end{aligned}
\end{equation}

Let us calculate the transformation law of the energy-momentum tensor.
From the Ward-Takahashi identity and the result of $T(z)T(w)$, we find that the transformation law of the energy-momentum tensor under the conformal transformation $z \to z+\epsilon(z)$ is given by
\begin{equation}\label{eq:Tinf}
\delta T(z) = - 2 \del\epsilon(z) T(z) -\epsilon(z)\del T(z) - \fr{c}{12} \del\del\del \epsilon(z).
\end{equation}
The first two terms are the same as the transformation law of a primary field (\ref{eq:primarytrans}).
The last term reflects the conformal anomaly.
By integrating this, we obtain the following expression for a finite conformal transformation $z \to w$:
\begin{equation}\label{eq:Ttrans}
T'(w) = \pa{\fr{dw}{dz}}^{-2}\pa{T(z) -\fr{c}{12} \{w,z \} }.
\end{equation}
Here, $\{f(z),z\}$ is the Schwarzian derivative which is defined as follows:
\footnote{
Using the following property of the Schwarzian derivative,
\begin{equation}
\{w,z\} = -\pa{\fr{dw}{dz}}^2\{z,w\},
\end{equation}
some references describe the transformation law of the energy-momentum tensor as
\begin{equation}
T'(w) = \pa{\fr{dw}{dz}}^{-2}T(z) +\fr{c}{12} \{z,w \}.
\end{equation}
}
\begin{equation}
\{f(z),z\} \equiv \fr{f'''}{f'} - \fr{3}{2} \fr{(f'')^2}{(f')^2}.
\end{equation}
This Schwarzian derivative vanishes when $f$ is a global conformal transformation.
This implies that the energy-momentum tensor behaves like a primary field under global conformal transformations.
Such a field, which behaves like a primary field only under global conformal transformations, is called a {\it quasi-primary field}.

Finally, using the transformation law of the energy-momentum tensor (\ref{eq:Ttrans}), let us consider the physical interpretation of the central charge.
While considering global conformal transformations, no macroscopic scale appears in the theory.
However, some local conformal transformations introduce a macroscopic scale.
When such a macroscopic scale is added to the system, a conformal anomaly appears.
For example, consider the following conformal transformation:
\begin{equation}
z\to w =\fr{L}{2\pi} \ln z.
\end{equation}
This transformation corresponds to mapping the angular coordinate to the space $S_L^1$ compactified with circumference $L$, and the radial coordinate to $\bb{R}$.
In other words, this transformation maps the plane $\bb{C}\cup\{\infty \}$ to the cylinder $S_L^1 \times \bb{R}$.
This transformation introduces a macroscopic scale of circumference $L$.
Correspondingly, the energy-momentum tensor changes as follows:
\begin{equation}\label{eq:Tcyl}
T_{\text{cyl}}(w) = \pa{\fr{2\pi}{L} }^2 \pa{ T_{\text{pl}}(z)z^2 -\fr{c}{24} }.
\end{equation}
From this result, we can calculate the vacuum expectation value of the energy-momentum tensor on the cylinder, obtaining the Casimir energy proportional to the central charge
\begin{equation}\label{eq:Casimir}
\braket{T_{\text{cyl}}} = -\fr{c\pi^2}{6L^2}.
\end{equation}
In general, as in this example, the central charge appears as a quantity characterizing how the system responds to the breaking of symmetry due to the introduction of a macroscopic scale.
For further explanation on this, refer to Section 5.4 of the Yellow book \cite{Francesco2012}.

\subsection{Measurement of Central Charge}\label{subsec:EE}
In the previous section, we introduced the central charge as a quantity characterizing the breaking of scale invariance.
Suppose that we have found a critical system through experiments or numerical simulations.
How can we measure the central charge of this critical system?
To measure the central charge, we need a quantity that does not depend on the details of the system other than the central charge itself.
In fact, such a quantity is given by the \und{entanglement entropy} (EE) for the vacuum state \cite{Calabrese2004}.
Entanglement entropy was originally introduced in the context of quantum information theory,
but it is also widely used in the context of physics as it is useful for investigating the properties of quantum many-body systems, particularly their non-local properties.
For example, the entanglement entropy can measure the global property of whether a given state is a pure state.
\footnote{
A state $\rho$ is said to be a {\it pure state} if it can be expressed as
\begin{equation}
\rho=\ket{\phi}\bra{\phi}.
\end{equation}
States other than pure states are called {\it mixed states},
which has the following form,
\begin{equation}
\rho = \sum_{i=1}^N p_i \ket{\phi_i}\bra{\phi_i}, \ \ \ N > 1.
\end{equation}
}
Additionally, while correlation functions depend on the choice of fields, entanglement entropy is defined universally,
which is useful for comparing a property of different theories.
\footnote{
In the sense that it is universally defined in different theories,
the partition function (not limited to genus 1) also serves a similar role.
In fact, the calculation of entanglement entropy is essentially equivalent to calculating a special higher-genus partition function called the {\it replica partition function} \cite{Calabrese2009}.
However, there is a well-known fact that this replica partition function can be regarded as a multi-point correlation function on the sphere in an orbifold CFT,
making the calculation of entanglement entropy somewhat easier than that of general higher-genus partition functions.
}

Entanglement entropy measures the amount of quantum correlation (i.e. entanglement) between two parts when a system is divided into two, where the Hamiltonian is decomposed as
\begin{equation}\label{eq:Hdec}
\ca{H} = \ca{H}_A \otimes \ca{H}_B.
\end{equation}
When the reduced density matrix $\rho_A$ is defined for the total density matrix $\rho=\ket{\psi}\bra{\psi}$ as follows,
\begin{equation}
\rho_A = \tr_B \rho,
\end{equation}
the entanglement entropy is defined as the von Neumann entropy of the reduced density matrix:
\begin{equation}
S_A = - \tr_A \rho_A \log \rho_A.
\end{equation}

Consider the entanglement entropy for a connected subsystem $A$ of the vacuum state.
When the length of the subsystem $A$ is $l$, the following universal formula is known to hold in 2D CFT \cite{Calabrese2004}:
\begin{equation}
S_A = \fr{c}{3} \ln l + const.
\end{equation}
The constant part depends on the details of the system and the measurement method.
As seen from this formula,
by measuring the entanglement entropy for various $l$ and determining the coefficient of the logarithmic part,
we can determine the central charge.
This can actually be implemented through Monte Carlo simulations.

\subsection{Operator Product Expansion}\label{subsec:OPE}

Unlike in standard field theory, Lagrangian does not appear at all in the calculations of CFT.
\footnote{
The strength of the Lagrangian method lies in the fact that all correlation functions are provided by the simple object, i.e. ``Lagrangian".
However, it is important to note that this method assumes the existence of a (simple) Lagrangian.
Not all CFTs have a simple Lagrangian description, and some do not have a Lagrangian description at all (e.g. orbifold CFTs).
}
In other words, in CFT, all correlation functions are calculated solely from symmetry and self-consistency, without using the standard QFT methods like perturbative methods.
The key to this approach is the \und{Operator Product Expansion} (OPE).
OPE is a method to approximate the product of two local fields that are very close to each other as a sum of local fields.
\begin{tcolorbox}[title=Operator Product Expansion]
   \begin{equation}
\phi_i(x) \phi_j(0) = \sum_p C_{ijp}(x) \phi_p (0).
\end{equation} 
\end{tcolorbox}
\noindent
Here, $C_{ijp}$ are called \und{OPE coefficients}.
This method can be used even without conformal symmetry, but it is generally not very useful.
However, with conformal symmetry, OPE becomes significantly more manageable.
In this section, we will explain why this is the case.
We strongly recommend readers who want to know the details to read \cite{Pappadopulo2012}.

Consider a CFT in two-dimensional complex coordinates.
First, using scale invariance (i.e. dimensional analysis), we can determine the functional form of $C_{ijp}(z,\bz)$.
\begin{equation}
C_{ijp}(z,\bz) = const. \times z^{h_p-h_i-h_j} \bz^{\bar{h}_p-\bar{h}_i-\bar{h}_j}.
\end{equation}
Next, we decompose the sum by Verma modules.
\begin{equation}\label{eq:preope}
\phi_i(z,\bz) \phi_j(0) = \sum_p \sum_{\{\bm{n},\bar{\bm{n}}\}} C_{ijp \{\bm{n},\bar{\bm{n}}\}} z^{h_p+N-h_i-h_j} \bz^{\bar{h}_p+\bar{N}-\bar{h}_i-\bar{h}_j} \phi^{\{\bm{n},\bar{\bm{n}}\}}_p (0).
\end{equation}
Here, $p$ labels the primary and contains both holomorphic and anti-holomorphic information $(h_p, \bar{h}_p)$.
$\bm{n}$ is a label for descendants $\bm{n} \equiv (n_1, n_2, \cdots n_k)$, and $N$ is the level of the descendant.
Also, we define
\begin{equation}
 \phi^{\{\bm{n},\bar{\bm{n}}\}}_p (0)\ket{0}
 \equiv
 L_{-\bm{n}}\ket{h_p} \otimes \bar{L}_{-\bar{\bm{n}}}\ket{\bar{h}_p}.
\end{equation}
Here, $\ket{h} \otimes \ket{\bar{h}}$ and $\ket{h,\bar{h}}$ have the same meaning.
Since descendant fields are constructed by acting the Virasoro algebra on primary fields,
knowing the information of primary fields should allow us to know the information of descendant fields.
Therefore, it is expected that the following decomposition can be made using a universal function $\beta$:
\begin{equation}
  C_{ijp \{\bm{n},\bar{\bm{n}}\}} = \beta(h_i, h_j, h_p; \bm{n}) \beta(\bar{h}_i, \bar{h}_j, \bar{h}_p; \bar{\bm{n}}) C_{ijp}.
\end{equation}
In short, the claim is that the contribution from descendant fields in the OPE can be expressed as the contribution from primary fields acted upon by some function determined solely by conformal symmetry.
We will see below that this is indeed correct.
In the following, for simplicity, we consider the OPE coefficients of level 1 descendants, but the OPE coefficients of any descendants can be obtained with the same procedure.
Rewrite (\ref{eq:preope}) using the state-operator correspondence as follows:
\begin{equation}
\phi_i(z,\bz) \ket{h_j} \otimes \ket{\bar{h}_j} = \sum_p \sum_{\{\bm{n},\bar{\bm{n}}\}} C_{ijp \{\bm{n},\bar{\bm{n}}\}} z^{h_p+N-h_i-h_j} \bz^{\bar{h}_p+\bar{N}-\bar{h}_i-\bar{h}_j}
L_{-\bm{n}}\ket{h_p} \otimes \bar{L}_{-\bar{\bm{n}}}\ket{\bar{h}_p}.
\end{equation}
Since different eigenstates are orthogonal to each other,
multiplying from the left by $\bra{h_k}L_1 \otimes \bra{\bar{h}_k}$ leads to
\begin{equation}\label{eq:opecov}
\bra{h_k}L_1 \otimes \bra{\bar{h}_k} \phi_i(z,\bz) \ket{h_j} \otimes \ket{\bar{h}_j}
=
C_{ijk\{1,0\}}z^{h_k+1-h_i-h_j} \bz^{\bar{h}_k-\bar{h}_i-\bar{h}_j} \bra{h_k}L_1 L_{-1} \ket{h_k} \braket{\bar{h}_k | \bar{h}_k}.
\end{equation}
From the OPE of primary fields (\ref{eq:primaryope}) and the Ward-Takahashi identity (\ref{eq:WT}),
we obtain
\begin{equation}
[L_n,\phi_i(z,\bz)] = h_i(n+1) z^n \phi_i(z,\bz) + z^{n+1} \del \phi_i(z,\bz),
\end{equation}
and using this, the left-hand side of (\ref{eq:opecov}) is given by
\begin{equation}
h_kC_{ijk} z^{h_k+1-h_i-h_j} \bz^{\bar{h}_k-\bar{h}_i-\bar{h}_j}.
\end{equation}
Also, from the Virasoro algebra (\ref{eq:Virasoro}), the right-hand side of (\ref{eq:opecov}) is given by
\begin{equation}
2h_kC_{ijk \{(1),0\}} z^{h_k+1-h_i-h_j} \bz^{\bar{h}_k-\bar{h}_i-\bar{h}_j}.
\end{equation}
Thus, we obtain
\begin{equation}
C_{ijk \{(1),0\}} = \fr{1}{2} C_{ijk}.
\end{equation}
Similarly, the OPE coefficients of any descendant fields can be calculated solely from conformal symmetry.

So far, we have explained that thanks to conformal symmetry, the structure of the OPE becomes relatively simple.
This is indeed an important property, but there is another essential property to discuss the usefulness of OPE in CFT.
Let us consider calculating $n$-point correlation functions using OPE,
\begin{equation}\label{eq:opeblock}
\braket{\phi_i \phi_j \cdots}
=
\sum_p C_{ijp} 
\sum_{\{\bm{n},\bar{\bm{n}} \}}   \beta(h_i, h_j, h_p; \bm{n}) \beta(\bar{h}_i, \bar{h}_j, \bar{h}_p; \bar{\bm{n}})   z^{h_p+N-h_i-h_j}   \bz ^{\bar{h}_p+\bar{N}-\bar{h}_i-\bar{h}_j} 
\braket{  \phi_p^{  \{ \bm{n}, \bar{\bm{n}} \}  }  \cdots }.
\end{equation}
As seen from this equation,
using OPE, the $n$-point correlation function can be expressed in terms of $(n-1)$-point correlation functions.
The most important point here is that in CFT, this OPE expansion is not an asymptotic expansion but \und{a series expansion with a finite radius of convergence}.

The key to the finite radius of convergence in CFT lies in the state-operator correspondence.
Let us explain this below.
As explained in Section \ref{subsec:stateoperator},
one can construct states from local operators,
\begin{equation}
\ket{\Psi} = \phi_i(x) \phi_j(y) \ket{0}.
\end{equation}
Any state in the Hilbert space can be expanded in terms of energy eigenstates.
\begin{equation}
\ket{\Psi} = \sum_p C_p(x-y) \ket{E_p}.
\end{equation}
Here, the energy eigenstate $\ket{E_p}$ is defined through the radial quantization centered at $y$.
Due to the properties of the Hilbert space, this expansion converges.
This holds in general QFT.
The crucial point is that in CFT, one can reconstruct local operators from states,
\begin{equation}
\ket{\Psi} = \sum_p C_p(x-y) \ca{O}_p(y) \ket{0},
\ \ \ \ \ \ca{O}_p(y)\ket{0} = \ket{E_p}.
\end{equation}
As one quickly realizes, this expansion in terms of local operators is the OPE itself.
Thus, from the equivalence of OPE and the energy eigenstate expansion, and the convergence of the energy eigenstate expansion,
the convergence of OPE can be proven.
Furthermore, as seen from this proof, the radius of convergence is the distance to the nearest insertion point among all other local operators excluding $\phi_i(x)$.
In other words, the condition for convergence is, letting $\{ z_i \}$ be the insertion points of the remaining operators, as follows:
\begin{equation}
\abs{x-y} < \min_{i} \abs{z_i - y}.
\end{equation}

From this convergence of OPE, we obtain the following two important consequences:
\begin{itemize}

\item
As long as the OPE coefficients are known, any $n$-point correlation function can be calculated to any desired precision.
In general QFT, the OPE is an asymptotic expansion, so the OPE calculation method was not mainstream.
However, in CFT, the OPE becomes the common calculation method, which can be understood in this section.

\item
OPE satisfies associativity.
When expanding $\braket{ \phi_i(x) \phi_j(y) \phi_k(z) \cdots }$
in terms of $\abs{x-y}$ and in terms of $\abs{y-z}$,
one notices that there is an overlap in the radius of convergence.
This leads to the following \und{OPE associativity},
\begin{equation}
\pa{ \phi_i(x) \phi_j(y)} \phi_k(z) = \phi_i(x) \pa{ \phi_j(y) \phi_k(z)}.
\end{equation}
The program for determining the OPE coefficients with this OPE associativity is called the \und{conformal bootstrap}.
This conformal bootstrap provides a non-perturbative calculation method that is entirely different from the traditional method based on Lagrangian,
bringing about remarkable progress in the understanding of field theory.

\end{itemize}

\subsection{Four-Point Function and Conformal Bootstrap}

Here, let us briefly explain the conformal bootstrap.
The detailed explanation will be lengthy, so we will save it for later,
and here we aim to give a sense of what it is about.

\begin{figure}[t]
 \begin{center}
  \includegraphics[width=10.0cm,clip]{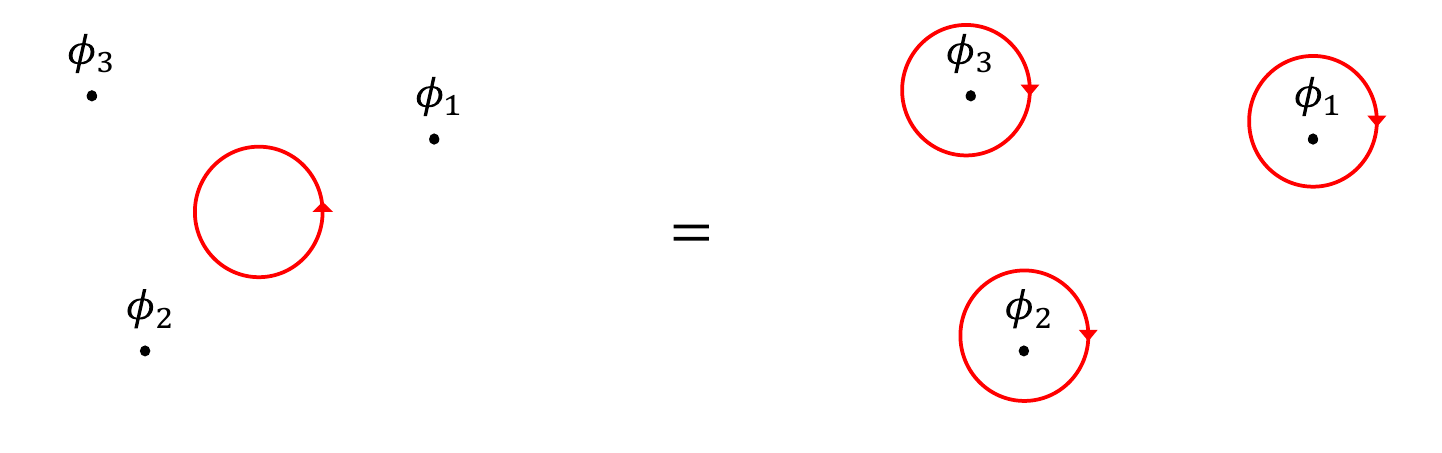}
 \end{center}
 \caption{Moving the topological operator corresponding to Virasoro generator.}
 \label{fig:move}
\end{figure}

The simplest example of the conformal bootstrap is obtained by imposing the OPE associativity on a four-point function on the plane.
Here, before discussing the four-point function, let us first consider $N<4$ point functions.
Correlation functions with three points or less are trivial in the sense that their functional form is completely determined by conformal symmetry.
This can be immediately seen from the OPE,
\begin{equation}
\bra{h_i, \bar{h}_i} \phi_j(z,\bz ) \ket{h_k, \bar{h}_k} = C_{ijk} z^{h_i-h_j-h_k} \bz ^{\bar{h}_i -\bar{h}_j -\bar{h}_k}.
\end{equation}
More directly, it can also be derived as follows:
\begin{equation}\label{eq:GWT1}
\begin{aligned}
\bra{0} X  L_{-n} \ket{0}
&= -\sum_i \oint_{z_i} \fr{dz}{2 \pi i} z^{1-n} \braket{T(z) X} \\
&=  -\sum_i \oint_{z_i} \fr{dz}{2 \pi i} z^{1-n}  \pa{\fr{h_i}{(z-z_i)^2}+\fr{1}{z-z_i} \del_{z_i}}\braket{X} \\
&=\sum_i \pa{ \fr{(n-1)h_i}{z_i^n} - \fr{1}{z_i^{n-1}}\del_{z_i}     } \braket{X}.
\end{aligned}
\end{equation}
Here, in the first equation, we moved the Virasoro generator as shown in Figure \ref{fig:move}.
In the second equation, we used the OPE (\ref{eq:primaryope}) between the primary field and the energy-momentum tensor.
In the third equation, we used the residue theorem.
Recall here that the vacuum is invariant under global conformal transformations,
\begin{equation}\label{eq:GWT2}
L_n \ket{0} = 0, \ \ \ \ \ \text{if} \ n=-1,0,1.
\end{equation}
Combining (\ref{eq:GWT1}) and (\ref{eq:GWT2}), we obtain the following three equations:
\begin{equation}\label{eq:GWT}
\begin{aligned}
\sum_i   \del_{z_i} \braket{X}  &= 0, \\
\sum_i \pa{ h_i + z_i\del_{z_i}     } \braket{X}  &= 0,  \\
\sum_i \pa{ 2h_i z_i + z_i^2 \del_{z_i}     } \braket{X}  &= 0. 
\end{aligned}
\end{equation}
Since this only looks at the part of the conformal Ward-Takahashi identity that pertains to global conformal transformations,
it is sometimes called the {\it global Ward-Takahashi identity}.
The global Ward-Takahashi identity for correlation functions with three points or less can be solved exactly.
\begin{equation}\label{eq:GWTres}
\begin{aligned}
\braket{\phi_i(z_i, \bz _i)} &= 0, \\
\braket{ \phi_i(z_i, \bz _i) \phi_j (z_j, \bz _j) } &= \fr{\delta_{ij}}{ z_{ij}^{2h_i} \bz _{ij}^{2\bar{h}_i} }, \\
\braket{  \phi_i(z_i, \bz _i) \phi_j (z_j, \bz _j)  \phi_k (z_k, \bz _k) } &= C_{ijk} z_{ij}^{ h_k-h_i-h_j  }  z_{jk}^{ h_i-h_j-h_k  } z_{ik}^{ h_j-h_i-h_k  }  \times (\text{anti-holomorphic part}).
\end{aligned}
\end{equation}
Here, $z_{ij} \equiv z_i-z_j$.
Although the four-point function is not determined by conformal symmetry alone,
using the OPE (\ref{eq:opeblock}), which expresses it as a sum of three-point functions, we obtain the following expression:
\begin{equation}
\begin{aligned}
\braket{  \phi_4(z_4, \bz _4)   \phi_3 (z_3, \bz _3)  \phi_2 (z_2, \bz _2)  \phi_1(z_1, \bz _1)   }
&= \prod_{i>j} z_{ij}^{\fr{h}{3} -h_i - h_j  }  \bz _{ij}^{\fr{\bar{h}}{3} -\bar{h}_i - \bar{h}_j  } G^{21}_{34}(z,\bz ),\\
G^{ji}_{kl}(z,\bz )
\equiv \braket{  \phi_l| \phi_k (1, 1)  \phi_j (z, \bz )  |\phi_i   }
&= \sum_p C_{ijp}C_{klp}  \ca{F}^{ji}_{kl}(h_p|z) \overline{ \ca{F}^{ji}_{kl}(h_p|z)}.
\end{aligned}
\end{equation}
Here, $h\equiv\sum_{i=1}^4$. $z$ is called the {\it cross ratio}, defined as $z \equiv \fr{z_{12} z_{34}}{ z_{13} z_{24}}$.
$\ca{F}^{ji}_{kl}(h_p|z)$ is a function of $\{ h_i, h_j, h_k, h_l, h_p, c, z \}$ called the \und{Virasoro block}, defined as follows:
\footnote{
Unlike in two dimensions, in higher dimensions only global conformal symmetry exists. Thus operators get organized into primaries and descendants based on the global conformal group. The function capturing the contribution of descendants, given that of a global conformal primary in the intermediate channel, to the four-point function, is called the global conformal block.
To distinguish it from the global conformal block, what is organized by Virasoro symmetry is often called the Virasoro block, not just the conformal block.
}
\begin{tcolorbox}[title=Virasoro block]
    \begin{equation}\label{eq:Virdef0}
\ca{F}^{ji}_{kl}(h_p|z) \equiv z^{h_p-h_i-h_j} \sum_{\bm{n}} \beta(h_i, h_j, h_p, \bm{n}) z^N \fr{ \braket{\phi_l|\phi_k(1,1)  L_{-\bm{n}}| \phi_p}   }{ \braket{\phi_l|\phi_k(1,1) | \phi_p}  }.
\end{equation}
\end{tcolorbox}
\noindent
The Virasoro block plays the role of organizing the contributions of descendant fields.
The theory dependence from the three-point functions cancels out between the denominator and numerator, so the Virasoro block is a function independent of the theory.
Just as we expanded the solutions to the Schrödinger equation using special functions in quantum mechanics,
the correlation functions in CFT are spanned by a functional basis called Virasoro blocks.
The expansion coefficients of these correlation functions are the OPE coefficients.

The radius of convergence for the $\phi_1\phi_2$ OPE in the four-point function $\braket{  \phi_4| \phi_3 (1, 1)  \phi_2 (z, \bz )  |\phi_1   }$ is $0<\abs{z}<1$.
On the other hand, the radius of convergence for the $\phi_2\phi_3$ OPE in the four-point function transformed by the conformal transformation $z \to 1-z$, $\braket{  \phi_4| \phi_1 (1, 1)  \phi_2 (1-z, 1-\bz )  |\phi_3   }$, is $0<\abs{1-z}<1$.
Since there is overlap in these radii of convergence, the following equality holds:
\begin{tcolorbox}[title=Conformal bootstrap equation (Crossing equation)]
\begin{equation}
 \sum_p C_{12p}C_{34p}  \ca{F}^{21}_{34}(h_p|z) \overline{ \ca{F}^{21}_{34}(h_p|z)}
 =
  \sum_p C_{32p}C_{14p}  \ca{F}^{23}_{14}(h_p|1-z) \overline{ \ca{F}^{23}_{14}(h_p|1-z)}.
\end{equation}
\end{tcolorbox}
\noindent
This is called the crossing equation, representing the associativity of the OPE.
The crossing equation is a non-trivial equation for the OPE coefficients,
and the approach of determining the OPE coefficients by solving this equation is called the \und{conformal bootstrap}.
To distinguish the Virasoro block expansions appearing on the left-hand side and right-hand side of this equation,
the Virasoro blocks on the left-hand side are called the {\it s-channel}, and those on the right-hand side are called the {\it t-channel}.

The conformal bootstrap equation is invariant under the rescaling of fields and OPE coefficients:
\begin{itemize}

\item
$ \phi_i \to \lambda_i \phi_i$ ,

\item
$C_{ijk} \to \kappa C_{ijk} $.

\end{itemize}
Therefore, the solution to the conformal bootstrap has the ambiguity due to the rescaling.
For CFTs including the vacuum state, this ambiguity is generally removed by choosing the normalization as follows:
\begin{equation}\label{eq:normalization}
\begin{aligned}
\braket{\phi|\phi}&=1, \\
C_{i i \bb{I}} &= 1.
\end{aligned}
\end{equation}

Finally, a note on notation. The overline indicates that all parameters are anti-holomorphic.
\begin{equation}
\overline{ \ca{F}^{21}_{34}(h_p|z)} \equiv \ca{F}^{\bar{2}\bar{1}}_{\bar{3}\bar{4}}(\bar{h}_p|\bz ).
\end{equation}
To avoid long expressions, it is common to abbreviate the expression (holomorphic) $\times$ (anti-holomorphic) as the absolute value.
For example, many references use the following abbreviated notation:
\begin{equation}
 \abs{\ca{F}^{21}_{34}(h_p|z)}^2 \equiv \ca{F}^{21}_{34}(h_p|z) \overline{ \ca{F}^{21}_{34}(h_p|z)}.
\end{equation}
However, note that the anti-holomorphic part is not necessarily the complex conjugate of the holomorphic part and can take independent values.

\newsavebox{\boxs}
\sbox{\boxs}{\includegraphics[width=3cm]{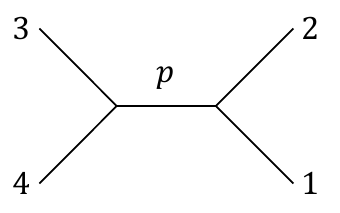}}
\newlength{\sw}
\settowidth{\sw}{\usebox{\boxs}} 

\newsavebox{\boxsd}
\sbox{\boxsd}{\includegraphics[width=2.5cm]{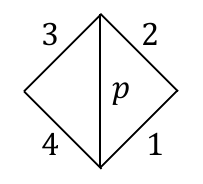}}
\newlength{\sdw}
\settowidth{\sdw}{\usebox{\boxsd}} 

\newsavebox{\boxt}
\sbox{\boxt}{\includegraphics[width=2.2cm]{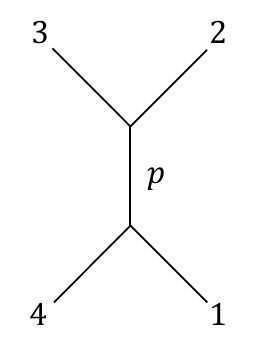}}
\newlength{\tw}
\settowidth{\tw}{\usebox{\boxt}}

Furthermore, due to the analogy with Feynman diagrams,
the Virasoro block is sometimes represented by the following graph:
\begin{equation}
\ca{F}^{21}_{34}(h_p|z) \equiv \parbox{\sw}{\usebox{\boxs}} \equiv \parbox{\sdw}{\usebox{\boxsd}}.
\end{equation}
The second graph is the dual graph of the first one (a graph where the vertices correspond to the faces of the original graph).
Using this diagrammatic notation, the conformal bootstrap equation can be expressed as follows:
\begin{equation}
 \sum_p C_{12p}C_{34p}  \abs{\parbox{\sw}{\usebox{\boxs}} }^2
 =
  \sum_p C_{32p}C_{14p}  \abs{\parbox{\tw}{\usebox{\boxt}}}^2  .
\end{equation}

\subsection{Belavin-Polyakov-Zamolodchikov Equation}\label{subsec:BPZ}

Recall the process of removing null states when constructing the irreducible representation of the Virasoro algebra in terms of Verma modules.
A similar process applies to the calculation of correlation functions,
leading to the requirement that correlation functions involving null states vanish.
This is called \und{Belavin-Polyakov-Zamolodchikov (BPZ) equation}\cite{Belavin1984}.
This requirement is generally independent of the Ward-Takahashi identity.

As the simplest example of the BPZ equation, consider the equation obtained from the level 1 null state,
\begin{equation}\label{eq:zero3pt}
\braket{ \chi_{11}(z) \phi_1(z_1) \phi_2(z_2)  } = 0.
\end{equation}
Correlation functions involving descendant fields can generally be written as derivatives of correlation functions of primary fields.
\begin{equation}
\begin{aligned}
\braket{ (L_{-n} \phi_1)(z_1) \phi_2(z_2), \cdots  }
&=\fr{1}{2\pi i} \oint_{z_1}  \fr{dw}{(w-z_1)^{n-1}} \braket{ T(w) \phi_1(z_1) \phi_2 (z_2) \cdots} \\
&= -\fr{1}{2\pi i}  \sum_{i=2}^k \oint_{z_i}  \fr{dw}{(w-z_1)^{n-1}}  \pa{ \fr{h_i}{(w-z_i)^2}+\fr{\del_i}{w-z_i  }}\braket{ \phi_1(z_1) \phi_2 (z_2) \cdots} \\
&= \sum_{i=2}^k  \pa{ \fr{(n-1)h_i}{(z_i-z_1)^n} - \fr{\del_i}{(z_i-z_1)^{n-1}  }}\braket{ \phi_1(z_1) \phi_2 (z_2) \cdots} \\
& \equiv \ca{L}_{-n} \braket{ \phi_1(z_1) \phi_2 (z_2) \cdots} .
\end{aligned}
\end{equation}
In the first equation, we used (\ref{eq:taylorT}); in the second equation, we used (\ref{eq:primaryope}); and in the third equation, we moved the integral contour (see Figure \ref{fig:move}).
More generally, the following holds,
\begin{equation}
\braket{ (L_{-{n_1}} L_{-{n_2}} \cdots \phi_1)(z_1) \phi_2(z_2), \cdots  }
= \ca{L}_{-{n_1}} \ca{L}_{-{n_2}}  \cdots \braket{ \phi_1(z_1) \phi_2 (z_2) \cdots} .
\end{equation}
Using this result, (\ref{eq:zero3pt}) can be rewritten as follows:
\begin{equation}
\del_z \pa{(z-z_1)^{h_2-h_1} (z-z_2)^{h_1-h_2} (z_1-z_2)^{-h_1-h_2}}=0.
\end{equation}
Here, we used $h_{11}=0$. This equation implies that $\braket{\phi_{11}(z) \phi_1(z_1)\phi_2(z_2)  }  = \braket{ \phi_1(z_1)\phi_2(z_2)  }$ is nonzero only when $h_1=h_2$. This result is consistent with the global Ward-Takahashi identity result (\ref{eq:GWTres}).

As a more non-trivial example, consider the BPZ equation obtained from the level 2 null state (\ref{eq:LevelTwo}),
\begin{equation}
\braket{\chi_{21}(z) \phi_1(z_1) \phi_2(z_2)} = \pa{\ca{L}_{-2}+\fr{1}{b^2}\ca{L}_{-1}^2} \braket{ \phi_{21}(z) \phi_1(z_1) \phi_2(z_2)}=0. 
\end{equation}
This equation means that $\braket{ \phi_{21}(z) \phi_1(z_1) \phi_2(z_2)}$ is nonzero only under the following condition:
\begin{equation}\label{eq:secondBPZ}
\lambda_1=\lambda_2\pm \fr{b}{2}, \ \ \ \ \ \ h_i \equiv \fr{Q^2}{4} -\lambda_i^2.
\end{equation}
This implies that the primary fields appearing in the $\phi_{21} \phi_1$ OPE are restricted to only two types of primary fields that satisfy the equation (\ref{eq:secondBPZ}). This represents a constraint on the dynamical data, i.e. the OPE, a restriction that cannot be obtained from the Ward-Takahashi identity.

The BPZ equation is a powerful tool for determining CFT data, but it is important to note that it can only be applied in special theories where the spectrum includes the zeros of the Kac determinant $\{ h_{rs} \}$.
\footnote{
A theory with a spectrum composed of $\{ h_{rs} \}$, called the minimal model, can be regarded as one where all information is determined solely by symmetry. Consequently, the BPZ equation, which is a consequence of symmetry, determines the correlation functions, which are generally dynamical objects.
}
In CFTs with $c > 1$, $h_{rs}$ is non-positive, therefore, no fields other than the vacuum lead to the BPZ equation in unitary CFTs with $c>1$. As a result, to investigate the CFT data of $c > 1$ unitary CFTs, other methods, such as conformal bootstrap, must be employed.

\section{Virasoro Blocks and Conformal Bootstrap}\label{sec:Virasoro}

Whether calculating correlation functions or solving the conformal bootstrap,
it is essential to understand the details of the Virasoro blocks, which form the functional basis.
However, Virasoro blocks are extremely complicated special functions that cannot be expressed in closed form,
requiring various techniques to understand their properties.
Here, we will introduce some particularly important ones.

\subsection{OPE $=$ Inserting the Identity Operator}

\begin{figure}[t]
 \begin{center}
  \includegraphics[width=15.0cm,clip]{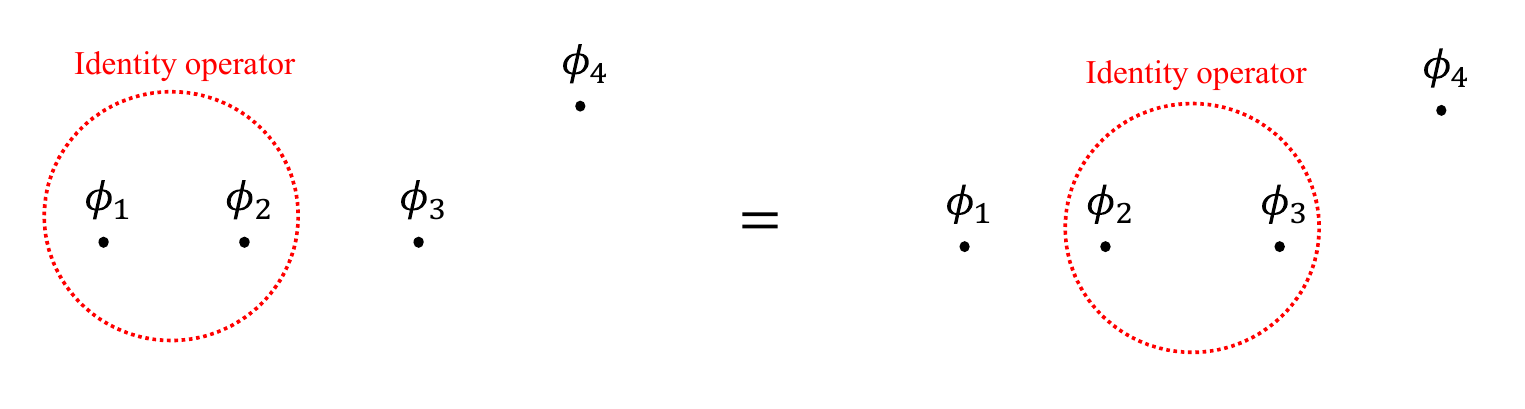}
 \end{center}
 \caption{Conformal bootstrap equation from two different insertions of the identity operator.}
 \label{fig:id}
\end{figure}

OPE involves expanding the product of two local operators as a sum of single local operators.
This ``mapping" can be constructed using the identity operator
\begin{equation}
1 = \sum_{p: \text{primary}} \Pi_{h_p} \otimes \bar{\Pi}_{\bar{h}_p},
\end{equation}
as
\begin{equation}\label{eq:OPEid}
\sum_{p: \text{primary}} \Pi_{h_p} \otimes \bar{\Pi}_{\bar{h}_p} \phi_i(z_i, \bz _i) \phi_j(z_j, \bz _j) \ket{0}.
\end{equation}
Here, $\Pi_{h_p}$ is the projection operator onto the Verma module $\ca{V}(c,h_p)$.
\begin{equation}
\Pi_{h_p} \equiv \sum_{ \substack{ \bm{n}, \bm{m} \\ N=M }} \ket{h_p, \bm{n}} [G^{(N)}_{c,h_p}]_{\bm{n}, \bm{m}}^{-1} \bra{h_p, \bm{m}}.
\end{equation}
In fact, (\ref{eq:OPEid}) is a sum of local operators.
\begin{equation}
\sum_{p: \text{primary}} \sum_{ \substack{ \bm{n}, \bar{\bm{n}} }} a_{p,\bm{n},\bar{\bm{n}}} \ket{h_p, \bm{n}} \otimes \ket{\bar{h}_p, \bar{\bm{n}}}
=
\sum_{p: \text{primary}} \sum_{ \substack{ \bm{n}, \bar{\bm{n}} }} a_{p,\bm{n},\bar{\bm{n}}} 
L_{-\bm{n}} \bar{L}_{-\bar{\bm{n}}}\phi_p(0,0)\ket{0},
\end{equation}
where the expansion coefficients are defined as follows:
\begin{equation}
a_{p,\bm{n},\bar{\bm{n}}} \equiv \sum_{ \substack{ \bm{m} \\ M=N }} \sum_{ \substack{ \bar{\bm{m}} \\ \bar{M}=\bar{N} }} 
[G^{(N)}_{c,h_p}]_{\bm{n}, \bm{m}}^{-1} [G^{(\bar{N})}_{c,\bar{h}_p}]_{\bar{\bm{n}}, \bar{\bm{m}}}^{-1}
\bra{h_p, \bm{m}} \otimes \bra{\bar{h}_p, \bar{\bm{m}}} \phi_i(z_i, \bz _i) \phi_j(z_j, \bz _j)\ket{0}.
\end{equation}
In other words, performing the OPE is equivalent to inserting the identity operator.
From this perspective, the conformal bootstrap equation simply becomes a relation that connects two different ways to insert the identity operator (see Figure \ref{fig:id}).
While this may seem trivial, the fact that the time axes are taken differently, meaning the quantization methods differ,
makes it a non-trivial relation between different quantizations.
As a result, it becomes a non-trivial relation concerning the states on the Hilbert spaces associated with each quantization.
Considering the OPE as the insertion of the identity operator is more convenient for dealing with general conformal bootstrap equations,
so this lecture mainly adopts this viewpoint.

Inserting the identity operator into a four-point function naturally yields an expansion into each Verma module.
\begin{equation}
\braket{h_4,\overline{h}_4|\phi_3(1,1)\phi_2(z,\overline{z})|h_1,\overline{h}_1}
=\braket{h_4,\overline{h}_4|\phi_3(1,1) \PA{ \sum_p  \Pi_{h_p} \otimes \bar{\Pi}_{\bar{h}_p} } \phi_2(z,\overline{z})|h_1,\overline{h}_1}.
\end{equation}
This is precisely the expansion of the correlation function in terms of Virasoro blocks.
Here, the Virasoro block is interpreted as the partial wave projected onto a single Verma module.
\begin{equation}
\ca{F}^{21}_{34}(h_p|z) \overline{ \ca{F}^{21}_{34}(h_p|z)} \propto \braket{h_4,\overline{h}_4|\phi_3(1,1) \Pi_{h_p}\otimes \bar{\Pi}_{\bar{h}_p} \phi_2(z,\overline{z})|h_1,\overline{h}_1}.
\end{equation}
This partial wave is universal, except for a constant factor.
Since the constant part is proportional to the OPE coefficients, it cannot be determined from conformal symmetry alone.
By normalizing the Virasoro block as follows:
\begin{equation}\label{eq:F0}
 \ca{F}^{21}_{34}(h_p|z) \ar{z\to 0} z^{h_p-h_1-h_2},
\end{equation}
the relationship between the Virasoro block and the projection of the correlation function onto the Verma module can be written as:
\begin{equation}
\ca{F}^{21}_{34}(h_p|z) \overline{ \ca{F}^{21}_{34}(h_p|z)} = \fr{\braket{h_4,\overline{h}_4|\phi_3(1,1) \Pi_{h_p} \otimes \bar{\Pi}_{\bar{h}_p} \phi_2(z,\overline{z})|h_1,\overline{h}_1}}{C_{12p} C_{34p} }.
\end{equation}

\subsection{Notation}
In studying the properties of Virasoro blocks,
various notations are introduced.
Here, we summarize this notation for reference.

\hrulefill
\begin{description}

\item[Primary Field ($\nu, \bar{\nu}$):]
\begin{equation}
\lim_{z,\bz  \to 0} \phi_i (z,\bz ) \ket{0} = \ket{\nu_i \otimes \bar{\nu}_i}.
\end{equation}
In this notation, descendant states within the Verma module $\ca{V}_{h_i}$ are described as follows:
\begin{equation}\label{eq:standard}
\ket{\nu_{i,\bm{m}}}=L_{-\bm{m}}\ket{\nu_i}
\equiv L_{-m_1}\dots L_{-m_j}\ket{\nu_i}.
\end{equation}
Here, $\bm{m}$ is an ordered set $\bm{m}=\{m_1,m_2,\dots,m_j \in \bb{N} |m_1\leq m_2 \leq \dots \leq m_j \text{ and } M=m_1+m_2+\cdots+m_j \}$.
Descendant fields defined by the state-operator correspondence are denoted as $\phi_{i,\bm{m}}$.
In the following, we use $\nu_i$ to denote primary states, while $\xi_i$ denotes general states.

\item[Three-Point Block $\rho(\xi_1,\xi_2,\xi_3|z)$:] \mbox{}\\
For states $\xi_i$ in the Verma module $\ca{V}_{h_i}$, the three-point block is defined as follows:
\begin{equation}
\braket{\xi_1, \bar{\xi}_1| \phi_2(z,\bz )| \xi_3,\bar{\xi}_3 }=\rho(\xi_1,\xi_2,\xi_3|z)\rho(\bar{\xi}_1,\bar{\xi}_2,\bar{\xi}_3|\bz )C_{123}.
\end{equation}
Assuming $\xi_i$ are eigenstates of $L_0$,
\begin{equation}
L_0\xi_i=\Delta(\xi_i) \xi_i.
\end{equation}
Here, we explicitly denote the eigenvalue dependence on the eigenstate to cover the most general situations.
From the Ward-Takahashi identity for dilatation, we get the following $z$ dependence:
\begin{equation}\label{eq:3blockz}
\rho(\xi_1,\xi_2,\xi_3|z)=z^{\D(\xi_1)-\D(\xi_2)-\D(\xi_3)}\rho(\xi_1,\xi_2,\xi_3|1).
\end{equation}
By definition, the three-point block for primary fields $\nu_i$ is given by
\begin{equation}
\rho(\nu_1,\nu_2,\nu_3|1)=1.
\end{equation}
The key point is that the three-point block is defined by removing the OPE coefficients, the only parameters that depend on the theory, from the three-point function.
Thus, by definition, the three-point block is merely a function that does not depend on the theory.

\item[Gram Matrix $G^{(N)}_{c,h}:$]
\begin{equation}
\br{G^{(N)}_i}_{\bm{n} \bm{m}}=\braket{\nu_{i,\bm{n}}|\nu_{i,\bm{m}} }.
\end{equation}
Here, $N=M$ must be satisfied.
Since the central charge remains constant within a single CFT,
we omit the central charge from the Gram matrix's subscript and retain only the primary field's label $i$.

\end{description}

\hrulefill \\
Using the above notation, the Virasoro block can be expressed as follows:
\begin{equation}\label{eq:defVir}
\begin{aligned}
\ca{F}^{21}_{34}(h_p|z) =\sum_{\substack{ \bm{n}, \bm{m}\\  N=M }} \rho(\nu_4, \nu_3, \nu_{p,\bm{n}}|1)
\br{G_{p}^{(N)}}_{\bm{n}, \bm{m}}^{-1} \rho(\nu_{p,\bm{m}}, \nu_{2}, \nu_1|z).
\end{aligned}
\end{equation}
This expression is equivalent to the definition of the Virasoro block (\ref{eq:Virdef0}) that appeared in Section \ref{sec:CFT}.
In older literature such as the Yellow book \cite{Francesco2012}, the Virasoro block is defined by (\ref{eq:Virdef0}),
but the expression (\ref{eq:defVir}) where the Gram matrix explicitly appears is more convenient for calculations.
Therefore, in practical 2D CFT research, the expression (\ref{eq:defVir}) is used in most cases.

As seen from the expression (\ref{eq:defVir}),
the sum over states $\nu_{p,\bm{n}} \in V(c,h_p)$ within the Verma module is very complicated,
making the calculation of Virasoro blocks generally difficult.
This often becomes the bottleneck in solving the conformal bootstrap for 2D CFT.
Conversely, since the Virasoro block contains much information about the theory,
clarifying the properties of the Virasoro block often directly leads to understanding the universal properties of the theory.
Additionally, the fact that the parameters identifying a 2D CFT are very few is due to the benefit of having much information packed into the infinite sum in the Virasoro block.
Therefore, the complexity of the Virasoro block can be seen as both a disadvantage and an advantage of 2D CFT.

\subsection{Brute Force Approach}\label{subsec:brute}
Using the Ward-Takahashi identity,
we obtain the following formula for the three-point blocks:
\begin{equation}\label{eq:3ptBlock}
\rho(L_{-n}\xi_3, \xi_2, \xi_1|z) = \rho(\xi_3, \xi_2, L_n \xi_1|z) + \sum^{l(n)}_{m=-1}
\left(
    \begin{array}{c}
       n+1   \\
       m+1   \\
    \end{array}
  \right)
  z^{n-m}
\rho(\xi_3, L_m\xi_2, \xi_1|z).
\end{equation}
Here, when $n\geq -1$, $l(n)=n$, otherwise $l(n)=\infty$. 
Similarly, the Ward-Takahashi identity also provides the following formulas:
\begin{equation}
\begin{aligned}
\rho(\xi_3, L_{-1} \xi_2, \xi_1|z) &=\del_z \rho(\xi_3,\xi_2,\xi_1|z),\\
\rho(\xi_3, L_{-n} \xi_2, \xi_1|z) &=
\sum_{m=0}^{\infty}
\left(
    \begin{array}{c}
       n-2+m   \\
       n-2   \\
    \end{array}
  \right)
  \pa{
  z^m \rho(L_{n+m}\xi_3, \xi_2, \xi_1|z)
  +(-1)^n z^{1-n-m} \rho(\xi_3, \xi_2, L_{m-1}\xi_1 |z)
  }
  , \ \ \ \ \ n \geq 2.
\end{aligned}
\end{equation}
By combining these formulas to convert raising operators one by one into lowering operators,
the Virasoro block can be calculated in a finite number of steps.
This procedure is essentially the same as the calculation of $\beta$ performed in Section \ref{subsec:OPE}.

Can this procedure be used to calculate the Virasoro block with high precision?
From (\ref{eq:3ptBlock}), the calculation of the level $N$ three-point block is expected to be completed in polynomial time with respect to $N$.
Furthermore, when the dimension of the subspace at level $N$ is $d_N$, it is necessary to perform the above three-point block calculation $d_N$ times.
The Gram matrix also appears in the definition of the Virasoro block, and its calculation can be made in polynomial time with respect to $d_N$ \cite{Whalen2014}.
The calculation of the inverse matrix is known to be $O(d_N^3)$.
Therefore, the overall computation time will be polynomial with respect to $d_N$.
However, in reality, it is not practical to perform this on a laptop.
This is because the dimension of the representation space increases exponentially as the level $N$ increases, as indicated by the Hardy-Ramanujan formula (\ref{eq:HR}).
In general, exponential algorithms are only feasible for small numbers.
\footnote{
In general, a laptop can perform about $10^6 \sim 10^8$ operations (such as $+$ or $\times$) per second.
Therefore, for an $O(2^N)$ exponential algorithm, even $N=60$ would take many years.
}
In this case, only a few initial terms (low-level terms) of the Virasoro block can be computed.
Such coarsely calculated Virasoro blocks are often insufficient for 2D CFT research.

To avoid exponential growth in computation,
it is necessary to perform level $N$ calculations with a computation cost independent of the dimension of the representation space.
This efficient algorithm was realized by Zamolodchikov \cite{Zamolodchikov1984,Zamolodchikov1987}.
This computational method is widely used not only in numerical calculations, but also in analytical calculations and the discovery of new dualities.
Therefore, it will be introduced in this lecture.

\subsection{Zamolodchikov Recursion Relation}
The properties of the inverse Gram matrix are fundamentally important in the calculation of Virasoro blocks.
For convenience, let us summarize the properties of the Gram matrix here:

\begin{description}

\item[Kac Formula:] \mbox{}\\
This was introduced in Section \ref{subsec:Gram},
\begin{equation}
\det   G^{(N)}_{c,h} \propto \prod^N_{\substack{r,s=1\\ 1\leq rs \leq N  }} \pa{h-h_{rs}}^{p(N-rs)}.
\end{equation}
Here,
\begin{equation}\label{eq:hrs}
\begin{aligned}
h_{rs}\equiv \fr{Q^2}{4} - \fr{\pa{rb + \fr{s}{b}  }^2 }{4}, \ \ \ \ c\equiv 1+6Q^2, \ \ \ \ Q\equiv \pa{b+\fr{1}{b}}.
\end{aligned}
\end{equation}
Below, we generally write the primary state with conformal dimension $h_{rs}$ as $\nu_{rs}$.
Among the descendant states of $\nu_{rs}$, we denote null states specifically as $\chi_{rs}$.

\item[Poles of the Inverse Gram Matrix:] \mbox{}\\
The inverse Gram matrix has simple poles at $h=h_{rs}$.
It has no other poles.
Alternatively, when viewed as a function of $c$ with $h$ fixed,
when the solution of (\ref{eq:hrs}) is denoted by $c_{rs}$,
the inverse Gram matrix has simple poles at $c=c_{rs}$.
It has no other poles.

\end{description}

The key idea is to focus on the fact that the Virasoro block is a meromorphic function of $c$ (i.e. an analytic function with only isolated singularities) and to perform a Mittag-Leffler expansion.
The Mittag-Leffler expansion represents a function as an expansion around its discrete set of poles $\{z_i \}$,
\begin{equation}
f(z) = c_0 + \sum_{i=1}^\infty \fr{c_i}{z-z_i}.
\end{equation}
Here, $c_i$ are constants, with $c_{i\geq1}$ being residues.
For this expansion to be valid, the following conditions must be satisfied:
\begin{equation}\label{eq:MLcond}
\abs{\fr{f(z)}{z}} \ar{z \to \infty} 0.
\end{equation}
The Virasoro block satisfies this condition,
and its limit can be written as
\begin{equation}\label{eq:Fcinf}
\lim_{c\to \infty} \ca{F}^{21}_{34}(h_p|z) = z^{h_p-h_1-h_2} {}_2F_1(h_p+h_2-h_1,h_p+h_3-h_4;2h_p;z),
\end{equation}
where ${}_2 F_1$ are the hypergeometric functions.
Although this section focuses on the Mittag-Leffler expansion of the Virasoro block,
the large $c$ Virasoro block is important for various reasons,
so its derivation will be detailed in Section \ref{subsec:GCB}.

Let us determine the coefficients of the Mittag-Leffler expansion.
This corresponds to calculating the following residue,
\begin{equation}
\lim_{h \to h_{rs}} (h-h_{rs})\ca{F}^{21}_{34}(h|z).
\end{equation}
Since the three-point block does not have a pole,
the only surviving contribution in the limit $h \to h_{rs}$ comes from the inverse Gram with poles at $ h_{rs}$.
In particular, only the components of the inverse of the Gram matrix with the null descendant label survive. Roughly speaking, this is because the norm of the null descendant is zero.
\footnote{
Let us consider constructing the inverse matrix from the cofactor matrix.
Recall that the cofactor $A_{\bm{n}, \bm{m}}$ is composed of the determinant of the minor matrix obtained by removing the $\bm{n}$-th row and the $\bm{m}$-th column.
When the label $\bm{n}$ or $\bm{m}$ of the cofactor matrix corresponds to a null descendant,
the order of the zero decreases by the extent to which the null descendant is excluded in constructing the cofactor.
In other words, the elements of the inverse of the Gram matrix have poles in $\{ h_{rs} \}$ when $\bm{n}$ or $\bm{m}$ is a null descendant.
Otherwise, the order of the zero of the cofactor matches that of the Kac determinant, and thus no pole appears.
}
Thus, we obtain the following result for the residue,
\begin{equation}
\lim_{h \to h_{rs}} (h-h_{rs})\ca{F}^{21}_{34}(h|z)
=A_{rs}^c \sum_{N=M} z^{h_{rs}-h_1-h_2+rs+N}
\rho(\nu_4, \nu_3, L_{-\bm{n}}\chi_{rs}|1) [G^{(N)}_{c, h_{rs}+rs}]^{-1}_{\bm{n}, \bm{m}}
\rho(L_{-\bm{m}}\chi_{rs},\nu_2,\nu_1|1).
\end{equation}
Here, we use the fact (\ref{eq:null}) that in the limit $h \to h_{rs}$, $\chi_{rs}$ behaves like a primary field.
Although the norm of $\chi_{rs}$ becomes zero in the limit $h \to h_{rs}$,
we are multiplying by $h-h_{rs}$, so
\begin{equation}\label{eq:Ars}
\lim_{h \to h_{rs}} \fr{ \braket{ \chi_{rs} | \chi_{rs}}  }{h-h_{rs}} \equiv \pa{A^c_{rs}}^{-1}
\end{equation}
remains. $A_{rs}^c$ can be calculated from the Virasoro algebra and is written as follows,
\begin{equation}
A_{rs}^c = \fr{1}{2} \prod_{m=1-r}^r \prod_{n=1-s}^s \pa{ mb + \fr{n}{b}  }^{-1}, \ \ \ (m,n)\neq 0, (r,s).
\end{equation}
Moreover, it is known that the three-point block involving a null state is given by the following {\it fusion polynomial},
\begin{equation}
\begin{aligned}
\rho(\nu_1, \nu_2, \chi_{rs})
=\rho(\chi_{rs}, \nu_2, \nu_1)
=P^{rs}_c
  \left[
    \begin{array}{c}
       h_1   \\
       h_2   \\
    \end{array}
  \right]
\equiv
\prod^{r-1}_{\substack{ p=1-r \\ p+r=1 \mod 2  }}
\prod^{s-1}_{\substack{ q=1-s \\ q+s=1 \mod 2  }}
\pa{\lambda_1 + \lambda_2 + \lambda_{p,q}}
\pa{\lambda_1 - \lambda_2 + \lambda_{p,q}}.
\end{aligned}
\end{equation}
Here, $\lambda_i$ is defined as follows,
\begin{equation}
h_i \equiv \fr{Q^2}{4}- \lambda_i^2.
\end{equation}
Also, $\lambda_{p,q}$ corresponds to $\lambda$ for the zeros $h_{pq}$ of the Kac determinant,
\begin{equation}
h_{pq} = \fr{Q^2}{4} - \lambda_{p,q}^2, \ \ \ \ \ \lambda_{p,q} = \fr{1}{2}\pa{pb+\fr{q}{b}}.
\end{equation}
Finally, for the calculation of the three-point block involving the null descendant field,
the three-point block has the following ``factorization",
\begin{equation}
\rho(L_{-\bm{n}}\chi_{rs},\nu_2,\nu_1|1)
=
\rho(L_{-\bm{n}}\nu_{h_{rs}+rs},\nu_2,\nu_1|1)
\rho(\chi_{rs},\nu_2,\nu_1|1).
\end{equation}
Here, $\nu_{h_{rs}+rs}$ is a primary field with conformal dimension $h_{rs}+rs$ (normalized by (\ref{eq:Ars})).
Combining these, we obtain the following result,
\begin{equation}\label{eq:resF}
\lim_{h \to h_{rs}} (h-h_{rs})\ca{F}^{21}_{34}(h|z)
=
A^c_{rs}
P^{rs}_c
  \left[
    \begin{array}{c}
       h_1   \\
       h_2   \\
    \end{array}
  \right]
P^{rs}_c
  \left[
    \begin{array}{c}
       h_4   \\
       h_3   \\
    \end{array}
  \right]
  \ca{F}^{21}_{34}(h_{rs}+rs|z).
\end{equation}
Let us rewrite this as a function of $c$ using (\ref{eq:hrs}),
\begin{equation}
\text{Res}_{c=c_{rs}(h)}{\ca{F}_c}^{21}_{34}(h|z)
= - \fr{\del c_{rs}(h)}{\del h} 
A^{c_{rs}}_{rs}
P^{rs}_{c_{rs}}
  \left[
    \begin{array}{c}
       h_1   \\
       h_2   \\
    \end{array}
  \right]
P^{rs}_{c_{rs}}
  \left[
    \begin{array}{c}
       h_4   \\
       h_3   \\
    \end{array}
  \right]
  {\ca{F}_{c_{rs}}}^{21}_{34}(h_{rs}+rs|z).
\end{equation}
Here, the dependence of the Virasoro block on the central charge is written explicitly.

Finally, the Mittag-Leffler expansion of the Virasoro block is given as follows.
\begin{equation}\label{eq:c-recursion}
\begin{aligned}
{\ca{F}_c}^{21}_{34}(h|z)
&=
z^{h-h_1-h_2} {}_2F_1(h+h_2-h_1,h+h_3-h_4;2h;z) \\
&-
\sum_{r \geq 2, s \geq 1} \fr{1}{c-c_{rs}(h)}
\fr{\del c_{rs}(h)}{\del h} 
A^{c_{rs}}_{rs}
P^{rs}_{c_{rs}}
  \left[
    \begin{array}{c}
       h_1   \\
       h_2   \\
    \end{array}
  \right]
P^{rs}_{c_{rs}}
  \left[
    \begin{array}{c}
       h_4   \\
       h_3   \\
    \end{array}
  \right]
  {\ca{F}_{c_{rs}}}^{21}_{34}(h+rs|z).
\end{aligned}
\end{equation}
Let us recall that the Virasoro block has the following expansion in terms of $z$.
\begin{equation}
\begin{aligned}
\ca{F}^{21}_{34}(h_p|z) 
&=\sum_{\substack{ \bm{n} , \bm{m}\\  N=M  }}  \rho(\nu_4, \nu_3, \nu_{p,\bm{n}}|1)
\br{G_{p}^{(N)}}_{\bm{n}, \bm{m}}^{-1} \rho(\nu_{p,\bm{m}}, \nu_{2}, \nu_1|z)\\
&=\sum_{\substack{ \bm{n} , \bm{m}\\  N=M  }}  \rho(\nu_4, \nu_3, \nu_{p,\bm{n}}|1)
\br{G_{p}^{(N)}}_{\bm{n}, \bm{m}}^{-1} \rho(\nu_{p,\bm{m}}, \nu_{2}, \nu_1|1)z^{h_p-h_1-h_2+N} \\
&\equiv z^{h_p-h_1-h_2} \sum_N \ca{F}^{(N)}_{c,h_p} z^N.
\end{aligned}
\end{equation}
If we know the coefficients $\ca{F}^{(N)}_{c,h_p} $ of this expansion up to $N=k$ for any $c$ and $h_p$,
by substituting this into (\ref{eq:c-recursion}),
we can determine $\ca{F}^{(N+2)}_{c,h_p} $ for any $c$ and $h_p$.
In other words, (\ref{eq:c-recursion}) derives a recurrence relation for the coefficients of the Virasoro block,
\begin{equation} 
\ca{F}^{(N)}_{c,h} = \sum_{k>1}^{N} \sum_{\substack{ r,s \in \bb{N}\\ rs=k }} \alpha^{rs}_{c_{rs},h} \ca{F}^{(N-k)}_{c_{rs},h+rs} + \beta^{(N)}_h.
\end{equation}
Here, we define
\begin{equation}
\begin{aligned}
\alpha^{rs}_{c_{rs},h}
&\equiv
-\fr{\del c_{rs}(h)}{\del h} 
A^{c_{rs}}_{rs}
P^{rs}_{c_{rs}}
  \left[
    \begin{array}{c}
       h_1   \\
       h_2   \\
    \end{array}
  \right]
P^{rs}_{c_{rs}}
  \left[
    \begin{array}{c}
       h_4   \\
       h_3   \\
    \end{array}
  \right],  \\
\beta^{(N)}_h
&\equiv
[z^N]{}_2F_1(h+h_2-h_1,h+h_3-h_4;2h;z),
\end{aligned}
\end{equation}
where, $[z^N]f(z)$ denotes the coefficient of $z^N$ in the series expansion of $f(z)$.

Since the number of pairs $(r,s)$ that satisfy $k=rs$ is given by the number of divisors of $k$, there are $O(k \ln k)$ such pairs.
By summing these $O(k \ln k)$ terms for each $k=2,\dots,n$,
we need $O(n^2 \ln n)$ operations to calculate $\ca{F}^{(n)}_{c,h} $.
Since we sequentially calculate for $n=1,\dots N-1$ to obtain $\ca{F}^{(N)}_{c,h} $,
we finally perform $O(N^3 \ln N)$ operations.
\footnote{
By storing the frequently used calculation results in memory,
we avoid repeated calculations.
Such an algorithm is called {\it dynamic programming}.
}
As this is a polynomial-time algorithm,
it is sufficiently efficient compared to the exponential algorithm introduced in Section \ref{subsec:brute}.
Specifically, while the exponential algorithm takes an astronomical amount of time to compute the $z^{100}$ term,
the polynomial algorithm mentioned above can compute the $z^{1000}$ term in a realistic time (as demonstrated in the author's research \cite{Kusuki2017,Das2021,Kusuki2018}).

\subsection{Virasoro Characters and Modular Bootstrap}
So far, we have looked at the Virasoro block expansion of the four-point function on the Riemann sphere.
In fact, the Virasoro block expansion can be performed for $N$-point functions on a Riemann surface of arbitrary genus.
Recalling that the conformal bootstrap is a relation between different ways of expanding a correlation function in the Virasoro block basis,
we notice that the conformal bootstrap equations can be obtained from various correlation functions other than the four-point function.
\footnote{
Not all of these conformal bootstrap equations are independent,
so we do not need to solve all of them.
In fact, it is known that solving the conformal bootstrap for the four-point function on the sphere and the one-point function on the torus is sufficient to determine all the CFT data.
}
Here, as an example of the Virasoro block expansion other than the four-point function, we consider the Virasoro block expansion of the torus zero-point function, i.e. the partition function.

\begin{figure}[t]
 \begin{center}
  \includegraphics[width=12.0cm,clip]{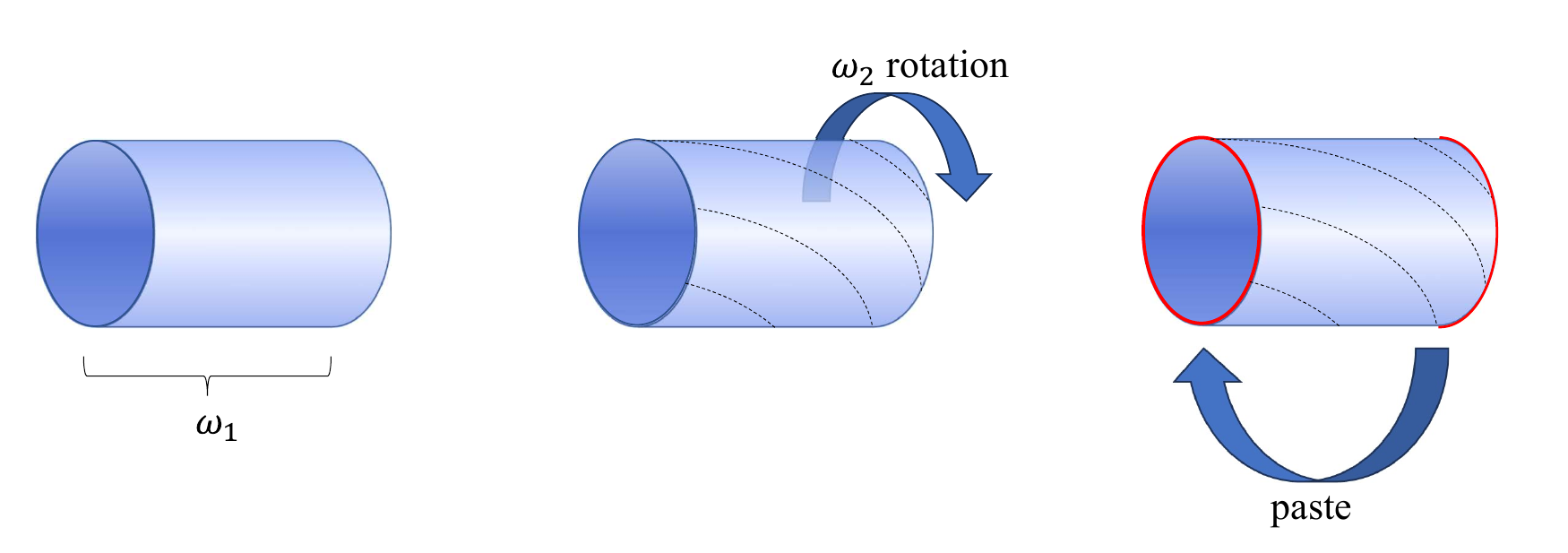}
 \end{center}
 \caption{Construction of the torus partition function.
 (Left) Imaginary time evolution by $\omega_1$ with $H=2\pi\pa{L_0+\bar{L}_0-\fr{c}{12}}$.
 (Middle) Rotation by $\omega_2$ with $P=2\pi i(L_0-\bar{L}_0)$.
 (Right) The torus partition function is created by joining both ends (i.e. taking the trace).
 }
 \label{fig:cooking}
\end{figure}

The partition function can be constructed from a cylinder evolved in imaginary time by $\omega_1$ with $H=2\pi\pa{L_0+\bar{L}_0-\fr{c}{12}}$ (see the left of Figure \ref{fig:cooking}).
Since we are now considering time evolution on a cylinder with period $1$,
we add the Casimir energy $-2\pi \fr{c}{12}$ to the Hamiltonian as seen in (\ref{eq:Tcyl}).
To create a torus, we need to join both ends of the cylinder,
and in doing so, we can twist the cylinder (see the middle of Figure \ref{fig:cooking}).
This twist is realized by rotating by $\omega_2$ with $P=2\pi i(L_0-\bar{L}_0)$.
Finally, the torus is completed by joining both ends of the cylinder (i.e. taking the trace) (see the right of Figure \ref{fig:cooking}).
Setting $(\tau,\bar{\tau}) \equiv (-\omega_2+i\omega_1, -\omega_2-i\omega_1)$,
the partition function can be written as follows,
\begin{equation}
Z(\tau) \equiv \tr_{\ca{H}} q^{L_0-\fr{c}{24}} \bar{q}^{\bar{L}_0-\fr{c}{24}}, \ \ \ \ \ \ q\equiv \ex{2\pi i \tau}, \bar{q} \equiv \ex{-2\pi i \bar{\tau}}.
\end{equation}
Here, $\ca{H}$ is the Hilbert space of the CFT (\ref{eq:hilbert}).
The parameter $\tau$ of the Riemann surface is called the {\it moduli}.

There are countless ways to take the time axis on the torus, and they should all be equivalent.
The mapping that shifts to a different choice of the time axis is called the {\it modular transformation}.
The modular transformation is generated by the following two transformations.

\begin{figure}[t]
 \begin{center}
  \includegraphics[width=15.0cm,clip]{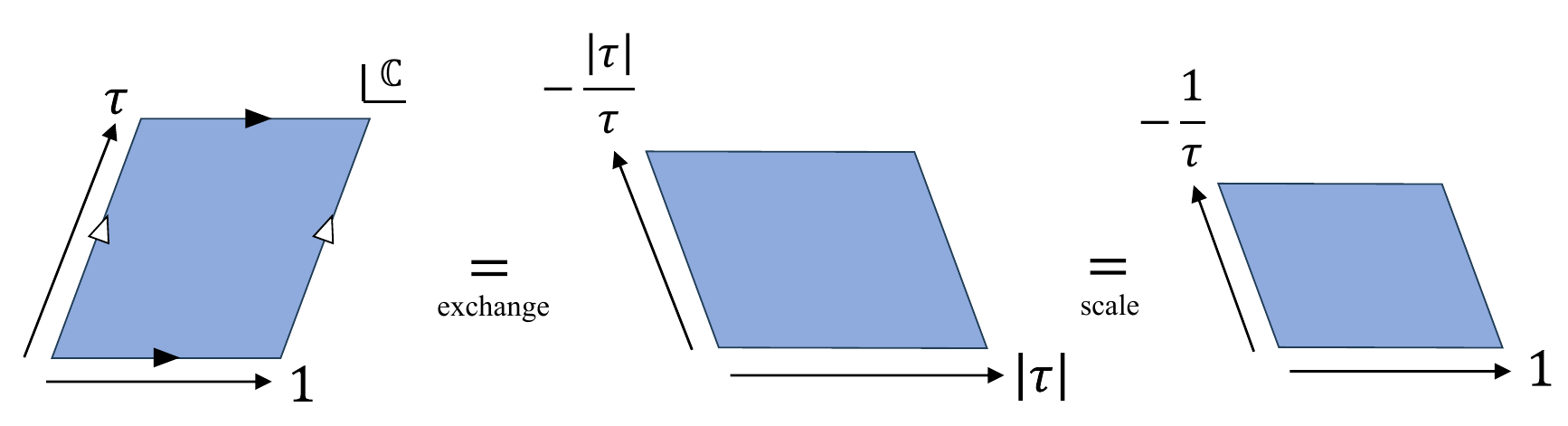}
 \end{center}
 \caption{
A diagram illustrating the modular $S$ transformation.
As shown in the left figure, the torus is represented by two periods $ \{1, \tau\} $.  
We refer to $1$ and $\tau$ as the angular cycle and the thermal cycle, respectively.  
The modular $S$ transformation corresponds to exchanging the angular cycle and the thermal cycle.  
The torus transformed by the modular $S$ transformation is depicted in the right figure, which is equivalent to the torus in the left figure up to a rescaling.  
Due to scale invariance, the path integrals on these two tori yield the same result.  
This is called the modular $S$ invariance of the partition function. 
 }
 \label{fig:swap}
\end{figure}

\begin{figure}[t]
 \begin{center}
  \includegraphics[width=12.0cm,clip]{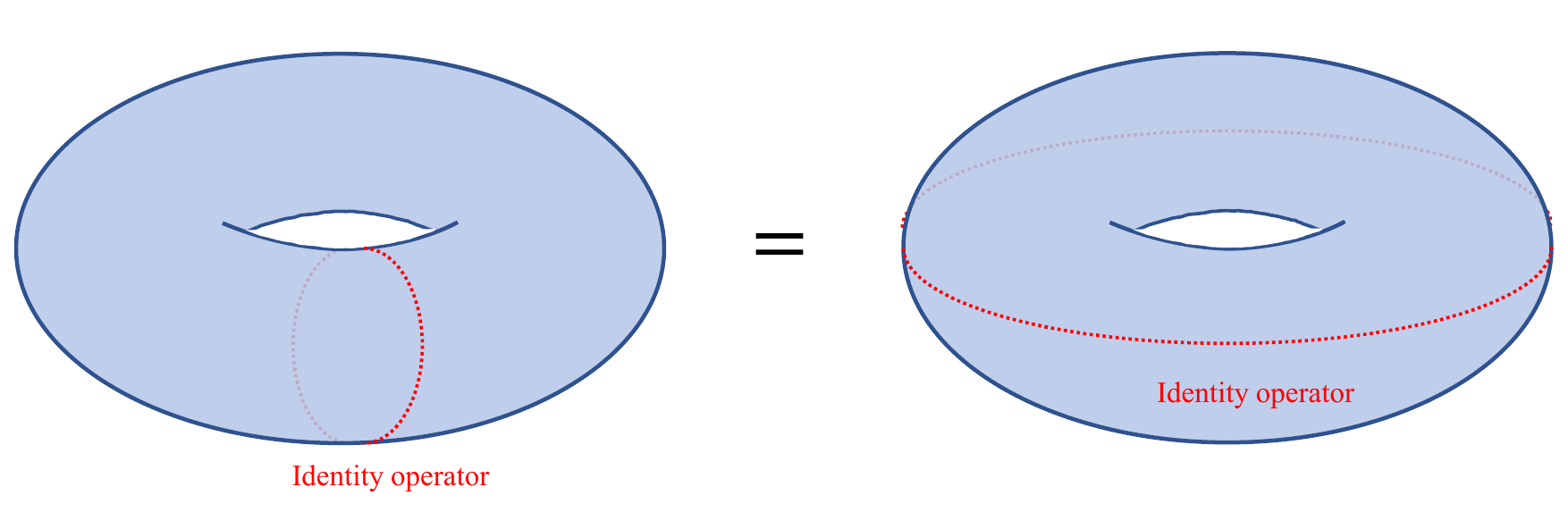}
 \end{center}
 \caption{Modular bootstrap equation.
 This figure shows that the result of the calculation does not change depending on which cycle the identity operator is inserted along.
 }
 \label{fig:modular}
\end{figure}

\begin{description}

\item[$T$ Transformation]\mbox{}\\
\begin{equation}
T: \tau \to \tau+1.
\end{equation}
Let us see how the partition function changes under this transformation,
\begin{equation}
Z(\tau+1) = \tr_{\ca{H}} q^{L_0-\fr{c}{24}} \bar{q}^{\bar{L}_0-\fr{c}{24}} \ex{2\pi i (L_0-\bar{L}_0)}.
\end{equation}
Since the eigenvalues of $L_0-\bar{L}_0$ correspond to the spin,
the invariance under the modular $T$ transformation implies that the spin must be an integer.
\footnote{
In this discussion, we have assumed the CFT is bosonic. For fermionic CFTs, the situation is slightly more subtle. Fermionic theories have half-integer spin and need a spin structure to be defined. On the torus, this means specifying periodic/anti-periodic boundary conditions along both the time and space circles. The four boundary conditions transform as a triplet and a singlet under the modular group. This more intricate structure can potentially lead to more powerful constraints from modular invariance \cite{Benjamin2020b}.
}

\item[$S$ Transformation]\mbox{}\\
\begin{equation}
S: \tau \to -\fr{1}{\tau}.
\end{equation}
Roughly speaking, this corresponds to the exchange of two periods of the torus, that is, the angular cycle and the thermal cycle (see Figure \ref{fig:swap}).
This exchange does not change the result of the path integral on the torus.
When expressing the partition function in the operator formalism,
we insert the identity operator,
and by swapping the time axis and the space axis, it changes which cycle of the torus the identity operator is inserted along (see Figure \ref{fig:modular}).
However, the result of the calculation does not change depending on how the identity operator is inserted,
so these are equivalent,
\begin{equation}\label{eq:modularS}
Z(\tau) = Z\pa{-\fr{1}{\tau}}.
\end{equation}
This is exactly the torus partition function version of the conformal bootstrap equation in the same spirit as Figure \ref{fig:id}.

\end{description}

\noindent
By combining the $S$ and $T$ transformations,
we can derive various conformal bootstrap equations from the partition function.
These are specifically called \und{modular bootstrap equations}.
While the modular $T$ invariance merely implies the integrality of the spin,
the modular $S$ invariance non-trivially constrains the spectrum.
Therefore, when referring to the modular bootstrap equation,
it generally means (\ref{eq:modularS}).

The partition function can be expressed using Virasoro characters as follows,
\begin{equation}
Z(\tau) = \sum_p \chi_p(\tau) \overline{\chi_p(\tau)}.
\end{equation}
Here, the Virasoro character is a special function defined by
\begin{equation}
\chi_i(\tau) \equiv \tr_{\ca{V}_i} q^{L_0 -\fr{c}{24}}= \sum_n d(n) q^{h_i+n-\fr{c}{24}},
\end{equation}
where $d(n)$ is the number of descendant states at level $n$ in the (irreducible) Verma module.
If the Verma module does not contain null states, $d(n)=p(n)$.
Note that this function is a special function determined solely by the Virasoro algebra,
independent of the details of the theory.
While (\ref{eq:defVir}) refers to the Virasoro block for the four-point function on the Riemann sphere,
the Virasoro character is the Virasoro block for the zero-point function on the torus.

Here are some comments on the modular bootstrap equation:
\begin{itemize}

\item
The modular bootstrap equation constrains the spectrum,
but it does not uniquely determine the theory.
In fact, there are examples of different CFTs having the same partition function \cite{Schellekens1993}.
To identify the theory (especially, the OPE coefficients of the theory), we need to solve the bootstrap equation for the four-point function.

\item
The modular bootstrap is not a subset of the conformal bootstrap for four-point functions but an independent condition.
It is known that solving the conformal bootstrap for the four-point function and the one-point function on the torus is sufficient to completely determine the theory \cite{Moore1989}.
The reason for this is simply that from the fusion transformation of the four-point Virasoro block and the modular transformation of the torus one-point block,
one can construct any fusion transformation (transformation between bases of functions for $N$-point functions on any Riemann surface).
This will be explained in Section \ref{sec:GenOPE}.

\item
Unlike the conformal bootstrap equations for four-point functions or one-point functions on the torus,
the modular bootstrap equation does not involve OPE coefficients.
In this sense,
the modular bootstrap equation is much easier to handle.
Therefore, when investigating the possibility of the existence or universal properties of a given theory,
one often considers the modular bootstrap first.
In fact, many notable constraints have been obtained just by considering the modular bootstrap.

\item
While the conformal bootstrap equation for four-point functions is easily generalized to higher dimensions,
the higher-dimensional version of the modular bootstrap equation is somewhat tricky.
Only recently has the higher-dimensional modular bootstrap equation been developed \cite{Benjamin2024}.

\end{itemize}

\subsection{Virasoro Block on Pillow Geometry}

Here, we describe the method of handling the four-point function on a cylinder.
This method is useful when considering the convergence and approximation of the Virasoro block,
and it has been applied in many studies.

\begin{figure}[t]
 \begin{center}
  \includegraphics[width=12.0cm,clip]{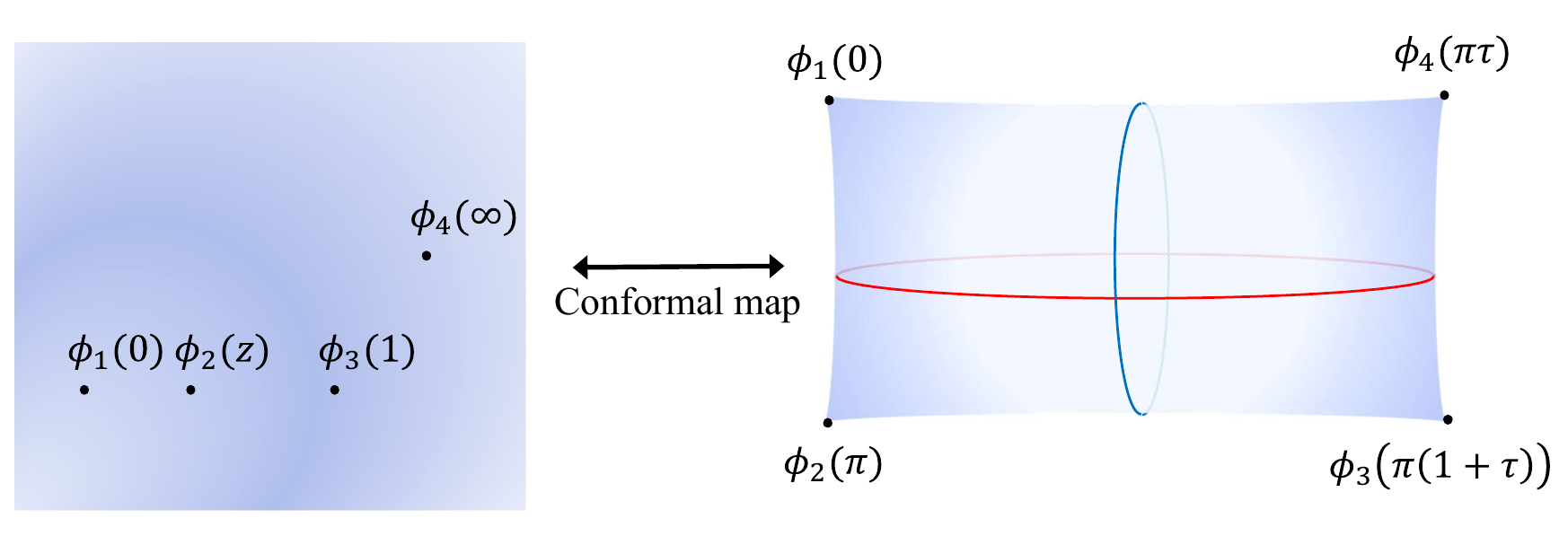}
 \end{center}
 \caption{Pillow geometry is the space $T^2/\bb{Z}_2$ obtained by quotienting a torus by $\bb{Z}_2$.
 Here, we consider a conformal transformation that maps the four points on the Riemann sphere to the corners of the pillow geometry.}
 \label{fig:pillow}
\end{figure}

\begin{figure}[t]
 \begin{center}
  \includegraphics[width=12.0cm,clip]{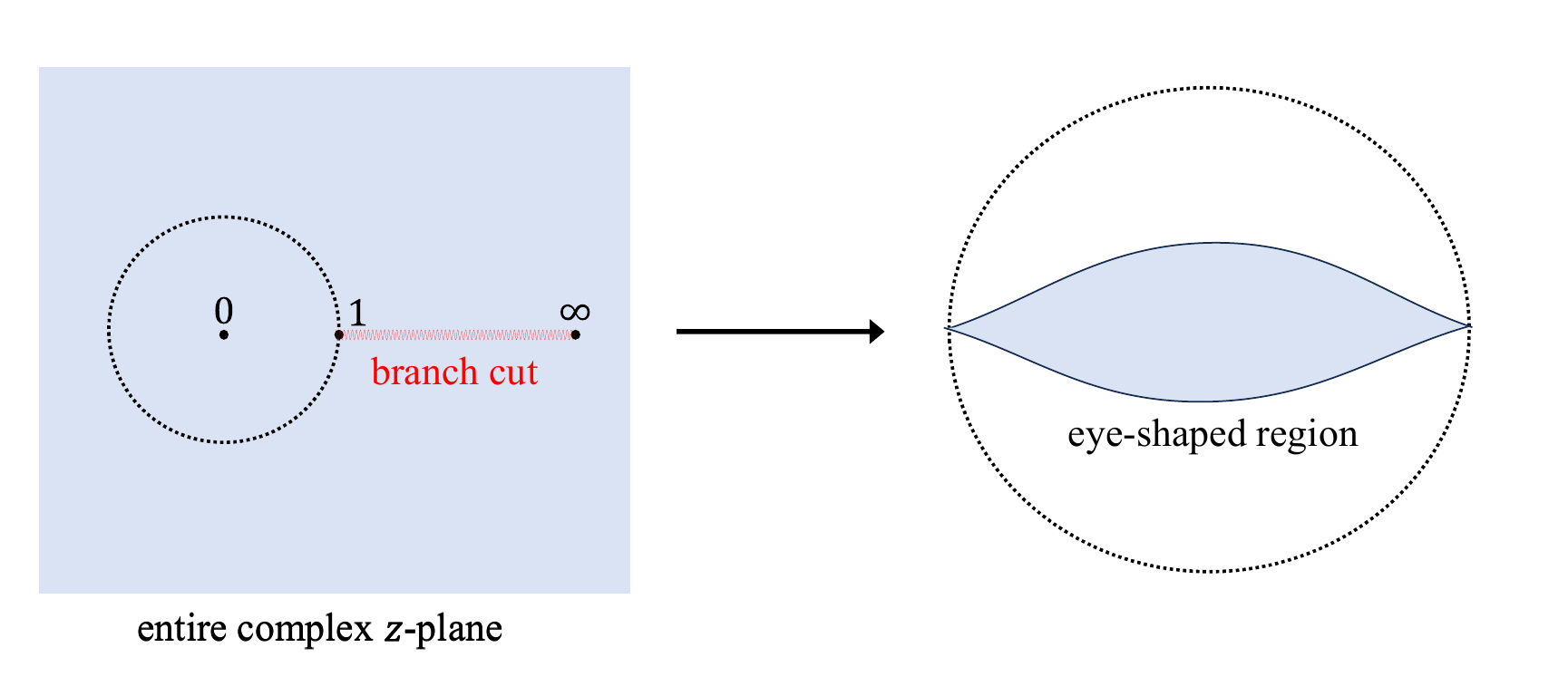}
 \end{center}
 \caption{The entire $z$-plane with a branch cut along $[1,\infty)$ maps to the interior of a unit circle in the $q$-plane, specifically to the eye-shaped region shown in blue.}
 \label{fig:eye}
\end{figure}

The space obtained by quotienting a torus by $\bb{Z}_2$ (denoted as $T^2/\bb{Z}_2$) is called {\it pillow geometry}.
We represent the complex coordinates on a torus using the following identification.
\begin{equation}
w \sim w+2\pi \sim w+2\pi\tau.
\end{equation}
The pillow metric is obtained by the $\bb{Z}_2$ identification $w \sim -w$.
We consider mapping the four-point function on the Riemann sphere to the pillow metric.
Such a conformal transformation is given by the Weierstrass elliptic function (or its inverse function) \cite{olver2010nist}.
\footnote{
This conformal transformation is also used in studies of 2nd-Renyi mutual information (e.g. in the author's papers \cite{Kusuki2020, KudlerFlam2022}).
}
This conformal transformation maps the four points on the Riemann sphere to the corners of the pillow metric (see Figure \ref{fig:pillow}).
The moduli $\tau$ is expressed as a function of the cross ratio $z$ as follows,
\begin{equation}
\tau= i\fr{K(1-z)}{K(z)}.
\end{equation}
Here, $K(z)$ is the complete elliptic integral of the first kind,
\footnote{
In the particle physics field of research, this definition is mainly used (presumably because Mathematica is based on this definition).
On the other hand, in mathematical textbooks, the following definition is often used for the complete elliptic integral of the first kind,
\begin{equation}
K(z)\equiv \int^{\fr{\pi}{2}}_0 \fr{d\theta}{\sqrt{1-z^2 \sin^2 \theta}}.
\end{equation}
Be careful not to confuse results based on different definitions.
}
\begin{equation}
K(z)\equiv \int^{\fr{\pi}{2}}_0 \fr{d\theta}{\sqrt{1-z \sin^2 \theta}} =\fr{\pi}{2} {}_2 F_1 \pa{\fr{1}{2},\fr{1}{2},1;z}.
\end{equation}
The inverse function can also be expressed as follows,
\begin{equation}\label{eq:invz}
z=\pa{ \fr{\theta_2(\tau)}{\theta_3(\tau)}} ^4,
\end{equation}
where the Jacobi theta functions $\theta_i$ are defined using the {\it elliptic nome} $q\equiv \ex{\pi i \tau}$ as follows,
\footnote{
Note that $q$ here is the elliptic nome $q\equiv \ex{\pi i \tau}$.
When using $q\equiv\ex{2\pi i \tau}$ defined for the partition function, the expression changes slightly.
}
\begin{equation}
\begin{aligned}
\theta_2(\tau) &=  \sum_{n \in \bb{Z}+\fr{1}{2}} q^{n^2}, \\
\theta_3(\tau) &=  \sum_{n \in \bb{Z}} q^{n^2}, \\
\theta_4(\tau) &=  \sum_{n \in \bb{Z}} (-1)^n q^{n^2}. \\
\end{aligned}
\end{equation}
In the pillow metric, the Virasoro block is interpreted as a propagator on the cylinder,
\footnote{
Since $\phi_p$ is interpreted as an intermediate state,
$h_p$ is sometimes called the {\it internal weight}.
In contrast, $h_i$ is called the {\it external weight}.
}
\begin{equation}
\braket{\phi_{34}^p|q^{L_0-\fr{c}{24}}| \phi_{12}^p }  = \sum_{n=0}^\infty a_n q^{h_p+n-\fr{c}{24}}, \ \ \ \ \ q\equiv \ex{\pi i \tau}.
\end{equation}
Here, $\ket{\phi_{12}^p}$ is the state created by inserting the operators $\phi_1$ and $\phi_2$ at the corners of the pillow and being projected onto the Verma module $\ca{V}_p$.
As seen from this expression, in the pillow geometry, the $z$-expansion is replaced by the $q$-expansion.
Adding the conformal factor for the conformal transformation from the Riemann sphere to the pillow geometry,
the Virasoro block is expressed as follows,
\begin{equation}
\ca{F}^{21}_{34}(h_p|z)=
\pa{ \theta_3(\tau) }^{ \fr{c}{2}-4(h_1+h_2+h_3+h_4)   } z^{\fr{c}{24}-h_1-h_2} (1-z)^{\fr{c}{24}-h_3-h_4}
\braket{\phi_{34}^p|q^{L_0-\fr{c}{24}}| \phi_{12}^p } .
\end{equation}
One of the advantages of this expression is that it is expanded in $q$.
The radius of convergence for the $q$-series expansion is 1,
\footnote{
One may have encountered the statement ``the convergence of the Virasoro block is not proven".
As seen in the proof of the convergence of the OPE,
we implicitly assume that the state space is a Hilbert space.
Without this assumption, we can consider the problem of whether the $z$- or $q$-series expansion in the Virasoro block converges, where the series expansion coefficients are functions of $\{h_i, h_j, h_k, h_l, h_p, c \}$ determined by the Virasoro algebra.
This is unproven.
It converges in all analytically known examples and generally converges numerically,
so it should not be a problem when solving physical problems.
}
covering the entire complex plane.
In particular, the entire $z$-plane is mapped only to the eye-shaped region in the $q$-plane (see Figure \ref{fig:eye}),
so the $q$-expansion converges better than the $z$-expansion.

When expressing the Virasoro block in terms of the elliptic nome, it is known that the limit $h_p \to \infty$ can be expressed in a simple form as follows,
\begin{equation}
\braket{\phi_{34}^p|q^{L_0-\fr{c}{24}}| \phi_{12}^p }
\ar{h_p \to \infty }
\pa{16q}^{h_p-\fr{c}{24}} \prod^\infty_{n=1}(1-q^{2n})^{-\fr{1}{2}} = \pa{16q}^{h_p-\fr{c}{24}} q^{\fr{1}{24}} \eta(\tau)^{-\fr{1}{2}},
\end{equation}
where the Dedekind eta function is defined as
\begin{equation}\label{eq:Dedekind}
\eta(\tau) \equiv q^{\fr{1}{12}} \prod_{n=1}^\infty \pa{1-q^{2n}}.
\end{equation}
The Dedekind eta function has the following relationship with the Jacobi theta functions,
\begin{equation}
\eta(\tau)^3 =\fr{1}{2} \theta_2(\tau) \theta_3(\tau) \theta_4(\tau).
\end{equation} 
Using this property along with (\ref{eq:invz}) and $\theta_2^2+\theta_4^2=\theta_3^2$, we obtain
\begin{equation}
\eta(\tau)^{-\fr{1}{2}} = 16^{\fr{1}{24} }\theta_3(\tau)^{-\fr{1}{2}} \pa{z(1-z)}^{-\fr{1}{24}}.
\end{equation}
Thus, the limit $h_p \to \infty$ of the Virasoro block can be written as follows,
\begin{equation}\label{eq:large-h}
\ca{F}^{21}_{34}(h_p|z) \ar{h_p \to \infty} 
\pa{16q}^{h_p-\fr{c-1}{24}} 
\pa{ \theta_3(\tau) }^{ \fr{c-1}{2}-4(h_1+h_2+h_3+h_4)   } z^{\fr{c-1}{24}-h_1-h_2} (1-z)^{\fr{c-1}{24}-h_3-h_4}.
\end{equation}
If we express the Virasro block as
\begin{equation}
\ca{F}^{21}_{34}(h_p|z) = 
\pa{16q}^{h_p-\fr{c-1}{24}} 
\pa{ \theta_3(\tau) }^{ \fr{c-1}{2}-4(h_1+h_2+h_3+h_4)   } z^{\fr{c-1}{24}-h_1-h_2} (1-z)^{\fr{c-1}{24}-h_3-h_4}
H^{21}_{34}(h_p|q),
\end{equation}
by applying (\ref{eq:MLcond}) to $H^{21}_{34}(h_p|q)$ and using (\ref{eq:resF}), we obtain the following Mittag-Leffler expansion,
\begin{tcolorbox}[title=Zamolodchikov $h$-recursion relation]
\begin{equation}
H^{21}_{34}(h_p|q) =
1 + 
\sum_{r, s \geq 1} \fr{q^{rs}}{h-h_{rs}}
A^{c}_{rs}
P^{rs}_{c}
  \left[
    \begin{array}{c}
       h_1   \\
       h_2   \\
    \end{array}
  \right]
P^{rs}_{c}
  \left[
    \begin{array}{c}
       h_4   \\
       h_3   \\
    \end{array}
  \right]
  H^{21}_{34}(h_{rs}+rs|q).
\end{equation}
\end{tcolorbox}
\noindent
This expansion in $h$ is called the \und{Zamolodchikov $h$-recursion relation},
and in contrast, (\ref{eq:c-recursion}) is called the \und{Zamolodchikov $c$-recursion relation}.
As explained in Figure \ref{fig:eye}, the $q$-expansion converges better than the $z$-expansion,
providing a better approximation for the Virasoro block.
Therefore, unless there is a special reason, we use the Zamolodchikov $h$-recursion relation.
Code for calculating the Virasoro block using the Zamolodchikov $h$-recursion relation is provided in \cite{Chen2017}.

\subsection{Large-$c$ Virasoro Block}\label{subsec:GCB}

In this section, we derive the large-$c$ limit of the Virasoro block introduced in (\ref{eq:Fcinf}),
\begin{tcolorbox}[title=Large $c$ Virasoro block]
\begin{equation}
\lim_{c\to \infty} \ca{F}^{21}_{34}(h_p|z) = z^{h_p-h_1-h_2} {}_2F_1(h_p+h_2-h_1,h_p+h_3-h_4;2h_p;z).
\end{equation}
\end{tcolorbox}
\noindent
Recall that the Virasoro block is obtained by inserting the projection operator onto the Verma module
\begin{equation}
\Pi_{h_p} \equiv \sum_{ \substack{ \bm{n}, \bm{m}  \\ N=M } }  \ket{h_p, \bm{n}}  [G^{(N)}_{c,h_p}]_{\bm{n}, \bm{m}  }^{-1} \bra{h_p, \bm{m}}
\end{equation}
into the four-point function.
We consider how this projection operator simplifies in the $c \to \infty$ limit.

First, recall that the inverse matrix of $A$ can be written using the cofactor matrix $\ti{A}$ as follows,
\begin{equation}
A^{-1} = \fr{1}{\det A} \ti{A}.
\end{equation}
From this, for the element $[G^{(N)}_{c,h_p}]_{\bm{n}, \bm{m}  }^{-1}$ of the inverse matrix to be non-zero in the large $c$ limit,
the cofactor $[\ti{G}^{(N)}_{c,h_p}]_{\bm{n}, \bm{m}  }$ must have the same maximum power of $c$ as the determinant $\det G^{(N)}_{c,h_p}$.
Since the diagonal elements of $G^{(N)}_{c,h_p}$ have the highest power of $c$,
the maximum power of $c$ in $\det G^{(N)}_{c,h_p}$ is the sum of the maximum powers of the diagonal elements.
For the cofactor to have the same maximum power as the determinant, the elements removed in constructing the cofactor must be $O(c^0)$.
This occurs only when the labels $(\bm{n},\bm{m})$ of the matrix elements are as follows,
\begin{equation}
\bm{n}=\bm{m}=(-1,-1,\cdots,-1).
\end{equation}
This is because when $\bm{n}$ contains $L_{n\leq-2}$, calculating its norm using the Virasoro algebra inevitably produces a central term ($=O(c)$).
In this case, removing elements with the highest power of 1 or more in constructing the cofactor $[\ti{G}^{(N)}_{c,h_p}]_{\bm{n}, \bm{m}  }$
results in the cofactor having a lower maximum power than the determinant.
Thus, in the large $c$ limit, the only non-zero elements of the inverse Gram matrix are given by
\begin{equation}
\lim_{c \to \infty}  [G^{(N)}_{c,h_p}]_{\bm{n}, \bm{n}  }^{-1} = \fr{1}{  \braket{h| L_1^N L_{-1}^N    |h}  } = \fr{1}{N!(2h)_N}.
\end{equation}
Here, $(a)_n\equiv \fr{\Gamma(a+n)}{\Gamma(a)}$ is the Pochhammer symbol.
Additionally, using the Ward-Takahashi identity, we can show
\begin{equation}
\rho( L_{-1}^n \nu, \nu_2, \nu_1|1 ) = (h+h_2-h_1)_n.
\end{equation}
Using these, we obtain the following expression,
\begin{equation}\label{eq:GCB}
\begin{aligned}
\lim_{c\to \infty}
\ca{F}^{21}_{34}(h_p|z)
&=\sum_{\substack{  \bm{n}=\underbrace{(-1,-1,\cdots,-1)}_{N}  }}  \rho(\nu_4, \nu_3,\nu_{p, \bm{n}}|1)
\br{G_{p}^{(N)}}_{\bm{n}, \bm{n}}^{-1} \rho(\nu_{p,\bm{n}}, \nu_{2}, \nu_1|z) \\
&= \sum_N \fr{ (h_p+h_3-h_4)_N (h_p+h_2-h_1)_N}{N! (2h_p)_N}z^{h_p+N-h_1-h_2} \\
&=z^{h_p-h_1-h_2} {}_2F_1(h_p+h_2-h_1,h_p+h_3-h_4;2h_p;z).
\end{aligned}
\end{equation}
This implies that in the large $c$ limit, only the $L_{-1}^N$ descendant states contribute.
Since $L_{-1}$ is the raising operator of the global conformal algebra, this is the global conformal block.

The motivation for deriving the large $c$ limit of the Virasoro block in this section was to obtain the Zamolodchikov $c$-recursion relation (\ref{eq:c-recursion}).
However, the application of the large $c$ Virasoro block goes beyond the efficient computation of Virasoro blocks.
In fact, the large $c$ limit corresponds to the classical limit in the context of the AdS/CFT correspondence,
making the large $c$ Virasoro block extremely useful for studying semiclassical gravity.
This will be explained in the section on the AdS/CFT correspondence (Section \ref{sec:AdSCFT}).

\section{Classification of CFTs and Conformal Bootstrap}\label{sec:classification}
So far, the discussion has been abstract, so one may not have seen the physics clearly.
Therefore, in this section, we will review the results while touching on specific examples of CFTs.

\subsection{Minimal Models}
When the spectrum of a CFT consists of a finite number of irreducible representations of the Virasoro algebra,
the CFT is called a \und{minimal model}.
Minimal models are labeled by two coprime natural numbers $(p,q)$ and are denoted as $\ca{M}(p,q)$.
The main CFT data of $\ca{M}(p,q)$ are as follows:

\begin{description}

\item[Central Charge]\mbox{}\\
\begin{equation}
c=1-6\fr{(p-q)^2}{pq}, \ \ \ \ \ p>q \in \{ 2,3,\cdots\}.
\end{equation}

\item[Spectrum]\mbox{}\\
\begin{equation}
\ca{H} = \bigoplus_{\substack{ 1\leq r < q \\ 1 \leq s < p }} \ca{V}_{h_{rs}} \otimes \bar{\ca{V}}_{h_{rs}}.
\end{equation}
Here, $h_{rs}$ are the zeros of the Kac determinant (\ref{eq:kaczero}),
\begin{equation}\label{eq:kaczero2}
\begin{aligned}
h_{rs}\equiv \fr{Q^2}{4} - \fr{\pa{rb + \fr{s}{b}}^2}{4}, \ \ \ \ \ c\equiv 1+6Q^2, \ \ \ \ Q\equiv \pa{b+\fr{1}{b}}.
\end{aligned}
\end{equation}
In other words, all primary fields appearing in a minimal model are degenerate.
\footnote{
Generally, there are infinitely many primary fields that can appear in the OPE,
but in minimal models, since all primary fields are degenerate,
the BPZ equation significantly restricts the primary fields that can appear in the OPE,
realizing the finiteness of the minimal model.
}
\end{description}

\subsubsection{Conformal Bootstrap in Minimal Models}
The simplest example of minimal models is the Ising model $\ca{M}(4,3)$.
The Ising model has the following three primary fields:
\begin{equation}
\begin{aligned}
&\bb{I}: (h_{\bb{I}}, \bar{h}_{\bb{I}}) = (h_{11}, h_{11})=  (0,0), \\
&\sigma: (h_\sigma, \bar{h}_\sigma) = (h_{12}, h_{12}) =\pa{\fr{1}{16}, \fr{1}{16}}, \\
&\epsilon: (h_\epsilon, \bar{h}_\epsilon) = (h_{21},h_{21}) =\pa{ \fr{1}{2}, \fr{1}{2} } .
\end{aligned}
\end{equation}
Let us evaluate a four-point function using the conformal bootstrap in this theory.
From the BPZ equation, the four-point function of $\sigma$ satisfies the following equation,
\begin{equation}\label{eq:BPZIsing}
\pa{\ca{L}_{-2} - \fr{4}{3} \ca{L}_{-1}^2} \braket{ \sigma (\infty) \sigma (1) \sigma(z,\bz) \sigma(0)} = 0.
\end{equation}
The following Virasoro blocks are obtained as two linearly independent solutions to this equation,
\footnote{
One may think that these can be directly derived from the Zamolodchikov recursion relation.
However, when one tries it, divergent terms appear when the central charge is rational.
Since the Virasoro block itself does not diverge, each divergence must cancel out nicely.
Therefore, it is theoretically possible to obtain the Virasoro blocks for rational central charges by taking the appropriate limit from the Zamolodchikov recursion relation.
However, due to technical difficulties, this remains an open problem.

In actual numerical calculations, the Virasoro block for rational central charges is prepared approximately by slightly shifting from the rational value.
}
\begin{equation}\label{eq:IsingVir}
\begin{aligned}
\ca{F}^{\sigma \sigma}_{\sigma \sigma}(\bb{I}|z) =\fr{1}{\sqrt{2}}\fr{\sqrt{1+\sqrt{1-z}}}{\pa{z(1-z)}^{\fr{1}{8}}}, \\
\ca{F}^{\sigma \sigma}_{\sigma \sigma}(\epsilon|z) =\fr{2}{\sqrt{2}}\fr{\sqrt{1-\sqrt{1-z}}}{\pa{z(1-z)}^{\fr{1}{8}}}. \\ 
\end{aligned}
\end{equation}
As we saw in Section \ref{sec:Virasoro}, generally, the Virasoro block cannot be written in a closed form.
However, when the Verma module contains null descendant states,
the process of removing invariant subspaces from the Verma module reduces the number of descendant states in the irreducible Verma module.
As a result, the expression of the Virasoro block, which is a function that organizes the contributions from descendant states, can sometimes be simplified.
In this case, the reason the Virasoro block of the Ising model could be written in a closed form is precisely because of this.
Moreover, the fact that the solution space is spanned by only two Virasoro blocks is due to the unique property of minimal models (i.e. finite spectrum).
In general, correlation functions are described by infinite sums or integrals of Virasoro blocks.

In terms of these Virasoro blocks, the four-point function can be expressed as follows,
\begin{equation}
 \braket{ \sigma (\infty) \sigma (1) \sigma(z,\bz) \sigma(0)} =  \abs{\ca{F}^{\sigma \sigma}_{\sigma \sigma}(\bb{I}|z)}^2+
C_{\sigma \sigma \epsilon}^2 \abs{\ca{F}^{\sigma \sigma}_{\sigma \sigma}(\epsilon|z)}^2.
\end{equation}
Note that the coefficient for the vacuum block is given by $1$ because of the normalization (\ref{eq:normalization}).
The conformal bootstrap equation for this correlation function is
\begin{equation}\label{eq:IsingBootstrap}
\abs{\ca{F}^{\sigma \sigma}_{\sigma \sigma}(\bb{I}|z)}^2+ C_{\sigma \sigma \epsilon}^2 \abs{\ca{F}^{\sigma \sigma}_{\sigma \sigma}(\epsilon|z)}^2
=
\abs{\ca{F}^{\sigma \sigma}_{\sigma \sigma}(\bb{I}|1-z)}^2+ C_{\sigma \sigma \epsilon}^2 \abs{\ca{F}^{\sigma \sigma}_{\sigma \sigma}(\epsilon|1-z)}^2.
\end{equation}
Since $\ca{F}^{\sigma \sigma}_{\sigma \sigma}(\bb{I}|1-z)$ and $\ca{F}^{\sigma \sigma}_{\sigma \sigma}(\epsilon|1-z)$ are also the solutions to the same BPZ equation (\ref{eq:BPZIsing}),
they should be expandable in terms of $\ca{F}^{\sigma \sigma}_{\sigma \sigma}(\bb{I}|z)$ and $\ca{F}^{\sigma \sigma}_{\sigma \sigma}(\epsilon|z)$.
There is a transformation corresponding to the change of the basis of the Virasoro blocks as follows,
\footnote{
It can be written in the form of a sum only when the Verma module $\ca{V}_p$ is degenerate.
In general, it becomes an integral transformation.
This will be explained later.
}
\begin{tcolorbox}[title=Fusion transformation]
    \begin{equation}
\ca{F}^{21}_{34}(h_p|z) = \sum_q 
 {\bold F}_{h_p, h_q} 
   \left[
    \begin{array}{cc}
    h_2   & h_1  \\
     h_3  &   h_4\\
    \end{array}
  \right]
  \ca{F}^{23}_{14}(h_q|1-z).
\end{equation}
\end{tcolorbox}
\noindent
Such a transformation is called a \und{fusion transformation}, and the coefficients $ {\bold F}_{h_p, h_q} $ are called \und{fusion matrices}.
The fusion matrices for the Virasoro blocks appearing in the Ising model (\ref{eq:IsingVir}) can be obtained using the following identities:
\begin{equation}
\begin{aligned}
\sqrt{1+\sqrt{1-z}}&=\fr{1}{\sqrt{2}}\pa{\sqrt{1+\sqrt{z}}+\sqrt{1-\sqrt{z}}},\\
\sqrt{1-\sqrt{1-z}}&=\fr{1}{\sqrt{2}}\pa{\sqrt{1+\sqrt{z}}-\sqrt{1-\sqrt{z}}}.
\end{aligned}
\end{equation}
From these identities, the fusion transformations are given by
\begin{equation}
\begin{aligned}
\ca{F}^{\sigma \sigma}_{\sigma \sigma}(\bb{I}|z)=\fr{1}{\sqrt{2}}\ca{F}^{\sigma \sigma}_{\sigma \sigma}(\bb{I}|1-z)+\fr{1}{2\sqrt{2}}\ca{F}^{\sigma \sigma}_{\sigma \sigma}(\epsilon|1-z), \\
\ca{F}^{\sigma \sigma}_{\sigma \sigma}(\epsilon|z)=\fr{2}{\sqrt{2}}\ca{F}^{\sigma \sigma}_{\sigma \sigma}(\bb{I}|1-z)-\fr{1}{\sqrt{2}}\ca{F}^{\sigma \sigma}_{\sigma \sigma}(\epsilon|1-z).
\end{aligned}
\end{equation}
Substituting these into the conformal bootstrap equation (\ref{eq:IsingBootstrap}) and comparing the coefficients, we obtain
\begin{equation}
C_{\sigma \sigma \epsilon} = \fr{1}{2}.
\end{equation}
Similarly, all the CFT data of the Ising model can be determined by the conformal bootstrap.
For reference, the CFT data of the Ising model are listed in Appendix \ref{app:Ising}.

\subsubsection{Modular Bootstrap in Minimal Models}

The partition function of a minimal model can be written as follows:
\begin{equation}
Z(\tau, \bar{\tau}) = \sum_{(h,\bar{h}) \in \ca{S}}  M_{h,\bar{h}} \chi_h(\tau) \bar{\chi}_{\bar{h}} (\bar{\tau}).
\end{equation}
An important point is that $\ca{S}$ is a finite discrete set.
Let us see what can be obtained from this assumption and the modular bootstrap alone.
To consider inequalities, let us set $\tau = -\bar{\tau} \in i\bb{R}$.
Recalling that an irreducible Verma module is obtained by removing the invariant subspaces from the original Verma module, we obtain the following inequality:
\begin{equation}
\chi_h(\tau)=q^{h-\fr{c}{24}}\sum_{n\geq 0} d(n) q^n \leq q^{h-\fr{c}{24}}\sum_{n\geq 0} p(n) q^n.
\end{equation}
Here, we set $q\equiv\ex{2\pi i \tau} $.
The right-hand side of this inequality is a formal power series (i.e. generating function) with coefficients representing the number of level $n$ descendants that can be constructed from $L_{-1}, L_{-2}, L_{-3}, \cdots$.
As a warm-up, let us consider the generating function for the number of level $n$ descendants that can be constructed from only $L_{-k}$. Since there is only one way to construct a level $n$ state using $L_{-k}^m$ when $n=mk \ \ (n \in \bb{Z}_{\geq0})$, we have
\begin{equation}
1+q^k+q^{2k}+q^{3k}+\cdots = \frac{1}{1-q^k}.
\end{equation}
Similarly, the formal power series with coefficients representing  the number of level $n$ descendants that can be constructed from $L_{-1}, L_{-2}, L_{-3}, \cdots$ is
\begin{equation}
\prod_{k=1}^{\infty} \frac{1}{1-q^k} = q^{-\frac{1}{24}}\eta(\tau).
\end{equation}
Here, $\eta(\tau)$ is the Dedekind eta function defined in (\ref{eq:Dedekind}). Since we are using the notation $q=e^{2\pi i \tau}$, the expression differs slightly from (\ref{eq:Dedekind}). Thus, we obtain
\begin{equation}
\sum_{n \geq 0} p(n) q^n = q^{-\frac{1}{24}}\eta(\tau).
\end{equation}
Using these results and the modular bootstrap equation, we obtain the following inequality:
\begin{equation}
\sum_{(h,\bar{h}) \in \ca{S}}  M_{h,\bar{h}}\chi_h \pa{-\fr{1}{\tau}} \bar{\chi}_{\bar{h}} \pa{\fr{1}{\tau}}
\leq
\fr{1}{\eta(\tau)^2}
\sum_{(h,\bar{h}) \in \ca{S}} M_{h,\bar{h}} q^{h+\bar{h}-\fr{c-1}{12}}.
\end{equation}
Taking the limit $\tau \to i 0^+$ of this inequality,
\begin{equation}
\beta^{-1} \ex{-2 \fr{(2\pi)^2}{\beta} \pa{h_{\text{min}}-\fr{c-1}{24}}} \leq   \sum_{(h,\bar{h}) \in \ca{S}}  M_{h,\bar{h}}.
\end{equation}
Here, for convenience, we introduced the temperature $\beta\equiv 2\pi i \tau \in \bb{R}$.
$h_{\text{min}}$ is the minimum conformal dimension appearing in the spectrum.
We also used the property of the Dedekind eta function, $\eta\pa{-\fr{1}{\tau}} = \sqrt{-i \tau} \eta(\tau)$.
Since the assumption that the number of irreducible representations appearing in the spectrum is finite implies that the right-hand side is finite, we obtain
\begin{equation}\label{eq:hmin}
h_{\text{min}} > \fr{c-1}{24}.
\end{equation}
In particular, assuming that the theory has a vacuum state, we have $h_{\text{min}}\leq 0$, thus $c<1$.

Next, we show that (\ref{eq:hmin}) is satisfied only when $c=1-6\fr{(p-q)^2}{pq}$.
Using the Virasoro algebra, we can obtain the following lemma about the Virasoro character:
\begin{equation}
\chi_h(\tau) \geq  \fr{q^{h-\fr{c-1}{24}}}{\eta(\tau)}(1-q), \ \ \ \ \ \text{if } c\neq 1-6\fr{(p-q)^2  }{pq}.
\end{equation}
From this inequality and the modular bootstrap equation in the limit $\tau \to i 0$, we obtain
\begin{equation}
\beta^{-3}\ex{-2 \fr{(2\pi)^2}{\beta} \pa{h_{\text{min}}-\fr{c-1}{24}}} \geq   \sum_{(h,\bar{h}) \in \ca{S}}  M_{h,\bar{h}}.
\end{equation}
Consequently, we obtain
\begin{equation}
h_{\text{min}} < \fr{c-1}{24},
\end{equation}
which contradicts (\ref{eq:hmin}).
Thus, from the modular bootstrap, we obtain the following conclusion:
\begin{tcolorbox}
    When the spectrum is composed of a finite number of irreducible representations of the Virasoro algebra, $c<1$ is always satisfied, and in particular, the central charge takes the following discrete values:
\begin{equation}
c=1-6\fr{(p-q)^2}{pq}, \ \ \ \ \ p>q \in \{2,3,\cdots\}.
\end{equation}
\end{tcolorbox}

\subsubsection{Comments on Minimal Models}
Finally, here are some comments on minimal models.
\begin{itemize}

\item
Minimal models are unitary only when $\abs{p-q}=1$.
This can be confirmed as follows.
Substituting the central charge expression for minimal models into (\ref{eq:kaczero2}), we obtain
\begin{equation}
h_{rs} = \fr{(pr-qs)^2 -(p-q)^2}{4pq}.
\end{equation}
By Bézout's identity, for coprime integers $p$ and $q$, there exist integers $x$ and $y$ that satisfy the following,
\begin{equation}
xp+yq=1.
\end{equation}
Therefore, for the theory to be unitary, the following inequality must be satisfied,
\begin{equation}
h_{\mathrm{min}}=h_{xy} = \fr{1-(p-q)^2}{4pq} \geq 0.
\end{equation}
This is satisfied only when $\abs{p-q}=1$.

\item
Specifying $p, q$ is enough to specify the chiral algebra, but it is not enough to specify the full CFT. From modular invariance, one can classify all possible minimal model CFTs. This classification is known as the \und{ADE classification}:
\begin{itemize}

\item $A$-series: the diagonal modular invariant, which exists for any coprime integers $p, q \geq 2$.

\item $D$-series: the non-diagonal modular invariant, which exists for any even $q \geq 6$ and odd $p$ (or vice versa), with $p, q$ coprime. 

\item $E$-series: the exceptional (non-diagonal) modular invariant, which exists for $p$ or $q=12, 18, 30$
\end{itemize}

Note that if we assume the theory to be unitary, the $A$-series always exists; the $D$-series exists for all values of $c<1$ except for $c=\frac12, \frac{7}{10}$; and the E-series only exists for $(p,p+1) = (11, 12), (12, 13), (17, 18), (18,19), (29, 30), (30, 31)$. For unitary minimal models, the first $D$-series minimal model is the three-states Potts model, which occurs at $c=\frac 45$.

\item 
The $A$-series minimal models are believed to be the IR fixed points under RG flow of the relevant deformation $\phi^{2m}$ of the $c=1$ free theory \cite{Zamolodchikov1986b}. By now there is strong evidence for this conjecture. For example for a supersymmetric generalization of this flow, there is strong evidence in \cite{Witten1993}. There is even a very recent generalizations to RG flow of nonunitary theories in \cite{Klebanov2022, Tanaka2024, Katsevich2024}.

\item
The fewer the number of irreducible representations included in the spectrum,
the easier it is to compute physical quantities.
Therefore, when investigating the behavior of a certain physical quantity,
one often starts by examining minimal models (especially the Ising model).

\item 
On the other extreme, the $p\rightarrow\infty$ limit for unitary minimal models is subtle and interesting. The central charge approaches $1$ from below. The spectrum approaches that of Liouville CFT, which we will discuss in the next spectrum. However, the three point functions behave differently, and were studied by Runkel and Watts \cite{Runkel2001}. Moreover, even though at any finite $p$, the theory is completely rational, the spectrum exhibits some features of chaotic CFTs in what is known as the spectral form factor \cite{Benjamin2018}. One piece of intuition for this is that the number of primary operators becomes very large, so the rationality of the conformal dimensions because harder to see at large $p$.

\end{itemize}

\subsection{Liouville CFT}

In Liouville CFT, it is convenient to use the following notation,
\begin{equation}\label{eq:Qb}
c=1+6Q^2, \ \ \ \ \ Q=b+\fr{1}{b}, \ \ \ \ \ h_i=\a_i(Q-\a_i).
\end{equation}
Here, $Q$ is called the {\it background charge}, $b$ is the {\it Liouville coupling}, and $\alpha_i$ is the {\it Liouville momentum}.
\begin{description}

\item[Central Charge]\mbox{}\\
Liouville CFT can be defined for any complex number $c \in \bb{C}$.
In particular, when the central charge is a real number $c>1$, Liouville CFT is a unitary CFT.

\item[OPE Coefficients]\mbox{}\\
The OPE coefficients of Liouville CFT are known as the DOZZ formula \cite{Dorn1994, Zamolodchikov1996},
\begin{equation}\label{eq:DOZZ}
C_{\alpha_1, \alpha_2, \alpha_3}
=
\frac{
\left(b^{\frac{2}{b}-2b} \lambda\right)^{Q-\alpha_1 -\alpha_2- \alpha_3} \Upsilon_b'(0)\Upsilon_b(2\alpha_1)\Upsilon_b(2\alpha_2)\Upsilon_b(2\alpha_3)
}{
\Upsilon_b(\alpha_1+\alpha_2+\alpha_3-Q)\Upsilon_b(\alpha_1+\alpha_2-\alpha_3)\Upsilon_b(\alpha_2+\alpha_3-\alpha_1)
\Upsilon_b(\alpha_3+\alpha_1-\alpha_2)
}.
\end{equation}
Here, $\lambda$ is an arbitrary normalization constant, and $\Upsilon_b(z)$ is defined using the Barnes double Gamma function $\Gamma_b(x)$ 
(see (\ref{eq:DefG}) in Appendix \ref{subapp:Gamma}) as follows:
\footnote{
Using (\ref{eq:DerG}), it can be shown that
\begin{equation}
\Upsilon_b'(0) = \fr{2\pi}{\Gamma_b(Q)^2}.
\end{equation}
Some references rewrite the DOZZ formula using this.
}
\begin{equation}
\Upsilon_b(x) \equiv \frac{1}{\Gamma_b(x) \Gamma_b(Q-x)}.
\end{equation}
By definition, $\Upsilon_b(x)$ has the following reflection symmetry:
\begin{equation}
\Upsilon_b(x) = \Upsilon_b(Q-x).
\end{equation}
An important feature of the function $\Upsilon_b(x)$ is the following:
\begin{equation}
\Upsilon_b(x) = 0 \ \ \ \ \ \ \ \text{for }  x \in \left(-b\mathbb{N}-\frac{1}{b}\mathbb{N}\right) \cup \left(Q+b\mathbb{N}+\frac{1}{b}\mathbb{N}\right).
\end{equation}
In particular, these zeros are simple zeros (of order one).  
Reflecting these zeros, the DOZZ formula also exhibits simple zeros and poles.

As stated in (\ref{eq:normalization}), the solution to the conformal bootstrap has an ambiguity up to a multiplicative constant.  
In Liouville CFT, the following normalization is adopted:
\begin{equation}
\Res{\alpha_1+\alpha_2+\alpha_3=Q} C_{123} = 1.
\end{equation}
Adopting the normalization differs from (\ref{eq:normalization}) because Liouville CFT has a continuous spectrum, as explained below.

\item[Normalization]\mbox{}\\

The two-point function can be defined from the $\alpha_2 \to 0$ limit of the DOZZ formula.  
Expanding the DOZZ formula around $\alpha_2 = \epsilon \ll 1$ gives
\begin{equation}\label{eq:twoDOZZ}
C_{\alpha_1, \epsilon, \alpha_3}
=
\frac{2 \epsilon B_{\alpha_1}}
{(\alpha_1-\alpha_3+\epsilon)(\alpha_3-\alpha_1+\epsilon)}
+
\frac{2\epsilon}{(Q-\alpha_1-\alpha_3+\epsilon)(\alpha_1+\alpha_3-Q+\epsilon)}. 
\end{equation}
Here, $B_{\alpha}$ is defined as:
\begin{equation}\label{eq:B}
B_{\alpha} \equiv \left(b^{\frac{2}{b}-2b}  \lambda\right)^{Q-2\alpha} \frac{\Upsilon_b(Q-2\alpha)}{\Upsilon_b(2\alpha-Q)}.
\end{equation}
From this expression, $C_{\alpha_1, \epsilon, \alpha_3}$ diverges in the limit $\epsilon \to 0$ when $\alpha_3 = \alpha_1$ or $\alpha_3 = Q-\alpha_1$,  
and vanishes otherwise.  
Therefore, the two-point function is defined only in the distributional sense.

We consider defining the norm on the Hilbert space of Liouville CFT.  
First, from the DOZZ formula, which satisfies $\left(C_{\alpha_1, \alpha_2, \alpha_3}\right)^* = C_{\alpha_1^*, \alpha_2^*, \alpha_3^*}$,  
the Hermitian conjugate of the fields in Liouville CFT is given by:
\begin{equation}
[V_{\alpha_i}(z,\bar{z})]^\dagger = \bar{z}^{-2h_i} z^{-2h_i} V_{\alpha_i^*}(1/\bar{z},1/z).
\end{equation}
Thus, the inner product should be defined by
\begin{equation}
\braket{\alpha_i|\alpha_j} = \lim_{\epsilon\to 0} C_{\alpha_i^*,\epsilon, \alpha_j}.
\end{equation}
From the positivity of the norm, it follows that $\alpha \in \frac{Q}{2} + i \mathbb{R}_+$ or $\alpha \in [0,\frac{Q}{2})$.  
The upper limit $\frac{Q}{2}$ in the second case is imposed using the following reflection formula for the DOZZ formula:
\begin{equation}
C_{\alpha_1, \alpha_2, \alpha_3} = B_{\alpha_1} C_{Q-\alpha_1,\alpha_2,\alpha_3}.
\end{equation}
The restriction $\alpha \in \frac{Q}{2} + i \mathbb{R}_+$ or $\alpha \in [0,\frac{Q}{2})$ on the Liouville momentum can also be understood from the reality of the conformal dimension $h_i = \alpha_i(Q-\alpha_i)$.
Furthermore, when defining the scalar product using (\ref{eq:twoDOZZ}) in the distributional sense,  
it is necessary to take $\Im \epsilon \neq 0$ so that the regularization parameter $\epsilon$ does not lie on the integration contour $\alpha \in [0,\frac{Q}{2})$.  
This breaks the positivity of the norm, resulting in the inability to define well-defined states for $\alpha \in [0,\frac{Q}{2})$.  
Finally, the spectrum of Liouville CFT is given as follows:

\item[Spectrum]\mbox{}\\
\begin{equation}
\ca{S} = \int_{\fr{Q}{2}+i0}^{\fr{Q}{2}+i \infty} d\alpha \ca{V}_\alpha \otimes \bar{\ca{V}}_\alpha.
\end{equation}
In terms of conformal dimension, $\alpha \in \fr{Q}{2}+i\bb{R}_{+}$ corresponds to
\begin{equation}
h \in \fr{c-1}{24}+\bb{R}_{+}.
\end{equation}

This spectrum has the following important features:
\begin{itemize}

\item
continuous spectrum

\item
irrational CFT

\item
diagonal CFT
\end{itemize}

\item[Lagrangian Formulation]\mbox{}\\
Liouville CFT can be described by the following local action,
\begin{equation}
S[\phi] = \fr{1}{4\pi} \int d^2 x \sqrt{g}
\pa{
g^{\mu \nu}\del_\mu \phi \del_\nu \phi + QR\phi + \lambda' \ex{2b\phi}
}.
\end{equation}
Here, $R$ is the Ricci scalar.
$Q$ and $b$ are the same as those that appeared in the notation of Liouville CFT (\ref{eq:Qb}).
$\lambda'$ is related to $\lambda$, which appears in the OPE coefficient, as follows.
\begin{equation}\label{eq:lprime}
\lambda'=4\fr{\Gamma(1-b^2)}{\Gamma(b^2)}\lambda^b.
\end{equation}
The coupling constant is sometimes expressed as $\lambda' = 4\pi \mu$ (e.g. \cite{Teschner2001}).
A primary field with conformal dimension $h=\alpha(Q-\alpha)$ can be expressed using the Liouville field,
\begin{equation}
V_\alpha \equiv \ex{2\alpha\phi}.
\end{equation}
Correlation functions are symmetric under the following transformations:
\begin{itemize}

\item
$b \leftrightarrow \fr{1}{b}$,

\item
$\alpha \leftrightarrow Q-\alpha$.

\end{itemize}
While this is obvious from the CFT perspective,
it is non-trivial in the Lagrangian formulation.

\end{description}

\subsubsection{Conformal Bootstrap in Liouville CFT}\label{subsubsec:Bootstrap}

An important assumption in solving the conformal bootstrap in Liouville CFT is that it is a diagonal CFT.
In general, the spectrum is $(h,\bar{h}) \in \bb{C}^2$, but in the case of a diagonal CFT, there is only one degree of freedom $h \in \bb{C}$,
which significantly reduces the number of equations needed to determine the CFT data.
In fact, the CFT data of a diagonal CFT can be uniquely determined from the conformal bootstrap of the four-point function \cite{Ribault2014}.
Let us see this in detail below.

When solving the conformal bootstrap, it is sometimes convenient to use the fusion transformation.
To see this, let us rewrite the conformal bootstrap equation using the fusion transformation.
When the Verma module is originally irreducible, the fusion transformation is expressed as an integral transformation \cite{Ponsot2001, Ponsot1999, Teschner2001},
\begin{tcolorbox}[title=Fusion transformation]
    \begin{equation}\label{eq:Fusion}
\begin{aligned}
\ca{F}^{21}_{34}(h_{\a_s}|z)=\int_{\fr{Q}{2}+i0}^{\fr{Q}{2}+i\infty} d\a_t {\bold F}_{\a_s, \a_t} 
   \left[
    \begin{array}{cc}
    \a_2   & \a_1  \\
     \a_3  &   \a_4\\
    \end{array}
  \right]
  \ca{F}^{23}_{14}(h_{\a_t}|1-z).
\end{aligned}
\end{equation}
\end{tcolorbox}
\noindent
The details of the fusion matrix ${\bold F}_{\a_s, \a_t}$ are described in Appendix \ref{app:FM}.
Using this fusion transformation, the conformal bootstrap can be rewritten as follows,
\begin{equation}
\begin{aligned}
&\int_{\fr{Q}{2}+i0}^{\fr{Q}{2}+i\infty} d\a_p \fr{C_{12p}C_{34p}}{B_p}
{\bold F}_{\a_p, \a_q} 
   \left[
    \begin{array}{cc}
    \a_2   & \a_1  \\
     \a_3  &   \a_4\\
    \end{array}
      \right]
\ca{F}^{21}_{34}(p|z)
 \bar{\ca{F}}^{23}_{14}(p|1-\bz ) \\
&= 
\int_{\fr{Q}{2}+i0}^{\fr{Q}{2}+i\infty} d\alpha_q \fr{C_{14q}C_{23q}}{B_q}
{\bold F}_{\a_q, \a_p} 
   \left[
    \begin{array}{cc}
    \a_2   & \a_3  \\
     \a_1  &   \a_4\\
    \end{array}
      \right]
\ca{F}^{21}_{34}(q|z)
 \bar{\ca{F}}^{23}_{14}(q|1-\bz ).
\end{aligned}
\end{equation}
Now, since both sides are expanded in the same functional basis, by comparing the coefficients, we obtain the following relation,
\begin{equation}\label{eq:Fbootstrap}
\fr{C_{12p}C_{34p}}{B_p}
{\bold F}_{\a_p, \a_q} 
   \left[
    \begin{array}{cc}
    \a_2   & \a_1  \\
     \a_3  &   \a_4\\
    \end{array}
      \right]
=
\fr{C_{14q}C_{23q}}{B_q}
{\bold F}_{\a_q, \a_p} 
   \left[
    \begin{array}{cc}
    \a_2   & \a_3  \\
     \a_1  &   \a_4\\
    \end{array}
      \right].
\end{equation}
This is equivalent to the conformal bootstrap equation, but since the fusion matrix is somewhat simpler than the Virasoro block, this expression can sometimes be easier to handle.

\begin{figure}[t]
 \begin{center}
  \includegraphics[width=10.0cm,clip]{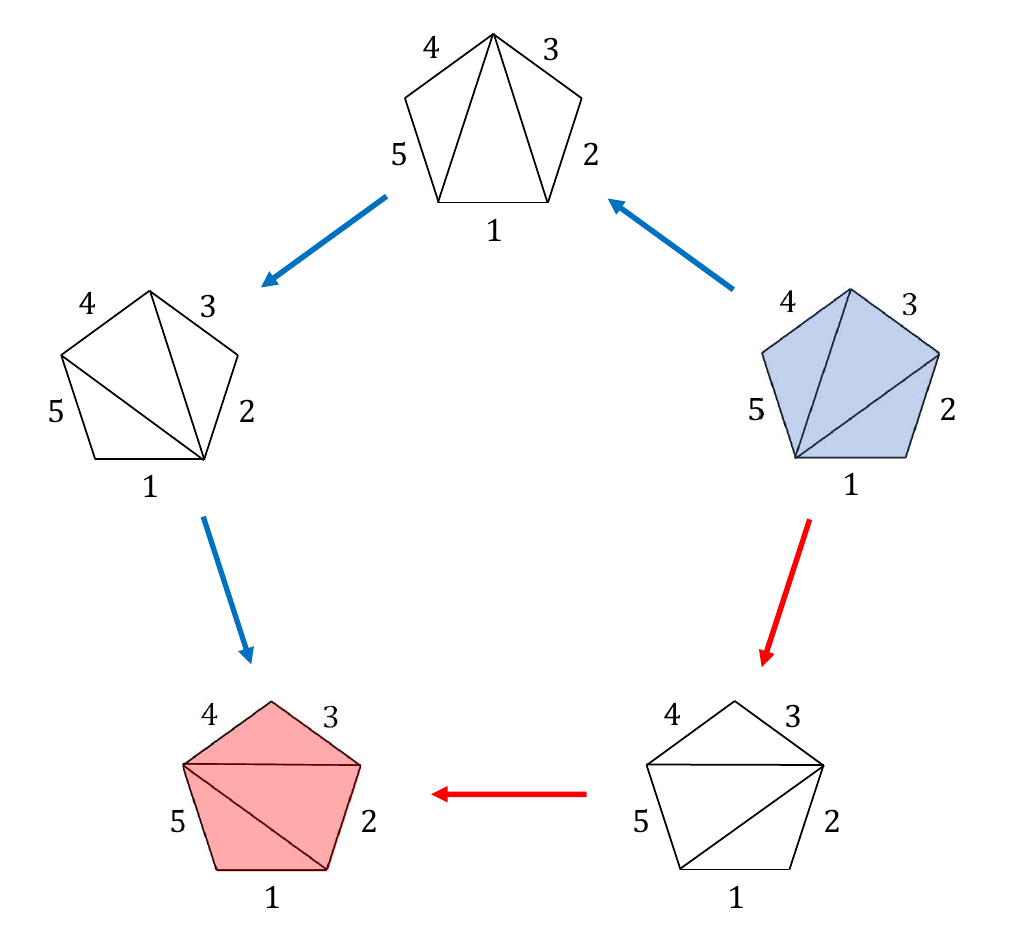}
 \end{center}
 \caption{Illustration of the derivation of the Pentagon identity using the dual graph representation of the Virasoro block. 
 There are two distinct ways to transform the block from the blue graph to the red graph using the fusion transformation, as shown in the figure.
 The Pentagon identity is obtained from the requirement that these two multiple fusion transformations must be equivalent.
 }
 \label{fig:pentagon}
\end{figure}
\noindent
Repeating the fusion transformation multiple times can return to the original Virasoro block, and the consistency of this fusion transformation leads to the following {\it pentagon identity} (See Figure \ref{fig:pentagon}),
\begin{equation}
\int_{\fr{Q}{2}+i0}^{\fr{Q}{2}+i\infty} d\alpha_s
{\bold F}_{\a_q, \a_s} 
   \left[
    \begin{array}{cc}
    \a_k   & \a_b  \\
     \a_j  &   \a_p \\
    \end{array}
   \right]
{\bold F}_{\a_p, \a_l} 
   \left[
    \begin{array}{cc}
    \a_s   & \a_b  \\
     \a_i  &   \a_a \\
    \end{array}
   \right]
{\bold F}_{\a_s, \a_r} 
   \left[
    \begin{array}{cc}
    \a_j   & \a_k  \\
     \a_i  &   \a_l \\
    \end{array}
   \right]
=
{\bold F}_{\a_p, \a_r} 
   \left[
    \begin{array}{cc}
    \a_j   & \a_q  \\
     \a_i  &   \a_a \\
    \end{array}
   \right]
{\bold F}_{\a_q, \a_l} 
   \left[
    \begin{array}{cc}
    \a_k   & \a_b  \\
     \a_r  &   \a_a \\
    \end{array}
   \right].
\end{equation}
To relate this to the fusion matrix appearing on the left-hand side of (\ref{eq:Fbootstrap}), we perform the following relabeling,
\begin{equation}
(a,b,q,r,i,j,k,l) \to (3,p,2,q,3,2,1,4).
\end{equation}
Furthermore, setting $p=\bb{I}$, we can show
\begin{equation}
{\bold F}_{\a_2, \a_s} 
   \left[
    \begin{array}{cc}
    \a_1   & \a_p  \\
     \a_2  &   0 \\
    \end{array}
   \right]
   =\delta_{\alpha_s, \alpha_p}.
\end{equation}
Substituting this into the pentagon identity allows us to remove the integral over $\alpha_s$,
\begin{equation}\label{eq:F1234}
{\bold F}_{0, \a_4} 
   \left[
    \begin{array}{cc}
    \a_p   & \a_p  \\
     \a_3  &   \a_3 \\
    \end{array}
   \right]
{\bold F}_{\a_p, \a_q} 
   \left[
    \begin{array}{cc}
    \a_2   & \a_1  \\
     \a_3  &   \a_4 \\
    \end{array}
   \right]
=
{\bold F}_{0, \a_q} 
   \left[
    \begin{array}{cc}
    \a_2   & \a_2  \\
     \a_3  &   \a_3 \\
    \end{array}
   \right]
{\bold F}_{\a_2, \a_4} 
   \left[
    \begin{array}{cc}
    \a_1   & \a_p  \\
     \a_q &   \a_3 \\
    \end{array}
   \right].
\end{equation}
Furthermore, to relate this to the fusion matrix appearing on the right-hand side of (\ref{eq:Fbootstrap}),
we consider the relabeling $p \leftrightarrow q$ and $1 \leftrightarrow 3$ in (\ref{eq:F1234}),
\begin{equation}\label{eq:F1432}
{\bold F}_{0, \a_4} 
   \left[
    \begin{array}{cc}
    \a_q   & \a_q  \\
     \a_1  &   \a_1 \\
    \end{array}
   \right]
{\bold F}_{\a_q, \a_p} 
   \left[
    \begin{array}{cc}
    \a_4   & \a_1  \\
     \a_3  &   \a_2 \\
    \end{array}
   \right]
=
{\bold F}_{0, \a_p} 
   \left[
    \begin{array}{cc}
    \a_2   & \a_2  \\
     \a_1  &   \a_1 \\
    \end{array}
   \right]
{\bold F}_{\a_2, \a_4} 
   \left[
    \begin{array}{cc}
    \a_1   & \a_p  \\
     \a_q &   \a_3 \\
    \end{array}
   \right].
\end{equation}
Here, we used the fact that the fusion matrix is symmetric under the following permutations,
\begin{equation}
{\bold F}_{\a_p, \a_q} 
   \left[
    \begin{array}{cc}
    \a_2   & \a_1  \\
     \a_3  &   \a_4 \\
    \end{array}
   \right]
=
{\bold F}_{\a_p, \a_q} 
   \left[
    \begin{array}{cc}
    \a_1   & \a_2  \\
     \a_4  &   \a_3 \\
    \end{array}
   \right]
=
{\bold F}_{\a_p, \a_q} 
   \left[
    \begin{array}{cc}
    \a_3   & \a_4  \\
     \a_2  &   \a_1 \\
    \end{array}
   \right].
\end{equation}
Finally, since (\ref{eq:F1234}) and (\ref{eq:F1432}) share the common factor
$
{\bold F}_{\a_2, \a_4} 
   \left[
    \begin{array}{cc}
    \a_1   & \a_p  \\
     \a_q &   \a_3 \\
    \end{array}
   \right],
$
we obtain
\begin{equation}
\begin{aligned}
&{\bold F}_{0, \a_4} 
   \left[
    \begin{array}{cc}
    \a_p   & \a_p  \\
     \a_3  &   \a_3 \\
    \end{array}
   \right]
 {\bold F}_{0, \a_p} 
   \left[
    \begin{array}{cc}
    \a_2   & \a_2  \\
     \a_1  &   \a_1 \\
    \end{array}
   \right]
{\bold F}_{\a_p, \a_q} 
   \left[
    \begin{array}{cc}
    \a_2   & \a_1  \\
     \a_3  &   \a_4 \\
    \end{array}
   \right] \\
&=
{\bold F}_{0, \a_4} 
   \left[
    \begin{array}{cc}
    \a_q   & \a_q  \\
     \a_1  &   \a_1 \\
    \end{array}
   \right]
{\bold F}_{0, \a_q} 
   \left[
    \begin{array}{cc}
    \a_2   & \a_2  \\
     \a_3  &   \a_3 \\
    \end{array}
   \right]
{\bold F}_{\a_q, \a_p} 
   \left[
    \begin{array}{cc}
    \a_4   & \a_1  \\
     \a_3  &   \a_2 \\
    \end{array}
   \right].
\end{aligned}
\end{equation}
We find that this is the solution to (\ref{eq:Fbootstrap}).
The fusion matrix is generally complicated, but in this case, we can use the following result,
\begin{equation}
\begin{aligned}
{\bold F}_{0, \a_t} 
   \left[
    \begin{array}{cc}
    \a_1   & \a_1  \\
     \a_4  &   \a_4 \\
    \end{array}
   \right]
&=
\lim_{\a_s=0}
\lim_{\a_2=\a_1}
\lim_{\a_3=\a_4}
{\bold F}_{\a_s, \a_t} 
   \left[
    \begin{array}{cc}
    \a_2   & \a_1  \\
     \a_3  &   \a_4 \\
    \end{array}
   \right] \\
&\propto
\fr{1}{
\Upsilon_b(\alpha_1+\alpha_4+\alpha_t-Q)\Upsilon_b(\alpha_1+\alpha_4-\alpha_t)\Upsilon_b(\alpha_4+\alpha_t-\alpha_1)
\Upsilon_b(\alpha_t+\alpha_1-\alpha_4).
}
\end{aligned}
\end{equation}
Therefore, the solution to the conformal bootstrap equation for a diagonal CFT without degenerate fields is
\begin{equation}
C_{123}
\propto
\fr{1}{
\Upsilon_b(\alpha_1+\alpha_2+\alpha_3-Q)\Upsilon_b(\alpha_1+\alpha_2-\alpha_3)\Upsilon_b(\alpha_2+\alpha_3-\alpha_1)
\Upsilon_b(\alpha_3+\alpha_1-\alpha_2).
}.
\end{equation}
The conformal bootstrap equation is invariant under field redefinitions, so the ambiguity due to normalization constants cannot be removed.
Moreover, since the spectrum of Liouville CFT does not contain the vacuum state, the ambiguity cannot be removed by the standard normalization (\ref{eq:normalization}).
Instead, we adopt the following normalization,
\begin{equation}
\Res{\alpha_1+\alpha_2+\alpha_3=Q} C_{123}=1.
\end{equation}
Under this normalization, the OPE coefficient can be written as follows,
\begin{tcolorbox}[title=DOZZ formula]
   \begin{equation}
C_{123}
=
\fr{
\pa{b^{\fr{2}{b}-2b} \lambda}^{Q-\alpha_1 -\alpha_2- \alpha_3} \Upsilon_b'(0)\Upsilon_b(2\alpha_1)\Upsilon_b(2\alpha_2)\Upsilon_b(2\alpha_3)
}{
\Upsilon_b(\alpha_1+\alpha_2+\alpha_3-Q)\Upsilon_b(\alpha_1+\alpha_2-\alpha_3)\Upsilon_b(\alpha_2+\alpha_3-\alpha_1)
\Upsilon_b(\alpha_3+\alpha_1-\alpha_2)
}.
\end{equation} 
\end{tcolorbox}
\noindent
This is the DOZZ formula (\ref{eq:DOZZ}).
The overall factor $\pa{b^{\fr{2}{b}-2b} \lambda}^{Q-\alpha_1 -\alpha_2- \alpha_3}$ comes from the convention in the Lagrangian formulation.

\subsubsection{Comments on Liouville CFT}
\begin{itemize}

\item
Liouville CFT is often studied as an example of a solvable ICFT.
However, it has unique properties that differ from typical physical systems, such as having a continuous spectrum and the Hilbert space not containing the vacuum state.

\item
Since the central charge can be arbitrary and the conformal dimensions are continuous,
Liouville CFT correlation functions include conformal blocks over a very wide range of parameters.
Therefore, the correlation functions of Liouville CFT are sometimes used to study the properties of Virasoro blocks.

\item
The Hilbert space of Liouville CFT does not contain the vacuum state in the standard sense.  
However, appropriately restricting the dual space makes it possible to "define" a vacuum state.  
Once the vacuum state is defined in this way, the state-operator correspondence can also be established.  
Some references describe this as "the state-operator correspondence is broken in Liouville CFT", in the sense that the relationship between eigenstates in the Hilbert space and local operators differs slightly from the standard interpretation.  
For more details on the state-operator correspondence in Liouville CFT, see \cite{Teschner2001}.

\end{itemize}

\subsection{Free Boson CFT}

A free boson CFT is a CFT without interactions. It is often defined using a Lagrangian, but it can also be defined without a Lagrangian as a CFT with $\hat{\mathfrak{u}}_1$ symmetry.

\begin{description}

\item[Central Charge]\mbox{}\\
A free boson CFT can be defined with any complex central charge $c$. However, the most commonly considered case is the $c=1$ free boson CFT. When $c=1$, it has the feature that it can be compactified on a circle of any radius $S^1$.

\item[Lagrangian Formulation]\mbox{}\\
A free boson CFT can be described by the following local action:
\begin{equation}\label{eq:FreeAction}
S[\phi] = \fr{1}{4\pi} \int d^2 x \sqrt{g} \pa{ g^{\mu \nu} \del_\mu \phi \del_\nu \phi + QR\phi}.
\end{equation}
Here, $R$ is the Ricci scalar, and $Q \in \bb{C}$ is the background charge. The primary fields can be expressed using the bosonic field $\phi$ as
\begin{equation}
V_\alpha=\ex{2\alpha \phi}.
\end{equation}

\item[OPE]\mbox{}\\
Due to the affine $\hat{\mathfrak{u}}_1$ symmetry, the momentum $\alpha$ is conserved. This imposes a strong constraint on the OPE:
\begin{equation}\label{eq:FreeOPE}
V_{\alpha_1, \bar{\alpha}_1}(z_1)  V_{\alpha_2, \bar{\alpha}_2}(z_2) = \abs{(z_1-z_2)^{h_3-h_1-h_2}}^2V_{\alpha_3, \bar{\alpha}_3}(z_2) + (\text{descendants}) .
\end{equation}
Here, $\alpha_3=\alpha_1+\alpha_2$. In general, various primary fields can appear in the OPE, but in the free boson CFT, the additive property of momentum ensures that only one primary field appears. This non-interacting nature is why it is called a ``free" boson CFT. Thanks to the simplification of the OPE, the correlation functions of the free boson CFT can be calculated very easily. This systematic calculation method is known as {\it Wick's theorem}.

\item[Spectrum]\mbox{}\\
Several different spectra are allowed for the free boson CFT. These will be explained in detail in Section \ref{subsubsec:spectrum}.

\end{description}

\subsubsection{Affine Lie Algebra}

A characteristic feature of the free boson CFT is the presence of the following conserved currents:
\begin{equation}
J=\del \phi, \ \ \ \ \ \ \bar{J} = \bar{\del}\phi.
\end{equation}
These currents satisfy the conservation laws:
\begin{equation}
\del \bar{J} = \bar{\del} J = 0.
\end{equation}
These conserved currents generate the abelian affine Lie algebra $\hat{\mathfrak{u}}_1$.
The OPE of the conserved currents is
\begin{equation}
J(z)J(w) = -\fr{1}{2(z-w)^2}.
\end{equation}
Expressing $J(z)$ in terms of its modes,
\begin{equation}
J(z) = \sum_{n \in \bb{Z}} \fr{J_n}{z^{n+1}},
\end{equation}
we obtain the $\hat{\mathfrak{u}}_1$ commutation relations:
\begin{equation}
[J_n, J_m] = \fr{1}{2} n \delta_{n+m,0}.
\end{equation}
From this $\hat{\mathfrak{u}}_1$ algebra, we can construct the Virasoro algebra as a subalgebra:
\begin{equation}
\begin{aligned}
L_n &= -\sum_{m \in \bb{Z}} J_{n-m}J_m + Q(n+1)J_n,  \ \ \ \ n \neq 0, \\
L_0 &= -2 \sum_{m=1}^{\infty} J_{-m} J_m - J_0^2 +QJ_0.
\end{aligned}
\end{equation}
These satisfy the Virasoro commutation relations with $c=1+6Q^2$.
The commutation relation between the Virasoro algebra and the $\hat{\mathfrak{u}}_1$ algebra is
\begin{equation}
[L_n,J_m] = -nJ_{n+m} - \fr{Q}{2}n(n+1)\delta_{n+m,0}.
\end{equation}
Let us construct the energy-momentum tensor from the Virasoro algebra,
\begin{equation}
T(z) =\sum_{n \in \bb{Z}} \fr{L_n}{z^{n+2}} = -(JJ)(z)-Q\del J(z).
\end{equation}
Here, $(AB)(z)$ denotes the normal ordered product, defined as
\begin{equation}
(AB)(z) = \fr{1}{2\pi i} \oint_z \fr{dw}{w-z}A(w)B(z).
\end{equation}
This energy-momentum tensor matches the one obtained from the action (\ref{eq:FreeAction}).
When a CFT has affine Lie algebra symmetry, it is often more convenient to consider the irreducible representations of the affine Lie algebra instead of the Virasoro algebra. The irreducible representation decomposition of the affine Lie algebra introduces the concept of {\it affine primary states},
\begin{equation}
J_{n>0}\ket{\alpha}=0, \ \ \ \ \ J_0 \ket{\alpha} =\alpha \ket{\alpha}.
\end{equation}
Since the Virasoro algebra is a subalgebra of the affine Lie algebra, affine primary fields are also (Virasoro) primary fields. Conversely, (Virasoro) primary fields are not necessarily affine primary fields. From the affine primary fields, we can define affine descendant fields similarly to the (Virasoro) descendant fields. To distinguish the Verma module $\ca{V}$ of the Virasoro algebra, we denote the Verma module composed of the affine primary field $V_\alpha$ and its descendant fields as $\ca{U}_\alpha$.

Rewriting in terms of OPE, $\hat{\mathfrak{u}}_1$ primary field are defined as follows,
\begin{equation}
J(z)V_\alpha(w) = \fr{\alpha V_{\alpha}}{z-w}.
\end{equation}
The (local) $\hat{\mathfrak{u}}_1$ Ward-Takahashi identity for the $\hat{\mathfrak{u}}_1$ primary fields is given by,
\begin{equation}
\braket{J(z) \prod_i^{N} V_{\alpha_i}(z_i)  }
=
\sum_{i=1}^N \fr{\alpha_i}{z-z_i}
\braket{\prod_i^{N} V_{\alpha_i}(z_i)  }.
\end{equation}
Moreover, from the global Ward-Takahashi identity, we obtain the charge conservation law,
\begin{equation}
\sum_i^N \alpha_i = Q.
\end{equation}
Consider the following general form of OPE,
\begin{equation}
V_{\alpha_1, \bar{\alpha}_1}(z_1)  V_{\alpha_2, \bar{\alpha}_2}(z_2) = \sum_p C_{12p} \abs{(z_1-z_2)^{h_p-h_1-h_2}}^2V_{\alpha_p, \bar{\alpha}_p}(z_2) + (\text{descendants}) .
\end{equation}
By inserting $J(z)$ on both sides of this OPE and using the Ward-Takahashi identity, we obtain the following selection rule,
\begin{equation}
\alpha_p=\alpha_1+\alpha_2.
\end{equation}
In other words, the OPE of the free boson CFT has additivity (\ref{eq:FreeOPE}).

Similar to Section \ref{sec:CFT}, various theorems can be derived from the representation theory of affine Lie algebras,
but explaining them would require a vast number of pages and would deviate from the purpose of these lecture notes,
so the details are left to textbooks, such as the Yellow book \cite{Francesco2012}.

\subsubsection{Spectrum}\label{subsubsec:spectrum}
Define the free boson CFT as a CFT with $\hat{\mathfrak{u}}_1$ symmetry.
Due to the additivity of the OPE, when there exists a primary field $\bm{\alpha} = (\alpha, \bar{\alpha}) \in \bb{C}^2$,
$n\bm{\alpha} \ \ (n \in \bb{N})$ must also exist in the spectrum.
Moreover, all these primary fields must satisfy the integrality of spin,
\begin{equation}\label{eq:SpinInt}
h-\bar{h} = (\alpha -\bar{\alpha})(Q-\alpha- \bar{\alpha}) \in \bb{Z}.
\end{equation}
This condition imposes strong constraints on the spectrum of the free boson CFT \cite{Ribault2014}.

As a result, the possible spectrum is as follows:

\begin{description}

\item[non-compactified free boson]\mbox{}\\
\begin{equation}
\ca{S} = \int_{Q\bb{R}} d \alpha \ca{U}_\alpha \otimes \bar{\ca{U}}_\alpha.
\end{equation}
The integral contour is in the same direction as $Q \in \bb{C}$.
When $Q=0$, the integral contour can be any straight line with an arbitrary angle.
However, it is unitary only when $Q=0$ and the integral contour is along the imaginary axis.
As seen from the spectrum, the non-compactified boson is an ICFT,
especially possessing a continuous spectrum.

\item[compactified free boson]\mbox{}\\
\begin{equation}\label{eq:Compact}
\ca{S}_R = \bigoplus_{(n,w) \in \bb{Z}^2}
\ca{U}_{\fr{i}{2}\pa{\fr{n}{R}+Rw}} \otimes \bar{\ca{U}}_{\fr{i}{2}\pa{\fr{n}{R}-Rw}}.
\end{equation}
When $ Q \neq 0 $, the spin generally does not satisfy the integer condition (\ref{eq:SpinInt}).
\footnote{
In fact, even when $Q \neq 0$, it is possible to satisfy the integer spin condition by choosing a special value for $R \in \bb{C}$.
This means that the compactification is possible even for $c \neq 1$.
For readers interested in the details, please refer to \cite{Ribault2014}.
}
Therefore, this spectrum is only possible when $Q=0$.
This spectrum is parameterized by $R\in\bb{C}$.
In particular, when $R \in \bb{R}$, it becomes a unitary theory.
This spectrum has a symmetry $R \leftrightarrow \fr{1}{R}$, which is called the {\it $T$-duality}.

This theory is called the {\it compactified boson}.
Starting from the Lagrangian formulation,
consider the canonical quantization of the free boson on a cylinder of period $2\pi$.
Let us identify $\phi$ with $\phi+2\pi R$ (compactification to $S^1$ with radius $R$).
In this case, the following periodic boundary condition can be imposed on the free boson field,
\begin{equation}
\phi(x+2\pi,t) = \phi(x,t) + 2\pi w R,
\ \ \ \ w\in \bb{Z}.
\end{equation}
Since this represents how many times the free boson wraps around $S^1_R$,
$w$ is called the {\it winding number}.
Imposing periodic boundary conditions quantizes the momentum, which corresponds to the label $n$.
The spectrum of this compactified boson CFT exactly matches (\ref{eq:Compact}).
For details, refer to Section 6.3.5 of the Yellow book \cite{Francesco2012}.

\end{description}

\subsubsection{Comments on Free Boson CFT}

\begin{itemize}

\item
The most significant feature of the free boson CFT is the absence of interactions.
In other words, the OPE has a very simple structure as shown in (\ref{eq:FreeOPE}).
Thanks to this simplification, various physical quantities can be easily calculated in the free boson CFT.
For these reasons, when discovering a property expected to hold in any CFT, one often starts by verifying whether it holds in the free boson CFT.

\item
There is a $\bb{Z}_2$ outer automorphism ($J \to -J$) in $\hat{\mathfrak{u}}_1$.
By gauging this $\bb{Z}_2$ symmetry, a new CFT can be constructed.
In general, CFTs obtained by gauging (also called orbifolding) are called {\it orbifold CFTs}.
For details, refer to Section 17.B of the Yellow book \cite{Francesco2012}.

\item
As a generalization of the $S^1$ compactification of the free boson CFT,
one can consider the $T^N$ compactification of $N$ copies of free boson CFTs.
This series of CFTs parameterized by the moduli of the $N$-dimensional torus $T^N$ is called the {\it Narain CFT}.
Due to the discovery that the ``averaged" CFT of Narain CFT over the moduli is equivalent to $U(1)$ Chern-Simons gravity,
it is also actively studied in the context of AdS/CFT correspondence \cite{Maloney2020,AfkhamiJeddi2021,Benjamin2022}.

\item
When $R^2 \in \bb{Q}$, it is known that the $\hat{\mathfrak{u}}_1$ symmetry is enlarged (see Section 14.4.4 of the Yellow book \cite{Francesco2012}).
Considering the representations of this extended $\hat{\mathfrak{u}}_1$ symmetry, it can be shown that they can be described by a finite number of irreducible representations.
In other words, when $R^2 \in \bb{Q}$, the compactified boson CFT becomes an RCFT.
Based on this fact, there are studies examining how certain physical quantities change during the process of changing the radius $R$ (called exactly marginal deformation),
to compare the behavior of physical quantities in ICFT and RCFT.

\end{itemize}

\subsection{Wess-Zumino-Witten Model}
We have stated that the free boson CFT is a CFT with $\hat{\mathfrak{u}}_1$ symmetry,
but more generally, we can consider a CFT with level $k$ affine Lie algebra $\hat{\mathfrak{g}}$ symmetry,
\begin{equation}
[J^a_n, J^b_m] = \sum_c i f_{abc} J^c_{n+m} + kn\delta_{ab} \delta_{n+m,0},
\end{equation}
where $J^a_n$ are the $n$-th mode of the current.
By generalizing $\hat{\mathfrak{u}}_1$ to non-abelian, new terms with non-trivial structure constants $f_{abc}$ appear.
The energy-momentum tensor can be constructed in almost the same way as in the case of the free boson,
\begin{equation}
T(z) \sim (JJ)(z).
\end{equation}
This is called the {\it Sugawara construction}.
It can be shown that this involves a Virasoro algebra as a subalgebra.
The CFT constructed from this energy-momentum tensor is called $\hat{G}_k$ \und{ Wess-Zumino-Witten (WZW) model}.
The WZW model consists of a finite number of $\hat{\mathfrak{g}}$ affine primary fields,
and hence is an RCFT.
However, note that while the representation of the affine Lie symmetry is described by a finite number of irreducible representations,
it does not mean that the representation of the Virasoro symmetry is described by a finite number of irreducible representations.
In general, there are infinitely many Virasoro primary fields in the Hilbert space of the WZW model.

Properly defining the WZW model requires an explanation of the affine Lie algebra,
but as this would deviate from the purpose of this lecture, it is not introduced here.
Interested readers should refer to Chapter 15 of the Yellow book \cite{Francesco2012}.

\subsection{RCFT and ICFT}

For an irreducible representation $R_i$ of the maximal current algebra $\ca{A}$ of a CFT, the character is defined as follows,
\begin{equation}
\chi_i(\tau) \equiv \tr_{R_i} q^{L_0-\fr{c}{24}}.
\end{equation}
In particular, when the maximal current algebra is the Virasoro algebra, this is the Virasoro character.
The partition function can be written as
\begin{equation}
Z(\tau, \bar{\tau}) = \sum_{i, \bar{i}} M_{i, \bar{i}} \chi_i(\tau) \bar{\chi}_{\bar{i}}(\bar{\tau}).
\end{equation}
When this partition function can be written as a finite sum, the CFT is called an \und{RCFT}.
All CFTs other than RCFT are called \und{ICFT}.

Below are some comments on RCFT and ICFT.
\begin{itemize}

\item
The central charge and conformal dimensions of an RCFT are always rational numbers \cite{Vafa1988b}.
\footnote{
It is not known whether the converse is true (as far as the author knows).
}

\item
A complete classification of RCFTs is not known.
The RCFTs known so far are of two types:
\begin{itemize}

\item
Coset construction, gauging, and tensor products of WZW models

\item
Rational points on the moduli space under exactly marginal deformations (such as Narain CFT)
\end{itemize}

\item
The proper construction of ICFTs is more difficult compared to that of RCFTs.
Examples of exactly constructed ICFTs include the free boson CFT and Liouville CFT introduced in this section.

\item
The CFTs appearing in the AdS/CFT correspondence are ICFTs.
In particular, the CFT dual to the simplest gravity without matter is considered to be a unitary compact ICFT (i.e. a unitary CFT with a discrete spectrum and the vacuum state) with only Virasoro symmetry.
This ICFT will be explained in the following sections.

\end{itemize}

\section{AdS/CFT Correspondence}\label{sec:AdSCFT}

This section will explain the AdS/CFT correspondence.
Since this lecture does not assume specialized knowledge of general relativity,
only the claims directly related to this lecture will be addressed.

\subsection{Anti-de Sitter Space}

AdS spacetime is a solution to the Einstein equation with a negative cosmological constant and the maximum symmetry.
\footnote{Having the maximum symmetry means that the number of independent Killing vectors is the maximum possible, that is, when the dimension of the space is $d$, there are $\fr{d(d+1)}{2}$ Killing vectors.
Intuitively, this means there are $d$ Killing vectors generating translations in the $d$ directions and $\fr{d(d-1)}{2}$ Killing vectors generating rotations around a point, totaling $\fr{d(d+1)}{2}$.
A familiar example is the symmetry of Minkowski space.
From the point of view of the hyperboloid (\ref{eq:con}), it is clear that AdS$_d$ has the symmetry $\mathrm{SO}(2,d-1)$.
Therefore, the number of independent Killing vectors is $\fr{d(d+1)}{2}$, the same as the number of its generators, confirming that AdS$_d$ indeed has the maximum symmetry.}
There are multiple ways to represent AdS$_{d+1}$ spacetime.
The simplest introduction is to realize it as a hyperboloid embedded in the space $\bb{R}^{2,d}$,
\begin{equation}
d s^2=-{d X_0}^2-{X_{d+1}}^2+{d X_1}^2+{d X_2}^2+\cdots +{d X_{d}}^2.
\end{equation}
The hyperboloid is given by
\begin{equation}\label{eq:con}
{X_0}^2+{X_{d+1}}^2={X_1}^2+{X_2}^2+\cdots +{X_{d}}^2+l^2,
\end{equation}
where $l$ is called the {\it AdS radius}.
To solve the constraint (\ref{eq:con}), the following coordinates are useful,
\begin{equation}\label{eq:cor}
\begin{aligned}
X_0&=l \cosh\rho \cos t, \\
X_{d+1}&=l \cosh\rho \sin t, \\
X_k&=l \sinh\rho \Omega_k, \\
\end{aligned}
\end{equation}
where $\Omega_k$ represents the coordinates of the $d-1$ dimensional unit sphere.
The metric of AdS$_{d+1}$ can be expressed as follows,
\begin{equation}
d s^2=l^2(-\cosh^2 \rho d t^2+ d \rho^2+\sinh^2 \rho {d \Omega_{d-1}}^2).
\end{equation}
This coordinate system covers a wider range $t \in [-\infty, \infty]$ than the original hyperbolic space $t \in [0,2\pi]$, hence it is called the {\it global coordinates}.
As a metric representing the global coordinates, the following metric is also often used,
\begin{equation}\label{eq:GAdS2}
ds^2= -\pa{ 1+\fr{r^2}{l^2}}d\tau^2 + \fr{dr^2}{\pa{ 1+\fr{r^2}{l^2}}} + r^2  {d \Omega_{d-1}}^2.
\end{equation}
This metric can be obtained by setting $r=l \sinh \rho$ and $\tau=lt$.

In the context of AdS/CFT, it is sometimes convenient to use the following {\it Poincaré coordinates},
\begin{equation}
d s^2=l^2\pa{\fr{d z^2-d t^2+d \bm{x}^2}{z^2}}.
\end{equation}
This can be obtained from the following coordinate transformation,
\begin{equation}\label{eq:GtoP}
\begin{aligned}
X_0&=\fr{z}{2}\pa{1+\fr{l^2+\bm{x}^2-x_0^2}{z^2}}, \\
X_i&=\fr{lx_i}{z}, \\
X_{d}&=\fr{z}{2}\pa{1-\fr{l^2-\bm{x}^2+x_0^2}{z^2}}, \\
X_{d+1}&=\fr{lx_0}{z}.
\end{aligned}
\end{equation}
This coordinate system has an asymptotic boundary at $z=0$.
Therefore, one considers either $z>0$ or $z<0$. 
If we take $z>0$,
\begin{equation}
\fr{X_0-X_{d}}{l^2}=\fr{1}{z}>0 \ \  \Leftrightarrow \ \  \cos t >\Omega_{d-1} \tanh \rho.
\end{equation}
Thus, the Poincaré coordinates cover only a part of the global coordinates.

\subsection{AdS/CFT Correspondence}

The AdS/CFT correspondence refers to the following conjecture \cite{Maldacena1997, Witten1998, Gubser1998},
\begin{tcolorbox}[title=AdS/CFT]
Quantum gravity on AdS${}_{d+1}$ is equivalent to CFT${}_{d}$.
\end{tcolorbox}
\noindent
Since a CFT is a field theory that can be treated non-perturbatively,
the AdS/CFT correspondence can be regarded as a non-perturbative formulation of quantum gravity.

Evidence of the AdS/CFT correspondence comes from the fact that both share the same symmetry $\mathrm{SO}(2,d)$.
In this section, by the detailed comparison between AdS and CFT, we will see how the AdS/CFT correspondence indeed works nicely.

\subsubsection{Virasoro Symmetry}

Specifically, let us focus on AdS${}_3$/CFT${}_2$ and see if the Virasoro symmetry can be reproduced from the gravity side.
Three-dimensional gravity with a negative cosmological constant is described by the following action:
\begin{equation}
S  =   \fr{1}{16 \pi G_N} \int_M dx^3 \  \sqrt{g} \pa{R  + \fr{2}{l^2} }  + \fr{1}{8\pi G_N} \int_{\del M} dx^2 \ \sqrt{h} \pa{K - \fr{1}{l}}.
\end{equation}
The first term is the Einstein-Hilbert action and the second term is the Gibbons-Hawking term on the asymptotic boundary.
$G_N$ is the Newton gravitational constant, $g$ is the metric, and $R$ is the scalar curvature.
$h$ is the induced metric on the boundary $\del M$ of $M$.
$K \equiv h^{ab} K_{ab}$ is defined as the trace of the extrinsic curvature $K_{ab}$, which is given by the following definition:
\footnote{
Using Gaussian normal coordinates
\begin{equation}
ds^2 = dr^2 + h_{ab} dx^a dx^b,
\end{equation}
the extrinsic curvature can be expressed in a simple form as:
\begin{equation}
K_{ab} = \frac{1}{2} \frac{\partial h_{ab}}{\partial r}.
\end{equation}
}
\begin{equation}
K_{ab} \equiv \nabla_a n_b.
\end{equation}
Here, $n$ denotes a unit vector normal to $\partial M$.
This action, which does not include any matter, is the simplest AdS gravity and is called the {\it pure AdS}.
The energy-momentum tensor for this action is defined as follows \cite{Balasubramanian1999}:
\begin{equation}\label{eq:TGrav}
T^{ij} \equiv -\fr{4\pi}{\sqrt{h}} \fr{\delta S}{\delta h_{ij}}.
\end{equation}
Below, we show that the transformation law of this stress tensor reproduces the Virasoro symmetry on the CFT side.

The transformations corresponding to the Virasoro symmetry on the gravity side should be those that do not change the metric near the asymptotic boundary.
To see this, expand the metric (\ref{eq:GAdS2}) with $z=\tau + l\phi, \bz =\tau-l\phi$, and $r\to \infty$,
\begin{equation}
ds^2 \simeq \fr{l^2}{r^2}dr^2 + \fr{r^2}{l^2} dz d\bz .
\end{equation}
The possible asymptotic AdS metric consistent with this asymptotics has the following form:
\begin{equation}\label{eq:Metric}
ds^2 = \fr{l^2}{r^2}dr^2 + \fr{r^2}{l^2} dz d\bz  +h_{zz}dz^2 + h_{\bz  \bz } d\bz ^2 + 2h_{z \bz } dz d\bz .
\end{equation}
This metric does not include all possible terms, but other possible terms can be eliminated by coordinate transformations.
By substituting this metric into (\ref{eq:TGrav}), we find that the energy-momentum tensor can be expressed as follows:
\begin{equation}\label{eq:T=h}
T_{zz} = -\fr{1}{4lG_N} h_{zz}, \ \ \ \ \ 
T_{\bz \bz } = -\fr{1}{4lG_N} h_{\bz \bz }.
\end{equation}
Here, we used the following Einstein equation (i.e. the equation of motion for the gravitational action):
\begin{equation}
h_{z\bz }=0, \ \ \ \ \ \del h_{\bz \bz } = \bar{\del} h_{zz} = 0.
\end{equation}

The infinitesimal coordinate transformations that preserve the form of the metric (\ref{eq:Metric}) can be expressed using arbitrary functions $\epsilon(z), \bar{\epsilon}(\bz )$ as follows:
\begin{equation}
\begin{aligned}
z &\rightarrow z+\e(z)-\fr{{l}^4}{2r^2}\bar{\e}''(\bz ),\\
\bz &\rightarrow \bz +\bar{\e}(\bz )-\fr{{l}^4}{2r^2}\e''(z),\\
r &\rightarrow r-\fr{r}{2}\e'(z)-\fr{r}{2}\bar{\e}'(\bz ).
\end{aligned}
\end{equation}
This transformation corresponds exactly to the infinitesimal conformal transformation (\ref{eq:epsilon}) of the CFT on the asymptotic boundary.
Under this coordinate transformation, the metric changes as follows:
\begin{equation}
\delta h_{zz} = -2 \epsilon'(z) h_{zz} -\epsilon(z) h'_{zz} + \fr{l^2}{2} \epsilon'''(z). 
\end{equation}
By converting this to the energy-momentum tensor using (\ref{eq:T=h}), we obtain
\begin{equation}
\delta T_{zz} = - 2 \epsilon'(z) T_{zz} -\epsilon(z)T'_{zz} - \fr{l}{8 G_N} \epsilon'''(z).
\end{equation}
This matches the transformation law of the energy-momentum tensor obtained earlier (\ref{eq:Tinf}).
Comparing the anomaly terms, we find that the central charge is given by:
\begin{tcolorbox}[title=Brown-Henneaux central charge]
    \begin{equation}\label{eq:Brown}
c=\fr{3l}{2G_N}.
\end{equation}
\end{tcolorbox}
\noindent
This result is called the \und{Brown-Henneaux central charge}, named after the authors who have shown the equivalence between the asymptotic symmetry of AdS and the Virasoro symmetry \cite{Brown1986}.

The Brown-Henneaux central charge shows that the classical limit $G_N \to 0$ corresponds to $c \to \infty$.
In other words,
\begin{tcolorbox}
Einstein gravity is dual to a large $c$ CFT.
\end{tcolorbox}
\noindent
This CFT is called the \und{holographic CFT}.
It is important to note that there is still no ``precise definition" of the holographic CFT.
The CFT data of the holographic CFT are still unknown,
and revealing its CFT data is one of the important challenges in the AdS/CFT correspondence.
An approach that is useful when only partial information such as large $c$ is known is the conformal bootstrap.

\subsubsection{GKPW Dictionary}

The basic objects in QFT are correlation functions.
What is the gravitational counterpart of correlation functions in the AdS/CFT correspondence?
This is given by the Gubser-Klebanov-Polyakov-Witten (GKPW) dictionary \cite{Gubser1998, Witten1998}:
\begin{tcolorbox}[title=GKPW dictionary]
    \begin{equation}\label{eq:dic}
Z_{\text{grav}}[\phi_i]=\Braket{\ex{i\sum_i \int d^{d}x O_i(x) J_i(x)}}_{\text{CFT}}.
\end{equation}
\end{tcolorbox}
\noindent
Here, each symbol is defined as follows:
\begin{equation}\label{eq:label}
\begin{aligned}
O_i &\cdots \text{Field in CFT} \\
\phi_i &\cdots \text{Field in gravity} \\
J_i &\cdots \text{Boundary condition for } \phi_i: \phi_i|_{\text{boundary}}\propto J_i \\
Z_{\text{grav}}[\phi_i] &\cdots \text{Partition function of quantum gravity}
\end{aligned}
\end{equation}

Let us first focus on the path integral on the gravity side on the left-hand side.
Since AdS space has a boundary, boundary conditions must be specified to define the path integral.
These are $J_i$ in (\ref{eq:label}).
In Poincaré coordinates, this boundary condition is expressed as follows:
\begin{equation}\label{eq:coupling}
\phi_i(z,x)\ar{z \to 0} z^{d-\D}J_i(x).
\end{equation}
If we consider the equation of motion for a scalar field $\phi_i$ with mass $m$, the following equation is obtained in the limit $z\to0$:
\begin{equation}\label{eq:ScalarMass}
l^2 m^2 =\D(\D-d), \ \ \ \ \ \D=\fr{d}{2}+\sqrt{\fr{d^2}{4}+m^2R^2}.
\end{equation}
In fact, this $\D$ is the conformal dimension of the field $O_i$ in the CFT.
The scale transformation on the CFT side corresponds to the following on the gravity side:
\begin{equation}
x_i \to \l x_i, \ \ \ z \to \l z.
\end{equation}
From this fact and (\ref{eq:coupling}), we can understand the behavior of $J_i(x)$ under the renormalization group.
For the coupling term
\begin{equation}
\int d^{d}x O_i(x) J_i(x)
\end{equation}
to be invariant under the renormalization group, the conformal dimension of $O_i(x)$ must be $\Delta$.

The right-hand side of (\ref{eq:dic}) is the generating functional of the CFT,
and the correlation functions of the CFT are obtained as follows:
\begin{equation}
\braket{O_1(x_1)\cdots O_n(x_n)}_{\text{CFT}}=\fr{\d^n}{\d J_1 (x_1)\cdots \d J_n(x(n))}\Braket{\ex{i\sum_i \int \mathrm{d}^{d}x O_i(x) J_i(x)}}_{\text{CFT}}.
\end{equation}

As seen from the GKPW dictionary, the CFT appears as a ``boundary condition" of the quantum gravity on AdS,
so it is often explained that ``the CFT lives on the asymptotic boundary of AdS".
While this is convenient for intuitive interpretation,
it should be noted that the entire AdS quantum gravity corresponds to the CFT,
and the CFT does not simultaneously exist on AdS.

\subsubsection{Energy}\label{subsubsec:energy}

In this section, we will compare the energy in AdS and CFT.
Let us first calculate the ADM mass of the global AdS geometry.
The ADM mass in AdS is obtained by integrating the boundary energy-momentum tensor \cite{Balasubramanian1999},
\begin{equation}\label{eq:MADM}
M_{\mathrm{ADM}} = \int^{2\pi}_0 d \phi T_{\tau \tau}.
\end{equation}
The boundary energy-momentum tensor for the global AdS geometry is given by
\begin{equation}
T_{\tau \tau} = -\fr{1}{16 \pi G_N},
\end{equation}
so the ADM mass is
\begin{equation}
M_{\mathrm{ADM}} = -\fr{1}{8G_N}.
\end{equation}
Using the Brown-Henneaux central charge, we obtain
\begin{equation}
E=lM_{\mathrm{ADM}}=-\fr{c}{12}.
\end{equation}
It can be confirmed that this ADM mass of the vacuum AdS matches the vacuum energy of the CFT, i.e. the Casimir energy (\ref{eq:Casimir}).

Next, consider a more non-trivial case, Euclidean conical defect geometry:
\begin{equation}\label{eq:conical}
ds^2= \pa{ 1+\fr{r^2}{l^2}}d\tau^2 + \fr{dr^2}{\pa{ 1+\fr{r^2}{l^2}}} + r^2  d\phi^2, \ \ \ \ \ \phi \in [0,2\pi\chi).
\end{equation}
This is the same as the global AdS metric (\ref{eq:GAdS2}),
but the range of $\phi$ is different.
Since this geometry has a deficit angle of $\delta\phi \equiv 2\pi(1-\chi)$,
it has a singularity at the origin.
Moreover, since there is no horizon, this is a naked singularity.
This naked singularity can be interpreted as a point particle.
In fact, the conical defect geometry is an on-shell solution when the following point particle action is added to the Einstein-Hilbert action,
\begin{equation}
S_{\mathrm{particle}}=-m\int_\Gamma \sqrt{\gamma}.
\end{equation}
Here, $\Gamma$ represents the worldline of the point particle.
Note that the mass $m$ of the point particle is related to the deficit angle as
\begin{equation}\label{eq:phim}
\delta \phi  = 2\pi(1-\chi) = 8\pi G_N m.
\end{equation}
Let us evaluate the ADM mass of this geometry.
Recall that CFT states are defined on the cylinder with period of $2\pi$.  
As can be seen from the expression (\ref{eq:MADM}), the ADM mass is not a scalar quantity, therefore, in order to compare the ADM mass with the energy of a CFT state,  it is necessary to evaluate the ADM mass in coordinates whose asymptotic boundary is the cylinder with period of $2\pi$.  
For this purpose, we perform the following coordinate transformation to map the metric (\ref{eq:conical}) into the appropriate coordinate,
\begin{equation}
\tau \to \fr{\tau}{\chi}, \ \ \ \ \phi \to \fr{\phi}{\chi}.
\end{equation}
Under this coordinate transformation, the metric takes the following form:
\begin{equation}
ds^2= \pa{ \chi^2+\fr{r^2}{l^2}}d\tau^2 + \fr{dr^2}{\pa{ \chi^2+\fr{r^2}{l^2}}} + r^2  d\phi^2, \ \ \ \ \ \phi \in [0,2\pi).
\end{equation}
Also, the energy-momentum tensor changes as follows:
\begin{equation}
T_{\tau \tau} = -\fr{\chi^2 }{8G_N}.
\end{equation}
Thus, the ADM mass is given by
\begin{equation}\label{eq:conicalADM}
E=lM_{\mathrm{ADM}}=-\fr{c}{12} \chi^2.
\end{equation}
Using (\ref{eq:phim}), (\ref{eq:conicalADM}), and the relation $E=2h-\fr{c}{12}$ between the energy of the scalar field in the CFT on the cylinder with period $2\pi$ and the conformal dimension, we obtain the following relation between the mass $m$ of the point particle and the conformal dimension $h$,
\begin{equation}
m=\fr{c}{6}\pa{1-\sqrt{1-\fr{24h}{c}}}.
\end{equation}
This equation can be regarded as an extension of (\ref{eq:ScalarMass}) to $h=O(c)$.
In fact, these two equations match in the regime $1\ll h \ll c$.

Finally, let us consider the black hole solution.
The solution corresponding to the black hole with mass $M$ is given by the Banados-Teitelboim-Zanelli (BTZ) metric:
\begin{equation}
ds^2 = -\pa{\fr{r^2}{l^2}-8G_N M}dt^2+ \fr{dr^2}{\fr{r^2}{l^2}-8G_N M} +r^2 d\phi^2.
\end{equation}
This solution can be regarded as an analytic continuation of the deficit angle of the conical defect geometry to $\chi^2 \to -8G_N M$.
Thus, the ADM mass of the BTZ black hole can be obtained by the analytic continuation of  (\ref{eq:conicalADM}),
\begin{equation}
M_{\mathrm{ADM}}=M.
\end{equation}
In other words, the mass of the BTZ black hole is the ADM mass itself.

Since the conical defect geometry has a negative ADM mass and the BTZ black hole has a positive ADM mass,
using the relation $\Delta-\fr{c}{12}=E$, we obtain the following conclusion:
\begin{tcolorbox}
\begin{equation}
\begin{aligned}
\text{BTZ\ BH} &= \text{primary state with } \Delta>\fr{c}{12}, \\
\text{conical\ defect} &= \text{primary state with } \Delta < \fr{c}{12}.
\end{aligned}
\end{equation}
\end{tcolorbox}
\noindent
From this fact, $\fr{c}{12}$ is called the {\it BTZ threshold}.
Note that since a descendant state is created by the action of Virasoro generators, a descendant state corresponds to an excitation caused by a diffeomorphism (a boundary graviton) on the gravity side.

\subsubsection{Thermodynamic Quantities}\label{subsubsec:thermal}

The GKPW dictionary for the partition function is the simplest example,
\begin{equation}
Z_{\mathrm{CFT}}(\beta) = Z_{\mathrm{grav}}(\beta).
\end{equation}
Here, the left-hand side is the partition function of the CFT defined on the manifold $S^1_{2\pi} \times S^1_{\beta}$,
and is expressed as $Z_{\mathrm{CFT}}(\beta) \equiv \tr \ \ex{-\beta H_{\mathrm{CFT}}}$.
$Z_{\mathrm{grav}}(\beta)$ is the quantity evaluated by the gravitational path integral on AdS whose asymptotic boundary is given by $S^1_{2\pi} \times S^1_{\beta}$,
and in particular, in the semiclassical approximation, it can be written as follows using the on-shell Euclidean action,
\begin{equation}
Z_{\mathrm{grav}}(\beta) = \ex{-S_{E}[g_1]} + \ex{-S_{E}[g_2]} + \cdots,
\end{equation}
where $g_i$ are solutions to the Einstein equation.

The agreement of these partition functions implies that the thermodynamic quantities on both sides match.
On the gravity side, by evaluating the on-shell Euclidean action, we can get the free energy $F\equiv-\beta^{-1} \ln Z$.
Thus, we can evaluate the thermodynamic quantities including the black hole entropy.
In fact, it can be shown that the black hole entropy calculated in this way matches the entropy calculated on the CFT side (i.e. Cardy formula).
The calculation details on the CFT side are explained in Section \ref{subsec:Cardy}.

The semiclassical approximation of the gravitational path integral is given by
\begin{equation}\label{eq:Zgrav}
Z_{\mathrm{grav}}(\beta)  = \ex{\fr{c}{12}\beta} + \ex{\fr{c}{12} \fr{(2\pi)^2}{\beta}} + \cdots ,
\end{equation}
where the two solutions are the thermal AdS and the BTZ black hole.
In the classical limit $c \to \infty$, it is approximated by the larger of these two terms.
This implies that there is a first-order phase transition at the critical temperature $\beta = 2\pi$,
\begin{equation}
\begin{aligned}
\ln Z_{\mathrm{grav}}&=\left\{
    \begin{array}{ll}
    \fr{c}{12}\beta   ,& \text{if }\beta <2\pi   ,\\
     \fr{c}{12} \fr{(2\pi)^2}{\beta}  ,& \text{if }\beta > 2\pi   .\\
    \end{array}
  \right.\\
\end{aligned}
\end{equation}
This phase transition is called the Hawking-Page transition \cite{Hawking1983}.

\subsubsection{Maloney-Witten-Keller Partition Function}

The two terms appearing in the semiclassical approximation of the gravitational path integral (\ref{eq:Zgrav}) are exactly related by the modular $S$ transformation.
This reflects the modular invariance of $Z_{\mathrm{CFT}}(\beta)$.
From this expression of the gravitational path integral, one might guess that the partition function of the CFT dual to pure AdS can be written in the following simple form,
\begin{equation}
Z(\tau,\bar{\tau}) = \abs{\chi_{\bb{I}}(\tau)}^2 + \abs{\chi_{\bb{I}}\pa{-\fr{1}{\tau}}}^2.
\end{equation}
However, although this is invariant under the modular $S$ transformation, it is not invariant under all modular transformations including the modular $T$ transformation.
Nevertheless, if it can be updated to a modular invariant partition function in some way,
it may be possible to reveal the CFT dual to the pure AdS.

This attempt to concretely construct a CFT dual to the pure AdS was initiated by Maloney-Witten-Keller \cite{Maloney2007,Keller2014}.
The idea of MWK is to sum over contributions with modular parameters obtained from modular transformations to realize modular invariance. In fact, each of the terms in the modular sum corresponds to a different Euclidean classical saddle \cite{Maldacena1998}, so in this way, the partition function is a sum over classical solutions.
\begin{equation}
Z_{\mathrm{MWK}}(\tau, \bar{\tau}) = \sum_{\gamma \in \mathrm{PSL}(2,\bb{Z})} \abs{\chi_{\bb{I}}(\gamma \tau)}^2.
\end{equation}
Here, $\mathrm{PSL}(2,\bb{Z})$ is the projective special linear group,
\begin{equation}
 \mathrm{PSL}(2,\bb{Z})
 = \left\{
 \left(
    \begin{array}{cc}
     a  & b  \\
     c  &  d \\
    \end{array}
  \right)
  \in \mathrm{M}_2(\bb{Z}) \ | \ ad-bc=1
  \right\}
  /m\sim m' :\Leftrightarrow m=-m'
  .
\end{equation}
For the modular invariance to be satisfied, this partition function must be convergent,
but in fact, this sum (called the Poincaré series) diverges.
MWK proposes a method to appropriately remove this divergence,
but the MWK partition function obtained in this way has the following problems \cite{Maloney2007, Benjamin2019}:
\begin{enumerate}

\item
 continuous spectrum.

\item
 non-unitary.

\end{enumerate}
Regarding the second problem, it has been proposed that it can be resolved by adding other contributions (e.g. Poincaré series of non-vacuum characters) to the MWK partition function \cite{Keller2014,Benjamin2019,Benjamin2020,Maxfield2020}.
Further advancing such improvements, ultimately constructing a unitary compact MWK partition function is one of the important challenges in the study of AdS/CFT.

\subsection{Correlation Functions in Holographic CFT}

In general, the correlation functions of a CFT are expressed as (infinite) sums of complicated special functions i.e. the Virasoro blocks,
which makes extracting physics from correlation functions complicated.
However, in holographic CFTs, the expression for correlation functions simplifies under certain assumptions.
The mechanism of simplification can be roughly described as follows:
\begin{itemize}

\item
In the large $c$ limit, from the viewpoint of the saddle point approximation,
the correlation function can be approximated by a single Virasoro block.

\item
In the large $c$ limit, the expression of the Virasoro block simplifies.

\end{itemize}
Below, we explain the details of the simplification of the Virasoro block.

\subsubsection{Heavy-Light Virasoro Block}\label{subsubsec:HHLLblock}

Here, we explain the simplification of the Virasoro block in the large $c$ limit.
In fact, we have already seen the simplification in a special case in Section \ref{subsec:GCB}.
For convenience, we repost (\ref{eq:GCB}) below,
\begin{equation}\label{eq:GCB2}
\begin{aligned}
\lim_{c\to \infty}
\ca{F}^{21}_{34}(h_p|z)
&=\sum_{\substack{  \bm{n}=\underbrace{(-1,-,1,\cdots,-1)}_{N}  }}
\rho(\nu_4, \nu_3,\nu_{p, \bm{n}}|0)
\br{G_{p}^{(N)}}_{\bm{n}, \bm{n}}^{-1} \rho(\nu_{p,\bm{n}}, \nu_{2}, \nu_1|z) \\
&= \sum_N \fr{ (h_p+h_3-h_4)_N (h_p+h_2-h_1)_N}{N! (2h_p)_N}z^{h_p+N-h_1-h_2} \\
&=z^{h_p-h_1-h_2} {}_2F_1(h_p+h_2-h_1,h_p+h_3-h_4;2h_p;z).
\end{aligned}
\end{equation}
Recall that when deriving this large $c$ Virasoro block,
we took the $c \to \infty$ limit while keeping all parameters (i.e. $\{ h_i \}$ and $z$) fixed.
Here, we generalize this and consider the Virasoro block in the $c \to \infty$ limit under the following assumptions:
\begin{equation}
h_3, h_4 = O(c), \ \ \ \ h_1, h_2, \abs{h_3-h_4}, h_p, z = O(c^0).
\end{equation}
Under these assumptions, the first equation in (\ref{eq:GCB2}) (i.e. the assertion that $L_{n<-1}$ descendants do not contribute in the large $c$ limit) no longer holds.
This is because the 3-point block $\rho(\nu_4, \nu_3,\nu_{p, \bm{n}}|0)$ becomes $O(c)$,
and the suppression by $\fr{1}{c}$ from the inverse Gram matrix is canceled out.
We explain how to resolve this problem below.

For simplicity, let $h_1=h_2\equiv h_L$ and $h_3=h_4\equiv h_H$.
$H$ stands for ``Heavy" and $L$ stands for ``Light".
In this case, by the Ward-Takahashi identity,
\begin{equation}\label{eq:corrT}
\braket{ \phi_H(\infty) \phi_H(1) T(z) \phi_h(0) } = C_{HHh} \left( \frac{h_H}{(1-z)^2} + \frac{h}{(1-z) z^2} \right).
\end{equation}
Recalling that the Virasoro generator corresponds to the Laurent coefficients of $T(z)$,
it is clear that the reason $\rho(\nu_4, \nu_3,\nu_{p, \bm{n}}|0)$ is $O(c)$ is essentially because the first term in (\ref{eq:corrT}) is proportional to $h_H=O(c)$.

\begin{figure}[t]
 \begin{center}
  \includegraphics[width=8.0cm,clip]{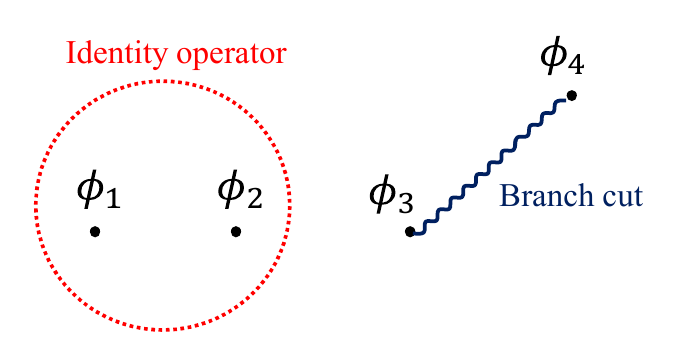}
 \end{center}
 \caption{Insertion of the identity operator into the four-point function on the $w$ coordinates. Note the presence of the branch cut.}
 \label{fig:branch}
\end{figure}

We consider eliminating this term proportional to $c$ by a conformal transformation.
The key is the transformation law (\ref{eq:Ttrans}) of the energy-momentum tensor $T(z)$:
\begin{equation}
T'(w) = \pa{\fr{dw}{dz}}^{-2}\pa{T(z) -\fr{c}{12} \{w,z \} }.
\end{equation}
Consider the following conformal transformation:
\begin{equation}\label{eq:HLmap}
z \to w =1- (1-z)^\alpha,  \ \ \ \ \  \alpha \equiv \sqrt{1 - 24 \frac{h_H}{c} }.
\end{equation}
Under this transformation, the Schwarzian derivative term exactly cancels the $c$-proportional term in (\ref{eq:corrT}):
\begin{equation}\label{eq:OOTO}
\braket{ \phi_H(\infty) \phi_H(1) T(w) \phi_h(0) } = C_{HHh} \left( h \,  \frac{ 1-z }{z^2} \right).
\end{equation}
With this fact in mind, we consider evaluating the Virasoro block on the $w$ coordinates.
The Virasoro block on the $w$ coordinates is also defined by inserting a projection operator onto the Verma module (see Figure \ref{fig:branch}).
In this case, the Virasoro block can be expressed as follows:
\begin{equation}\label{eq:HHLL1}
\begin{aligned}
\lim_{c\to \infty}
\ca{F}^{21}_{34}(h_p|z)
&=\sum_{\substack{  \bm{n}=\underbrace{(-1,-,1,\cdots,-1)}_{N}  }}
\ti{\rho}(\nu_4, \nu_3,\nu_{p, \bm{n}}|0)
\br{G_{p}^{(N)}}_{\bm{n}, \bm{n}}^{-1}
\rho(\nu_{p,\bm{n}}, \nu_{2}, \nu_1|w).
\end{aligned}
\end{equation}
The three-point block $\rho(\nu_{p,\bm{n}}, \nu_{2}, \nu_1|w)$ can be calculated in the same way as in the plane coordinates.
On the other hand, since the three-point block $\ti{\rho}(\nu_4, \nu_3,\nu_{p, \bm{n}}|0)$ includes a branch cut as seen in Figure \ref{fig:branch},
the action of the Virasoro algebra changes.
This is emphasized by adding a tilde.
As a result of the change in the action of the Virasoro algebra, $\ti{\rho}(\nu_4, \nu_3,\nu_{p, \bm{n}}|0)$ becomes $O(c^0)$.
This is a consequence of (\ref{eq:OOTO}).
It can be shown that the insertion of multiple energy-momentum tensors also results in $O(c^0)$.
Therefore, all contributions from $L_{n<-1}$ are suppressed by $\fr{1}{c}$ by the inverse Gram matrix.
Thus, only the contributions from descendants consisting of $L_{-1}$ remain, as shown in (\ref{eq:HHLL1}).

Next, we consider how $L_{-1}$ acts on the three-point block with the branch cut.
From (\ref{eq:OOTO}), we directly obtain the following result:
\begin{equation}
\ti{\rho}(\nu_4, \nu_3,\nu_{p, \bm{n}}|0) = \alpha^{-h_p} \pa{ h_p + \fr{h_3-h_4}{\alpha} }_N.
\end{equation}
Here, we have written a general expression including the case of $h_1 \neq h_2$ and $h_3 \neq h_4$.
The three-point block $\rho(\nu_{p,\bm{n}}, \nu_{2}, \nu_1|w)$ including the Jacobian factor is
\begin{equation}
\rho(\nu_{p,\bm{n}}, \nu_{2}, \nu_1|w) =  \alpha^{h_1+h_2}(1-w)^{\pa{1-\fr{1}{\alpha}}h_2}  (h_p+h_2-h_1)_N w^{h_p+N-h_1-h_2}.
\end{equation}
Thus, we finally obtain the following result \cite{Fitzpatrick2015}:
\begin{tcolorbox}[title=Heavy-Light Virasoro block]
    \begin{equation}\label{eq:HHLL}
\begin{aligned}
\lim_{c\to \infty}
\ca{F}^{21}_{34}(h_p|z)
&=\sum_{\substack{  \bm{n}=\underbrace{(-1,-,1,\cdots,-1)}_{N}  }}
\ti{\rho}(\nu_4, \nu_3,\nu_{p, \bm{n}}|0)
\br{G_{p}^{(N)}}_{\bm{n}, \bm{n}}^{-1} \rho(\nu_{p,\bm{n}}, \nu_{2}, \nu_1|z) \\
&= 
  \alpha^{-h_p+h_1+h_2}(1-w)^{\pa{1-\fr{1}{\alpha}}h_2}
\sum_N \fr{ \pa{h_p+\fr{h_3-h_4}{\alpha}}_N (h_p+h_2-h_1)_N}{N! (2h_p)_N} w^{h_p+N-h_1-h_2} \\
&=
(1-w)^{\pa{1-\fr{1}{\alpha}}h_2}
\pa{\fr{w}{\alpha}}^{h_p-h_1-h_2}
{}_2F_1\pa{h_p+h_2-h_1,h_p+\fr{h_3-h_4}{\alpha};2h_p;w}.
\end{aligned}
\end{equation}
\end{tcolorbox}
\noindent
Since $\abs{h_3-h_4}=O(c^0)$, $\alpha$ is defined similarly even for $h_3 \neq h_4$:
\begin{equation}
\alpha \equiv \sqrt{1 - 24 \frac{h_3}{c} } + O(c^{-1}) = \sqrt{1 - 24 \frac{h_4}{c} } + O(c^{-1}) .
\end{equation}
This expression for the Virasoro block in the large $c$ limit is called the \und{Heavy-Light Virasoro block} or \und{HHLL Virasoro block}.

\subsubsection{Heavy-Light Virasoro Block and Information Loss Problem}\label{subsec:HHLL}

\begin{figure}[t]
 \begin{center}
  \includegraphics[width=6.0cm,clip]{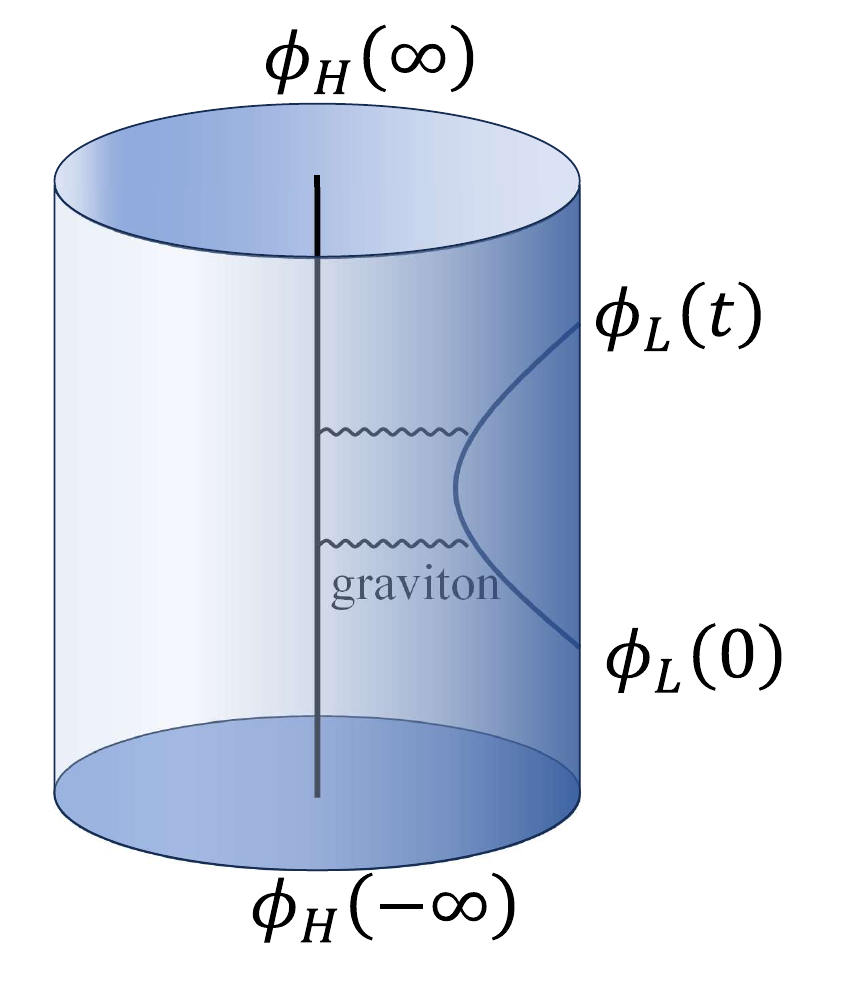}
 \end{center}
 \caption{Two-point correlation function in the black hole background geometry.}
 \label{fig:HHLL}
\end{figure}

The two-point function $\braket{\phi_L(t) \phi_L(0)}_{BH}$ in the black hole background geometry (see Figure \ref{fig:HHLL}) decays exponentially under time evolution.
Intuitively, this reflects the fact that the information swallowed by the black hole is permanently lost.
This contradicts the unitarity of quantum mechanics, and this contradiction is known as \und{information loss problem}.
Resolving this information loss problem is one of the major goals in quantum gravity.

In Euclidean time, the information loss problem appears as the periodicity of the correlation function.
Using the AdS/CFT correspondence, we consider understanding the mechanism of periodicity in the semiclassical approximation from the CFT.
The two-point function in the black hole microstate depicted in Figure \ref{fig:HHLL} corresponds to the following correlation function in the CFT:
\begin{equation}\label{eq:HHLLcorr}
\braket{ \phi_H(\infty) \phi_H(1) \phi_L(z,\bz ) \phi_L(0)  }.
\end{equation}
Although we would like to directly investigate this correlation function,
since the exact CFT data of holographic CFTs are unknown,
instead of examining the correlation function, we focus on the Virasoro block, which is a component of the correlation function.

The Virasoro block appearing in the Heavy-Light correlation function is precisely the Heavy-Light Virasoro block (\ref{eq:HHLL}) obtained earlier.
First, to match the setup in Figure \ref{fig:HHLL}, we consider the following coordinate transformation:
\begin{equation}
1-z=\ex{- t_E}.
\end{equation}
In this case, the Heavy-Light vacuum block becomes:
\begin{equation}
\begin{aligned}
\ca{F}^{LL}_{HH}(0|t_E)
&=\ex{ - h_L t_E} \ca{F}^{LL}_{HH}(0|z) \\
&= \pa{ \fr{\alpha}{2 \sinh \fr{\alpha t_E}{2}}  }^{2h_L}.
\end{aligned}
\end{equation}
The factor appearing on the right-hand side of the first line is the Jacobi factor.
Recall from Section \ref{subsubsec:thermal} that the state corresponding to the black hole had $h_H > \fr{c}{24}$.
The $\alpha = \sqrt{1 - 24 \frac{h_H}{c} }$ appearing in the Heavy-Light Virasoro block becomes purely imaginary above the BTZ threshold, i.e. $h_H > \fr{c}{24}$.
By setting $\alpha \equiv 2 \pi i T_H$, we obtain the following expression:
\begin{equation}\label{eq:HHLLthermal}
\ca{F}^{LL}_{HH}(0|t_E)= \pa{\fr{\pi T_H}{\sin \pa{\pi T_H t_E}}}^{2\Delta_L}.
\end{equation}
As can be seen from the expression, this Virasoro block has a periodicity parameterized by $T_E$ in the Euclidean time direction.
In other words, the expected ``periodicity" appears precisely above the BTZ threshold.
Furthermore, by the Wick rotation $t_E \to it$ on this expression, it can be observed that it decays exponentially as $\ex{-\pi T_H t}$ for $t \to \infty$.

In fact, this functional form is expected to appear.
The Euclidean black hole has periodicity in the time direction,
implying that the corresponding CFT is none other than a finite-temperature CFT.
The two-point function on a finite-temperature CFT at temperature $T_E$ is the two-point function on a cylinder with periodicity $T_E$ in the time direction.
As seen in the state-operator correspondence, the cylinder and the plane are connected by a conformal transformation,
so this two-point function on the cylinder can be calculated from the two-point function on the plane (\ref{eq:GWTres}).
Using the conformal transformation $z=1-\ex{ t_E}$, we obtain:
\begin{equation}
\braket{\phi_L(t_E) \phi_L (0)}_{T_H}=\pa{\fr{\pi T_H}{\sin \pa{\pi T_H t_E}}}^{2h_L}.
\end{equation}
In other words, the Heavy-Light vacuum block (\ref{eq:HHLLthermal}) is none other than the two-point function on the finite-temperature CFT at temperature $T_H$.
This result suggests that pure states with conformal dimensions above the BTZ threshold and the mixed states of finite temperature are indistinguishable in the semiclassical limit.

One consequence of the periodicity of the Heavy-Light vacuum block is the existence of infinitely many singularities $z=1-\ex{\fr{n}{T_H}} \ \ \ (n \in \bb{Z})$ \cite{Fitzpatrick2016a}.
The following translation is possible for the information loss problem:
\begin{quote}
Information loss problem $\ar{\mathrm{Euclidean}}$ Periodicity $\to$ Infinitely many singularities
\end{quote}
On the other hand, as is clear from the discussion of the convergence of the OPE, the Heavy-Light correlation function (\ref{eq:HHLLcorr}) should have no singularities other than the OPE singularities.
The reason why these ``forbidden singularities" appear is due to the large $c$ limit.

In this section, we have clarified how (the Euclidian version of) the information loss problem is reproduced in the semiclassical approximation on the CFT side.
The mechanism for solving the information loss problem should be understood by investigating the $\fr{1}{c}$ corrections in large $c$ CFT.
This direction of research is ongoing, as seen in \cite{Chen2017}, and continues to develop.

\subsubsection{Entanglement Entropy}

\begin{figure}[t]
 \begin{center}
  \includegraphics[width=10.0cm,clip]{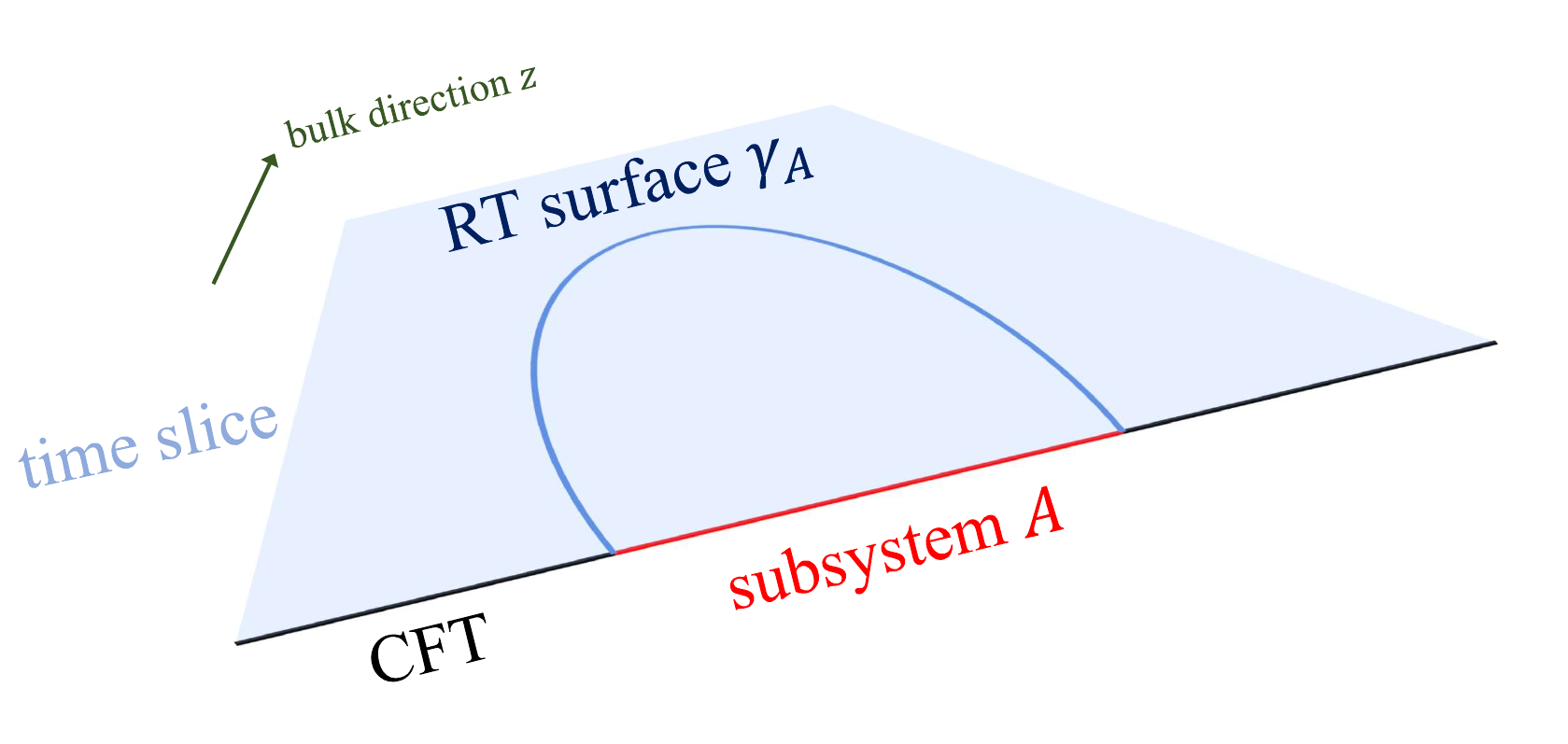}
 \end{center}
 \caption{Ryu-Takayanagi surface $\g_A$ corresponding to the entanglement entropy of subsystem $A$ on a constant-time slice of the CFT}
 \label{fig:RT}
\end{figure}

Here, as an application of the Heavy-Light Virasoro block,
we introduce the calculation of entanglement entropy.

In these lecture notes, entanglement entropy appeared in Section \ref{subsec:EE} as one method of measuring the central charge.
In the context of the AdS/CFT correspondence, this entanglement entropy plays an important role.
One reason is that the gravitational dual of entanglement entropy is given by a simple formula:
\begin{equation}\label{eq:RTformula}
S_A=\min_{\substack{\g_A \\ \del \g_A = \del A}} \fr{\text{Area}(\g_A)}{4G_N}.
\end{equation}
$\g_A$ is a surface homologous to the subsystem $A$ on the boundary CFT.
\footnote{
Roughly speaking, $A$ is homologous to $B$ if $A$ can be continuously deformed into $B$.
}
A depiction of $\g_A$ is provided in Figure \ref{fig:RT}.
This formula is known as the \und{Ryu-Takayanagi formula}, named after its discoverers \cite{Ryu2006}.
The minimal surface $\g_A$ is called the {\it Ryu-Takayanagi surface} (or RT surface).

\begin{figure}[t]
 \begin{center}
  \includegraphics[width=6.0cm,clip]{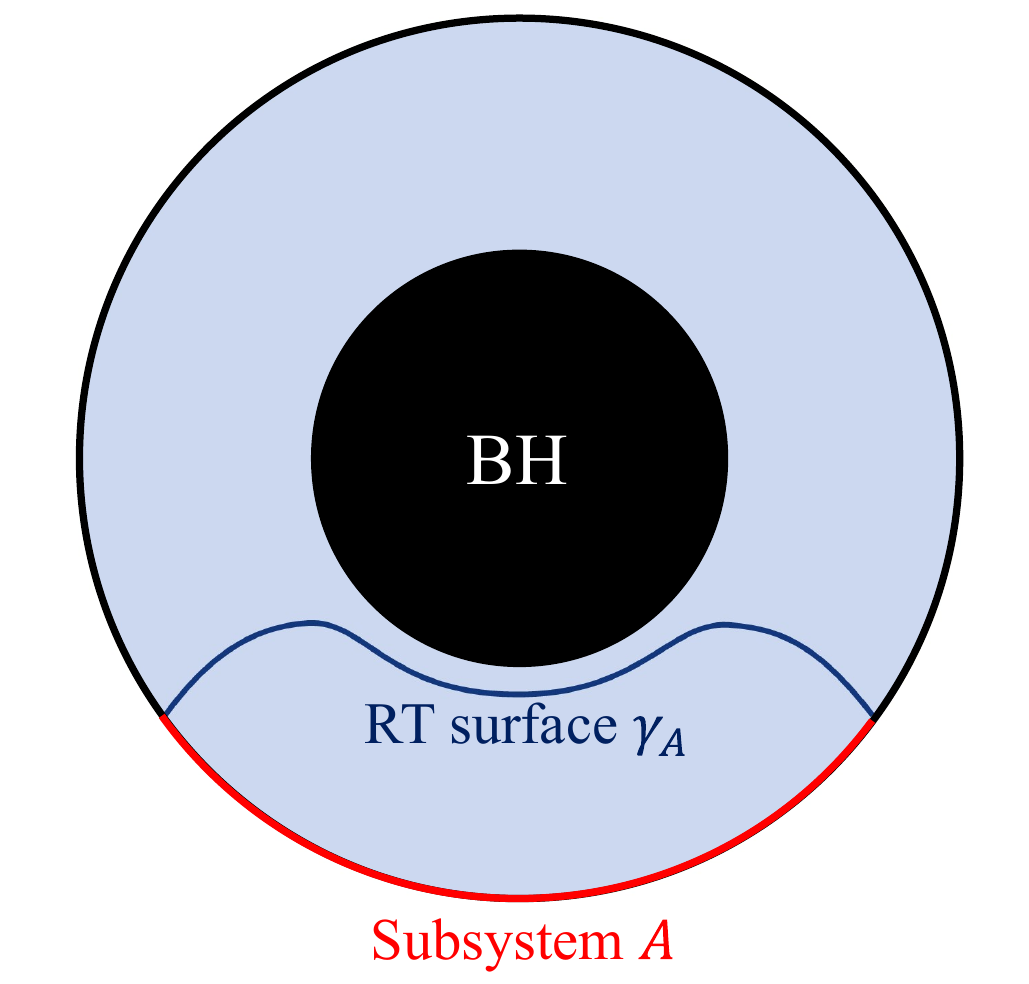}
 \end{center}
 \caption{RT surface in the BTZ black hole geometry}
 \label{fig:RT2}
\end{figure}

Let us compute the holographic entanglement entropy in the BTZ black hole geometry.
The RT surface is given by a surface as shown in Figure \ref{fig:RT2}.
Calculating the area of this surface in the BTZ black hole background, we obtain the following result:
\begin{equation}\label{eq:HSA}
\begin{aligned}
S_A &=\left\{
    \begin{array}{ll}
     \fr{c}{3} \log \fr{\sinh \pi T_H R }{\pi T_{H} \epsilon}, \ \ \ \ \   & \text{if \ }  0<R<\pi   ,\\
     \fr{c}{3} \log \fr{\sinh \pi T_H (2\pi-R) }{\pi T_{H} \epsilon}, \ \ \ \ \   & \text{if \ }  \pi<R<2\pi   .\\
    \end{array}
  \right.\\
\end{aligned}
\end{equation}
Here, we used the Brown-Henneaux formula $c=\fr{3l}{2G_N}$.
$R \in [0, 2\pi]$ is the length of the subsystem $A$.
$ 2\pi T_H  (= \sqrt{8M})$ is the temperature of the black hole,
and rewriting it using the conformal dimension, $2\pi i T_H = \alpha \equiv \sqrt{1-\fr{24}{c}h_H}$.
Since the AdS metric diverges at the asymptotic boundary,
the area of the surface extending from the asymptotic boundary also diverges.
Therefore, a cutoff parameter $\epsilon$ must be introduced near the asymptotic boundary to regularize the calculation.
\footnote{
A detailed explanation of how to introduce the cutoff can be found in \cite{Kusuki2023}.
}

The entanglement entropy for a state $\rho = \ket{\phi_H} \bra{\phi_H}$ can be calculated using the following formula \cite{Calabrese2004}:
\begin{equation}\label{eq:SA}
S_A = \lim_{n \to 1} \fr{1}{1-n} \log \braket{\phi_H^{\otimes n} |  \bar{\sigma}_n(R) \sigma_n(0)  | \phi_H^{\otimes n} },
\end{equation}
where $ \phi_H^{\otimes n}$ is a primary operator with conformal dimension $nh_H$,  
and the twist operator $\sigma_n$ is a primary operator with conformal dimension $h_n \equiv \frac{c}{24}\left(n-\frac{1}{n}\right)$.  
Since the information required for the calculations here is limited to the conformal dimensions,  
the details of the notation can be ignored.  
\footnote{
Let us denote the CFT as $\mathcal{M}$.  
The twist operator is defined as the endpoint of a $\mathbb{Z}_n$ symmetry defect in the tensor product $\mathcal{M}^{\otimes n}$.  
The correlation function $\braket{\phi_H^{\otimes n} |  \bar{\sigma}_n(R) \sigma_n(0)  | \phi_H^{\otimes n} }$  
is defined not on $\mathcal{M}$ but on $\mathcal{M}^{\otimes n}$,  
where $\phi_H^{\otimes n}$ is a tensor product of the primary operators defined on $\mathcal{M}^{\otimes n}$.  
Further details would deviate significantly from the purpose of these lecture notes, so they will not be covered here. See \cite{Asplund2015} for more details.
}

The CFT is defined on a cylinder $w = t_E + i\theta$ through the conformal transformation $z = 1 - \ex{-w}$,  
and the twist operators are inserted at $\theta = 0$ and $\theta = R$.  
Using the saddle-point approximation in the large $c$ limit,  
we can approximate the correlation function with a single Virasoro block.
$\phi_H$ represents a black hole microstate, which means $h_H = O(c)$.  
Additionally, the conformal dimension of $\sigma_n$ satisfies $h_n \ll 1$ when $n - 1 \ll 1$.  
Therefore, the correlation function appearing in (\ref{eq:SA}) can ultimately be approximated by the Heavy-Light vacuum block (\ref{eq:HHLLthermal}).
Substituting $t_E=iR$ into (\ref{eq:HHLLthermal}) and then into (\ref{eq:SA}), we obtain:
\begin{equation}\label{eq:SA2}
S_A \simeq \fr{c}{3} \log \fr{\sinh \pi T_H R }{\pi T_{H} \epsilon}, \ \ \ \ R\ll 2\pi.
\end{equation}
Here, the UV cutoff parameter $\epsilon$ is added through dimensional analysis.
\footnote{
To define entanglement entropy,
the Hilbert space needs to be decomposed as in (\ref{eq:Hdec}),
but the Hilbert space of QFT cannot be decomposed in principle.
Therefore, by introducing a lattice size cutoff parameter,
the QFT Hilbert space is approximately decomposed like a lattice system.
As a result, the entanglement entropy in QFT always includes a cutoff parameter.
How to specifically introduce the cutoff is discussed in \cite{Ohmori2015}.
A detailed explanation would be lengthy,
so here the UV cutoff parameter is heuristically determined by dimensional analysis.
In fact, it is known that the exact definition of the twist operator necessarily involves the cutoff parameter \cite{Lunin2001}.
}
The validity region of this result is limited because the saddle point approximation is used.
Although details are omitted here, by examining the correct choice of the saddle point and the validity region of the saddle point approximation,
the above result (\ref{eq:SA2}) is ultimately extended as follows \cite{Asplund2015}:
\begin{equation}
\begin{aligned}
S_A &=\left\{
    \begin{array}{ll}
     \fr{c}{3} \log \fr{\sinh \pi T_H R }{\pi T_{H} \epsilon}, \ \ \ \ \   & \text{if \ }  0<R<\pi   ,\\
     \fr{c}{3} \log \fr{\sinh \pi T_H (2\pi-R) }{\pi T_{H} \epsilon}, \ \ \ \ \   & \text{if \ }  \pi<R<2\pi   .\\
    \end{array}
  \right.\\
\end{aligned}
\end{equation}
This result beautifully matches (\ref{eq:HSA}).

The Ryu-Takayanagi formula has brought a quantum information-theoretic approach to quantum gravity,
establishing a new field that merges quantum gravity and quantum information.
Practically, it contributes to understanding the properties of quantum gravity by bringing inequalities known in quantum information theory into quantum gravity.

\subsubsection{Monodromy Method}

Here, we introduce another method for obtaining the Virasoro block in the large $c$ limit.
Since Liouville CFT has a continuous spectrum,
its correlation functions contain Virasoro blocks with a wide range of parameters.
The basic idea is to derive the expression for the Virasoro block by examining the correlation functions of Liouville CFT in the large $c$ limit.

Here, we focus on the application to the large $h$ Virasoro block (\ref{eq:large-h}),
but it is also applied to the Heavy-Light Virasoro block mentioned above and generalizations to $n$-point functions (e.g. \cite{Fitzpatrick2014, Asplund2015}).

\begin{enumerate}

\item {\it BPZ equation}\\
For the degenerate primary field $\Psi$ with conformal dimension $h_{21}=-\fr{1}{2}-\fr{3}{4}b^2$, we set up the BPZ equation (see Section \ref{subsec:BPZ}).
\footnote{
Although Liouville CFT does not have degenerate primary fields,
the analyticity of the correlation functions allows analytic continuation to $h_i \to h_{21}$,
and the correlation function after the analytic continuation satisfies the BPZ equation.
This method is also used in the derivation of the DOZZ formula.
}
\begin{equation}\label{eq:ODE}
\left[\frac{1}{b^2}\partial_z^2+\sum_{i=1}^4\left(\frac{h_i}{(z-z_i)^2}+\frac{1}{z-z_i}\partial_i\right)\right]
\langle\phi_4(z_4,\bz _4)\phi_3(z_3,\bz _3)\Psi(z,\bz )\phi_2(z_2,\bz _2)\phi_1(z_1,\bz _1)\rangle=0.
\end{equation}

\item {\it BPZ for each intermediate state}\\
We consider the large $c$ limit with $h_i=O(c)$.
In this limit, define $\Psi_p(z,\bz )$ as follows:
\begin{equation}
\Psi_p(z,\bz )
\equiv
\fr{\braket{\phi_4(z_4,\bz _4) \phi_3(z_3, \bz _3) \Psi(z, \bz ) \phi_{p,\bm{n}}(0,0)  }}
{\braket{\phi_4(z_4,\bz _4) \phi_3(z_3, \bz _3) \phi_{p,\bm{n}}(0,0)  }}.
\end{equation}
Here, it is important that $\Psi_p(z,\bz )$ does not depend on the label $\bm{n}$ of the descendant.
This is because,
by replacing the action of the Virasoro algebra on $\phi_p$ with the action on $\phi_4$, $\phi_3$, and $\Psi$ as in Figure \ref{fig:move},
the action of the Virasoro algebra on $\Psi$ is negligible compared to the other actions ($=O(c)$), 
since the conformal dimension of $\Psi$ is $O(c^0)$.

By using $\Psi_p$, the five-point function can be rewritten as follows:
\begin{equation}\label{eq:5dec}
\begin{aligned}
&\langle\phi_4(z_4,\bz _4)\phi_3(z_3,\bz _3)\Psi(z,\bz )\phi_2(z_2,\bz _2)\phi_1(z_1,\bz _1)\rangle\\
&= \sum_p \langle\phi_4(z_4,\bz _4)\phi_3(z_3,\bz _3)\Psi(z,\bz ) \pa{ \Pi_{h_p} \otimes \bar{\Pi}_{\bar{h}_p}}   \phi_2(z_2,\bz _2)\phi_1(z_1,\bz _1)\rangle \\
&= \sum_p \Psi_p(z,\bz ) C_{12p} C_{34p} \abs{\ca{F}^{21}_{34}(h_p|z_1,z_2,z_3,z_4)}^2.
\end{aligned}
\end{equation}
Here, $\Pi_{h_p}$ is the projection operator onto the Verma module $\ca{V}_{h_p}$.

Let us focus on the {\it monodromy}, which is a change of a function as we run around a singularity.
For example, the logarithmic function $\log z$ has a singularity at $z=0$, and therefore, it exhibits a nontrivial monodromy around $z=0$:
\begin{equation}
\log z \to \log z + 2 \pi i.
\end{equation}

The BPZ equation is invariant under a monodromy of $z_2$ around $z_1$.
Therefore, when expressing the solution of the BPZ equation as a linear combination of functions with different monodromies around $z_1$,
each function is expected to be a solution separately.
When decomposing the solution as in (\ref{eq:5dec}), the monodromy of $z_2$ around $z_1$ for each $p$ is given by the phase $2\pi (h_p-h_1-h_2)$ (see (\ref{eq:F0})),
so the following separate differential equation holds for each $p$:
\begin{equation}\label{eq:BPZ2}
\left[\frac{1}{b^2}\partial_z^2+\sum_{i=1}^4\left(\frac{h_i}{(z-z_i)^2}+\frac{1}{z-z_i}\partial_i\right)\right]
\Psi_p(z,\bz ) C_{12p} C_{34p} \abs{\ca{F}^{21}_{34}(h_p|z_1,z_2,z_3,z_4)}^2=0.
\end{equation}

\item {\it Ward--Takahashi identity}\\
Define the accessory parameters $C_i$ as follows:
\begin{equation}
C_i \equiv -b^2  \del_i \log \pa{\Psi_p(z,\bz ) C_{12p} C_{34p} \abs{\ca{F}^{21}_{34}(h_p|z_1,z_2,z_3,z_4)}^2} .
\end{equation}
By following the same procedure as the global Ward-Takahashi identity (\ref{eq:GWT}),
we obtain the following system of equations for the accessory parameters:
\begin{equation}
\begin{aligned}
&\sum_i C_i=0,  \\
&\sum_i \left(C_i z_i-\delta_i\right)=0,  \\
&\sum_i\left(C_i z_i^2-2\delta_i z_i\right)=0.
\end{aligned}
\end{equation}
Here, $\delta_i \equiv b^2 h_i$. Solving this system of equations and setting $(z_1,z_2,z_3,z_4) \to (0,x,1,\infty)$, we obtain:
\begin{equation}
\begin{aligned}
C_1&= -(\delta_1 +\delta_2 +\delta_3 - \delta_4) +C_2(x-1), \\
C_3 &= \delta_1 +\delta_2 +\delta_3 - \delta_4 -C_2 x, \\
C_4 &= 0.
\end{aligned}
\end{equation}
Substituting these into (\ref{eq:BPZ2}) and setting $(z_1,z_2,z_3,z_4) \to (0,x,1,\infty)$, we obtain the following differential equation:
\begin{equation}\label{eq:BPZ3}
\left[\partial_z^2+\frac{\delta_1}{z^2}+\frac{\delta_2}{(z-x)^2}+\frac{\delta_3}{(1-z)^2}+\frac{\delta_1+\delta_2+\delta_3-\delta_4}{z(1-z)}-\frac{C_2 x(1-x)}{z(z-x)(1-z)}\right]\Psi_p=0.
\end{equation}
Assume that the Virasoro block behaves as follows in the large $c$ limit:
\begin{equation}\label{eq:semif}
\mathcal{F}^{21}_{34}(p|x)\sim \ex{-\frac{c}{6} f_{cl}(p|x)  }.
\end{equation}
This assumption is justified by the expression of the correlation function in Liouville CFT.
Thus, the accessory parameter is expressed as follows:
\begin{equation}\label{eq:Acc}
C_2=\del_x f_{cl}(p|x)  .
\end{equation}

\item {\it Monodromy equation}\\
Now, the problem of finding the Virasoro block has been reduced to the problem of determining the accessory parameter $C_2$.
In the next step, we use the { \it monodromy method}, which is a method to determine the accessory parameter so that it reproduces the known monodromy of the solution.
Since $\Psi$ is a degenerate field with conformal dimension $h_{21}$, the BPZ equation gives the following OPE:
\begin{equation}
\Psi(z) \phi_p(0) \sim C_+ \phi_{+}z^{ \fr{1}{2} \pa{  1+\sqrt{1-4b^2h_p}}} \pa{1 + O(z)}   + C_-   \phi_{-}z^{ \fr{1}{2} \pa{  1-\sqrt{1-4b^2h_p}}} \pa{1 + O(z)} .
\end{equation}
$\phi_{\pm}$ are the possible primary operators in the OPE determined by the BPZ equation (\ref{eq:secondBPZ}),  
and $C_{\pm}$ are their respective OPE coefficients.  
Let the solutions to (\ref{eq:BPZ3}) corresponding to each intermediate state be $(\Psi_+, \Psi_-)$,  
and the solutions after one cycle of $\Psi$ be $(\Psi'_+, \Psi'_-)$.  
The relationship between these solutions can be expressed using the monodromy matrix $M$ as follows:  
\begin{equation}\label{eq:Monodromy}
\begin{aligned}
  \left(
    \begin{array}{c}
       \Psi'_+   \\
       \Psi'_-   \\
    \end{array}
  \right)
  =
  M
    \left(
    \begin{array}{c}
      \Psi_+    \\
       \Psi_-   \\
    \end{array}
  \right),
  \ \ \ \ \ \ \ 
  M=\begin{pmatrix}
\ex{i\pi  \left(  1+\sqrt{1-4b^2h_p} \right)}  & 0\\
0 & \ex{i\pi\left(1-\sqrt{1-4b^2h_p}\right)}
\end{pmatrix}.
\end{aligned}
\end{equation}
Thus, the accessory parameter $C_2$ is determined by the condition that the monodromy of the solutions to the differential equation is given by (\ref{eq:Monodromy}).

\item {\it WKB approximation}\\
To solve the monodromy method, it is necessary to specify the form of the solution to some extent initially.
Perturbation methods or WKB methods are used for this purpose.
Here, we use the WKB method under the condition $\delta_p \to \infty$.
Approximating the differential equation (\ref{eq:BPZ3}) under $\delta_i \ll \delta_p$, we obtain the following approximate solution:
\begin{equation}\label{eq:Msolution}
\Psi_p\simeq \exp\left[\pm\sqrt{x(1-x)C_2}\int_{z_0}^z\frac{dz'}{\sqrt{z'(1-z')(z'-x)}}\right].
\end{equation}
Imposing that this solution is consistent with the monodromy equation (\ref{eq:Monodromy}), we obtain the following equation for $C_2$:
\begin{equation}\label{eq:C2}
C_2\simeq -\frac{\pi^2 b^2 h_p}{x(1-x)K(x)^2}.
\end{equation}
From the relationship between the accessory parameter and the Virasoro block (\ref{eq:Acc}), we obtain
\begin{equation}
\mathcal{F}^{21}_{34}(h_p|x)\simeq \pa{16q}^{h_p},\ \ \ \ \ \ q(x)=\ex{-\pi \fr{K(1-x)}{K(x)}}.
\end{equation}
Furthermore, by solving up to higher orders, we finally obtain
\begin{tcolorbox}[title=Large $h$ Virasoro block]
    \begin{equation}
\mathcal{F}^{21}_{34}(h_p|x)\simeq(16q)^{h_p-\frac{c-1}{24}}
(\theta_3(q))^{\frac{c-1}{2}-4(h_1+h_2+h_3+h_4)}
x^{\frac{c-1}{24}-h_1-h_2}(1-x)^{\frac{c-1}{24}-h_2-h_3},
\ \ \ \ \ \ 
h_p \gg c
.
\end{equation}
\end{tcolorbox}
\noindent
In fact, it can be proven that this result holds for any $c$.
The proof is somewhat complicated, but interested readers should refer to \cite{Zamolodchikov1987}.
This large $h$ limit of the Virasoro block is the leading term of the Zamolodchikov $h$-recursion relation (\ref{eq:large-h}).

\end{enumerate}

\section{Holographic CFT and Modular Bootstrap}

In Section \ref{sec:AdSCFT}, we introduced the concept of a holographic CFT.
Although the details of holographic CFTs are not fully understood, there are several known properties, as discussed in Section \ref{sec:AdSCFT}.
For example:
\begin{itemize}

\item
$c \gg 1$ \\
Especially, $c > 1$ is important.

\item
irrational \\
This is because the number of primary fields is clearly not finite, as suggested by the fact that the number of states is given by the black hole entropy.

\item
compact: discrete spectrum including the vacuum state \\
This derives from the fact that the Einstein equation has the vacuum solution.
\footnote{
Generally, physical systems are expected to have a discrete spectrum, therefore, we implicitly assume a discrete spectrum from the start.
It might seem contradictory that black hole states are characterized by the continuous parameter $M$,
but this continuity is due to the large $c$ limit, where the differences between adjacent energy levels become almost indistinguishable.
}

\item
unitary

\end{itemize}
In situations where some properties of a given CFT are known,
the conformal bootstrap can be a powerful tool.
As an example, in this section, we introduce the application of the modular bootstrap to holographic CFTs.

\subsection{Cardy Formula and Black Hole Entropy}\label{subsec:Cardy}
The Cardy formula implies a universal property of the number of states at high energy \cite{Cardy1986a}.
It is a result derived from the modular invariance and simple calculations,
making it one of the widely known applications of the modular bootstrap equation.

\subsubsection{Derivation of Cardy Formula}

Assuming a unitary compact CFT, consider the following modular invariant partition function:
\begin{equation}
Z(\beta) \equiv \tr \ex{-\beta H} = \sum_{\Delta} d_\Delta \ex{-\beta (\Delta-\fr{c}{12})}.
\end{equation}
Here, the sum runs over all states, and $d_\Delta$ is the number of states with scaling dimension $\Delta=h+\bar{h}$.
Define the density of states as follows:
\begin{equation}
\rho(\Delta) = \sum_{p} d_{\Delta_p} \delta(\Delta - \Delta_p),
\end{equation}
where $p$ is a label for a state in the Hilbert space.
The partition function can be expressed in terms of the density of states as follows:
\begin{equation}
Z(\beta)= \int_0^{\infty} d \Delta \  \rho(\Delta) \ex{-\beta (\Delta-\fr{c}{12})}.
\end{equation}
The modular bootstrap equation is expressed as:
\begin{equation}
\int_0^{\infty} d \Delta \ \rho(\Delta) \ex{-\beta (\Delta-\fr{c}{12})}
=
\int_0^{\infty} d \Delta \ \rho(\Delta) \ex{-\fr{ (2\pi)^2 }{\beta} (\Delta-\fr{c}{12})}.
\end{equation}
Consider the $\beta \to 0$ limit of this modular bootstrap equation.
This limit corresponds to taking the high-temperature limit on the left-hand side and the low-temperature limit on the right-hand side,
hence it is also called {\it high-low temperature duality}.
In this limit, the right-hand side can be approximated by the contribution from the vacuum state ($\Delta=0$):
\begin{equation}
\int_0^{\infty} d \Delta \  \rho(\Delta) \ex{-\beta (\Delta-\fr{c}{12})}
\simeq
 \ex{\fr{ (2\pi)^2 }{\beta} \fr{c}{12}}.
\end{equation}
One finds that this is a Laplace transform.
Therefore, by the inverse Laplace transform, we obtain
\begin{equation}
\rho(\Delta) = \fr{1}{2\pi i } \int_{-i\infty}^{i \infty} d\beta \  \ex{\beta \pa{\Delta-\fr{c}{12}} + \fr{ (2\pi)^2 }{\beta} \fr{c}{12}}
\simeq \ex{2\pi \sqrt{ \fr{c}{3} \pa{\Delta-\fr{c}{12}} }}.
\end{equation}
The second expression comes from the saddle point approximation.
The saddle point is given by:
\begin{equation}
\beta_* = 2\pi \sqrt{\fr{c}{12} \fr{1}{\Delta-\fr{c}{12}}}.
\end{equation}
The approximation by the vacuum state is valid only for $\beta \ll 1$,
so this saddle point approximation is valid only for $\Delta \gg c$.
Thus, we obtain the following \und{Cardy formula}:
\begin{tcolorbox}[title=Cardy formula]
\begin{equation}\label{eq:CardyEntropy}
\log \rho(\Delta) \simeq 2\pi \sqrt{ \fr{c}{3} \pa{\Delta-\fr{c}{12}} }, \ \ \ \ \ \Delta \gg c.
\end{equation}
\end{tcolorbox}
\noindent

\subsubsection{Bekenstein-Hawking Entropy}
The Cardy formula plays an important role in AdS/CFT.
Consider the BTZ black hole metric with mass $M$,
\begin{equation}
ds^2 = -\pa{\fr{r^2}{l^2}-8G_N M}dt^2+ \fr{dr^2}{\fr{r^2}{l^2}-8G_N M} +r^2 d\phi^2.
\end{equation}
Here, $\phi \sim \phi+2\pi$. Also, $G_N$ is the Newton constant and $l$ is the AdS radius.
The entropy of this black hole is given by the Bekenstein-Hawking entropy formula:
\begin{equation}\label{eq:BHentropy}
S=\fr{\mathrm{Area}}{4G_N} = \fr{2\pi l}{4 G_N} \sqrt{8 G_N M},
\end{equation}
where $\mathrm{Area}$ represents the area of the event horizon of the black hole.
According to the AdS/CFT correspondence, the BTZ black hole with mass $M$ can be interpreted as a state with the following conformal dimension \cite{Kraus2008} (see also Section \ref{subsubsec:thermal}):
\begin{equation}
\Delta-\fr{c}{12} = Ml.
\end{equation}
With this relation, we can re-express the black hole entropy (\ref{eq:BHentropy}) as
\begin{equation}
S= 2 \pi \sqrt{\fr{c}{3} \pa{\Delta- \fr{c}{12}}}.
\end{equation}
Here, we used the Brown-Henneaux formula $c=\fr{3l}{2G_N}$.
Compared with the Cardy formula (\ref{eq:CardyEntropy}),
we see that the Bekenstein-Hawking entropy perfectly matches the Cardy formula.

Hawking linked the area of the black hole horizon with entropy due to its analogy with thermodynamics,
but it was extremely difficult at the time to clarify the statistical mechanics behind black hole thermodynamics.
The Cardy formula provided evidence that the BTZ black hole entropy indeed corresponds to counting quantum states.

\subsubsection{Rotating BTZ Black Hole}\label{subsubsec:rotating}
The Cardy formula can be generalized to cases with nonzero spin.
In the derivation of the Cardy formula above, we assumed $\beta=\bar{\beta}$, but now we consider removing this assumption,
\begin{equation}
Z(\beta, \bar{\beta}) \equiv \tr \ex{-\beta \pa{L_0-\fr{c}{24}} - \bar{\beta}\pa{\bar{L}_0-\fr{c}{24}}  } = \sum_{h, \bar{h}} d_{h, \bar{h}} \ex{-\beta \pa{h-\fr{c}{24}} - \bar{\beta}\pa{\bar{h}-\fr{c}{24} }}.
\end{equation}
Performing independent inverse Laplace transforms for $\beta, \bar{\beta} \to 0$, we ultimately obtain the following Cardy formula:
\begin{equation}\label{eq:CardySpin}
\log \rho(h, \bar{h}) \simeq 2\pi \sqrt{ \fr{c}{6} \pa{h-\fr{c}{24}} }   + 2\pi \sqrt{ \fr{c}{6} \pa{\bar{h}-\fr{c}{24}} },  \ \ \ \ \ h, \bar{h} \gg c.
\end{equation}

Let us compare this result with the gravity calculation.
The metric of the BTZ black hole with mass $M$ and angular momentum $J$ is given by
\begin{equation}
ds^2 = -\pa{\fr{r^2}{l^2}+ \fr{16G_N^2 J^2}{r^2}  -8G_N M  }dt^2+ \fr{dr^2}{\fr{r^2}{l^2} + \fr{16G_N^2 J^2}{r^2}  -8G_N M} +r^2 \pa{d\phi  -\fr{4G_N J}{r^2} dt}^2.
\end{equation}
The outer and inner horizons of this black hole are given by
\begin{equation}
r_{\pm}^2 = 4G_N M l^2 \pa{1 \pm \sqrt{1-\pa{\fr{J}{Ml}}^2}}.
\end{equation}
The relation between $(M,J)$ and the horizons is
\begin{equation}\label{eq:MJ}
M= \fr{r_+^2 + r_-^2}{8G_N l^2}, \ \ \ \ J= \fr{r_+ r_-}{4 G_N l}.
\end{equation}
According to the AdS/CFT correspondence, the BTZ black hole with mass $M$ and angular momentum $J$ corresponds to the following state \cite{Kraus2008}:
\begin{equation}\label{eq:MJ2}
h-\fr{c}{24} = \fr{1}{2}(Ml-J)= \fr{(r_+ + r_-)^2}{16G_N l}    , \ \ \ \ \ \bar{h}-\fr{c}{24}=\fr{1}{2}(Ml+J)  = \fr{(r_+ - r_-)^2}{16G_N l}.
\end{equation}
With these relations, the Bekenstein-Hawking entropy can be expressed as
\begin{equation}
S = \fr{2\pi r_+}{4G_N} = 2\pi \sqrt{ \fr{c}{6} \pa{h-\fr{c}{24}} }   + 2\pi \sqrt{ \fr{c}{6} \pa{\bar{h}-\fr{c}{24}} }.
\end{equation}
This perfectly matches (\ref{eq:CardySpin}).

\subsection{Hellerman Bound and Pure Gravity}\label{subsec:Hellerman}

Consider Einstein gravity without any matter (called pure gravity),
\begin{equation}
S_{EH}=\fr{1}{16\pi G_N} \int d^3x \ \sqrt{g} \pa{R+\fr{2}{l^2}}.
\end{equation}
The solutions to the Einstein equation include the thermal AdS (i.e. the empty AdS) and the BTZ black holes.
The CFT state corresponding to the empty AdS has a conformal dimension $\Delta=0$.
On the other hand, according to the AdS/CFT correspondence, the BTZ black hole with mass $M$ and angular momentum $J$ corresponds to the state given by (\ref{eq:MJ2}):
\begin{equation}
h-\fr{c}{24} = \fr{1}{2}(Ml-J), \ \ \ \ \ \bar{h}-\fr{c}{24}=\fr{1}{2}(Ml+J).
\end{equation}
The condition for the black hole to have a smooth horizon is simply written by $r_{\pm}\geq 0$.
This can be translated using the relation (\ref{eq:MJ}) as
\begin{equation}
\abs{J}\leq Ml.
\end{equation}
This condition is called the {\it cosmic censorship hypothesis}.
The cosmic censorship hypothesis implies that states interpreted as black holes are expected to satisfy the following inequality:
\begin{equation}\label{eq:CCH}
h \geq \fr{c}{24} , \ \ \ \ \ \ \bar{h}\geq \fr{c}{24}.
\end{equation}
Since no matter exists by assumption, the spectrum of the CFT dual to pure gravity must have a large gap proportional to $c$:
\begin{equation}\label{eq:Gap}
\Delta_1 - \Delta_0 = \fr{c}{12}.
\end{equation}
Here, $\Delta_i$ is the $i$-th excited state ($\Delta_0 < \Delta_1 < \cdots$).
Note that $\Delta_0 = 0$ for the empty AdS, and $\Delta_1 = \fr{c}{12}$ which is the lower bound on the black hole energy (\ref{eq:CCH}).

Pure gravity is classically allowed,
but it is not clear if it can exist when quantum effects are included.
One of the important tasks is examining the possibility of pure gravity as a consistent theory of quantum gravity,
and the modular bootstrap can help with this task.
Consider the following problem:
\begin{tcolorbox}
What is the upper bound on $\Delta_1 - \Delta_0$?
\end{tcolorbox}
\noindent
If the modular bootstrap provides an upper bound such that
\begin{equation}
\Delta_1 - \Delta_0 < \fr{c}{12}
\end{equation}
(without the equality), then pure gravity cannot satisfy (\ref{eq:Gap}) and thus cannot exist at the quantum level.
On the other hand, if pure gravity does exist at the quantum level,
since the gap in all existing CFTs is smaller than (\ref{eq:Gap}),
pure gravity is expected to saturate the gap bound:
\begin{equation}\label{eq:expectedGap}
\Delta_1 - \Delta_0 \leq \fr{c}{12}.
\end{equation}

This problem of exploring the upper bound of the gap was proposed by Hellerman \cite{Hellerman2009} and has been tackled by many papers such as \cite{Friedan2013} and \cite{Collier2016}. However, it remains an open problem. Hellerman's work \cite{Hellerman2009} successfully provided a weaker version of the upper bound (\ref{eq:expectedGap}), and subsequent research has essentially developed Hellerman's ideas. Therefore, here we will introduce the ideas from \cite{Hellerman2009}.

The core idea is to extract an infinite number of equations explicitly by expanding the modular bootstrap equation around the fixed point of the modular transformation, $\beta=2\pi$. Specifically, the following equations are obtained:
\begin{equation}
\left. \pa{ \beta \fr{\del}{\del \beta}}^N Z(\beta) \right|_{\beta=2\pi} = 0, \ \ \ \ N \in 2\bb{Z}_{\geq0}+1.
\end{equation}
As a warm-up, let us consider $N=1$,
\begin{equation}\label{eq:N=1}
\sum_{i} d_i \pa{\Delta_i-\fr{c}{12}} \ex{-2\pi \pa{\Delta_i-\fr{c}{12}}} = 0.
\end{equation}
Here, we set $\Delta_0 < \Delta_1 < \cdots$.
Assuming $\Delta_0 > \fr{c}{12}$,
all terms on the left side of (\ref{eq:N=1}) would be positive, and their sum would be nonzero, leading to a contradiction.
Therefore, the conformal dimension of the lowest energy state $\Delta_0$ must satisfy the following inequality:
\begin{equation}\label{eq:N=1WU}
\Delta_0  \leq \fr{c}{12}.
\end{equation}
For unitary compact CFTs, the vacuum state $\Delta_0=0$ satisfies this inequality,
so this inequality is not very useful, but it is sufficient to understand the basic idea of Hellerman.
The result of \cite{Hellerman2009} is simply an extension of this idea to higher $N$.

Consider the case of $N=3$.
For brevity, let $E_i \equiv \Delta_i - \fr{c}{12}$, then we obtain
\begin{equation}
\sum_i d_i \pa{E_i - 6\pi E_i^2 + 4\pi^2 E_i^3}\ex{-2\pi E_i}=0.
\end{equation}
From (\ref{eq:N=1}), the contribution of the first term is zero.
To apply the same logic as (\ref{eq:N=1WU}),
we want the lowest conformal dimension in the sum to be $\Delta_1$.
To do this, we consider removing the contribution of $i=0$ from the sum using (\ref{eq:N=1}),
\begin{equation}\label{eq:N=3}
\begin{aligned}
&\sum_i d_i E_i  \pa{ - 3 E_i + 2\pi E_i^2}\ex{-2\pi E_i} - \sum_i d_i E_i  \pa{ - 3 E_0 + 2\pi E_0^2}\ex{-2\pi E_i} \\
&=
\sum_i d_i E_i  (E_i-E_0)\pa{3-2\pi(E_i+E_0) } = 0.
\end{aligned}
\end{equation}
Assume the following:
\begin{equation}
3-2\pi(E_1+E_0) < 0.
\end{equation}
In this case, all terms on the left side of (\ref{eq:N=3}) would be negative, leading to a contradiction with zero.
Consequently, we obtain the upper bound on $\Delta_1$ so-called the \und{Hellerman bound},
\begin{tcolorbox}[title=Hellerman bound]
\begin{equation}\label{eq:Hellerman}
3-2\pi(E_1+E_0) \geq 0 \ \ \ \ \Leftrightarrow E_1\leq \fr{3}{2\pi} -E_0 \ \ \ \ \Leftrightarrow \Delta_1 \leq \fr{c}{6} + \fr{3}{2\pi}.
\end{equation}
\end{tcolorbox}

Here, note that $\Delta_1$ is the conformal dimension of the first excited state.
Since every CFT has the energy-momentum tensor with $\Delta=2$,
for large $c$, (\ref{eq:Hellerman}) is trivially satisfied.
In the AdS/CFT correspondence, descendant states correspond to excitations by diffeomorphisms (i.e. boundary gravitons),
and the empty AdS and the BTZ black holes correspond to primary states.
Therefore, the upper bound we are really interested in is the upper bound on the conformal dimension of the first excited \und{primary} state.
To provide an upper bound for the first excited primary state,
one can repeat the previous discussion after replacing the function basis with Virasoro characters:
\begin{equation}
Z(\beta) = \sum_{i: \mathrm{primary}} d_i \abs{\chi_i(\beta)}^2.
\end{equation}
Since Virasoro characters are more complicated than exponential functions,
the calculation is much harder, but the result is almost the same as (\ref{eq:Hellerman}).
Readers interested in the generalization to primary states should refer to \cite{Hellerman2009}.

It is natural to expect that considering higher $N$ will provide stronger upper bounds.
Achieving this by hand is almost impossible,
but \cite{Collier2016} does similar work numerically.
In \cite{Collier2016}, they provide the improved upper bound:
\begin{equation}
\Delta_1 < \fr{c}{8}.
\end{equation}
The study in this direction continues to develop even today;
to the best knowledge of the author,
the best analytical bound is given in \cite{Hartman2019}, and the best numerical bound is provided in \cite{AfkhamiJeddi2019}.

\subsection{Hartman-Keller-Stoica Bound and Black Hole Entropy}\label{subsubsec:HKS}

In Section \ref{subsec:Cardy}, we saw how the Cardy formula beautifully reproduces the Bekenstein-Hawking entropy. However, there is a crucial difference between the Cardy formula and the Bekenstein-Hawking entropy: their applicable ranges.
\begin{description}

\item[Cardy Formula]\mbox{}\\
\begin{equation}
S \simeq 2\pi \sqrt{ \fr{c}{3} \pa{\Delta-\fr{c}{12}} } \ \ \ \ \ (\Delta \gg c).
\end{equation}

\item[Bekenstein-Hawking Entropy]\mbox{}\\
\begin{equation}
S \simeq 2\pi \sqrt{ \fr{c}{3} \pa{\Delta-\fr{c}{12}} } \ \ \ \ \ (\Delta > \fr{c}{12}, \ \ c\gg 1).
\end{equation}

\end{description}
The Cardy formula holds for any unitary compact CFT, not limited to holographic CFTs. On the other hand, the Bekenstein-Hawking entropy formula is not expected to hold for general CFTs. Therefore, some specific properties of holographic CFTs must be responsible for the extension of the validity region of the Cardy formula. This condition for the extended validity region was partially investigated by Hartman, Keller, and Stoica \cite{Hartman2014}. This section explains their results.

Consider dividing the spectrum into two regions by a small positive real number $\epsilon \ll 1$,
\begin{equation}
Z_L(\beta) \equiv \sum_{-\fr{c}{12} \leq E \leq \epsilon} d_E \ex{ -\beta E},
\ \ \ \ \  
Z_H(\beta) \equiv \sum_{ \epsilon < E} d_E \ex{ -\beta E}.
\end{equation}
From the modular bootstrap equation, we have
\begin{equation}
Z_L(\beta) - Z_L \pa{ \fr{(2\pi)^2 }{\beta}}
=
Z_H \pa{ \fr{(2\pi)^2 }{\beta}} - Z_H(\beta).
\end{equation}
For $\beta>2\pi$, we obtain the following inequality,
\begin{equation}
Z_H(\beta) 
=
\sum_{ \epsilon < E}  \ex{\pa{ \fr{(2\pi)^2}{\beta} -\beta }E}  d_E \ex{ - \fr{(2\pi)^2}{\beta} E}
\leq
\varepsilon Z_H \pa{\fr{(2\pi)^2}{\beta}},
\end{equation}
where we define $\varepsilon  \equiv   \ex{\pa{ \fr{(2\pi)^2}{\beta} -\beta }\epsilon}$, which satisfies $0<\varepsilon <1$ for $\beta>2\pi$.
Subtracting $\varepsilon Z_H(\beta)$ from both sides and using the modular bootstrap equation, we get
\begin{equation}
(1-\varepsilon) Z_H(\beta)
\leq \varepsilon \pa{    Z_H \pa{\fr{(2\pi)^2}{\beta}} - Z_H(\beta) }
= \varepsilon \pa{    Z_L(\beta) - Z_L \pa{\fr{(2\pi)^2}{\beta}}  }
\leq \varepsilon Z_L(\beta).
\end{equation}
Adding $(1-\varepsilon)Z_L(\beta)$ to both sides, we finally obtain
\begin{equation}
Z(\beta) \leq \fr{ Z_L(\beta) }{1-\varepsilon}.
\end{equation}
By adding a trivial lower bound to this inequality, we get
\begin{equation}\label{eq:Sand}
\log Z_L(\beta) \leq \log Z(\beta) \leq \log Z_L(\beta) - \log(1-\varepsilon).
\end{equation}
Rewrite $Z_L(\beta)$ as
\begin{equation}
Z_L(\beta) = \ex{\beta \fr{c}{12}} \int^{\fr{c}{12} + \epsilon}_0 d \Delta \  \rho(\Delta) \ex{-\beta \Delta}.
\end{equation}
If the density of states satisfies
\begin{equation}\label{eq:HKS}
\rho(\Delta)  <  \ex{2\pi \Delta}, \ \ \ \ \Delta < \fr{c}{12}+\epsilon,
\end{equation}
then, for $\beta > 2\pi$, we have
\begin{equation}\label{eq:sparseness1}
\log \pa{ \int^{\fr{c}{12} + \epsilon}_0 d \Delta \  \rho(\Delta) \ex{-\beta \Delta} }= O(c^0)
\end{equation}
so in the large $c$ limit, the partition function can be approximated by the vacuum contribution,
\begin{equation}
\log Z_L(\beta)  \simeq \beta \fr{c}{12}.
\end{equation}
In (\ref{eq:Sand}), since $\log(1-\varepsilon) = O(c^0)$, by the squeeze theorem, we get
\begin{equation}
\log Z(\beta) \simeq \beta \fr{c}{12} \ \  \ \ \ \ \ (\beta>2\pi , \  c \gg 1).
\end{equation}
This result indicates that the vacuum approximation region extends from $\beta \gg 1$ to $\beta > 2\pi$ in the large $c$ limit, provided (\ref{eq:HKS}) is satisfied. This extension of the vacuum approximation region leads to the extended validity region of the Cardy formula as follows:
\begin{tcolorbox}[title=HKS formula]
    \begin{equation}\label{eq:Cardy2}
S \simeq 2\pi \sqrt{ \fr{c}{3} \pa{\Delta-\fr{c}{12}} } \ \ \ \ \ (\Delta > \fr{c}{6}, \ \ c\gg 1).
\end{equation}
\end{tcolorbox}
\noindent
The validity region $\Delta > \fr{c}{6}$ is smaller than the expected range $\Delta > \fr{c}{12}$, but it is significantly extended compared to $\Delta \gg c$. Thus, it is understood that the spectrum of holographic CFTs must at least satisfy (\ref{eq:HKS}). This upper bound on the density of states is called the \und{Hartman-Keller-Stoica (HKS) bound}. To achieve the extension of the validity region to $\Delta > \fr{c}{12}$, conditions stronger than the HKS bound are required, but the exact conditions that lead to $\Delta > \fr{c}{12}$ are still unknown.

Previously we have derived in the $c\to\infty$ limit 
\begin{equation}
\log Z(\beta)\sim \begin{cases}
\frac{c}{12} \beta \,, &\quad \beta>2\pi\,,\\
\frac{c}{12}\frac{(2\pi)^2}{\beta}\,, & \quad \beta<2\pi.
\end{cases}
\end{equation}
Thus, the black hole saddle dominates below $\beta=2\pi$ and gives rise to the black hole entropy formula \eqref{eq:Cardy2} in the microcanonical ensemble.  Above $\beta=2\pi$,  the thermal AdS dominates. 

Now what happens if we turn on both $\beta$ and $\bar{\beta}$ with $\beta\neq\bar{\beta}$?  The partition function $Z(\beta,\bar\beta)$ becomes sensitive to both $h$ and $\bar h$, particularly,  on the spectrum of twist, defined as $2\text{min}(h,\bar h)$.  In this setup, to derive the universality of free energy in the large $c$ limit,  it is not enough to assume the sparseness condition in \eqref{eq:HKS}.   On top of that,  one needs to assume the sparseness of the states with twist below $c/12$ i.e. 
\begin{equation}\label{eq:HKS2}
\rho(h,\bar h) < e^{4\pi\sqrt{h\bar h}}\,, \quad \text{min}(h,\bar h)\leq \fr{c}{24}.
\end{equation}
The above implies $\rho(h,\bar h)< e^{2\pi(h+\bar h)}$ and ensures
\begin{equation}
\log \pa{ \int_{0\leq \text{min}(h,\bar h)\leq c/24} d h\ d\bar h\  \rho(h,\bar h) \ex{-\beta h-\bar{\beta} \bar{h}} }= O(c^0)\,,\quad \beta\bar\beta>4\pi^2\,.
\end{equation}
 Under these assumptions, given in \eqref{eq:HKS} and \eqref{eq:HKS2},  HKS conjectured
\begin{equation}
\log Z(\beta,\bar\beta)\sim \begin{cases}
\frac{c}{24} (\beta+\bar\beta)\,, &\quad \beta\bar{\beta}>4\pi^2\,,\\
\frac{c}{24}\left(4\pi^2/\beta+4\pi^2/\bar\beta\right)\,, & \quad \beta\bar{\beta}<4\pi^2.
\end{cases}
\end{equation}
The proof of this conjecture is non-trivial and has only recently been proven in \cite{Dey:2024nje}. The above produces the Cardy formula for the black hole entropy in the following extended regime: 
\begin{equation}\label{eq:extended}
    \left(h-\frac{c}{24}\right)    \left(\bar h-\frac{c}{24}\right)>\pa{\frac{c}{24}}^2.
\end{equation}

\section{Holographic CFT and Conformal Bootstrap}

In this section, we consider CFTs satisfy the following conditions:
\begin{itemize}

\item
$c>1$

\item
compact: discrete spectrum including the vacuum state

\item
unitary

\item
twist gap: $\min(h,\bar{h})>0$ for any non-vacuum state \\
This particularly implies that there are no symmetries other than the Virasoro symmetry.

\end{itemize}
The last condition, the twist gap, is a newly added condition.
Note that a CFT satisfying these conditions is necessarily irrational.
In actual research on solving holographic CFTs using conformal bootstrap,
these assumptions are often made.
In the following, we will refer to CFTs with these properties as {\it pure CFT}.
\footnote{
In Section \ref{sec:classification}, we examined specific examples of CFT,
but none of them are pure CFTs.
For example, minimal models and WZW models are not irrational,
the free boson CFT lacks a twist gap,
and Liouville CFT is non-compact.
In fact, no pure CFTs have been found among CFTs whose exact CFT data are known.
Thus, constructing explicit examples of pure CFTs remains one of the significant challenges.
Recently, there has been notable progress on this problem.
Considering the RG flow from a tensor product of $N$ copies of unitary minimal models,
Antunes and Behan have provided strong evidence that a CFT obtained at its IR fixed point
satisfies the criteria for being a pure CFT \cite{Antunes2023,Antunes2024}.
}

The advantage of considering pure CFT is that all Verma modules except the vacuum are non-degenerate (see around (\ref{eq:lnln})).
This fact leads to the following simplification,
\begin{equation}
\chi_h(\tau) = \sum_n d(n) \ex{2\pi i \tau \pa{h-\fr{c}{24}}}
=  \sum_n p(n) \ex{2\pi i \tau \pa{h-\fr{c}{24}}}
= \fr{ q^{h-\fr{c-1}{24}}}{\eta(\tau)}
\ \ \ \ \ \ \text{if } h>0.
\end{equation}
Also, since the vacuum has a level $1$ null state,
\begin{equation}
\chi_0(\tau) = \fr{ q^{h-\fr{c-1}{24}}}{\eta(\tau)} (1-q).
\end{equation}
In this section, we explain how to analytically solve the conformal bootstrap in the asymptotic region using these facts.

\subsection{Warm-up: Cardy Formula}

In Section \ref{subsec:Cardy}, we derived the density of states in high-energy regime.
Let us generalize this to the density of high-energy \und{primary} states.
There are several ways to derive the Cardy formula for primary states,
a modern method uses the fusion transformations \cite{Kusuki2019,Kusuki2019a, Collier2019, Collier2019a}.

For convenience, we introduce the Liouville notations:
\begin{equation}
c\equiv1+6Q^2, \ \ \ Q \equiv b+\fr{1}{b}, \ \ \ h_i \equiv \alpha_i(Q-\alpha_i) = \fr{Q^2}{4} + \gamma_i^2.
\end{equation}
$\gamma$ is the imaginary part of the Liouville momentum $\alpha = \fr{Q}{2} + i \gamma$.
Since it is more convenient to use the momentum $\gamma$ rather than the Liouville momentum $\alpha$ to explicitly express modular transformation properties,
we use $\gamma$ in the following.

We write the modular bootstrap equation using the Virasoro character basis,
\begin{equation}
\int_0^\infty
d\gamma \ 
\int_0^\infty
d \bar{\gamma} \ 
\rho(\gamma, \bar{\gamma})
\abs{\chi_\gamma(\tau)}^2
=
\int_0^\infty
d\gamma \  
\int_0^\infty
d \bar{\gamma}  \ 
\rho(\gamma, \bar{\gamma})
\abs{\chi_\gamma\pa{ -\fr{1}{\tau} }}^2.
\end{equation}
Here, the density of primary states $\rho(\gamma, \bar{\gamma})$ is defined using the number of primary states $d_i$ as follows,
\begin{equation}\label{eq:DefRho}
\rho(\gamma, \bar{\gamma}) = \sum_i d_i \delta(\gamma - \gamma_i) \delta(\bar{\gamma} - \bar{\gamma}_i).
\end{equation}
Taking the $\tau, \bar{\tau} \to i 0$ limit of the modular bootstrap equation,
the right-hand side can be approximated by the contribution from the vacuum state,
\begin{equation}\label{eq:Modular1}
\int_0^\infty
d\gamma \  
\int_0^\infty
d \bar{\gamma}  \ 
\rho(\gamma, \bar{\gamma})
\abs{\chi_\gamma(\tau)}^2
\simeq
\abs{\chi_\bb{I} \pa{ -\fr{1}{\tau} }}^2, \ \ \ \ \ \tau, \bar{\tau}  \to i0.
\end{equation}
We consider re-expanding both sides in the same functional basis using the fusion transformation.
The Virasoro character with $c>1, h\geq0$ has the following $S$ modular transformation property,
\begin{equation}\label{eq:Strans}
\begin{aligned}
\chi_\mathbb{I} \pa{-\fr{1}{\tau}} &= \int^\infty_0 d \gamma' \ S_{\mathbb{I} \gamma'} \chi_{\gamma'} \pa{\tau},
\ \ \ S_{\mathbb{I} \gamma'} =4\sqrt{2} \sinh\pa{2\pi \gamma' / b} \sinh \pa{2\pi \gamma' b},\\
\chi_\gamma\pa{-\fr{1}{\tau}} &= \int^\infty_0 d \gamma' \ S_{\gamma \gamma'} \chi_{\gamma'} \pa{\tau},
\ \ \ S_{\gamma \gamma'} = 2\sqrt{2} \cos \pa{4\pi \gamma \gamma'}.
\end{aligned}
\end{equation}
$S_{\gamma \gamma'}$ is a fusion matrix, specifically called the {\it modular $S$ matrix}.
It should be noted that the above formula for the modular $S$ matrix only holds for non-degenerate Verma modules,
and it does not hold in minimal models.
Using this modular $S$ transformation to rewrite (\ref{eq:Modular1}), we obtain the following asymptotic equation,
\begin{equation}
\int_0^\infty
d\gamma \  
\int_0^\infty
d \bar{\gamma}  \ 
\rho(\gamma, \bar{\gamma})
\abs{\chi_\gamma(\tau)}^2
\simeq
\int_0^\infty
d\gamma \  
\int_0^\infty
d \bar{\gamma}  \ 
S_{\bb{I} \gamma}
S_{\bb{I} \bar{\gamma}}
\abs{\chi_\gamma(\tau)}^2.
\end{equation}
By comparing the coefficients, we get
\begin{equation}\label{eq:SS}
\rho(\gamma, \bar{\gamma})
\simeq 
S_{\bb{I} \gamma}
S_{\bb{I} \bar{\gamma}},
\ \ \ \ \ \gamma, \bar{\gamma} \gg Q.
\end{equation}
The asymptotic form of the modular $S$ matrix is given by
\begin{equation}
S_{\bb{I} \gamma} \simeq \sqrt{2} \ex{2\pi Q \gamma }= \sqrt{2} \ex{2 \pi \sqrt{\fr{c-1}{6} \pa{h-\fr{c-1}{24}}}}, \ \ \ \ \gamma, \bar{\gamma} \gg Q.
\end{equation}
Thus, (\ref{eq:SS}) can be rewritten in terms of the conformal dimensions as follows.
\footnote{
It may seem strange that the Cardy formula becomes a continuous spectrum even though we assumed a discrete spectrum for pure CFT.
This is because we ignored all approximation errors.
One way to handle this accurately is to interpret that this asymptotic formula treats the spectrum averaged over a width $2\delta$, $\rho_{\delta} (h) \equiv \int_{h-\delta}^{h+\delta} dh \rho(h)$,
and explicitly write the error due to this averaging.
This can be investigated using the Tauberian theorem, and it can be shown that the error in $\log \rho$ is $O(1)$\cite{Mukhametzhanov2019}.
}
\begin{tcolorbox}[title=Cardy formula for primary states]
\begin{equation}\label{eq:CardyVir}
\log \rho(h,\bar{h}) \simeq 2 \pi \sqrt{\fr{c-1}{6} \pa{h-\fr{c-1}{24}}} + 2 \pi \sqrt{\fr{c-1}{6} \pa{\bar{h}-\fr{c-1}{24}}}, \ \ \ \ \ h, \bar{h} \gg c.
\end{equation} 
\end{tcolorbox}
\noindent
While the original Cardy formula (\ref{eq:CardySpin}) represented the total number of states with conformal dimensions $(h,\bar{h})$,
The formula (\ref{eq:CardyVir}) represents the number of \und{primary} states with conformal dimensions $(h,\bar{h})$.
From this expression, we see that the only difference from the original Cardy formula is the shift $c \to c-1$.
In particular, both match in the limit $c \to \infty$.
The reason for this is clear when we recall that the number of descendant states is given by the Hardy-Ramanujan formula (\ref{eq:HR}).
Since the asymptotic form of the Hardy-Ramanujan formula does not depend on $c$, the number of descendants can be overwhelmingly small compared to the number of primaries in the large $c$ limit.
This is why the original Cardy formula (\ref{eq:CardySpin}) and the Cardy formula for primary fields (\ref{eq:CardyVir}) coincide in the large $c$ limit.

\subsection{Generalized Cardy Formula}\label{subsec:GenCardy}

Although the definition of pure CFT included ``twist gap",
the discussions so far have not actually used the assumption of twist gap at all.
What can we gain by assuming the twist gap?
This section explains that.

Twist is a parameter defined as follows,
\begin{equation}
t = 2\min(h, \bar{h}).
\end{equation}
When every primary state except the vacuum state in the CFT spectrum satisfies $t>0$, we say that the CFT has a twist gap.
This implies the absence of primary fields with conformal dimensions $(h>0,0)$ or $(0, \bar{h}>0)$ (called the {\it chiral primaries}).
Since currents are chiral primary, the presence of a twist gap particularly means the absence of symmetries other than the Virasoro symmetry.

If there is a twist gap, $\ca{V}_{\bb{I}}$ only couples with $\bar{\ca{V}}_{\bb{I}}$, allowing a vacuum approximation in the limit $\bar{\tau} \to i0$.
\begin{equation}
\int_0^\infty
d\gamma \ 
\int_0^\infty
d \bar{\gamma}  \ 
\rho(\gamma, \bar{\gamma})
\abs{\chi_\gamma(\tau)}^2
\simeq
\abs{\chi_\bb{I} \pa{ -\fr{1}{\tau} }}^2, \ \ \ \ \ \bar{\tau}  \to i0.
\end{equation}
The difference from (\ref{eq:Modular1}) is that the limit with respect to $\tau$ is not taken.
By comparing the coefficients, we get the Cardy formula again \cite{Kusuki2019, Collier2019},
\begin{tcolorbox}[title=Generalized Cardy formula]
   \begin{equation}\label{eq:CardyVir2}
\log \rho(h,\bar{h}) \simeq 2 \pi \sqrt{\fr{c-1}{6} \pa{h-\fr{c-1}{24}}} + 2 \pi \sqrt{\fr{c-1}{6} \pa{\bar{h}-\fr{c-1}{24}}}, \ \ \ \ \ \bar{h} \gg c.
\end{equation} 
\end{tcolorbox}
\noindent
This formula holds for any $h$.
In this sense, it is sometimes called the \und{generalized Cardy formula} \cite{Benjamin2019}.
Moreover, fixing $h$ and taking the limit $\bar{h} \to \infty$ is equivalent to taking the limit of infinite spin $s=\abs{h-\bar{h}} \to \infty$,
which is called the {\it large spin limit}.
In short, while the Cardy formula was for the density of states at high energy,
the generalized Cardy formula is for the density of states at high spin.

\subsection{Conformal Bootstrap of Four-Point Functions}

Here, we use the Liouville momentum $\alpha$ again.
Unlike Liouville CFT, the spectrum can include primary fields with conformal dimensions $h<\fr{c-1}{24}$.
Therefore, we analytically continue the range of possible values for the Liouville momentum to the following region:
\begin{equation}
\alpha \in \left\{ \fr{Q}{2} + i \bb{R}_+ \right\} \cup \left[0, \fr{Q}{2}  \right).
\end{equation}

Let us consider the conformal bootstrap equation for four-point functions.
As in (\ref{eq:DefRho}), we define the {\it OPE coefficient density} to rewrite the sum as an integral,
\begin{equation}
\sum_p C_{12p}C_{34p} \ca{F}^{21}_{34}(h_p|z)\overline{\ca{F}^{21}_{34}}(\bar{h}_p|\bz )
=
\int d\alpha \int d\bar{\alpha} \ca{C}^{21}_{34}(\alpha, \bar{\alpha}) \ca{F}^{21}_{34}(\alpha |z)\overline{\ca{F}^{21}_{34}}( \alpha |\bz ),
\end{equation}
where the OPE coefficient density $\ca{C}^{21}_{34}(\alpha, \bar{\alpha})$ is defined as follows using the number of primary states $d_i$,
\begin{equation}
\ca{C}^{21}_{34}(\alpha, \bar{\alpha}) \equiv \sum_i d_i C_{12i}C_{34_i} \delta(\alpha-\alpha_i) \delta(\bar{\alpha} -\bar{\alpha}_i).
\end{equation}
In this way, the conformal bootstrap equation is expressed as follows.
\begin{equation}
\int d\alpha \int d\bar{\alpha} \ca{C}^{21}_{34}(\alpha, \bar{\alpha}) \ca{F}^{21}_{34}(\alpha |z)\overline{\ca{F}^{21}_{34}}( \alpha |\bz )
=
\int d\alpha' \int d\bar{\alpha'} \ca{C}^{23}_{14}(\alpha', \bar{\alpha}') \ca{F}^{23}_{14}(\alpha' |1-z)\overline{\ca{F}^{23}_{14}}( \bar{\alpha}' |1-\bz ).
\end{equation}
The solution method is basically the same as for the Cardy formula.
In the following, to make the vacuum approximation, we consider the case ($1=4$ and $2=3$) where the vacuum is possible in the OPE.
With the vacuum approximation on the right-hand side of the conformal bootstrap equation in the limit $z, \bz  \to 1$,
we consider the following fusion transformation (\ref{eq:Fusion}),
\begin{equation}
\begin{aligned}
\ca{F}^{23}_{14}(h_{\alpha}|1-z)=\int_{\fr{Q}{2}+i0}^{\fr{Q}{2}+i \infty} d\alpha
{\bold F}_{\alpha, \alpha'} 
   \left[
    \begin{array}{cc}
    \a_2   & \a_3  \\
     \a_1  &   \a_4\\
    \end{array}
  \right]
  \ca{F}^{21}_{34}(h_{\alpha'}|z).
\end{aligned}
\end{equation}
Then we obtain the asymptotic conformal bootstrap equation,
\footnote{
${\bold F}_{\bb{I}, \alpha} $ can be simply obtained by analytically continuing ${\bold F}_{\alpha', \alpha} $ to $\alpha' \to i \fr{Q}{2}$.
As a side note, when performing a similar analytic continuation for the modular $S$ matrix, care must be taken to properly handle the divergence from null states.
}
\begin{equation}
\int d\alpha \int d\bar{\alpha} \ca{C}^{21}_{21}(\alpha, \bar{\alpha}) \abs{\ca{F}^{21}_{21}(\alpha |z)}^2
\simeq
\int d\alpha \int d\bar{\alpha} 
{\bold F}_{\bb{I}, \alpha} 
   \left[
    \begin{array}{cc}
    \a_2   & \a_2  \\
     \a_1  &   \a_1\\
    \end{array}
  \right]
{\bold F}_{\bb{I}, \bar{\alpha}} 
   \left[
    \begin{array}{cc}
    \bar{\a}_2   & \bar{\a}_2  \\
     \bar{\a}_1  &   \bar{\a}_2\\
    \end{array}
  \right]
\abs{\ca{F}^{21}_{21}(\alpha |z)}^2.
\end{equation}
Thus, we obtain the asymptotic formula for the OPE coefficient density,
\begin{equation}\label{eq:OPErho}
\ca{C}^{21}_{21}(\alpha, \bar{\alpha}) 
\simeq
{\bold F}_{\bb{I}, \alpha} 
   \left[
    \begin{array}{cc}
    \a_2   & \a_2  \\
     \a_1  &   \a_1\\
    \end{array}
  \right]
{\bold F}_{\bb{I}, \bar{\alpha}} 
   \left[
    \begin{array}{cc}
    \bar{\a}_2   & \bar{\a}_2  \\
     \bar{\a}_1  &   \bar{\a}_2\\
    \end{array}
  \right],
  \ \ \ \ \ \abs{\a}, \abs{\bar{\a}} \gg Q.
\end{equation}
Using this result and the density of states (\ref{eq:SS}), we can define the average value of the OPE coefficient as follows,
\begin{equation}\label{eq:OPEave}
\overline{\pa{C_{123}}^2} \equiv \fr{\ca{C}^{21}_{21}(\alpha_3, \bar{\alpha}_3)}{\rho(\alpha_3, \bar{\alpha}_3)}. 
\end{equation}
Since the holomorphic and anti-holomorphic parts are separated, it is convenient to introduce the quantity with only the holomorphic part extracted as follows,
\begin{equation}\label{eq:DefC0}
C_0 (\a_1,\a_2, \a_3)
\equiv
\fr{
{\bold F}_{\bb{I}, \a_3} 
   \left[
    \begin{array}{cc}
    \a_2   & \a_2  \\
     \a_1  &   \a_1\\
    \end{array}
  \right]
  }{
  S_{\bb{I} \a_3}},
\end{equation}
which satisfies $\overline{\pa{C_{123}}^2} = \abs{C_0 (\a_1,\a_2, \a_3)}^2$.

Let us consider how this quantity is specifically expressed.
In fact, it is known that the fusion matrix we are considering has a simple form,
and as a result, $C_0 (\a_1,\a_2, \a_3)$ can be expressed as follows \cite{Collier2019}.
\begin{equation}\label{eq:C0}
\begin{aligned}
C_0(\alpha_1, \alpha_2, \alpha_3)
&\equiv
\fr{\alpha_b(2Q)}{2\sqrt{2}\pi \alpha_b(Q)}
\fr{1}{ \prod_{i=1,2,3} \Gamma_b(2\alpha_i) \Gamma(2Q-2\alpha_i)}\\
&\times
\fr{ \Upsilon_b'(0) }{\Upsilon_b(\alpha_1+\alpha_2+\alpha_3-Q)\Upsilon_b(\alpha_1+\alpha_2-\alpha_3)\Upsilon_b(\alpha_2+\alpha_3-\alpha_1)\Upsilon_b(\alpha_3+\alpha_1-\alpha_2)}.
\end{aligned}
\end{equation}
This formula is similar to the DOZZ formula (\ref{eq:DOZZ}).
Recalling that the DOZZ formula was determined by the fusion matrix (see Section \ref{subsubsec:Bootstrap}),
this similarity is inevitable.
The relationship between the two is specifically expressed as follows:
\begin{tcolorbox}[title=Asymptotic formula for OPE coefficient density]
    \begin{equation}
C_0(\alpha_1, \alpha_2, \alpha_3) = \fr{C_{123}^{\mathrm{DOZZ}}}{\ca{E}  \prod_{i=1,2,3} \sqrt{S_{\bb{I}\gamma_i} B_i }}.
\end{equation}
\end{tcolorbox}
\noindent
Here, $S_{\bb{I} \g}$ is the modular $S$ matrix, and $B_i$ is the normalization constant of the two-point function in Liouville CFT (\ref{eq:B}).
Also, $\ca{E}$ is a constant independent of the conformal dimensions:
\begin{equation}
\ca{E} \equiv 2^{\fr{3}{4}}\pi \pa{b^{\fr{2}{b}-2b} \lambda}^{-\fr{Q}{2}} \fr{\Gamma_b(Q)}{\Gamma_b(2Q)}.
\end{equation}
The detailed derivation of this relationship is described in Appendix \ref{subapp:DOZZ}.

\subsection{Lightcone Conformal Bootstrap}

Let us apply what we did in the derivation of the generalized Cardy formula (Section \ref{subsec:GenCardy}) to the four-point conformal bootstrap \cite{Kusuki2019a, Collier2019}.
Again, we assume a twist gap.
In this case, the conformal bootstrap equation can be approximated as follows in the limit $\bz  \to 1$,
\begin{equation}\label{eq:FF}
\int d\alpha \int d\bar{\alpha} \ca{C}^{21}_{21}(\alpha, \bar{\alpha}) \abs{\ca{F}^{21}_{21}(\alpha |z)}^2
\simeq
\int d\alpha \int d\bar{\alpha}
{\bold F}_{\bb{I}, \alpha} 
   \left[
    \begin{array}{cc}
    \a_2   & \a_2  \\
     \a_1  &   \a_1\\
    \end{array}
  \right]
{\bold F}_{\bb{I}, \bar{\alpha}} 
   \left[
    \begin{array}{cc}
    \bar{\a}_2   & \bar{\a}_2  \\
     \bar{\a}_1  &   \bar{\a}_2\\
    \end{array}
  \right]
\abs{\ca{F}^{21}_{21}(\alpha |z)}^2.
\end{equation}
By comparing both sides, we get
\begin{equation}\label{eq:OPErho2}
\ca{C}^{21}_{21}(\alpha, \bar{\alpha}) 
\simeq
{\bold F}_{\bb{I}, \alpha} 
   \left[
    \begin{array}{cc}
    \a_2   & \a_2  \\
     \a_1  &   \a_1\\
    \end{array}
  \right]
{\bold F}_{\bb{I}, \bar{\alpha}} 
   \left[
    \begin{array}{cc}
    \bar{\a}_2   & \bar{\a}_2  \\
     \bar{\a}_1  &   \bar{\a}_2\\
    \end{array}
  \right],
  \ \ \ \ \ \abs{\bar{\a}} \gg Q.
\end{equation}
This seems correct at first glance,
but in fact, it is only valid when $\Re (\alpha_1 +\alpha_2)>\fr{Q}{2}$.
The reason is that when the fusion transformation is analytically continued to the region $\Re (\alpha_1 +\alpha_2)<\fr{Q}{2}$,
the poles of the fusion matrix cross the integral contour $\fr{Q}{2}+0i \sim \fr{Q}{2} + i \infty$.
As a result of the contribution of these residues appearing in the integral transformation (\ref{eq:FF}),
the OPE spectrum consists of not only a continuous spectrum with $h>\fr{c-1}{24}$ but also a discrete spectrum with $h<\fr{c-1}{24}$.
This is explained below.

\subsubsection{Poles of Fusion Matrix}

\begin{figure}[t]
 \begin{center}
  \includegraphics[width=12.0cm,clip]{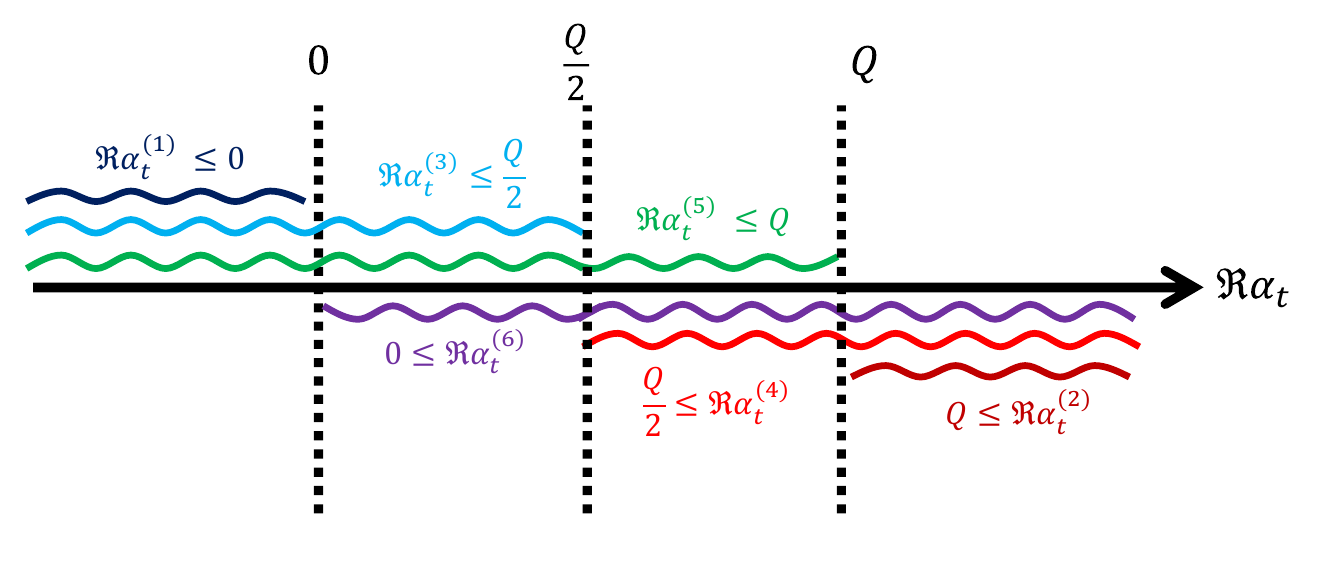}
 \end{center}
 \caption{The region where simple poles of the fusion matrix exist on the $\alpha_t$ plane}
 \label{fig:poles}
\end{figure}

Let us specifically look at the pole structure of the fusion matrix 
$
{\bold F}_{\bb{I}, \a_t} 
   \left[
    \begin{array}{cc}
    \a_2   & \a_2  \\
     \a_1  &   \a_1\\
    \end{array}
  \right]
  $.
The fusion matrix we are interested in is given by (\ref{eq:DefC0}) and (\ref{eq:C0}).
Since the Barnes double gamma function has simple poles at $x=-mb-\fr{n}{b}, \ \ (n,m \in \bb{Z}_{\geq 0})$,
the fusion matrix has simple poles at the following points on the $\alpha_t$ plane.
\begin{equation}\label{eq:Poles}
\begin{aligned}
\a_t^{(1)}&=(\a_1+\a_2)-Q-Q_{m,n},\\
\a_t^{(2)}&=-(\a_1+\a_2)+2Q+Q_{m,n},\\
\a_t^{(3)}&=\pm(\a_1-\a_2)-Q_{m,n},\\
\a_t^{(4)}&=\pm(\a_1-\a_2)+Q+Q_{m,n},\\
\a_t^{(5)}&=-(\a_1+\a_2)+Q-Q_{m,n},\\
\a_t^{(6)}&=\a_1+\a_2+Q_{m,n}.\\
\end{aligned}
\end{equation}
Here, $Q_{m,n}\equiv -mb-\fr{n}{b}, \ \ (n,m \in \bb{Z}_{\geq 0})$.
Since the Liouville momentum in unitary theories satisfies $0 \leq \Re \alpha_i \leq \fr{Q}{2}$,
the possible range of each pole location is restricted as shown in Figure \ref{fig:poles}.

\begin{figure}[t]
 \begin{center}
  \includegraphics[width=15.0cm,clip]{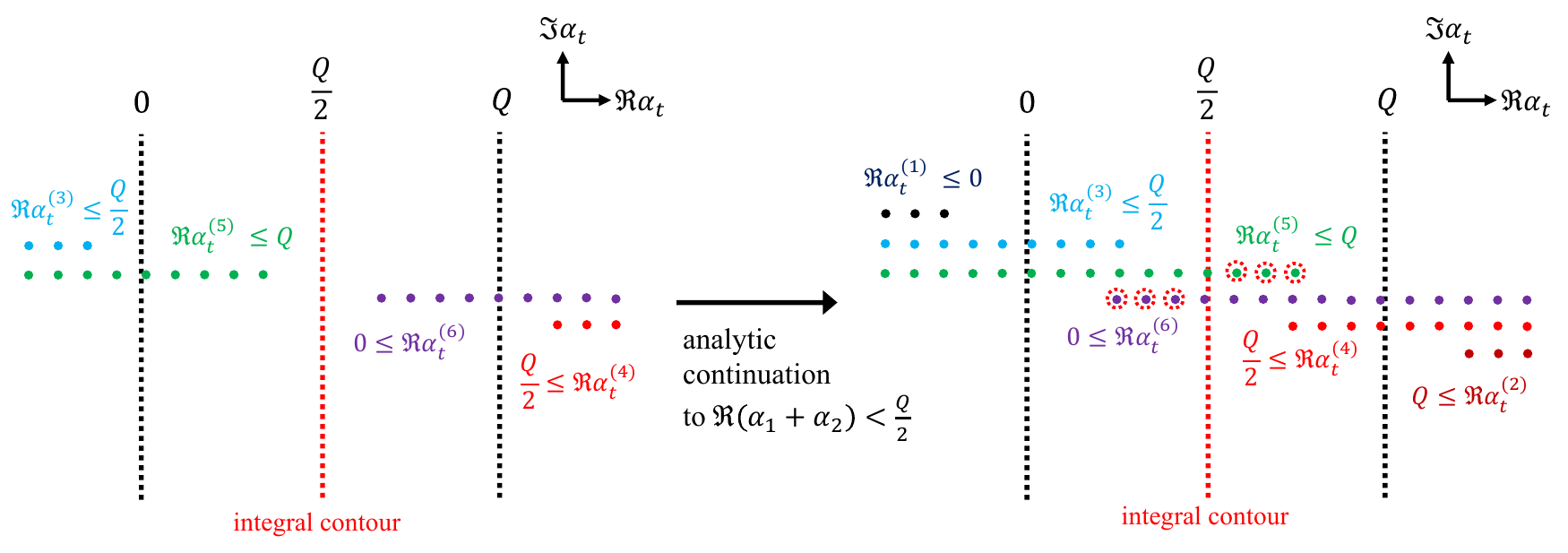}
 \end{center}
 \caption{When the fusion transformation is analytically continued to $\Re(\alpha_1+\alpha_2)<\fr{Q}{2}$,
the poles of the fusion matrix cross the integral contour,
resulting in the contribution of residues appearing in the integral transformation.}
 \label{fig:integral}
\end{figure}

From this pole structure,
when the fusion transformation is analytically continued to the region $\Re(\alpha_1+\alpha_2)<\fr{Q}{2}$,
it can be seen that the poles of the fusion matrix cross the integral contour,
resulting in the contribution of residues appearing in the integral transformation (see Figure \ref{fig:integral}).
Taking into account the contribution of these residues,
the OPE spectrum (\ref{eq:OPErho2}) is modified to include the following discrete spectrum,
\begin{equation}\label{eq:OPErho3}
\ca{C}^{21}_{21}(\alpha_m, \bar{\alpha}) 
\simeq
-2\pi
\Res{\alpha=\alpha_m}
{\bold F}_{\bb{I}, \alpha} 
   \left[
    \begin{array}{cc}
    \a_2   & \a_2  \\
     \a_1  &   \a_1\\
    \end{array}
  \right]
{\bold F}_{\bb{I}, \bar{\alpha}} 
   \left[
    \begin{array}{cc}
    \bar{\a}_2   & \bar{\a}_2  \\
     \bar{\a}_1  &   \bar{\a}_2\\
    \end{array}
  \right],
  \ \ \ \ \ \abs{\bar{\a}} \gg Q.
\end{equation}
Here, $\alpha_m \equiv \alpha_1 + \alpha_2 + mb \ \ \ \ (m \in \bb{Z}_{\geq 0})$.
Also, assuming $b\leq \fr{1}{b}$, since $\fr{1}{b}>\fr{Q}{2}$, the poles with $n>0$ do not contribute.

\subsubsection{Interpretation on the Gravity Side}

States created by two local operators correspond to two-particle systems on the gravity side.
The asymptotic formula for the OPE spectrum (\ref{eq:OPErho3}) predicts how the energy spectrum of two-particle systems with large spin is given,
\begin{tcolorbox}[title=OPE spectrum at large spin]
\begin{equation}\label{eq:Add}
h_m=h_1 + h_2 +m + \delta h_m, \ \ \ \ \ \delta h_m \equiv -2(\alpha_1+mb)(\alpha_2+bm) + m(m+1)b^2 <0, \ \ \ (m \in \bb{Z}_{\geq0}).
\end{equation}
\end{tcolorbox}
\noindent
The characteristic point of this energy spectrum is that in addition to the rest masses of the two particles $h_1+h_2$, there is a binding energy $\delta h_m$.
Recalling that the OPE spectrum was determined purely by kinematic information,
it will be clear that this binding energy comes from the exchanging of descendants between the two particles.
Moreover, recalling that the descendants correspond to gravitons on the gravity side,
we can see that this binding energy represents gravitational interaction.

It may seem strange that a large gravitational interaction exists between two particles in the large spin limit (equivalently the long-distance limit).
In fact, the intuition that gravitational interaction disappears in the large spin limit is correct in AdS${}_{d\geq 4}$ \cite{Komargodski2012, Fitzpatrick2013a}.
However, it can be shown that the gravitational potential does not decay as a function of distance in AdS${}_3$.
As a result, in the limit of infinite spin, finite gravitational interaction remains as in (\ref{eq:Add}).
Furthermore, the characteristic $\delta_m <0$ is also important.
This reflects the fact that gravitational interaction is attractive.

The continuous OPE spectrum (\ref{eq:OPErho2}) reflects the fact that states with $\min(h,\bar{h})>\fr{c}{24}$ correspond to the BTZ black holes.
The condition $\Re(\alpha_1+\alpha_2)>\fr{Q}{2}$ becomes the BTZ threshold when expressed as the lowest energy of the OPE spectrum,
\begin{equation}\label{eq:OPEBTZ}
h_0 = (\alpha_1+\alpha_2)(Q-\alpha_1-\alpha_2) > \fr{c-1}{24}.
\end{equation}

The most challenging part of deriving this asymptotic formula is extracting the necessary information from the complicated special function, i.e. the Virasoro block.
In \cite{Fitzpatrick2014}, they approximate the Virasoro block in the limit $c\to \infty$ and investigate the behavior near $z \to 1$,
identifying the leading term in $c$ for (\ref{eq:Add}).
Since the limit $(z,\bz ) \to (0,1)$ that appears in this derivation corresponds to the limit $t \to x$ when analytically continued to Lorentzian time as $z\equiv x+i\tau$,
it is called the {\it lightcone limit}.
In the present derivation, we only took the limit $\bz \to 0$ without taking the limit $z \to 1$,
but it is conventionally called the {\it lightcone conformal bootstrap}.

\subsection{General Conformal Bootstrap and Asymptotic Formula for OPE Coefficients}\label{sec:GenOPE}

\newsavebox{\boxZs}
\sbox{\boxZs}{\includegraphics[width=4cm]{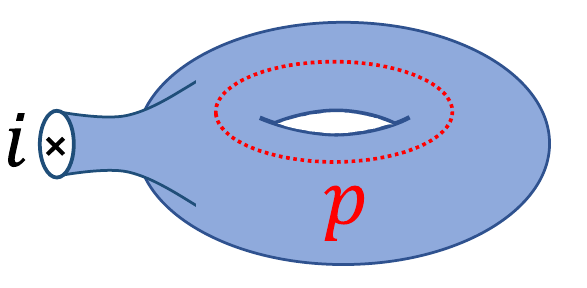}}
\newlength{\Zsw}
\settowidth{\Zsw}{\usebox{\boxZs}}

\newsavebox{\boxZt}
\sbox{\boxZt}{\includegraphics[width=4cm]{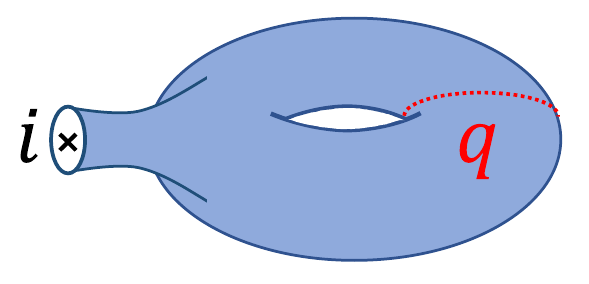}}
\newlength{\Ztw}
\settowidth{\Ztw}{\usebox{\boxZt}} 

\newsavebox{\boxFs}
\sbox{\boxFs}{\includegraphics[width=4cm]{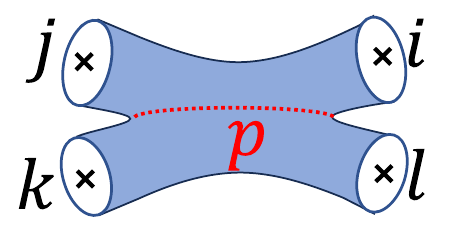}}
\newlength{\Fsw}
\settowidth{\Fsw}{\usebox{\boxFs}} 

\newsavebox{\boxFt}
\sbox{\boxFt}{\includegraphics[width=4cm]{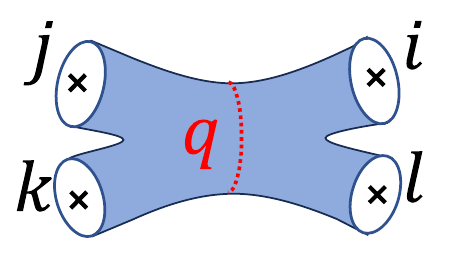}}
\newlength{\Ftw}
\settowidth{\Ftw}{\usebox{\boxFt}} 

\newsavebox{\boxFZss}
\sbox{\boxFZss}{\includegraphics[width=5cm]{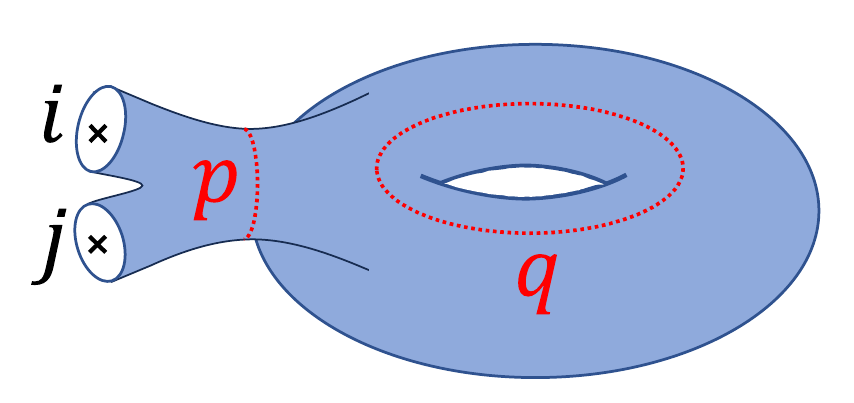}}
\newlength{\FZssw}
\settowidth{\FZssw}{\usebox{\boxFZss}} 

\newsavebox{\boxFZst}
\sbox{\boxFZst}{\includegraphics[width=5cm]{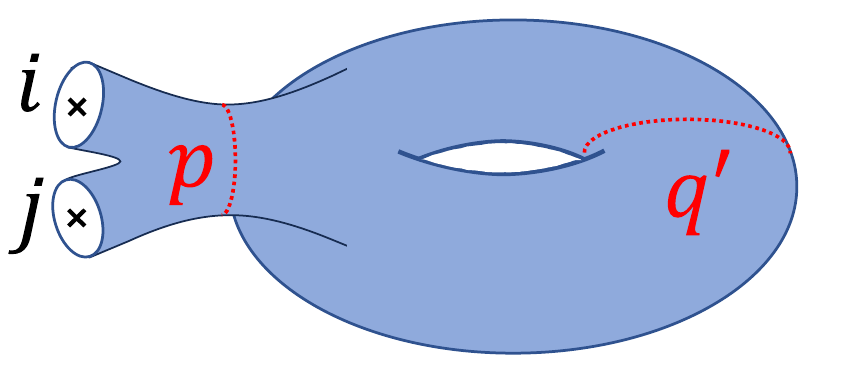}}
\newlength{\FZstw}
\settowidth{\FZstw}{\usebox{\boxFZst}} 

\newsavebox{\boxFZtt}
\sbox{\boxFZtt}{\includegraphics[width=5cm]{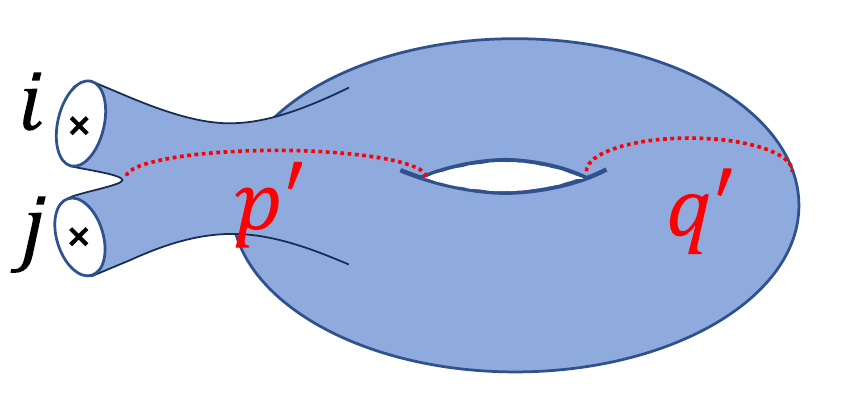}}
\newlength{\FZttw}
\settowidth{\FZttw}{\usebox{\boxFZtt}} 

\newsavebox{\boxtwos}
\sbox{\boxtwos}{\includegraphics[width=4cm]{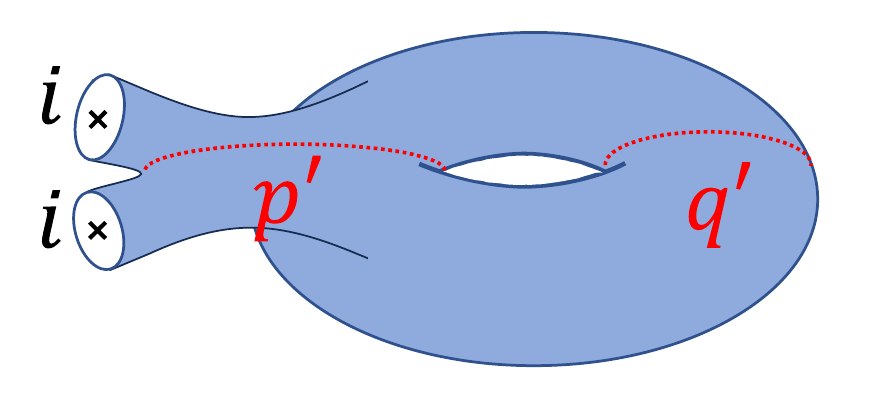}}
\newlength{\twosw}
\settowidth{\twosw}{\usebox{\boxtwos}} 

\newsavebox{\boxtwot}
\sbox{\boxtwot}{\includegraphics[width=4cm]{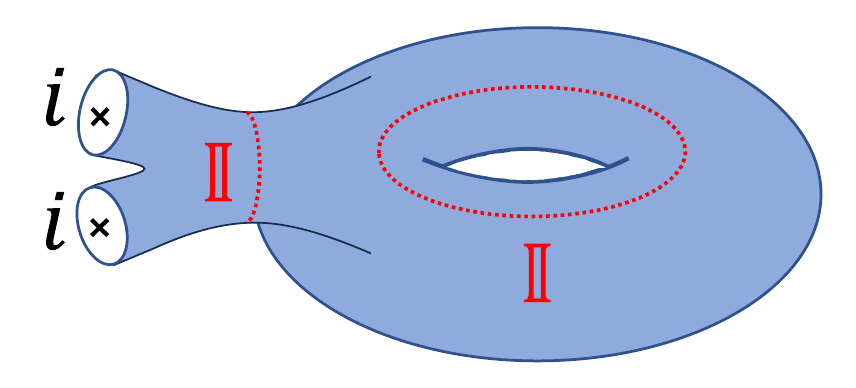}}
\newlength{\twotw}
\settowidth{\twotw}{\usebox{\boxtwot}} 

Here, we derive various asymptotic formulas for OPE coefficients in different asymptotic regions by considering more general conformal bootstrap equations.
While the derivation method using the inverse Laplace transformation as in Section \ref{subsec:Cardy} provided separate formulas for each asymptotic region \cite{Das2018, Kusuki2018, Kusuki2018a,Das2021,Kraus2016, Hikida2018, RomeroBermudez2018, Brehm2018, Cardy2017},
the derivation using the fusion transformation can handle them uniformly \cite{Collier2019a}.
The key to solving general conformal bootstrap equations using the fusion transformation is the Moore-Seiberg construction \cite{Moore1989}.
The Moore-Seiberg construction asserts that any fusion can be constructed from modular transformations of one-point functions on the torus and fusion transformations of four-point functions.
\begin{description}

\item[Modular transformation of one-point functions on the torus]\mbox{}\\
\begin{equation}
\parbox{\Zsw}{\usebox{\boxZs}}  =
\int^\infty_0 d \gamma_q 
S_{\gamma_p, \gamma_q}[\g_i]
\parbox{\Ztw}{\usebox{\boxZt}}.
\end{equation}

\item[Fusion transformation of four-point functions]\mbox{}\\
\begin{equation}
\parbox{\Fsw}{\usebox{\boxFs}}  =
\int^\infty_0 d \gamma_q
{\bold F}_{\g_p, \g_q} 
   \left[
    \begin{array}{cc}
    \g_j  & \g_i  \\
     \g_k  &   \g_l
    \end{array}
  \right]
\parbox{\Ftw}{\usebox{\boxFt}}.
\end{equation}
\end{description}
Here, the red lines represent the projection operator onto a Verma module.

As an example, we show how the fusion transformation of two-point functions on the torus is generated,
\begin{equation}
\begin{aligned}
&\parbox{\FZssw}{\usebox{\boxFZss}}  =
\int^\infty_0 d \gamma_{q'}
S_{\gamma_q, \gamma_{q'}}[\g_p]
\parbox{\FZstw}{\usebox{\boxFZst}} \\
&=
\int^\infty_0 d \gamma_{q'} 
\int^\infty_0 d \gamma_{p'}
S_{\gamma_q, \gamma_{q'}}[\g_p]
{\bold F}_{\g_p, \g_{p'}} 
   \left[
    \begin{array}{cc}
    \g_j  & \g_i  \\
     \g_{q'}  &  \g_{q'}
    \end{array}
  \right]
\parbox{\FZttw}{\usebox{\boxFZtt}}.
\end{aligned}
\end{equation}
From this fusion transformation and the following asymptotic conformal bootstrap equation
\begin{equation}
\begin{aligned}
\int^\infty_0 d \gamma_{p'}
\int^\infty_0 d \bar{\gamma}_{p'} 
\int^\infty_0 d \gamma_{q'}
\int^\infty_0 d \bar{\gamma}_{q'} 
\ca{C}_{ii} (\g_{p'}, \bar{\g}_{p'}, \g_{q'}, \bar{\g}_{q'}) 
&\abs{\parbox{\twosw}{\usebox{\boxtwos}}}^2 \\
&\simeq
\abs{
\parbox{\twotw}{\usebox{\boxtwot}}
}^2,
\end{aligned}
\end{equation}
we obtain the following asymptotic formula for the OPE coefficient density,
\begin{equation}
\ca{C}_{ii} (\g_p, \bar{\g}_p, \g_q, \bar{\g}_q) \simeq
\abs{
S_{\bb{I} \gamma_{q}}
{\bold F}_{\g_\bb{I}, \g_{p}} 
   \left[
    \begin{array}{cc}
    \g_i  & \g_i  \\
     \g_{q}  &  \g_{q}
    \end{array}
  \right]
  }^2
  , \ \ \ \ \ \g_p, \bar{\g}_p, \g_q, \bar{\g}_q \gg Q.
\end{equation}
Here, $\ca{C}_{ii} (\g_p, \bar{\g}_p, \g_q, \bar{\g}_q) $ is defined as follows:
\begin{equation}
\ca{C}_{ii}(\g_p, \bar{\g}_p, \g_q, \bar{\g}_q) \equiv \sum_{j, k} d_j d_k C_{ijk}^2 \delta(\g_p-\g_j) \delta(\g_q-\g_k) \delta(\bar{\g}_p-\bar{\g}_j) \delta(\bar{\g}_q-\bar{\g}_k)  .
\end{equation}
As with (\ref{eq:OPEave}), we define the average value of the OPE coefficient as follows:
\begin{equation}
\overline{C_{123}^2} \equiv \fr{\ca{C}_{11} (\g_2, \bar{\g}_2, \g_3, \bar{\g}_3)}{\rho(\gamma_2, \bar{\gamma_2}) \rho(\gamma_3, \bar{\gamma_3})}. 
\end{equation}
One can find that the asymptotic formula is exactly the same as (\ref{eq:DefC0}), except for the range of applications,
\begin{equation}
\overline{C_{123}^2} \simeq \abs{C_0(\alpha_1, \alpha_2, \alpha_3)}^2, \ \ \ \ \ \a_2, \bar{\a}_2, \a_3, \bar{\a}_3 \gg Q.
\end{equation}
Here, $\gamma$ is rewritten as the Liouville momentum $\alpha$ using $h_i \equiv \alpha_i(Q-\alpha_i) = \fr{Q^2}{4} + \gamma_i^2$.

By performing the same procedure with the genus 2 correlation function, the following asymptotic formula can be obtained,
\begin{equation}
\overline{C_{123}^2} \simeq \abs{C_0(\alpha_1, \alpha_2, \alpha_3)}^2, \ \ \ \ \ \a_1, \bar{\a}_1, \a_2, \bar{\a}_2, \a_3, \bar{\a}_3 \gg Q.
\end{equation}
Moreover, assuming a twist gap, the asymptotic formula for the OPE coefficients can be generalized as we derived the generalized Cardy formula in Section \ref{subsec:GenCardy}.
To summarize these discussions, all the asymptotic formulas for the OPE coefficients can be summarized as follows:
\begin{tcolorbox}[title=Unified asymptotic formula for OPE coefficients]
   \begin{equation}\label{eq:fullOPE}
\overline{C_{123}^2} \simeq \abs{C_0(\alpha_1, \alpha_2, \alpha_3)}^2, \ \ \ \ \ \max(\a_1, \bar{\a}_1, \a_2, \bar{\a}_2, \a_3, \bar{\a}_3) \gg Q.
\end{equation} 
\end{tcolorbox}
\noindent
In short,
the question ``What is universal and what depends on the theory in CFT data?",
has the answer ``All OPE coefficients in the high-energy region are universal."
\footnote{
Strictly speaking, it assumes $\a_1 \neq \a_2 \neq \a_3$.
If one considers a limit such as $\a_1 = \a_2 \to \infty$,
the asymptotic form of the OPE coefficients depends on the theory \cite{Kraus2016}.
}

Below, we summarize comments on this result.
\begin{itemize}

\item
As in the result of HKS (see Section \ref{subsubsec:HKS}),
it is expected that the applicable range of the asymptotic formula for the OPE coefficients (\ref{eq:fullOPE}) extends in holographic CFTs.
However, unlike the case of the Cardy formula,
it is already known that simply providing an upper bound on the OPE coefficients of light operators (i.e. an analog of the HKS bound) is insufficient \cite{Belin2017}.
Clarifying under what assumptions the applicability region expands remains an open question.

\item
It can be shown that (\ref{eq:fullOPE}) is consistent with the Eigenstate Thermalization Hypothesis (ETH).
This is considered to reflect the chaotic nature of holographic CFTs.
Refining this ETH in holographic CFTs is one of the challenges in this field.

\end{itemize}

\subsection{Application of Asymptotic OPE Coefficients: AdS/CFT and Averaging}

In this section, we discuss the role of the asymptotic formula for OPE coefficients in AdS/CFT.

It has been conjectured that
\begin{tcolorbox}
The holographic CFT is not a single CFT, but rather an average over a certain ensemble of CFTs (so-called {\it averaged CFT}).
\end{tcolorbox}
\noindent
This conjecture stems from the relationship between quantum gravity on AdS${}_2$ and random matrix theory\cite{Saad2019}.
Another basis for this conjecture comes from the following:
When considering an AdS spacetime with two asymptotic boundaries,
a wormhole solution exists connecting these two boundaries.
In other words, there is a correlation between the two asymptotic boundaries.
However, if we regard the asymptotic boundaries as CFTs,
then this should simply be the product of two CFTs,
and the asymptotic boundaries should be completely decoupled,
\begin{equation}\label{eq:factorization}
Z_{\mathrm{grav}}(\del M = \Sigma_1 \times \Sigma_2) \stackrel{?}{=}  Z_{\mathrm{CFT}}(\Sigma_1) \times Z_{\mathrm{CFT}}(\Sigma_2).  
\end{equation}
Here, $Z_{\mathrm{grav}}(\del M )$ represents the partition function of quantum gravity on AdS with the asymptotic boundary $\del M$,
and $Z_{\mathrm{CFT}}(\Sigma)$ represents the partition function of the CFT defined on the manifold $\Sigma$.
The issue that the wormhole solution contained in $Z_{\mathrm{grav}}(\del M )$ breaks the equality (\ref{eq:factorization}) is known as the \und{factorization problem}.
One proposed resolution to this issue is through averaging.
In other words, by averaging, correlations are induced between the two CFTs,
\begin{equation}
Z_{\mathrm{grav}}(\del M = \Sigma_1 \times \Sigma_2) = \overline{Z_{\mathrm{CFT}}(\Sigma_1) \times Z_{\mathrm{CFT}}(\Sigma_2)} .  
\end{equation}
However, the AdS/CFT constructed from a top-down approach via string theory clearly does not include any averaging operation.
Therefore, it is natural to consider that this averaging appears from the coarse-graining in the effective theory in the classical limit.
However, the mechanism remains unknown and is actively studied as one of the mysteries of the AdS/CFT correspondence.

In the above discussion, averaging was introduced abstractly,
but clarifying which specific averaging of CFTs produces the effective theory is likely one of the first tasks to be undertaken.
Thus, we consider the following as a concrete problem:
\begin{tcolorbox}
How are solutions beyond the thermal AdS and BTZ, particularly wormhole solutions, reproduced from averaged CFT?
\end{tcolorbox}
\noindent

Let us define the OPE coefficients of averaged CFT as random variables.
First, assume that the correlation between conical defects on AdS is only through gravitational interaction.
In other words, we consider pure AdS where there are no particles mediating between conical defects.
In the CFT language, this assumption translates to the following:
\begin{equation}
\overline{C_{ijk}} = 0.
\end{equation}
Next, since the OPE coefficients satisfy the conformal bootstrap equation,
they need to be consistent with (\ref{eq:fullOPE}) in particular.
Since we are now taking the large $c$ limit, the applicable range of the asymptotic formula for OPE coefficients extends to $\{ h_i \}=O(c)$.
However, as mentioned in Section \ref{subsubsec:HKS}, extending the applicable range for the asymptotic formula is technically challenging \cite{Michel2019,Das2021},
and the exact range of applicability remains unclear.
Therefore, we assume the following instead:
\footnote{
To avoid technical explanations, some assumptions have been omitted.  
Readers interested in the precise assumptions should refer to \cite{Chandra2022}.
}
\begin{quote}
(\ref{eq:fullOPE}) holds if all three weights satisfy $h > \fr{c}{32}$.
\end{quote}
Here, $\fr{c}{32}$ corresponds to the BTZ threshold of $\alpha_1$ when $\alpha_1=\alpha_2$ in (\ref{eq:OPEBTZ}).
In other words, we assume that (\ref{eq:fullOPE}) holds for the OPE coefficients corresponding to black hole formation.
This is a natural extension of the expectation (''expected'' HKS formula) that the Cardy formula holds up to the BTZ threshold.
Thus, we obtain the following conjecture for the variance of the OPE coefficients:
\begin{equation}\label{eq:OPEensemble}
\overline{ C_{ijk}C_{lmn}  } = \abs{C_0 (\alpha_i, \alpha_j, \alpha_k)}^2 \times \pa{ \delta_{il} \delta_{jm} \delta_{kn} \pm \mathrm{permutations}  }.
\end{equation}
Here, we assume that the OPE coefficients follow the Gaussian distribution.
It is important to note that the ensemble introduced here is an ensemble of CFT data, not an ensemble of CFTs.
Ultimately, it will be necessary to clarify what kind of ensemble of CFTs leads to the OPE coefficient random variables introduced here,
but for now, we assume the existence of some ensemble of CFTs that leads to the random variables (\ref{eq:OPEensemble}).
It is worth emphasizing that assuming the existence of such an ensemble is not so unnatural, given that any large $c$ CFT satisfies (\ref{eq:fullOPE}).

\newsavebox{\boxworma}
\sbox{\boxworma}{\includegraphics[width=4cm]{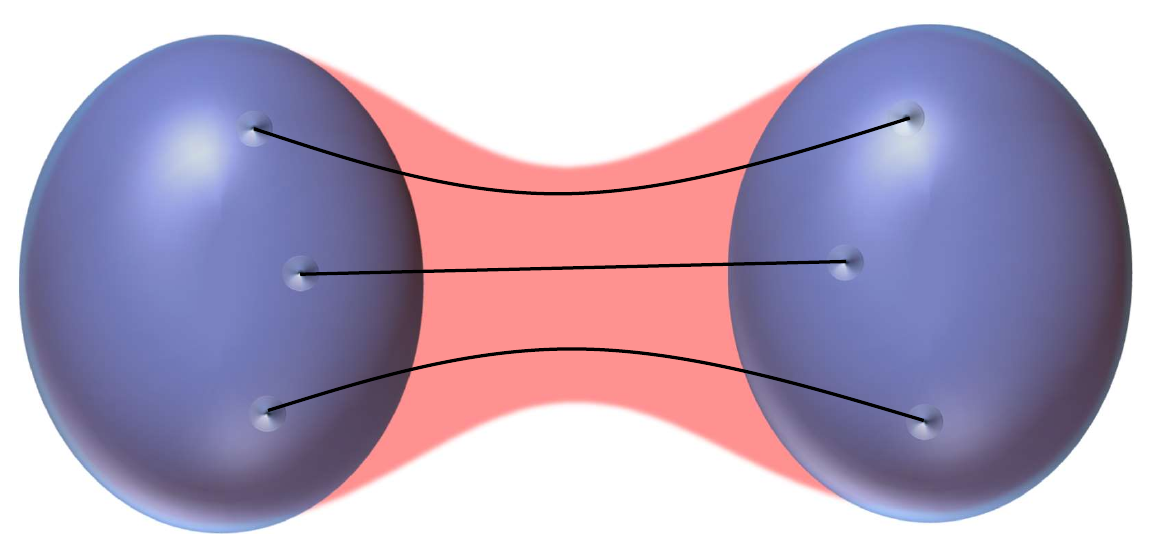}}
\newlength{\wormaw}
\settowidth{\wormaw}{\usebox{\boxworma}} 

\newsavebox{\boxwormb}
\sbox{\boxwormb}{\includegraphics[width=4cm]{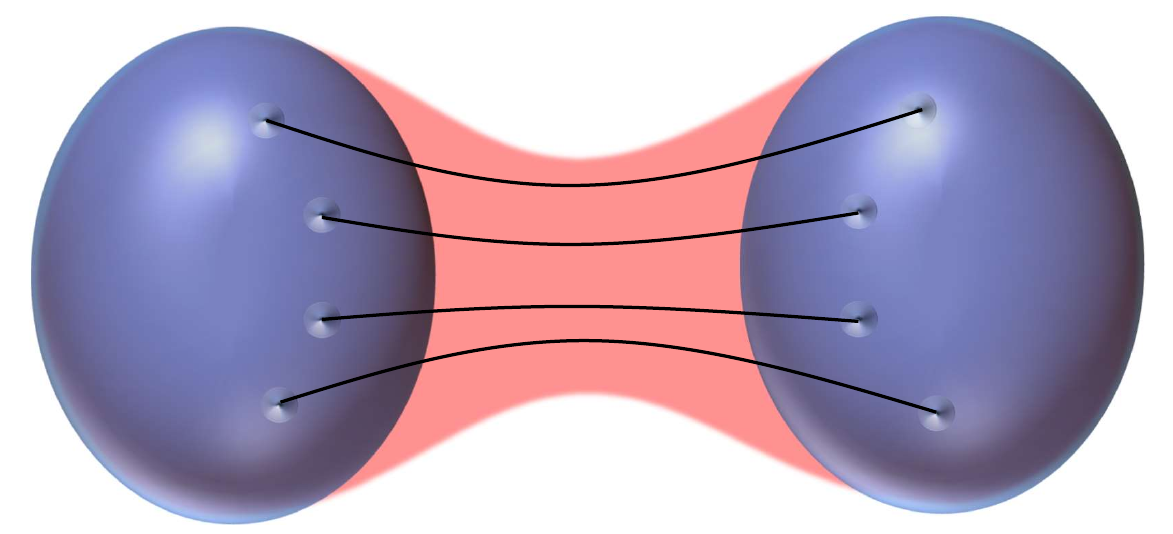}}
\newlength{\wormbw}
\settowidth{\wormbw}{\usebox{\boxwormb}}

Let us reproduce the wormhole solution between two asymptotic boundaries from CFT.
As a simple example, consider the case where the asymptotic boundaries are spheres.
A wormhole solution is possible when more than two local operators are inserted at the asymptotic boundary.
By calculating the Euclidean action of the wormhole solution with three conical defects, it can be shown that the following equality holds:
\begin{equation}
C_0 (\alpha_i, \alpha_j, \alpha_k)^2 = \parbox{\wormaw}{\usebox{\boxworma}}.
\end{equation}
In other words, the average of the product of three-point functions coincides with the wormhole solution.
Here, the black line in the diagram represents the conical defect propagating through the wormhole.
As a slightly nontrivial example, let us consider the four-point function,
\begin{equation}
\braket{\phi_4(\infty) \phi_3(1) \phi_2(z) \phi_1(0)}  \braket{\phi_4(\infty) \phi_3(1) \phi_2(z') \phi_1(0)}
= \sum_{p,q} C_{12p} C_{34p} C_{12q} C_{34q} \abs{\ca{F}^{21}_{34}(p|z) \ca{F}^{21}_{34}(q|z')}^2.
\end{equation}
Taking the average of the product of two correlation functions, only the term with $p=q$ remains due to the statistical properties of the OPE coefficients.
\begin{equation}
\overline{  \braket{\phi_4(\infty) \phi_3(1) \phi_2(z) \phi_1(0)}  \braket{\phi_4(\infty) \phi_3(1) \phi_2(z') \phi_1(0)} }
= \sum_p \overline{C_{12p}^2} \overline{C_{34p}^2}  \abs{\ca{F}^{21}_{34}(p|z) \ca{F}^{21}_{34}(q|z')}^2.
\end{equation}
The large $c$ limit of this result matches the calculation of the four-point wormhole solution in gravity.
\begin{equation}
\sum_p \overline{C_{12p}^2} \overline{C_{34p}^2}  \abs{\ca{F}^{21}_{34}(p|z) \ca{F}^{21}_{34}(q|z')}^2 = \parbox{\wormbw}{\usebox{\boxwormb}}.
\end{equation}

Similarly, it has been confirmed that wormhole solutions between genus 2 partition functions can also be computed from averaged CFT and are consistent with gravity-side calculations\cite{Chandra2022}.
In other words, the averaged CFT constructed here effectively describes pure AdS, and in this sense, we have successfully identified the "averaging" we were seeking.
It is a future challenge to further develop a microscopic understanding while confirming the physical significance of this toy model of averaged CFT,
ultimately leading to a better understanding of the factorization problem.

\section*{Acknowledgments}
I would like to thank Nathan Benjamin and Sridip Pal for numerous valuable discussions while preparing these lecture notes. I also thank Cyuan-Han Chang and Yu-ki Suzuki for their comments on the draft. Finally, I extend my heartfelt appreciation to all members of the conformal bootstrap community at Caltech for several enjoyable years of collaborative research.
YK is supported by the INAMORI Frontier Program at Kyushu University and JSPS KAKENHI Grant Number 23K20046.
These lecture notes are based on courses given at Meiji University.

\appendix

\section{Ising CFT data}\label{app:Ising}

\begin{description}
\item[conformal dimension]
\begin{equation}
(h_\sigma,\bar{h}_\sigma)=\pa{\fr{1}{16},\fr{1}{16}}, \ \ \ 
(h_\e,\bar{h}_\e)=\pa{\fr{1}{2},\fr{1}{2}}. \ \ \ 
\end{equation}
\item[OPE coefficient]
\begin{equation}
C_{\sigma \sigma 0} =C_{\e \e 0} =1, \ \ \ 
C_{\sigma \sigma \e} =\fr{1}{2}.
\end{equation}
\item[fusion matrix] (+permutation)
\begin{equation}\label{eq:fusionIsing}
\begin{aligned}
 {\bold F}_{0, 0}
   \left[
    \begin{array}{cc}
    \sigma   &  \sigma  \\
      \sigma  &    \sigma\\
    \end{array}
  \right]
&=2 {\bold F}_{0, \e}
   \left[
    \begin{array}{cc}
    \sigma   &  \sigma  \\
      \sigma  &    \sigma\\
    \end{array}
  \right]
=\fr{1}{2} {\bold F}_{\e, 0}
   \left[
    \begin{array}{cc}
    \sigma   &  \sigma  \\
      \sigma  &    \sigma\\
    \end{array}
  \right]
=- {\bold F}_{\e, \e}
   \left[
    \begin{array}{cc}
    \sigma   &  \sigma  \\
      \sigma  &    \sigma\\
    \end{array}
  \right]
=\fr{1}{\sqrt{2}}, \\
 {\bold F}_{0, 0}
   \left[
    \begin{array}{cc}
    \e   &  \e \\
      \e  &    \e \\
    \end{array}
  \right]
&=
 2{\bold F}_{0, \sigma}
   \left[
    \begin{array}{cc}
    \sigma   &  \sigma \\
      \e  &    \e \\
    \end{array}
  \right]
=
 \fr{1}{2}{\bold F}_{\sigma, 0}
   \left[
    \begin{array}{cc}
    \e   &  \sigma \\
      \e  &    \sigma \\
    \end{array}
  \right]
=- {\bold F}_{\sigma, \sigma}
   \left[
    \begin{array}{cc}
    \e   &  \sigma \\
      \sigma  &    \e \\
    \end{array}
  \right]
=1.
\end{aligned}
\end{equation}
\item[conformal block]
\begin{equation}\label{eq:blockIsing}
\ca{F}^{\sigma \sigma}_{\sigma \sigma}(0|z) =\fr{1}{\sqrt{2}}\fr{\sqrt{1+\sqrt{1-z}}}{\pa{z(1-z)}^{\fr{1}{8}}}, \ \ \ 
\ca{F}^{\sigma \sigma}_{\sigma \sigma}(\e|z) =\fr{2}{\sqrt{2}}\fr{\sqrt{1-\sqrt{1-z}}}{\pa{z(1-z)}^{\fr{1}{8}}}, \ \ \ 
\ca{F}^{\e \e}_{\e \e}(0|z) =\fr{1-z+z^2}{z(1-z)}.
\end{equation}
\end{description}

\section{Special Functions}\label{app:Special}

\subsection{Theta function}\label{subapp:Theta}

Here, let $q \equiv \ex{2\pi i \tau}$.

\begin{description}

\item[Definition]\mbox{}\\
\begin{equation}
\begin{aligned}
\theta_2(\tau) &=  \sum_{n \in \bb{Z}+\fr{1}{2}} q^{\fr{n^2}{2}}, \\
\theta_3(\tau) &=  \sum_{n \in \bb{Z}} q^{\fr{n^2}{2}}, \\
\theta_4(\tau) &=  \sum_{n \in \bb{Z}} (-1)^n q^{\fr{n^2}{2}}, \\
\eta(\tau) &= q^{\fr{1}{24}} \prod_{n=1}^\infty (1-q^n).
\end{aligned}
\end{equation}

\item[Modular $T$-transformation]\mbox{}\\
\begin{equation}
\begin{aligned}
\theta_2(\tau+1) &=  \ex{\fr{\pi}{4}i}\theta_2(\tau), \\
\theta_3(\tau+1) &=  \theta_4(\tau), \\
\theta_4(\tau+1) &=  \theta_3(\tau), \\
\eta(\tau+1) &= \ex{\fr{\pi}{12}i}\eta(\tau).
\end{aligned}
\end{equation}

\item[Modular $S$-transformation]\mbox{}\\
\begin{equation}
\begin{aligned}
\theta_2\pa{-\fr{1}{\tau}} &=  \sqrt{-i \tau}\theta_4(\tau), \\
\theta_3\pa{-\fr{1}{\tau}}&=  \sqrt{-i \tau}\theta_3(\tau)  , \\
\theta_4\pa{-\fr{1}{\tau}} &=  \sqrt{-i \tau}\theta_2(\tau) , \\
\eta\pa{-\fr{1}{\tau}} &= \sqrt{-i \tau}\eta(\tau).
\end{aligned}
\end{equation}

\item[Relation]\mbox{}\\
\begin{equation}
\begin{aligned}
&\eta(\tau)^3 = \fr{1}{2} \theta_2(\tau) \theta_3(\tau) \theta_4(\tau), \\
&\theta_3(\tau)^4 = \theta_2(\tau)^4 + \theta_4(\tau)^4 .
\end{aligned}
\end{equation}

\end{description}

\subsection{Barnes Double Gamma Function}\label{subapp:Gamma}

\begin{description}

\item[Definition]\mbox{}\\
\begin{equation}\label{eq:DefG}
\log \Gamma_b(x) \equiv \int^\infty_0 \fr{dt}{t} \pa{ \fr{\ex{-xt}-\ex{-\fr{Qt}{2}}}{(1-\ex{-bt})(1-\ex{-\fr{t}{b}})}-\fr{\pa{\fr{Q}{2}-x}^2\ex{-t}}{2}-\fr{\fr{Q}{2}-x}{t}}.
\end{equation}

\item[Asymptotics]\mbox{}\\
\begin{equation}\label{eq:DerG}
\Gamma_b(x) \ar{ x\to 0} \fr{\Gamma_b(Q)}{2\pi x} + O(x^0).
\end{equation}

\begin{equation}
\log \Gamma_b(x) \ar{\abs{x} \to \infty} -\fr{1}{2} x^2 \ln x + \fr{3}{4}x^2 + \fr{Q}{2}x \log x - \fr{Q}{2} x - \fr{Q^2+1}{12} \ln x + \ln \Gamma_0(b)+O(x^{-1}).
\end{equation}

\item[Poles]\mbox{}\\
$\Gamma_b(x) $ has simple poles at the following points:
\begin{equation}\label{eq:Poles2}
x=-mb-\fr{n}{b}, \ \ (n,m \in \bb{Z}_{\geq 0}).
\end{equation}
Also, it has no zeros.

\end{description}

\section{Liouville CFT}

\subsection{Notations in Liouville CFT}
Here, we summarize the notations and functions introduced for convenience in Liouville CFT.

\begin{description}

\item[Notation]\mbox{}\\
\begin{equation}
c=1+6Q^2, \ \ \ \ \ Q=b+\fr{1}{b}, \ \ \ \ \ h_i=\a_i(Q-\a_i) = \fr{Q^2}{4}+\gamma_i^2.
\end{equation}

\item[$\gamma$ function]\mbox{}\\
\begin{equation}
\gamma(x) \equiv \fr{\Gamma(x)}{\Gamma(1-x)}.
\end{equation}
This satisfies the following identity.
\begin{equation}
\gamma(x+1) = -x^2\gamma(x).
\end{equation}

\item[Upsilon function]\mbox{}\\
\begin{equation}
\Upsilon_b(x) \equiv \fr{1}{\Gamma_b(x) \Gamma_b(Q-x)}.
\end{equation}
This satisfies the following {\it shift identity},
\begin{equation}
\begin{aligned}
\fr{\Upsilon_b(x+b)}{\Upsilon_b(x)} &= b^{1-2bx}\gamma(bx), \\
\fr{\Upsilon_b\pa{x+\fr{1}{b}}}{\Upsilon_b(x)} &= b^{\fr{2x}{b}-1}\gamma\pa{\fr{x}{b}}. 
\end{aligned}
\end{equation}

\item[Double Sine function]\mbox{}\\
\begin{equation}\label{eq:DefS}
S_b(x) \equiv \fr{\Gamma_b(x) }{\Gamma_b(Q-x)}.
\end{equation}
This satisfies the following {\it shift identity},
\begin{equation}
\begin{aligned}
\fr{S_b(x+b)}{S_b(x)} &= 2\sin(\pi b x), \\
\fr{S_b\pa{x+\fr{1}{b}}}{S_b(x)} &= 2\sin \pa{\pi \fr{x}{b} }. \\
\end{aligned}
\end{equation}
In particular, the following holds,
\begin{equation}\label{eq:Shift}
\fr{S_b(2i\gamma+Q)}{S_b(2i\gamma)} = 4 \sinh\pa{2\pi \gamma / b} \sinh \pa{2\pi \gamma b} = \fr{S_{\bb{I} \gamma}}{\sqrt{2}}.
\end{equation}
Here, $S_{\bb{I} \gamma}$ is the modular $S$ matrix.

\item[Normalization]\mbox{}\\
\begin{equation}
\begin{aligned}
\braket{V_{\alpha_i}(z)V_{\alpha_j}(0)} &= \left(i B_{\alpha_i}\delta(\alpha_i - \alpha_j) + \delta(\alpha_i + \alpha_j)\right) \frac{1}{|z|^{4h_i}},\\
\Res{\alpha_1+\alpha_2+\alpha_3=Q} C_{123}  &= 1.
\end{aligned}
\end{equation}
Here, the normalization constant of the two-point function is defined as follows:
\begin{equation}\label{eq:AppB}
B_{\alpha} \equiv \left(b^{\frac{2}{b}-2b}  \lambda\right)^{Q-2\alpha} \frac{\Upsilon_b(Q-2\alpha)}{\Upsilon_b(2\alpha-Q)}.
\end{equation}
In addition, $B_{\alpha}$ satisfies the following reflection formula:
\begin{equation}
C_{\alpha_1, \alpha_2, \alpha_3} = B_{\alpha_1} C_{Q-\alpha_1, \alpha_2, \alpha_3}.
\end{equation}
Some references (e.g. \cite{Teschner2001}) rewrite $B_{\alpha}$ using the $\gamma$ function as:
\begin{equation}
B_{\alpha} = \frac{\left(\pi \mu \gamma(b^2)\right)^{\frac{Q-2\alpha}{b}}}{b^2} 
\frac{\gamma\left(2b\alpha-b^2\right)}{\gamma \left(2+\frac{1}{b^2} -\frac{2\alpha}{b}\right)}.
\end{equation}
Here, $4\pi \mu = \lambda'$ was used in conjunction with (\ref{eq:lprime}).

\end{description}

\subsection{DOZZ Formula and Fusion Matrix}\label{subapp:DOZZ}

The OPE coefficients in Liouville CFT are given by the DOZZ formula.
\begin{equation}
\begin{aligned}
C_{123}^{\mathrm{DOZZ}}
&=
\fr{
\pa{b^{\fr{2}{b}-2b} \lambda}^{Q-\alpha_1 -\alpha_2- \alpha_3}
\Upsilon_b'(0)\Upsilon_b(2\alpha_1)\Upsilon_b(2\alpha_2)\Upsilon_b(2\alpha_3)
}{
\Upsilon_b(\alpha_1+\alpha_2+\alpha_3-Q)\Upsilon_b(\alpha_1+\alpha_2-\alpha_3)\Upsilon_b(\alpha_2+\alpha_3-\alpha_1)
\Upsilon_b(\alpha_3+\alpha_1-\alpha_2)
}.
\end{aligned}
\end{equation}
On the other hand, the fusion matrix that appears in (\ref{eq:OPErho}) is expressed as
\begin{equation}
{\bold F}_{\bb{I}, \g_3} 
   \left[
    \begin{array}{cc}
    \g_2   & \g_2  \\
     \g_1  &   \g_1\\
    \end{array}
  \right]
  =
  S_{\bb{I}  \gamma_3} C_0 (\g_1,\g_2, \g_3).
\end{equation}
Here, $C_0 (\g_1,\g_2, \g_3)$ is defined as
\begin{equation}
\begin{aligned}
C_0(\alpha_1, \alpha_2, \alpha_3)
&\equiv
\fr{\Gamma_b(2Q)}{2\sqrt{2}\pi \Gamma_b(Q)}
\fr{1}{ \prod_{i=1,2,3} \Gamma_b(2\alpha_i) \Gamma(2Q-2\alpha_i)}\\
&\times
\fr{ \Upsilon_b'(0) }{\Upsilon_b(\alpha_1+\alpha_2+\alpha_3-Q)\Upsilon_b(\alpha_1+\alpha_2-\alpha_3)\Upsilon_b(\alpha_2+\alpha_3-\alpha_1)\Upsilon_b(\alpha_3+\alpha_1-\alpha_2)}.
\end{aligned}
\end{equation}
Let us calculate the ratio between the two,
\begin{equation}
\begin{aligned}
\fr{C_{123}^{\mathrm{DOZZ}}}{C_0(\alpha_1, \alpha_2, \alpha_3)}
=  2 \sqrt{2} \pi \pa{b^{\fr{2}{b}-2b} \lambda}^{Q-\alpha_1 -\alpha_2- \alpha_3} \fr{\Gamma_b(Q)}{\Gamma_b(2Q)} \prod_{i=1,2,3} \fr{\Gamma_b(2Q-2\alpha_i)}{\Gamma_b(Q-2\alpha_i)}.
\end{aligned}
\end{equation}
Here, the content of the product can be rewritten using the normalization factor of the two-point function (\ref{eq:AppB}) and the modular $S$ matrix as follows:
\begin{equation}
\fr{\Gamma_b(2Q-2\alpha)}{\Gamma_b(Q-2\alpha)}
= \sqrt{ \fr{S_b(2\a)}{S_b(2\a-Q)}} \sqrt{ \fr{\Upsilon_b(Q-2\a)}{\Upsilon_b(2\a-Q)}}
=\sqrt{\fr{S_{\bb{I}\gamma}}{\sqrt{2}}} \sqrt{ \fr{B_\alpha}{\pa{b^{\fr{2}{b}-2b} \lambda}^{Q-2\alpha}}}.
\end{equation}
In the second equation, we used (\ref{eq:Shift}).

Thus, the ratio between the two is expressed as
\begin{equation}
\fr{C_{123}^{\mathrm{DOZZ}}}{C_0(\alpha_1, \alpha_2, \alpha_3)} = \ca{E}  \prod_{i=1,2,3} \sqrt{S_{\bb{I}\gamma_i} B_i }.
\end{equation}
Here, the constant $\ca{E}$ independent of the conformal dimensions is defined by
\begin{equation}
\ca{E} \equiv 2^{\fr{3}{4}}\pi \pa{b^{\fr{2}{b}-2b} \lambda}^{-\fr{Q}{2}} \fr{\Gamma_b(Q)}{\Gamma_b(2Q)}.
\end{equation}

\section{Fusion Transformation} \label{app:FM}

The fusion transformation of four-point functions is expressed as follows \cite{Teschner2001}.
\begin{equation}\label{eq:fusiontrans}
\begin{aligned}
\ca{F}^{21}_{34}(h_{\a_s}|z)=\int_{\bb{S}} d \a_t {\bold F}_{\a_s, \a_t} 
   \left[
    \begin{array}{cc}
    \a_2   & \a_1  \\
     \a_3  &   \a_4\\
    \end{array}
  \right]
  \ca{F}^{23}_{14}(h_{\a_t}|1-z).
\end{aligned}
\end{equation}
The integral contour is from $\fr{Q}{2}+i0$ to $\fr{Q}{2}+i\infty$.
The integral kernel $ {\bold F}_{\a_s, \a_t} $ is called the {\it fusion matrix} or {\it crossing matrix}.
The exact form of the fusion matrix is expressed as follows \cite{Ponsot1999,Teschner2001},
\begin{equation}\label{eq:crossing}
\begin{aligned}
{\bold F}_{\a_s, \a_t} 
   \left[
    \begin{array}{cc}
    \a_2   & \a_1  \\
     \a_3  &   \a_4\\
    \end{array}
  \right]
=\fr{N(\a_4,\a_3,\a_s)N(\a_s,\a_2,\a_1)}{N(\a_4,\a_t,\a_1)N(\a_t,\a_3,\a_2)}
   \left\{
    \begin{array}{cc|c}
    \a_1   & \a_2    &  \a_s  \\
     \a_3  & \a_4    &  \a_t   \\
    \end{array}
  \right\}_b.
\end{aligned}
\end{equation}
Here, $N(\a_3,\a_2,\a_1)$ is defined by
\begin{equation}
N(\a_3,\a_2,\a_1) \equiv \fr{\G_b(2\a_1)\G_b(2\a_2)\G_b(2Q-2\a_3)}{\G_b(2Q-\a_1-\a_2-\a_3)\G_b(Q-\a_1-\a_2+\a_3)\G_b(\a_1+\a_3-\a_2)\G_b(\a_2+\a_3-\a_1)}.
\end{equation}
$ \left\{
    \begin{array}{cc|c}
    \a_1   & \a_2    &  \a_s  \\
     \a_3  & \bar{\a_4}    &  \a_t   \\
    \end{array}
  \right\}_b$
is the Racah--Wigner coefficient of the quantum group $U_q(sl(2,\bb{R}))$ and is expressed as follows \cite{Teschner2014},
\begin{equation}\label{eq:6j}
\begin{aligned}
&\left\{
    \begin{array}{cc|c}
    \a_1   & \a_2    &  \a_s  \\
     \a_3  & \a_4    &  \a_t   \\
    \end{array}
  \right\}_b\\
&= \fr{S_b(\a_1+\a_4+\a_t-Q)S_b(\a_2+\a_3+\a_t-Q)S_b(\a_3-\a_2-\a_t+Q)S_b(\a_2-\a_3-\a_t+Q)}{S_b(\a_1+\a_2-\a_s)S_b(\a_3+\a_s-\a_4)S_b(\a_3+\a_4-\a_s)}\\
&\times \abs{S_b(2\a_t)}^2 \int^{2Q+i \infty}_{2Q-i \infty} d u
\fr{S_b(u-\a_{12s})S_b(u-\a_{s34})S_b(u-\a_{23t})S_b(u-\a_{1t4})}{S_b(u-\a_{1234}+Q)S_b(u-\a_{st13}+Q)S_b(u-\a_{st24}+Q)S_b(u+Q)}.
\end{aligned}
\end{equation}
Here, $\bar{\a}\equiv Q-\a$, $\a_{ijk}\equiv \a_i+\a_j+\a_k$, and $\a_{ijkl}\equiv \a_i+\a_j+\a_k+\a_l$.
$\Gamma_b(x)$ is the Barnes double gamma function (\ref{eq:DefG}), and $S_b(x)$ is the double sine function (\ref{eq:DefS}).

Since the Barnes double gamma function has simple poles at $x=-mb-\fr{n}{b}, \ \ (n,m \in \bb{Z}_{\geq 0})$,
the fusion matrix has zeros and poles on both the $\alpha_s$ plane and the $\alpha_t$ plane.
In particular, when $\alpha_s=0$, the position of the simple poles on the $\alpha_t$ plane is given as follows (generalization of (\ref{eq:Poles})),
\begin{equation}
\begin{aligned}
\a_t^{(1)}&=(\a_2+\a_3)-Q-Q_{m,n},\\
\a_t^{(2)}&=-(\a_1+\a_4)+2Q+Q_{m,n},\\
\a_t^{(3)}&=\pm(\a_1-\a_4)-Q_{m,n},\\
\a_t^{(4)}&=\pm(\a_2-\a_3)+Q+Q_{m,n},\\
\a_t^{(5)}&=-(\a_2+\a_3)+Q-Q_{m,n}.\\
\a_t^{(6)}&=\a_1+\a_4+Q_{m,n}.\\
\end{aligned}
\end{equation}
Here, $Q_{m,n}\equiv -mb-\fr{n}{b}, \ \ (n,m \in \bb{Z}_{\geq 0})$.
Since the Liouville momentum in unitary theories satisfies $0 \leq \Re \alpha_i \leq \fr{Q}{2}$,
the possible range of each pole coordinate is restricted as shown in Figure \ref{fig:poles}.

\clearpage
\bibliographystyle{JHEP}
\bibliography{main}

\end{document}